\pdfoutput=1

\documentclass[11pt,twoside,a4paper,cmspaper,final,collab]{cms-tdr}
\def\svnVersion{70182:70891}\def\cmsCernNo{2011-102}\def\cmsCernDate{\today}\def\cmsMessage{Submitted to JINST}

\begin{document}\cmsNoteHeader{JME-10-011}

\hyphenation{had-ron-i-za-tion}
\hyphenation{cal-or-i-me-ter}
\hyphenation{de-vices}
\RCS$Revision: 70891 $
\RCS$HeadURL: svn+ssh://alverson@svn.cern.ch/reps/tdr2/papers/JME-10-011/trunk/JME-10-011.tex $
\RCS$Id: JME-10-011.tex 70891 2011-07-21 13:37:31Z alverson $
\newcommand{\calojets}{CaloJets}
\newcommand{\jptjets}{JPTJets}
\newcommand{\pfjets}{PFJets}

\newcommand{\ptave}{\ensuremath{p_{{T}}^{{ave}}}\xspace}
\newcommand{\sigmaA}{\ensuremath{\sigma_A}\xspace}
\newcommand{\ptfirst}{\ensuremath{p_{{T}}^{{Jet}1}}\xspace}
\newcommand{\ptthird}{\ensuremath{p_{{T}}^{{Jet}3}}\xspace}
\newcommand{\ptthirdmax}{\ensuremath{p_{{T,max}}^{{Jet}3}}\xspace}
\newcommand{\ptthirdraw}{\ensuremath{p_{{T,raw}}^{{Jet}3}}\xspace}
\newcommand{\ptthirdcorr}{\ensuremath{p_{{T,cor}}^{{Jet}3}}\xspace}
\newcommand{\ptthirdpara}{\ensuremath{p_{{T,rel}}^{Jet3,\parallel}}\xspace}
\newcommand{\ptrelthree}{\ensuremath{\ptthird/\ptave}\xspace}

\newcommand{\ptsecond}{\ensuremath{p_{{T}}^{{Jet}2}}\xspace}
\newcommand{\ptsecondmax}{\ensuremath{p_{{T,max}}^{{Jet}2}}\xspace}

\newcommand{\antikt}{Anti-$k_{T}$\xspace}
\newcommand{\aachen}{Cambridge/Aachen\xspace}
\newcommand{\siscone}{SISCone\xspace}
\newcommand{\itcone}{Iterative Cone\xspace}

\newcommand{\Rsoft}{\ensuremath{R_{\mathrm{soft}}}\xspace}
\newcommand{\ksoft}{\ensuremath{\mathrm{k}_{\mathrm{soft}}}\xspace}
\newcommand{\kpli}{\ensuremath{\mathrm{k}_{\mathrm{pli}}}\xspace}

\newcommand{\nhits}{\ensuremath{n^{90}_{\mathrm{hits}}}\xspace}
\newcommand{\fhpd}{\ensuremath{f_{\mathrm{HPD}}}\xspace}

\newcommand{\DeltaR}{\ensuremath{\Delta R}\xspace}
\newcommand{\DeltaRMax}{\ensuremath{\Delta R_{\mathrm{max}}}\xspace}
\newcommand{\Rch}{\ensuremath{R_{\mathrm{ch}}}\xspace}
\renewcommand{\pt}{\ensuremath{p_{{T}}}\xspace}
\newcommand{\ptmin}{\ensuremath{p^{{min}}_{{T}}}\xspace}
\newcommand{\ptcorr}{\ensuremath{p_{\mathrm{T}}^{\mathrm{corr}}}\xspace}
\newcommand{\pthat}{\ensuremath{\hat{p}_{\mathrm{T}}}\xspace}
\newcommand{\ptgen}{\ensuremath{p_{\mathrm{T}}^{\mathrm{gen}}}\xspace}
\newcommand{\ptref}{\ensuremath{p_{\mathrm{T}}^{\mathrm{REF}}}\xspace}
\newcommand{\etagen}{\ensuremath{\eta^{\mathrm{gen}}}\xspace}
\newcommand{\etamax}{\ensuremath{\eta^{\mathrm{max}}}\xspace}
\newcommand{\phigen}{\ensuremath{\varphi^{\mathrm{gen}}}\xspace}
\newcommand{\dphijj}{\ensuremath{\left|\Delta\varphi(\mathrm{j}_1\mathrm{j}_2)\right|}\xspace}

\def\eslash{\ensuremath{{\hbox{$E$\kern-0.6em\lower-.05ex\hbox{/}\kern0.10em}}}}
\def\vecmet{\mbox{$\vec{\eslash}_T$}\xspace} 
\def\MET{\mbox{$\eslash_T$}\xspace}
\def\met{\mbox{$\eslash_T$}\xspace} 
\def\mex{\mbox{$\eslash_x$}\xspace} 
\def\mey{\mbox{$\eslash_y$}\xspace} 

\newcommand{\etabarrel}{\,\ensuremath{0<|\eta|\leq1.4}\xspace}
\newcommand{\etaendcaps}{\,\ensuremath{1.4<|\eta|\leq2.6}\xspace}
\newcommand{\etatransition}{\,\ensuremath{2.6<|\eta|\leq3.2}\xspace}
\newcommand{\etaforward}{\,\ensuremath{3.2<|\eta|\leq4.7}\xspace}

\newcommand{\ipb}{\,pb^{-1}\xspace}
\newcommand{\ifb}{\,fb^{-1}\xspace}

\newcommand{\fasym}{\ensuremath{f_{\text{Asym}}}\xspace}
\newcommand{\fasymdata}{\ensuremath{f^{\text{Data}}_{\text{Asym}}}\xspace}
\newcommand{\fasymmc}{\ensuremath{f^{\text{MC}}_{\text{Asym}}}\xspace}
\newcommand{\fresp}{\ensuremath{f_{\text{Resp}}}\xspace}

\cmsNoteHeader{JME-10-011} 
\title{Determination of Jet Energy Calibration and Transverse Momentum Resolution in CMS}
\date{\today}
\abstract{
Measurements of the jet energy calibration and transverse momentum resolution in CMS are presented, performed with a data sample collected in proton-proton collisions at a centre-of-mass energy of 7\TeV, corresponding to an integrated luminosity of $36\pbinv$. The transverse momentum balance in dijet and $\gamma$/Z+jets events is used to measure the jet energy response in the CMS detector, as well as the transverse momentum resolution. The results are presented for three different methods to reconstruct jets: a calorimeter-based approach, the ``Jet-Plus-Track" approach, which improves the measurement of calorimeter jets by exploiting the associated tracks, and the ``Particle Flow" approach, which attempts to reconstruct individually each particle in the event, prior to the jet clustering, based on information from all relevant subdetectors.
}

\hypersetup{%
pdfauthor={CMS Collaboration},%
pdftitle={Determination of Jet Energy Calibration and Transverse Momentum Resolution in CMS},%
pdfsubject={CMS},%
pdfkeywords={CMS, jets, JetMET, energy scale, energy resolution}}

\maketitle

\clearpage

\section{Introduction}

Jets are the experimental signatures of quarks and gluons produced in high-energy processes such as hard scattering of partons in proton-proton collisions. The detailed understanding of both the jet energy scale and of the transverse momentum resolution is of crucial importance for many physics analyses, and it is an important component of the systematic uncertainty. This paper presents studies for the determination of the energy scale and resolution of jets, performed with the Compact Muon Solenoid (CMS) at the CERN Large Hadron Collider (LHC), on proton-proton collisions at $\sqrt{s}=7\TeV$, using a data sample corresponding to an integrated luminosity of $36\pbinv$.

The paper is organized as follows: Section~\ref{sec:detector} describes briefly the CMS detector, while Section~\ref{sec:jets} describes the jet reconstruction methods considered here. Sections~\ref{sec:data} and~\ref{sec:methods} present the data samples and the experimental techniques used for the various measurements. The jet energy calibration scheme is discussed in Section~\ref{sec:jec} and the jet transverse momentum resolution is presented in Section~\ref{sec:res}. 

\section{The CMS Detector}\label{sec:detector}

A detailed description of the CMS detector can be found elsewhere~\cite{CMS}. A right-handed coordinate system is used with the origin at the nominal interaction point (IP). The x-axis points to the center of the LHC ring, the y-axis is vertical and points upward, and the z-axis is parallel to the counterclockwise beam direction. The azimuthal angle $\phi$ is measured with respect to the x-axis in the xy-plane and the polar angle $\theta$ is defined with respect to the z-axis, while the pseudorapidity is defined as $\eta=-\ln\left[\tan\left(\theta/2\right)\right]$. The central feature of the CMS apparatus is a superconducting solenoid, of 6\,m internal diameter, that produces a magnetic field of 3.8\,T. Within the field volume are the silicon pixel and strip tracker and the barrel and endcap calorimeters ($|\eta| < 3$), composed of a crystal electromagnetic calorimeter (ECAL) and a brass/scintillator hadronic calorimeter (HCAL). Outside the field volume, in the forward region ($3 < |\eta| < 5$), there is an iron/quartz-fibre hadronic calorimeter. The steel return yoke outside the solenoid is instrumented with gaseous detectors used to identify muons. The CMS experiment collects data using a two-level trigger system, the first-level hardware trigger (L1)~\cite{PTDRI} and the high-level software trigger (HLT)~\cite{HLT}. 

\section{Jet Reconstruction}\label{sec:jets}

Jets considered in this paper are reconstructed using the anti-$k_T$ clustering algorithm~\cite{AKT} with a size parameter $R=0.5$ in the $y-\phi$ space. In some cases, jets with a size parameter $R=0.7$ are also considered. The clustering is performed by four-momentum summation. The rapidity $y$ and the transverse momentum \pt of a jet with energy $E$ and momentum $\vec{p}=(p_x,p_y,p_z)$ are defined as $y=\frac{1}{2}\ln\left(\frac{E+p_z}{E-p_z}\right)$ and $\pt=\sqrt{p_x^2+p_y^2}$ respectively. The inputs to the clustering algorithm are the four-momentum vectors of detector energy deposits or of particles in the Monte Carlo (MC) simulations. Detector jets belong to three types, depending on the way the individual contributions from subdetectors are combined: Calorimeter jets, Jet-Plus-Track jets and Particle-Flow jets.  

{\bf Calorimeter (CALO) jets} are reconstructed from energy deposits in the calorimeter towers. A calorimeter tower consists of one or more HCAL cells and the geometrically corresponding ECAL crystals. In the barrel region of the calorimeters, the unweighted sum of one single HCAL cell and 5x5 ECAL crystals form a projective calorimeter tower. The association between HCAL cells and ECAL crystals is more complex in the endcap regions. In the forward region, a different calorimeter technology is employed, using the Cerenkov light signals collected by short and long quartz readout fibers to aid the separation of electromagnetic and hadronic signals. A four-momentum is associated to each tower deposit above a certain threshold, assuming zero mass, and taking as a direction the tower position as seen from the interaction point.

{\bf Jet-Plus-Track (JPT) jets} are reconstructed calorimeter jets whose energy response and resolution are improved by incorporating tracking information, according to the Jet-Plus-Track algorithm~\cite{JME-09-002}. Calorimeter jets are first reconstructed as described above, and then charged particle tracks are associated with each jet, based on the spatial separation between the jet axis and the track momentum vector, measured at the interaction vertex, in the $\eta-\phi$ space. The associated tracks are projected onto the front surface of the calorimeter and are classified as \textit{in-cone} tracks if they point to within the jet cone around the jet axis on the calorimeter surface. The tracks that are bent out of the jet cone because of the CMS magnetic field are classified as \textit{out-of-cone} tracks. The momenta of charged tracks are then used to improve the measurement of the energy of the associated calorimeter jet: for \textit{in-cone} tracks, the expected average energy deposition in the calorimeters is subtracted and the momentum of the tracks is added to the jet energy. For \textit{out-of-cone} tracks the momentum is added directly to the jet energy. The Jet-Plus-Track algorithm corrects both the energy and the direction of the axis of the original calorimeter jet.

The {\bf Particle-Flow (PF) jets} are reconstructed by clustering the four-momentum vectors of particle-flow candidates. The particle-flow algorithm~\cite{PFT-09-001,PFT-10-002} combines the information from all relevant CMS sub-detectors to identify and reconstruct all visible particles in the event, namely muons, electrons, photons, charged hadrons, and neutral hadrons. Charged hadrons, electrons and muons are reconstructed from tracks in the tracker. Photons and neutral hadrons are reconstructed from energy clusters separated from the extrapolated positions of tracks in ECAL and HCAL, respectively. A neutral particle overlapping with charged particles in the calorimeters is identified as a calorimeter energy excess with respect to the sum of the associated track momenta. The energy of photons is directly obtained from the ECAL measurement, corrected for zero-suppression effects. The energy of electrons is determined from a combination of the track momentum at the main interaction vertex, the corresponding ECAL cluster energy, and the energy sum of all bremsstrahlung photons associated with the track. The energy of muons is obtained from the corresponding track momentum. The energy of charged hadrons is determined from a combination of the track momentum and the corresponding ECAL and HCAL energy, corrected for zero-suppression effects, and calibrated for the non-linear response of the calorimeters. Finally, the energy of neutral hadrons is obtained from the corresponding calibrated ECAL and HCAL energy. The PF jet momentum and spatial resolutions are greatly improved with respect to calorimeter jets, as the use of the tracking detectors and of the high granularity of ECAL allows resolution and measurement of charged hadrons and photons inside a jet, which together constitute $\sim$85\% of the jet energy. 

The {\bf Monte Carlo particle jets} are reconstructed by clustering the four-momentum vectors of all stable ($c\tau > 1$ cm) particles generated in the simulation. In particular, there are two types of MC particle jets: those where the neutrinos are excluded from the clustering, and those where both the neutrinos and the muons are excluded. The former are used for the study of the PF and JPT jet response in the simulation, while the latter are used for the study of the CALO jet response (because muons are minimum ionizing particles and therefore do not contribute appreciably to the CALO jet reconstruction).

The {\bf Particle-Flow missing transverse energy} ($\vecmet$), which is needed for the absolute jet energy response measurement, is reconstructed from the particle-flow candidates and is defined as $\vecmet=-\displaystyle\sum_{i}{\left(E_i\sin\theta_i\cos\phi_i\hat{\mathbf{x}}+E_i\sin\theta_i\sin\phi_i\hat{\mathbf{y}}\right)}=\mex\hat{\mathbf{x}}+\mey\hat{\mathbf{y}}$, where the sum refers to all candidates and $\hat{\mathbf{x}},\hat{\mathbf{y}}$ are the unit vectors in the direction of the x and y axes.

\section{Event Samples and Selection Criteria}\label{sec:data}

In this Section, the data samples used for the various measurements are defined. In all samples described below, basic common event preselection criteria are applied in order to ensure that the triggered events do come from real proton-proton interactions. First, the presence of at least one well-reconstructed primary vertex (PV) is required, with at least four tracks considered in the vertex fit, and with $|\text{z}(\mathrm{PV})|<24\cm$, where $\text{z}(\mathrm{PV})$ represents the position of the proton-proton collision along the beams. In addition, the radial position of the primary vertex, $\rho(\mathrm{PV})$, has to satisfy the condition $\rho(\mathrm{PV})<2\cm$.

Jet quality criteria (``Jet ID'') have been developed for CALO jets~\cite{JME-09-008} and PF jets~\cite{JME-10-003}, which are found to retain the vast majority ($>99\%$) of genuine jets in the simulation, while rejecting most of the misidentified jets arising from calorimeter and/or readout electronics noise in pure noise non-collision data samples: such as cosmic-ray trigger data or data from triggers on empty bunches during LHC operation. Jets used in the analysis are required to satisfy proper identification criteria.

\subsection{Zero Bias and Minimum Bias Samples}

The zero bias and minimum bias samples are used for the measurement of the energy clustered inside a jet due to noise and additional proton-proton collisions in the same bunch crossing (pile-up, or PU), as described in Section~\ref{sec:offset}. The zero bias sample is collected using a random trigger in the presence of a beam crossing. The minimum bias sample is collected by requiring coincidental hits in the beam scintillating counter~\cite{HLT} on either side of the CMS detector. 

\subsection{Dijet Sample}
\label{sec:jjsample}
The dijet sample is composed of events with at least two reconstructed jets in the final state and is used for the measurement of the relative jet energy scale and of the jet \pt resolution. This sample is collected using dedicated high-level triggers which accept the events based on the value of the average uncorrected \pt (\pt not corrected for the non-uniform response of the calorimeter) of the two CALO jets with the highest \pt (leading jets) in the event. The selected dijet sample covers the average jet $\pt$ range from $15\GeV$ up to around $1\TeV$.

\subsection{$\gamma+$jets Sample}

The $\gamma+$jets sample is used for the measurement of the absolute jet energy response and of the jet \pt resolution. This sample is collected with single-photon triggers that accept an event if at least one reconstructed photon has $\pt>15\GeV$. Offline, photons are required to have transverse momentum $\pt^{\gamma}>15\GeV$ and $|\eta|<1.3$. The jets used in the $\gamma+$jets sample  are required to lie in the $|\eta|<1.3$ region. The $\gamma+$jets sample is dominated by dijet background, where a jet mimics the photon. To suppress this background, the following additional photon isolation and shower-shape requirements~\cite{EGM-10-005} are applied:

\begin{itemize}

\item
\textbf{HCAL isolation}: the energy deposited in the HCAL within a cone of radius $R=0.4$ in the $\eta-\phi$ space, around the photon direction, must be smaller than $2.4\GeV$ or less than $5\%$ of the photon energy ($E_{\gamma}$);

\item
\textbf{ECAL isolation}: the energy deposited in the ECAL within a cone of radius $R=0.4$ in the $\eta-\phi$ space, around the photon direction, excluding the energy associated with the photon, must be smaller than $3\GeV$ or less than $5\%$ of the photon energy;

\item
\textbf{Tracker isolation}: the number of tracks in a cone of radius $R=0.35$ in the $\eta-\phi$ space, around the photon direction, must be less than three, and the total transverse momentum of the tracks must be less than $10\%$ of the photon transverse momentum;

\item
\textbf{Shower shape}: the photon cluster major and minor must be in the range of 0.15-0.35, and 0.15-0.3, respectively. Cluster major and minor are defined as second moments of the energy distribution along the direction of the maximum and minimum spread of the ECAL cluster in the $\eta-\phi$ space;

\end{itemize}

The selected $\gamma+$jets sample covers the $\pt^{\gamma}$ range from $15\GeV$ up to around $400\GeV$.

\subsection{$Z(\mu^+\mu^-)$+jets Sample}

The $Z(\mu^+\mu^-)$+jets sample is used for the measurement of the absolute jet energy response. It is collected using single-muon triggers with various \pt thresholds. Offline, the events are required to have at least two opposite-sign reconstructed global muons with $\pt>15\GeV$ and $|\eta^\mu|<2.3$ and at least one jet with $|\eta|<1.3$. A global muon is reconstructed by a combined fit to the muon system hits and tracker hits, seeded by a track found in the muon systems only. The reconstructed muons must satisfy identification and isolation requirements, as described in Ref.~\cite{EWK-10-002}. Furthermore, the invariant mass $M_{\mu\mu}$ of the two muons must satisfy the condition $70<M_{\mu\mu}<110\GeV$. Finally, the reconstructed Z is required to be back-to-back in the transverse plane with respect to the jet with the highest \pt: $|\Delta\phi(Z,jet)|>2.8 rad$. 

\subsection{$Z(e^+e^-)$+jets Sample}

The $Z(e^+e^-)$+jets sample is used for the measurement of the absolute jet energy response. It is collected using single-electron triggers with various \pt thresholds. Offline, the events are required to have at least two opposite-sign reconstructed electrons with $\pt>20\GeV$ in the fiducial region $|\eta|<1.44$ and $1.57<|\eta|<2.5$ and at least one jet with $|\eta|<1.3$. The reconstructed electrons must satisfy identification and isolation requirements, as described in Ref.~\cite{EWK-10-002}. Furthermore, the invariant mass $M_{ee}$ of the electron-positron pair must satisfy the condition $85<M_{ee}<100\GeV$. Finally, the reconstructed Z is required to be back-to-back in the transverse plane with respect to the jet with the highest \pt: $|\Delta\phi(Z,jet)|>2.7 rad$.

\section{Experimental Techniques}\label{sec:methods}

\subsection{Dijet \pt-Balancing}

The dijet \pt-balancing method is used for the measurement of the relative jet energy response as a function of $\eta$. It is also used for the measurement of the jet \pt resolution. The technique was introduced at the CERN p$\bar{\text{p}}$ collider (SP$\bar{\text{P}}$S)~\cite{spps} and later refined by the Tevatron experiments~\cite{jes_d0, jes_cdf}. The method is based on transverse momentum conservation and utilizes the \pt-balance in dijet events, back-to-back in azimuth. 

For the measurement of the relative jet energy response, one jet (barrel jet) is required to lie in the central region of the detector ($|\eta|<1.3$) and the other jet (probe jet) at arbitrary $\eta$. The central region is chosen as a reference because of the uniformity of the detector, the small variation of the jet energy response, and because it provides the highest jet \pt-reach. It is also the easiest region to calibrate in absolute terms, using $\gamma$+jet and Z+jet events. The dijet calibration sample is collected as described in Section~\ref{sec:jjsample}. Offline, events are required to contain at least two jets. The two leading jets in the event must be  azimuthally separated  by $\Delta \phi > 2.7 rad$, and one of them must lie in the $|\eta|<1.3$ region.  

The balance quantity $\mathcal{B}$ is defined as:

\begin{equation}
\mathcal{B}=\frac{\pt^{probe}-\pt^{barrel}}{\ptave},
\end{equation}

where \ptave is the average \pt of the two leading jets:

\begin{equation}
  \ptave = \frac{\pt^{barrel}+\pt^{probe}}{2}.
\end{equation}

The balance is recorded in bins of $\eta^{probe}$ and \ptave. In order to avoid a trigger bias, each \ptave bin is populated by events satisfying the conditions of the fully efficient trigger with the highest threshold.

The average value of the $\mathcal{B}$ distribution, $\langle \mathcal{B}\rangle$, in a given $\eta^{probe}$ and \ptave bin, is used to determine the relative response $\mathcal{R_\text{rel}}$: 

\begin{equation}
\mathcal{R_\text{rel}}(\eta^{probe},\ptave)=\frac{2 + \langle \mathcal{B}\rangle}{2 - \langle \mathcal{B}\rangle}.
\end{equation}

The variable $\mathcal{R_\text{rel}}$ defined above is mathematically equivalent to $\langle\pt^{probe}\rangle/\langle\pt^{barrel}\rangle$ for narrow bins of \ptave. The choice of \ptave minimizes the resolution-bias effect (as opposed to binning in $\pt^{barrel}$, which leads to maximum bias) as discussed in Section~\ref{sec:resbias} below. 

A slightly modified version of the dijet \pt-balance method is applied for the measurement of the jet \pt resolution. The use of dijet events for the measurement of the jet \pt resolution was introduced by the D0 experiment at the Tevatron~\cite{d0-asymmetry} while a feasibility study at CMS was presented using simulated events~\cite{JME-09-007}.

In events with at least two jets, the asymmetry variable $\mathcal{A}$ is defined as:

\begin{equation}
\mathcal{A} = \frac{\ptfirst - \ptsecond}{\ptfirst + \ptsecond},
\end{equation}

where \ptfirst and \ptsecond refer to the randomly ordered transverse momenta of the two leading jets. The variance of the asymmetry variable $\sigma_\mathcal{A}$ can be formally expressed as:

\begin{equation}
  \sigma_\mathcal{A}^2 = \left|\frac{\partial\mathcal{A}}{\partial\ptfirst}\right|^2\cdot\sigma^2(\ptfirst) +
  \left|\frac{\partial\mathcal{A}}{\partial\ptsecond}\right|^2\cdot\sigma^2(\ptsecond).
\end{equation}

If the two jets lie in the same $\eta$ region, $\pt\equiv\langle\ptfirst\rangle = \langle\ptsecond\rangle$ and $\sigma(\pt)\equiv\sigma(\ptfirst) = \sigma(\ptsecond)$. The fractional jet \pt resolution is calculated to be:

\begin{equation}
  \frac{\sigma(\pt)}{\pt} = \sqrt{2}\,\sigma_\mathcal{A}.
\end{equation}

The fractional jet \pt resolution in the above expression is an estimator of the true resolution, in the limiting case of no extra jet activity in the event that spoil the \pt balance of the two leading jets. The distribution of the variable $\mathcal{A}$ is recorded in bins of the average \pt of the two leading jets, $\ptave=\left(\ptfirst+\ptsecond\right)/2$, and its variance is proportional to the relative jet \pt resolution, as described above.

\subsection{$\gamma$/Z+jet \pt-Balancing}

The $\gamma$/Z+jet \pt-balancing method is used for the measurement of the jet energy response and the jet \pt resolution with respect to a reference object, which can be a $\gamma$ or a Z boson. The \pt resolution of the reference object is typically much better than the jet resolution and the absolute response $R_{abs}$ is expressed as: 

\begin{equation}
  R_{abs} = \frac{\pt^{jet}}{\pt^{\gamma,Z}}.
\end{equation}

The absolute response variable is recorded in bins of $\pt^{\gamma,Z}$. It should be noted that, because of the much worse jet \pt resolution, compared to the $\gamma$ or Z \pt resolution, the method is not affected by the resolution bias effect (see Section~\ref{sec:resbias}), as it happens in the dijet \pt-balancing method. Also, for the same reason, the absolute response can be defined as above, without the need of more complicated observables, such as the balance $\mathcal{B}$ or the asymmetry $\mathcal{A}$.

\subsection{Missing Transverse Energy Projection Fraction}

The missing transverse energy projection fraction (MPF) method (extensively used at the Tevatron~\cite{jes_d0}) is based on the fact that the $\gamma,Z$+jets events have no intrinsic \vecmet and that, at parton level, the $\gamma$ or Z is perfectly balanced by the hadronic recoil in the transverse plane:

\begin{equation}
\vec{\pt}^{\gamma,Z} + \vec{\pt}^{recoil} = 0.
\end{equation}

For reconstructed objects, this equation can be re-written as:

\begin{equation}
R_{\gamma,Z}\vec{\pt}^{\gamma,Z} + R_{recoil}\vec{\pt}^{recoil} = -\vecmet,
\end{equation}

where $R_{\gamma,Z}$ and $R_{recoil}$ are the detector responses to the $\gamma$ or Z and the hadronic recoil, respectively. 

Solving the two above equations for $R_{recoil}$ gives:

\begin{equation}
R_{recoil}= R_{\gamma,Z} +\frac{\vecmet \cdot \vec{\pt}^{\gamma,Z}}{(\pt^{\gamma,Z})^2}\equiv R_{MPF}.
\end{equation}

This equation forms the definition of the MPF response $R_{MPF}$. The additional step needed is to extract the jet energy response from the measured MPF response. In general, the recoil consists of additional jets, beyond the leading one, soft particles and unclustered energy. The relation $R_{leadjet}=R_{recoil}$ holds to a good approximation if the particles, that are not clustered into the leading jet, have a response similar to the ones inside the jet, or if these particles are in a direction perpendicular to the photon axis. Small response differences are irrelevant if most of the recoil is clustered into the leading jet. This is ensured by vetoing secondary jets in the selected back-to-back $\gamma,Z$+jets events.

The MPF method is less sensitive to various systematic biases compared to the $\gamma,Z$ \pt-balancing method and is used in CMS as the main method to measure the jet energy response, while the $\gamma,Z$ \pt-balancing is used to facilitate a better understanding of various systematic uncertainties and to perform cross-checks.

\subsection{Biases}

All the methods based on data are affected by inherent biases related to detector effects (e.g. \pt resolution) and to the physics properties (e.g. steeply falling jet \pt spectrum). In this Section, the two most important biases related to the jet energy scale and to the \pt resolution measurements are discussed: the resolution bias and the radiation imbalance.

\subsubsection{Resolution Bias}\label{sec:resbias}

The measurement of the jet energy response is always performed by comparison to a reference object. Typically, the object with the best resolution is chosen as a reference object, as in the $\gamma$/Z+jet balancing where the $\gamma$ and the Z objects have much better \pt resolution than the jets. However, in other cases, such as the dijet \pt-balancing, the two objects have comparable resolutions. When such a situation occurs, the measured relative response is biased in favor of the object with the worse resolution. This happens because a reconstructed jet \pt bin is populated not only by jets whose true (particle-level) \pt lies in the same bin, but also from jets outside the bin, whose response has fluctuated high or low. If the jet spectrum is flat, for a given bin the numbers of true jets migrating in and out are equal and no bias is observed. In the presence of a steeply falling spectrum, the number of incoming jets with lower true \pt that fluctuated high is larger and the measured response is systematically higher. In the dijet \pt-balancing, the effect described above affects both jets. In order to reduce the resolution bias, the measurement of the relative response is performed in bins of \ptave, so that if the two jets have the same resolution, the bias is cancelled on average. This is true for the resolution measurement with the asymmetry method where both jets lie in the same $\eta$ region. For the relative response measurement, the two jets lie in general in different $\eta$ regions, and the bias cancellation is only partial.

\subsubsection{Radiation Imbalance}\label{sec:radbias}

The other source of bias is the \pt-imbalance caused by gluon radiation. In general, the measured \pt-imbalance is caused by the response difference of the balancing objects, but also from any additional objects with significant \pt. The effect can by demonstrated as follows: an estimator $\mathcal{R}^{meas}$ of the response of an object with respect to a reference object, is $\mathcal{R}^{meas}=\pt/\pt^{ref}$ where \pt and $\pt^{ref}$ are the measured transverse momenta of the objects. These are related to the true \pt ($p^{true}_{T,ref}$) through the true response: $\pt=R^{true}\cdot\pt^{true}$ and $\pt^{ref}=R^{true}_{ref}\cdot p^{true}_{T,ref}$. In the presence of additional hard objects in the event, $\pt^{true}=p^{true}_{T,ref}-\Delta\pt$, where $\Delta\pt$ quantifies the imbalance due to radiation.  By combining all the above, the estimator $\mathcal{R}^{meas}$ is expressed as: $\mathcal{R}^{meas}=R^{true}/R^{true}_{ref}\left(1-\Delta\pt/p^{true}_{T,ref}\right)$. This relation indicates that the \pt-ratio between two reconstructed objects is a good estimator of the relative response, only in the case where the additional objects are soft, such that $\Delta\pt/p_{T,ref}^{true}\rightarrow 0$. 

The above considerations are important for all \pt-balancing measurements presented in this paper (the dijet \pt-balancing and the $\gamma$/Z+jet \pt-balancing), both for the scale and the resolution determination. Practically, the measurements are performed with a varying veto on an estimator of $a^{true}=\Delta\pt/p_{T,ref}^{true}$ and then extrapolated linearly to $a^{true}=0$. For the dijet \pt-balancing, the estimator of $a^{true}$ is the ratio $\ptthird/\ptave$, while for the $\gamma,Z$+jet \pt-balancing it is the ratio $\ptsecond/\pt^{\gamma,Z}$.

\section{Jet Energy Calibration}\label{sec:jec}

\subsection{Overview of the Calibration Strategy}

The purpose of the jet energy calibration is to relate, on average, the energy measured for the detector jet to the energy of the corresponding true particle jet. A true particle jet results from the clustering (with the same clustering algorithm applied to detector jets) of all stable particles originating from the fragmenting parton, as well as of the particles from the underlying event (UE) activity. The correction is applied as a multiplicative factor $\mathcal{C}$ to each component of the raw jet four-momentum vector $p_\mu^{raw}$ (components are indexed by $\mu$ in the following):

\begin{equation}
  \label{eq:master}
  p_\mu^{cor} = \mathcal{C}\cdot p_\mu^{raw}.
\end{equation}

The correction factor $\mathcal{C}$ is composed of the offset correction $C_\text{offset}$, the MC calibration factor $C_\text{MC}$, and the residual calibrations $C_\text{rel}$ and $C_\text{abs}$ for the relative and absolute energy scales, respectively. The offset correction removes the extra energy due to noise and pile-up, and the MC correction removes the bulk of the non-uniformity in $\eta$ and the non-linearity in \pt. Finally, the residual corrections account for the small differences between data and simulation. The various components are applied in sequence as described by the equation below:

\begin{equation}
  \label{eq:jec_components}
  \mathcal{C} = C_\text{offset}(\pt^{raw})\cdot C_\text{MC}(\pt^{\prime},\eta)\cdot C_\text{rel}(\eta)\cdot C_\text{abs}(\pt^{\prime\prime}),
\end{equation}

where $\pt^{\prime}$ is the transverse momentum of the jet after applying the offset correction and $\pt^{\prime\prime}$ is the \pt of the jet after all previous corrections. In the following sections, each component of the jet energy calibration will be discussed separately.

\subsection{Offset Correction}\label{sec:offset}

The offset correction is the first step in the chain of the factorized corrections. Its purpose is to estimate and subtract the energy not associated with the high-\pt scattering. The excess energy includes contributions from electronics noise and pile-up. In CMS, three approaches are followed for the offset correction: the jet area, the average offset and the hybrid jet area methods.

\subsubsection{Jet Area Method}

Recent developments in the jet reconstruction algorithms have allowed a novel approach for the treatment of pile-up~\cite{PU_JET_AREAS,JET_AREAS}: for each event, an average \pt-density $\rho$ per unit area is estimated, which characterizes the soft jet activity and is a combination of the underlying event, the electronics noise, and the pile-up. The two latter components contaminate the hard jet energy measurement and need to be corrected for with the offset correction.

The key element for this approach is the jet area $A_j$. A very large number of infinitely soft four-momentum vectors (soft enough not to change the properties of the true jets) are artificially added in the event and clustered by the jet algorithm together with the true jet components. The extent of the region in the $y-\phi$ space occupied by the soft particles clustered in each jet defines the active jet area. The other important quantity for the pile-up subtraction is the \pt density $\rho$, which is calculated with the $k_T$ jet clustering algorithm~\cite{KT1,KT2,KT3} with a distance parameter $R=0.6$. The $k_T$ algorithm naturally clusters a large number of soft jets in each event, which effectively cover the entire $y-\phi$ space, and can be used to estimate an average \pt-density. The quantity $\rho$ is defined on an event-by-event basis as the median of the distribution of the variable ${\pt}_j/A_j$, where $j$ runs over all jets in the event, and is not sensitive to the presence of hard jets. At the detector level, the measured density $\rho$ is the convolution of the true particle-level activity (underlying event, pile-up) with the detector response to the various particle types.

Based on the knowledge of the jet area and the event density $\rho$, an event-by-event and jet-by-jet pile-up correction factor can be defined:

\begin{equation}
\label{eq:fastjet}
  C_\text{area}(\pt^{raw},A_j,\rho) = 1-\frac{\left(\rho-\langle\rho_\text{UE}\rangle\right)\cdot A_j}{\pt^{raw}}.
\end{equation}

\begin{figure}[ht!]
  \begin{center}
    \includegraphics[width=0.45\textwidth]{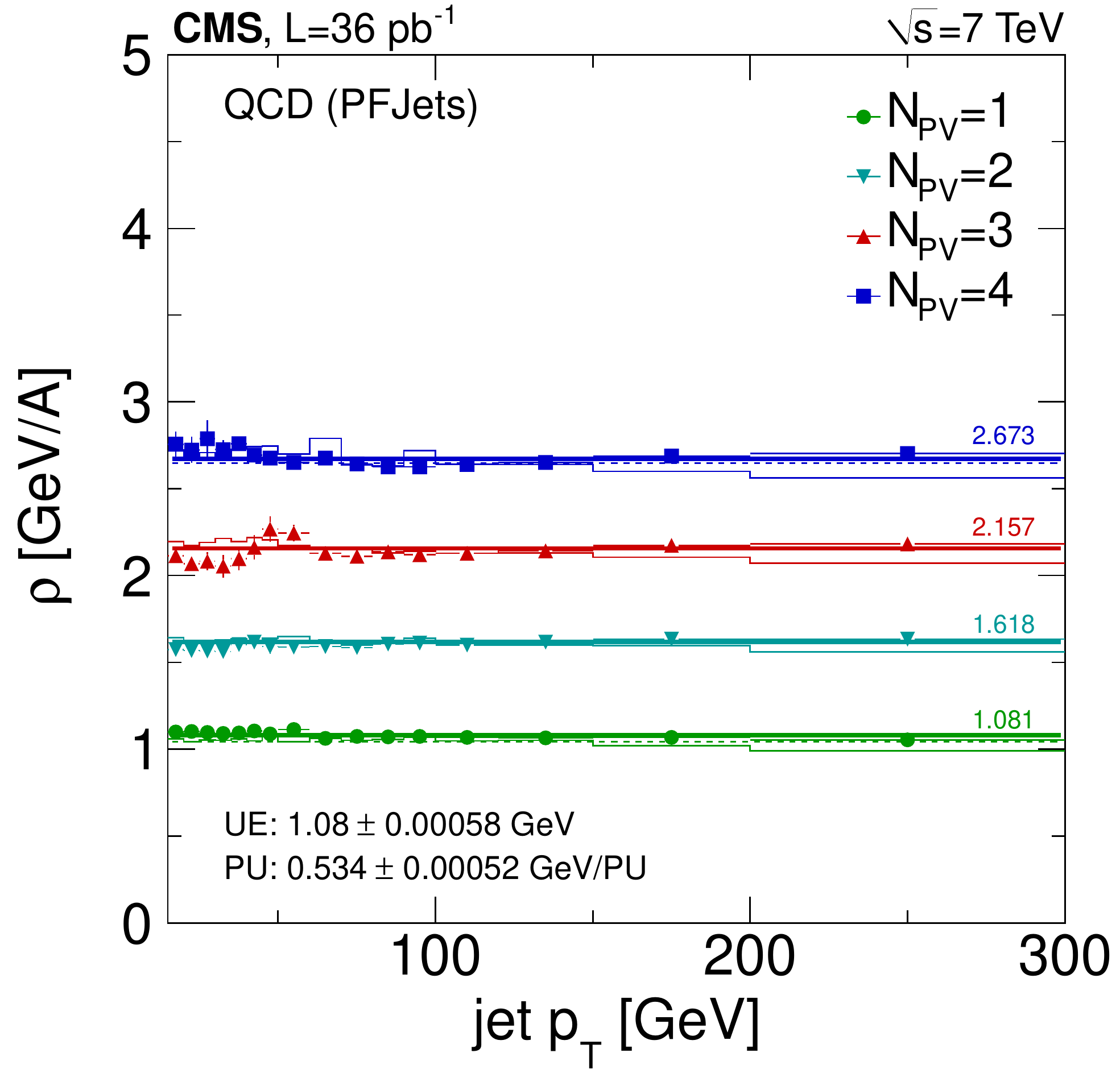}
    \caption{Pile-up and underlying event PF \pt-density $\rho$, as a function of the leading jet \pt in the QCD multijet sample for various pile-up conditions (here $N_\text{PV}$ denotes the number of reconstructed vertices, and A denotes the unit area in the $y-\phi$ space).}
    \label{fig:fastjet}
  \end{center}
\end{figure}

In the formula above, $\langle\rho_\text{UE}\rangle$ is the \pt-density component due to the UE and electronics noise, and is measured in events with exactly one reconstructed primary vertex (no pile-up). Figure~\ref{fig:fastjet} shows the PF \pt-density $\rho$, as a function of the leading jet \pt in QCD events and for various pile-up conditions. The fact that $\rho$ does not depend on the hard scale of the event confirms that it is really a measure of the soft jet activity. Finally, the density $\rho$ shows linear scaling properties with respect to the amount of pile-up.

\subsubsection{Average Offset Method}

\begin{figure}[ht!]
  \begin{center}
    \includegraphics[width=0.45\textwidth]{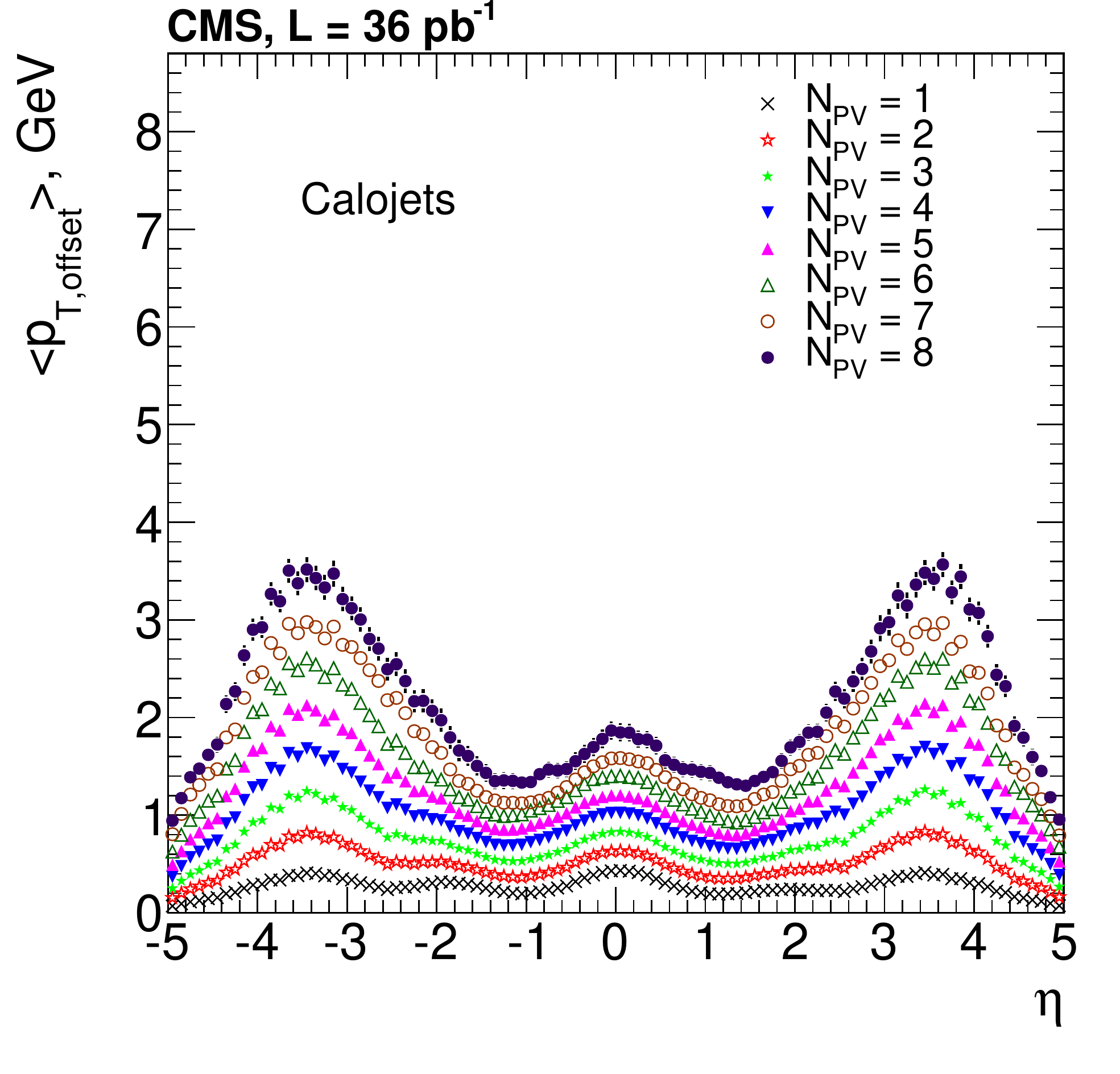}
    \includegraphics[width=0.45\textwidth]{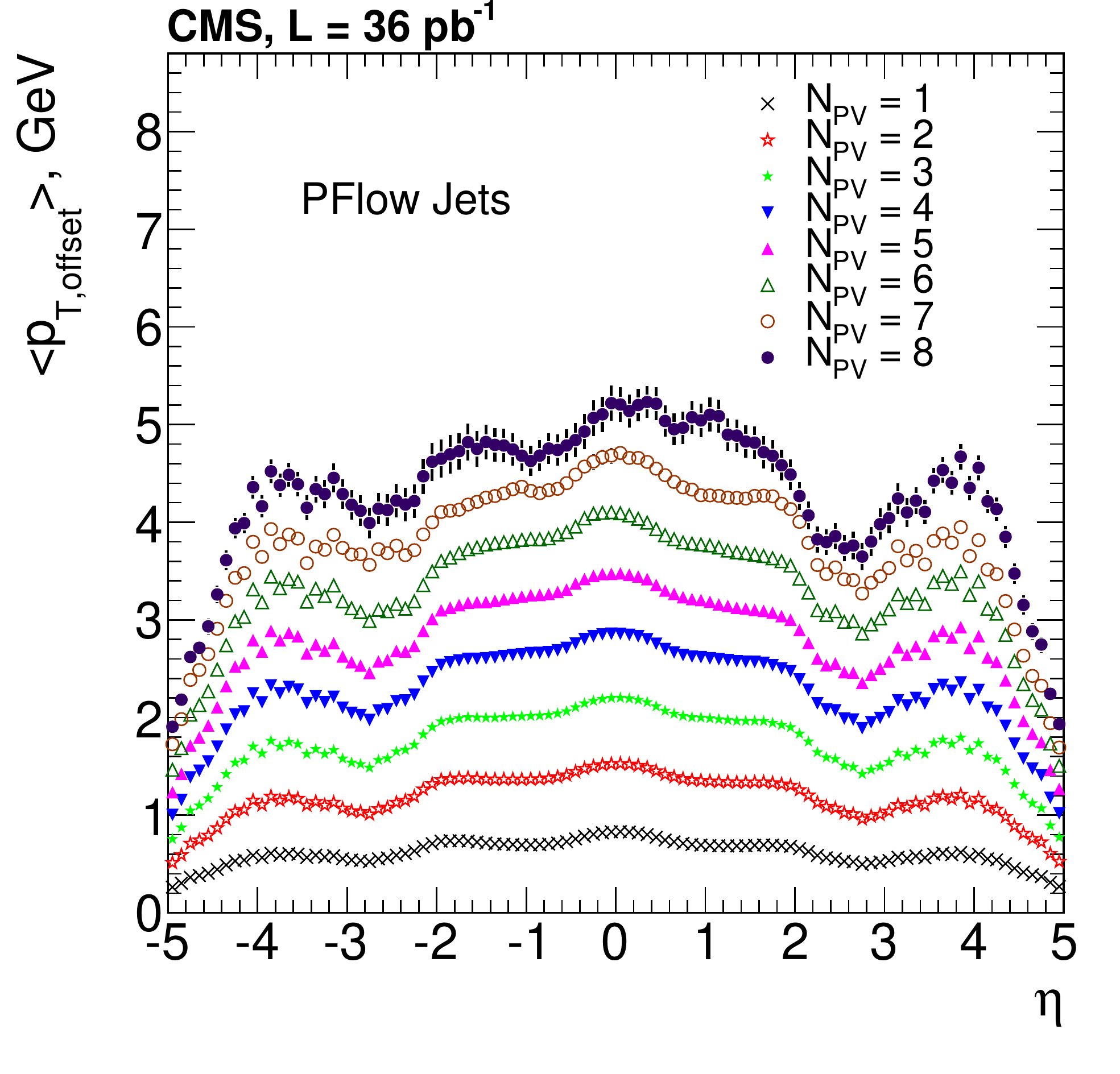}
    \caption{Average offset in \pt, as a function of $\eta$, measured in minimum bias events for different pile-up conditions (categorized according to the number $N_\text{PV}$ of reconstructed primary vertices). Left: CALO jets. Right: PF jets.}
    \label{fig:offset}
  \end{center}
\end{figure}

\begin{figure}[ht!]
  \begin{center}
    \includegraphics[width=0.45\textwidth]{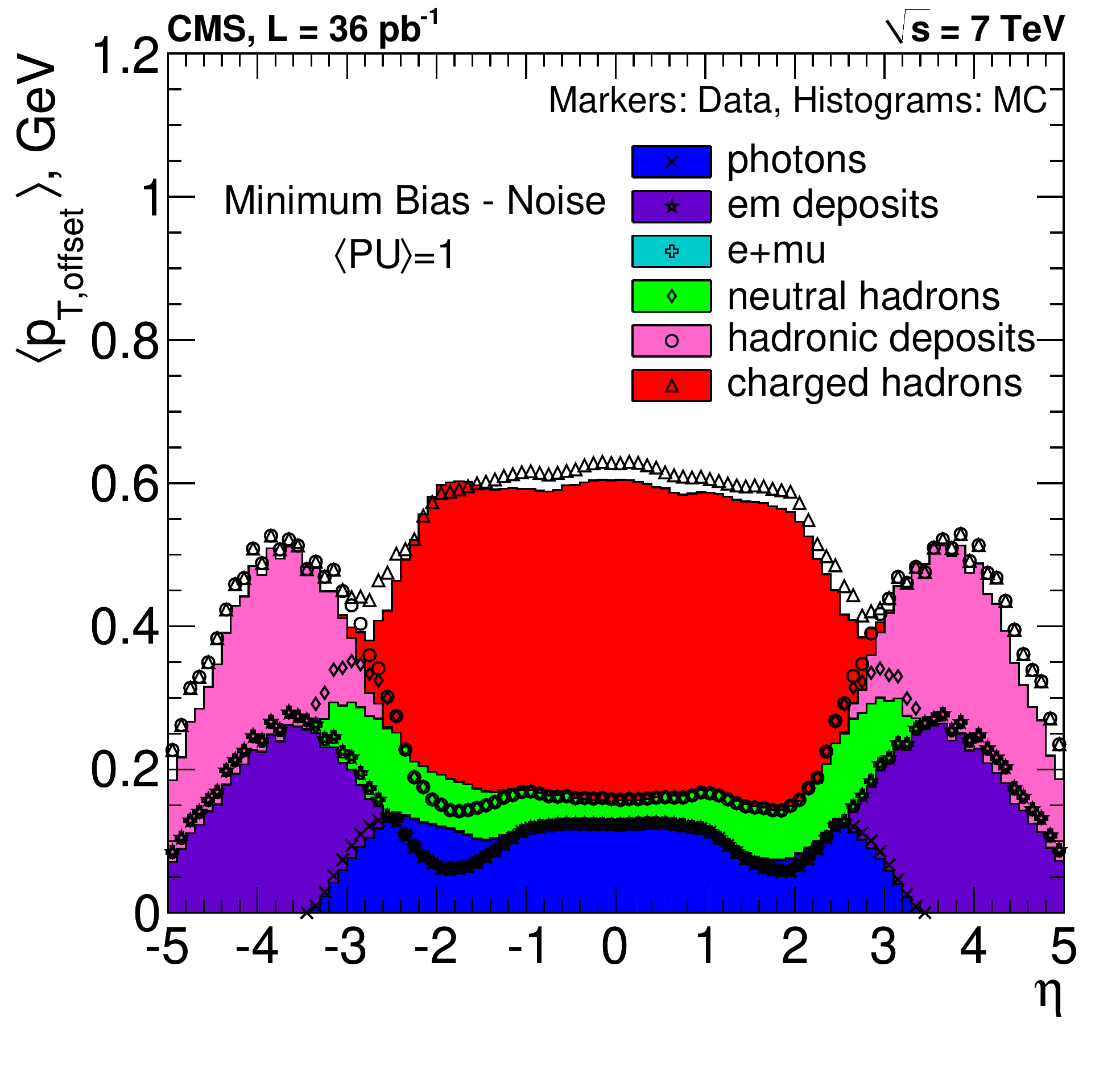}
    \caption{Breakdown of the average offset \pt, in terms of the PF candidates, as a function of $\eta$, for events with one PU interaction. Data are shown by markers and MC is shown as filled histograms.}
    \label{fig:PUcomposition}
  \end{center}
\end{figure}

\begin{figure}[ht!]
  \begin{center}
    \includegraphics[width=0.9\textwidth]{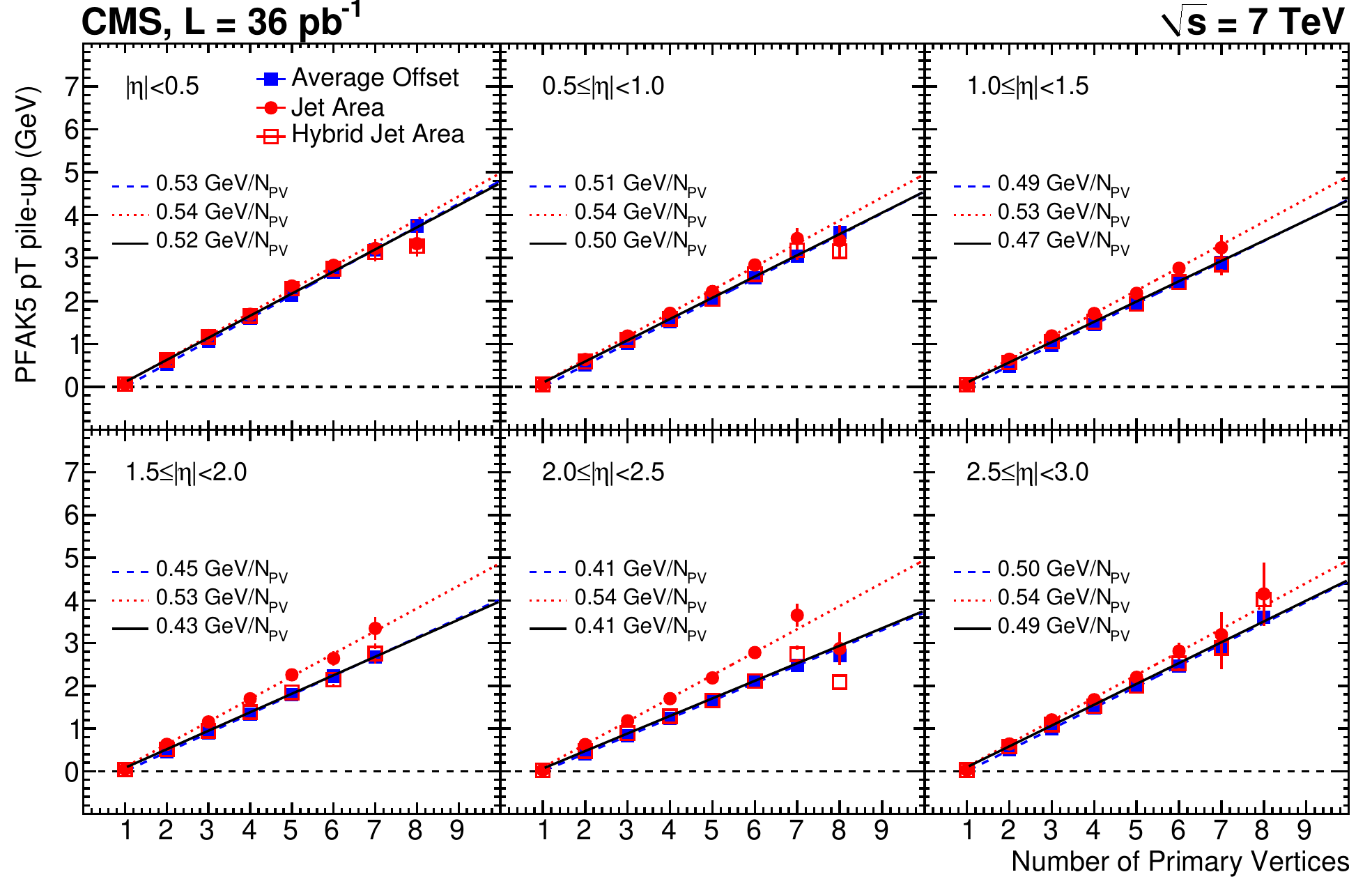}
    \caption{Average PF jet pile-up \pt, as a function of the number of reconstructed vertices ($N_\text{PV}$) for the jet area, the average offset, and the hybrid jet area methods in 6 different $\eta$ regions. In the y-axis title, PFAK5 denotes the PF jets reconstructed with the anti-$k_T$ algorithm with distance parameter $R=0.5$.}
    \label{fig:pileupPF}
  \end{center}
\end{figure}

The average offset method attempts to measure the average energy due to noise and pile-up, clustered inside the jet area, in addition to the energy associated with the jet shower itself. The measurement of the noise contribution is made in zero bias events by vetoing those that pass the minimum bias trigger. In the remaining events, the energy inside a cone of radius $R=0.5$ in the $\eta-\phi$ space is summed. The measurement is performed in cones centered at a specific $\eta$ bin and averaged across $\phi$. The noise contribution is found to be less than $250\MeV$ in \pt, over the entire $\eta$ range. The total average offset (over the entire dataset) is determined from inclusive zero bias events (with no veto on minimum bias triggers) and is classified according to the number of reconstructed vertices. Figure~\ref{fig:offset} shows the average offset \pt as a function of $\eta$ and for different pile-up conditions. The calorimetric offset \pt shows strong variations as a function of $\eta$, which follow the non-uniform particle response in the calorimeter, while for PF candidates, the offset \pt is more uniform versus $\eta$. The higher measured offset \pt for the PF-candidates is due to the much higher response with respect to the pure calorimetric objects. The observed $\eta$-asymmetry is related to calorimeter instrumental effects. For the highest number of vertices, in particular, the asymmetry is also of statistical nature (the adjacent points are highly correlated because at a given $\eta$ a large fraction of the energy in a cone of $R=0.5$ also ends up in overlapping cones). Figure~\ref{fig:PUcomposition} shows the breakdown, in terms of PF candidates, of the average offset \pt in events with one PU interaction, as measured in the data and compared to the MC prediction. The slight asymmetry observed in the MC is due to the asymmetric noise description in the specific version of the simulation. The average offset in \pt scales linearly with the number of reconstructed primary vertices, as shown in Fig.~\ref{fig:pileupPF}. The linear scaling allows the expression of the jet offset correction as follows:

\begin{equation}
  C_\text{offset}(\eta,\pt^{raw},N_\text{PV}) = 1-\frac{(N_\text{PV}-1)\cdot\mathcal{O}(\eta)}{\pt^{raw}},
\end{equation}

where $\mathcal{O}(\eta)$ is the average \pt due to one pile-up event, $\pt^{raw}$ is the \pt of the uncorrected jet, and $N_\text{PV}$ is the number of reconstructed primary vertices. The average offset method can be applied to jet algorithms that produce circular jets, while the quantity $\mathcal{O}(\eta)$ scales to larger cone sizes in proportion to the jet area. It should be noted that, in both the average offset subtraction and in the jet area method, the noise contribution and the UE are not subtracted. Because of the good description of the noise contribution in the simulation, the noise is taken into account with the MC-based correction.

\subsubsection{Hybrid Jet Area Method}

The measurement of the average offset presented in the previous paragraph confirms the $\eta$-dependence of the offset energy. This is explained by the fact that the measured offset is the convolution of the pile-up activity with the detector response. In order to take into account the $\eta$-dependence, a hybrid jet area method is employed:

\begin{equation}
\label{eq:hybrid}
  C_\text{hybrid}(\pt^{raw},\eta,A_j,\rho) = 1-\frac{(\rho-\langle\rho_\text{UE}\rangle)\cdot\beta(\eta)\cdot A_j}{\pt^{raw}}.
\end{equation}

In Eq.~(\ref{eq:hybrid}), the \pt density $\rho$ and the corresponding density due to the UE, $\langle\rho_{UE}\rangle$ are constants over the entire $\eta$ range. The multiplicative factor $\beta(\eta)$ corrects for the non-uniformity of the energy response and is calculated from the modulation of the average offset in \pt (Fig.~\ref{fig:offset}).

In the case of PF jets, the response variation versus $\eta$ is relatively small and the hybrid jet area method is found to be in excellent agreement with the average offset method. Figure~\ref{fig:pileupPF} shows the average offset in \pt as a function of the number of reconstructed primary vertices, for the three different methods (jet area, average offset, hybrid jet area). It can be seen that the differences between the jet area method and the average offset method are entirely due to the response dependence on $\eta$. The hybrid jet area method is chosen for the pile-up correction of PF jets. In the case of CALO jets, and also JPT jets (initially reconstructed as CALO jets), the response variation versus $\eta$ shows dramatic changes and neither the simple jet area nor the hybrid jet area methods are able to reproduce the average offset measurement. Therefore, for CALO jets, the average offset method is the one chosen for the pile-up correction.

\subsubsection{Offset Uncertainty}\label{sec:offset_unc}

\begin{figure}[ht!]
  \begin{center}
    \includegraphics[width=0.45\textwidth]{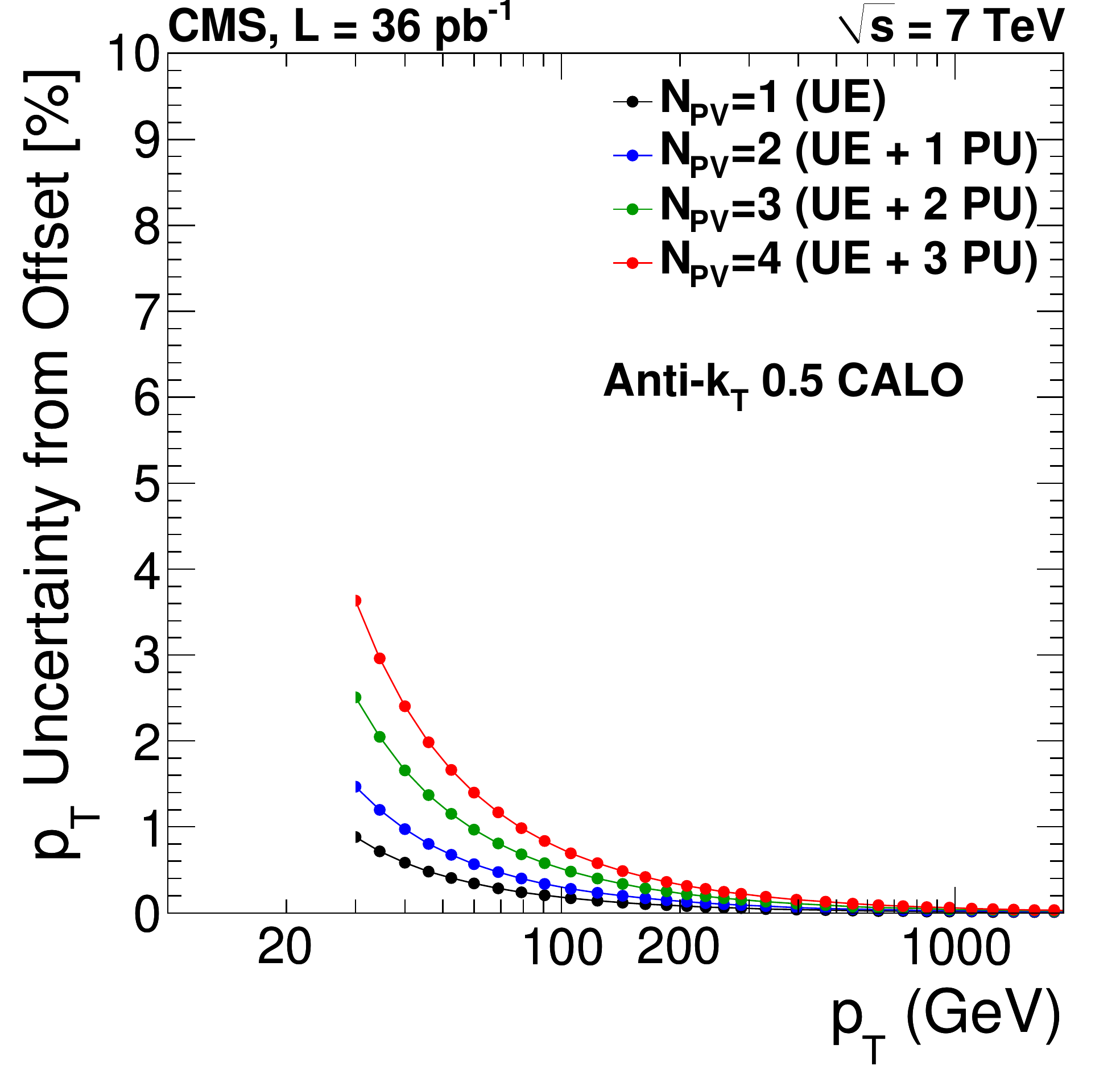}
    \includegraphics[width=0.45\textwidth]{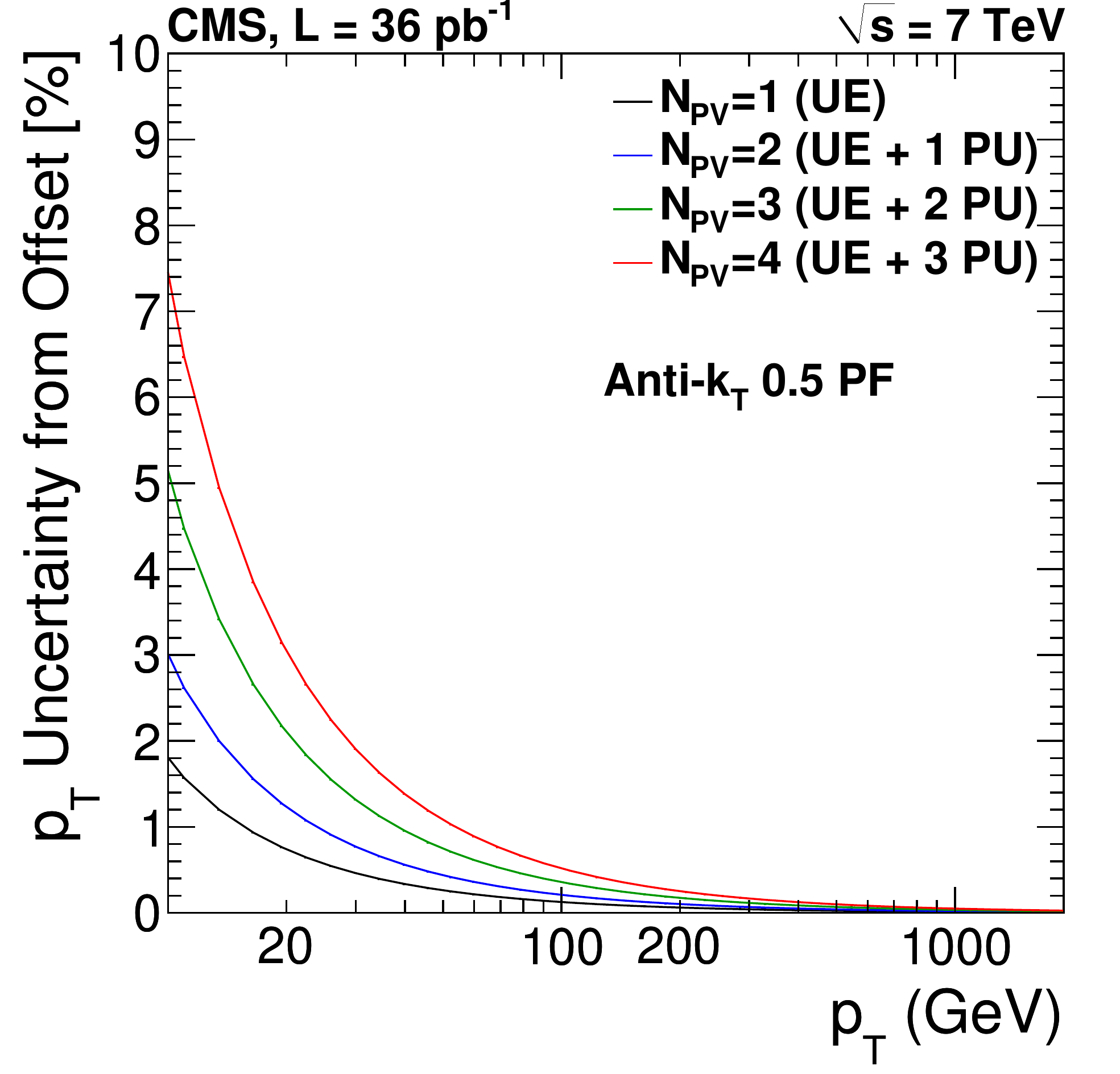}
    \caption{Offset jet-energy-correction uncertainty as a function of jet \pt. Left: CALO jets. Right: PF jets.}
    \label{fig:offsetUnc}
  \end{center}
\end{figure}

The uncertainty of the offset correction is quantified using the jet area method. Specifically, the quantities $\rho$ and $\langle\rho_\text{UE}\rangle$ in Eq.~(\ref{eq:fastjet}) are varied independently and the resulting shifts are added in quadrature. The event \pt-density $\rho$ uncertainty is estimated as $0.2\GeV$ per unit jet area and per pile-up event. This uncertainty is based on the maximum slope difference between the jet area and the average offset methods, and the residual non-closure in the average offset method. The UE \pt-density $\langle\rho_\text{UE}\rangle$ uncertainty is estimated as $0.15\GeV$ per unit jet area, based on the differences observed between the QCD multijet and Z+jets samples, and on the effective difference when applied in the inclusive jet cross-section measurement. Figure~\ref{fig:offsetUnc} shows the uncertainty of the offset correction, as a function of jet \pt and the number of primary vertices.

\subsection{Monte Carlo Calibration}

\begin{figure}[ht!]
  \begin{center}
    \includegraphics[width=0.45\textwidth]{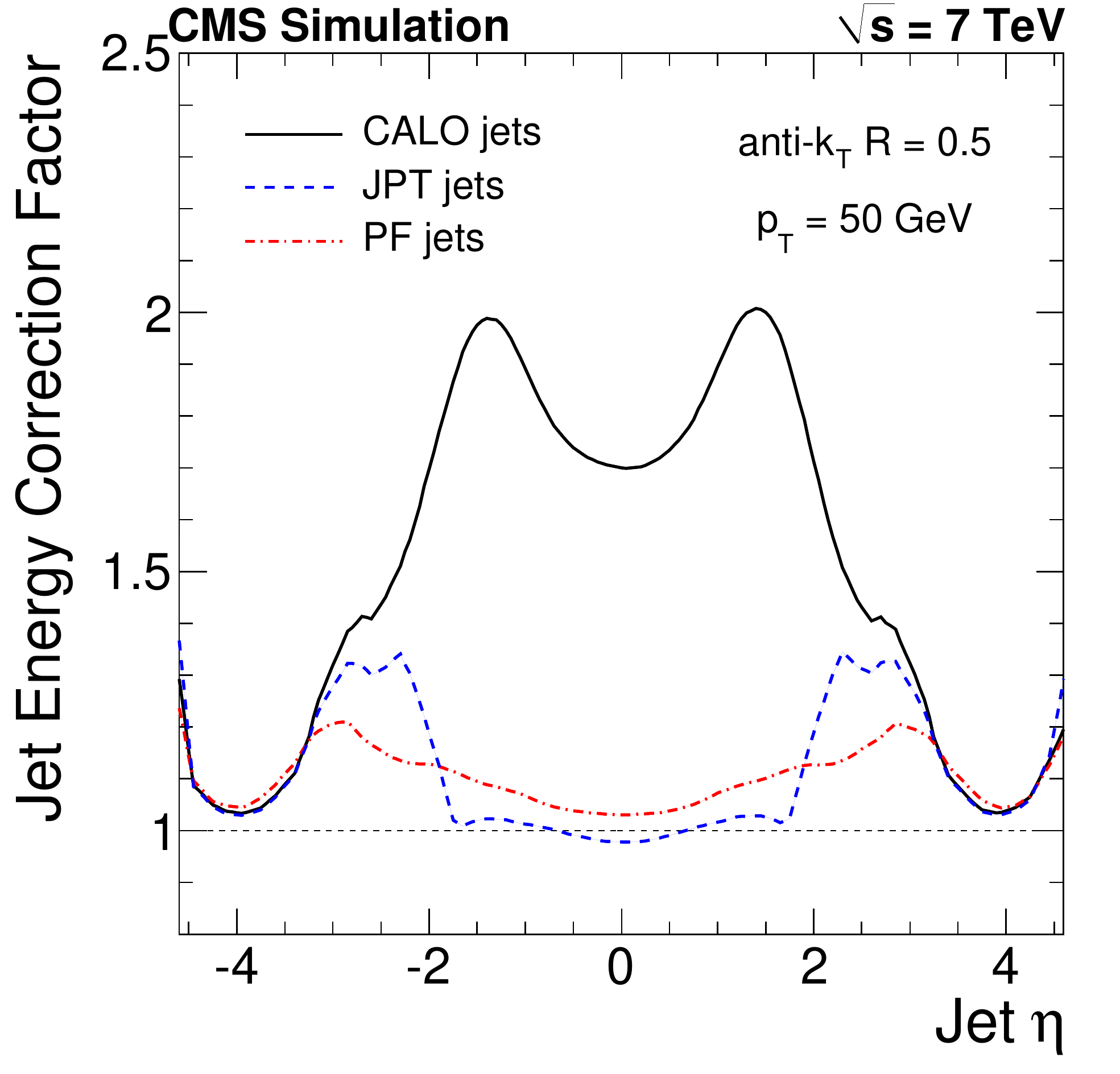}
    \includegraphics[width=0.45\textwidth]{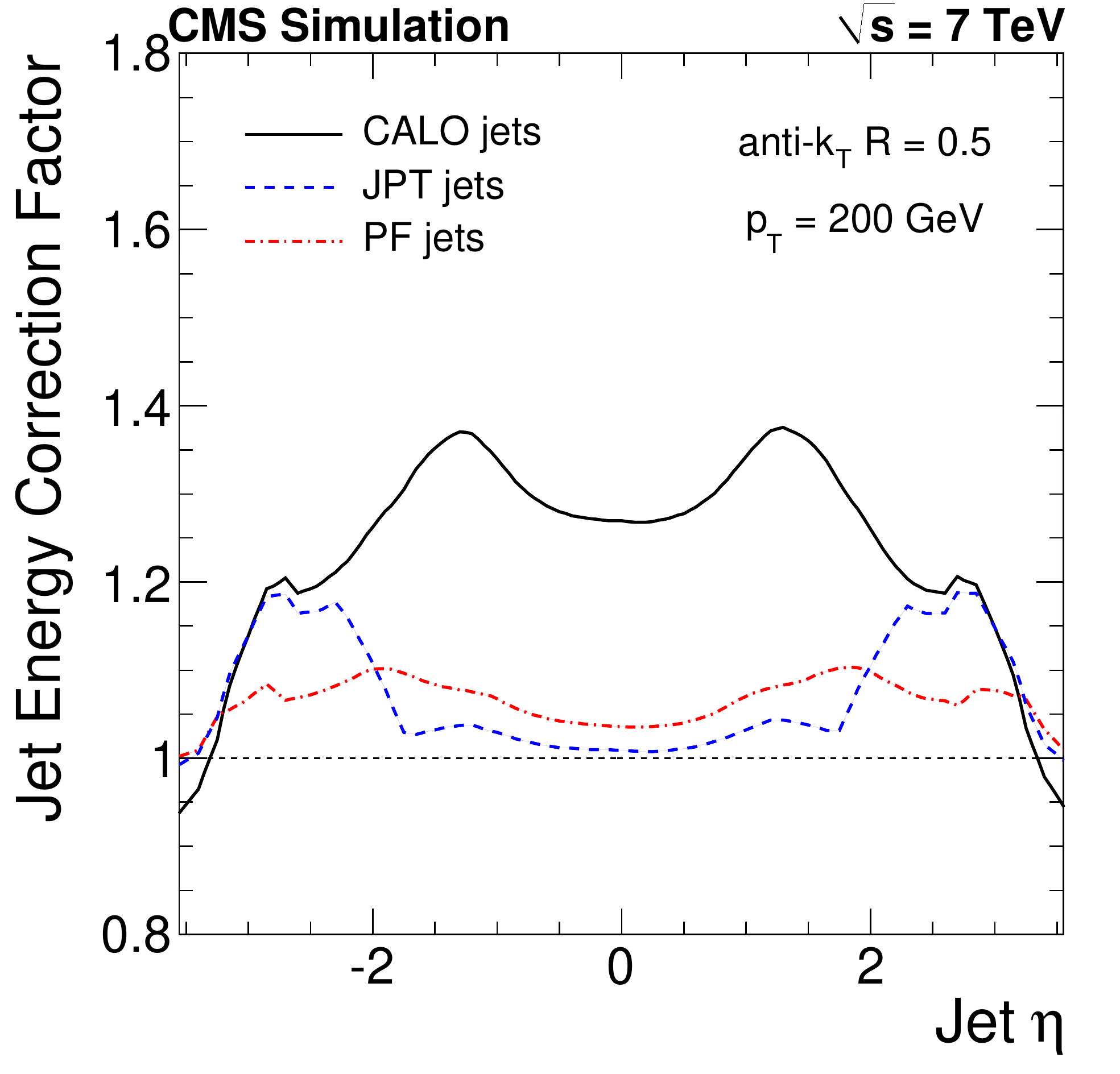}
    \caption{Monte Carlo jet-energy-correction factors for the different jet types, as a function of jet $\eta$. Left: correction factor required to get a corrected jet $\pt=50\GeV$. Right: correction factor required to get a corrected jet $\pt=200\GeV$.}
    \label{fig:mctruthVsEta}
  \end{center}
\end{figure}

\begin{figure}[ht!]
  \begin{center}
    \includegraphics[width=0.45\textwidth]{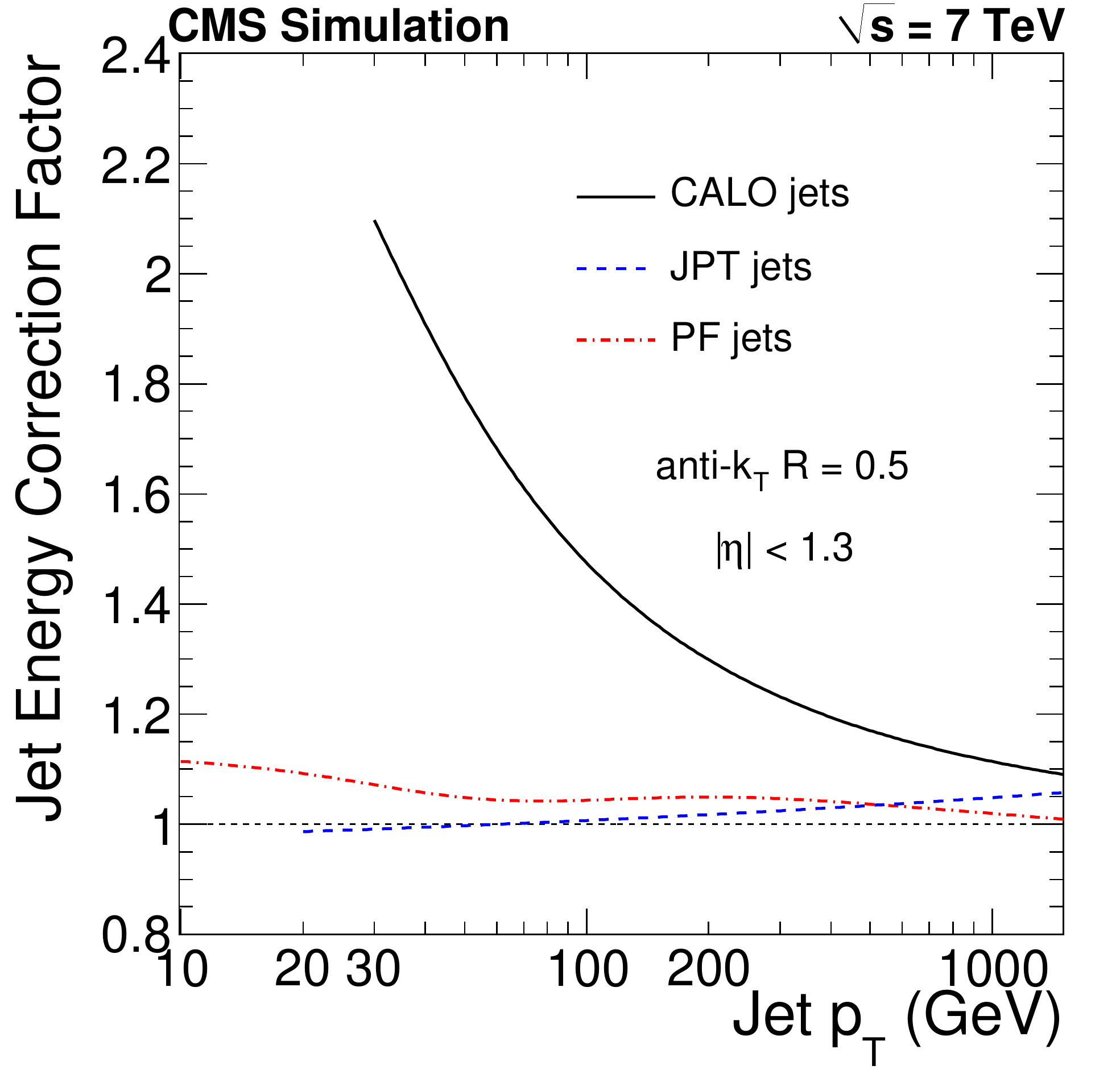}
    \caption{Monte Carlo jet-energy-correction factors for the different jet types, as a function of jet \pt.}
    \label{fig:mctruthVsPt}
  \end{center}
\end{figure}

The MC calibration is based on the simulation and corrects the energy of the reconstructed jets such that it is equal on average to the energy of the generated MC particle jets. Simulated QCD events are generated with {\sc PYTHIA}6.4.22~\cite{PYTHIA}, tune Z2 (the {\sc Z2} tune is identical to the {\sc Z1} tune described in~\cite{D6T} except that {\sc Z2} uses the CTEQ6L PDF, while {\sc Z1} uses CTEQ5L) and processed through the CMS detector simulation, based on {\sc GEANT4}~\cite{GEANT4}. The jet reconstruction is identical to the one applied to the data. Each reconstructed jet is spatially matched in the $\eta-\phi$ space with a MC particle jet by requiring $\Delta R<0.25$. In each bin of the MC particle transverse momentum $\pt^{gen}$, the response variable $\mathcal{R}=\pt^{reco}/\pt^{gen}$ and the detector jet $\pt^{reco}$ are recorded. The average correction in each bin is defined as the inverse of the average response $C_\text{MC}(\pt^{reco})=\frac{1}{<\mathcal{R}>}$, and is expressed as a function of the average detector jet \pt $<\pt^{reco}>$. Figure~\ref{fig:mctruthVsEta} shows the MC jet energy correction factor for the three jet types, vs. $\eta$, for different corrected jet \pt values. Figure~\ref{fig:mctruthVsPt} shows the average correction in $|\eta|<1.3$, as a function of the corrected jet \pt.

Calorimeter jets require a large correction factor due to the non-linear response of the CMS calorimeters. The structures observed at $|\eta|\sim 1.3$ are due to the barrel-endcap boundary and to the tracker material budget, which is maximum in this region. The fast drop observed in the endcap region $1.3<|\eta|<3.0$ is due to the fact that the jet energy response depends on energy rather than on jet \pt. For higher values of $|\eta|$ more energy corresponds to a fixed \pt value $E\approx\pt\cdot\cosh(\eta)$, which means that the jet response is higher and the required correction factor is smaller. The structure observed at $|\eta|\sim 3.0$ coincides with the boundary between the endcap and the forward calorimeters. Finally, in the region $|\eta|>4.0$, the jet energy response is lower because parts of the jets pointing toward this region extend beyond the forward calorimeter acceptance.

The track-based jet types (JPT and PF) require much smaller correction factors because the charged component of the jet shower is measured accurately in the CMS tracker which extends up to $|\eta|=2.4$. The fast rise of the correction factor for JPT jets in the region $2.0<|\eta|<2.5$ is explained by the fact that part of the jets lying in this region extends beyond the tracker coverage. For PF jets, the transition beyond the tracker acceptance is smoother because the PF candidates, which are input to the clustering of PF jets, are individually calibrated prior to the clustering. While both PF jets and JPT jets exploit the tracker measurements, the JPT jets require lower correction in the region $|\eta|<2.0$ because the tracker inefficiency is explicitly corrected for by the JPT algorithm.
In the forward region ($|\eta|>3.0$) all three jet types converge to simple calorimetric objects and therefore require almost identical corrections.

\begin{figure}[ht!]
  \begin{center}
    \includegraphics[width=0.32\textwidth]{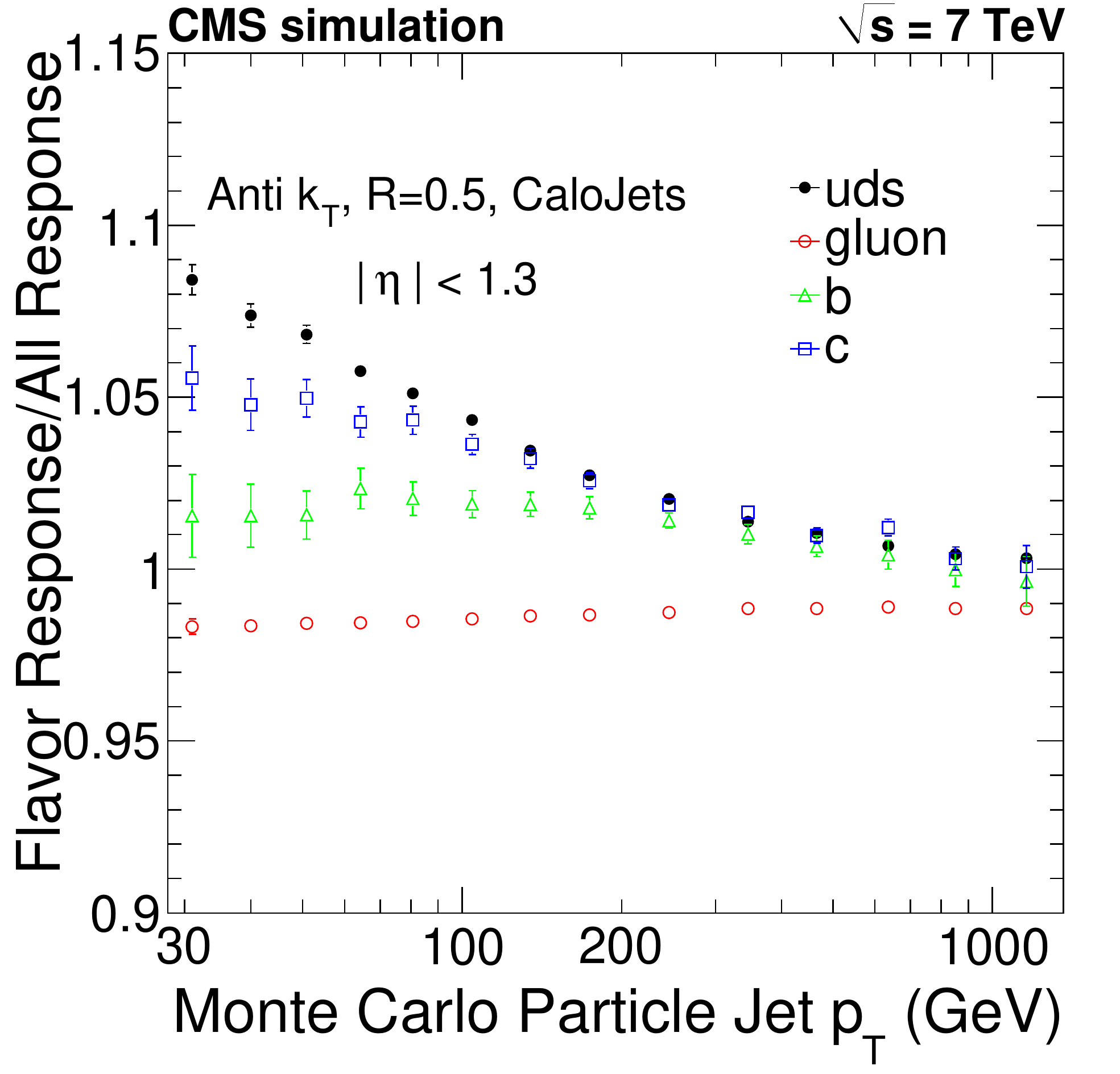}
    \includegraphics[width=0.32\textwidth]{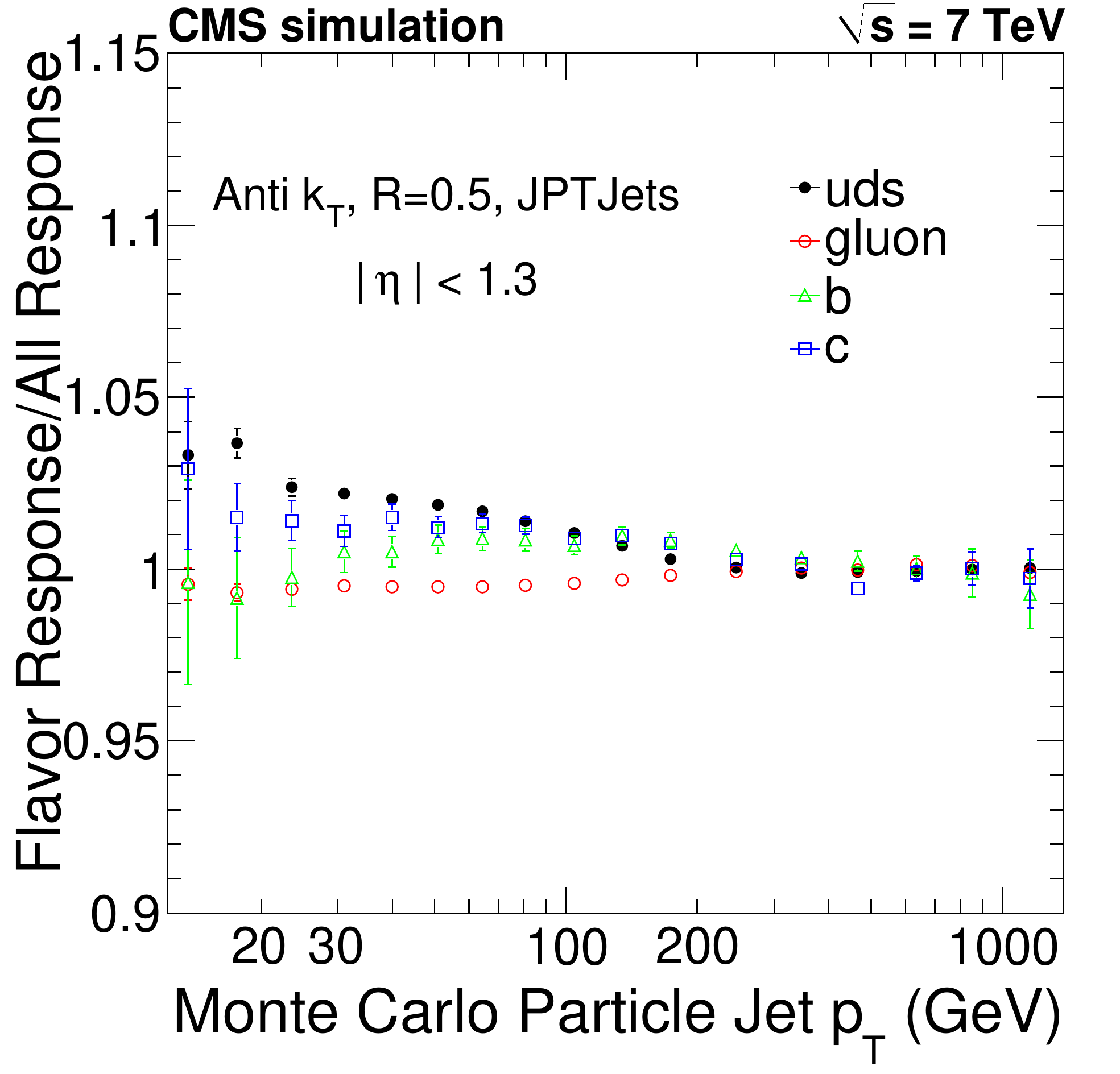}
    \includegraphics[width=0.32\textwidth]{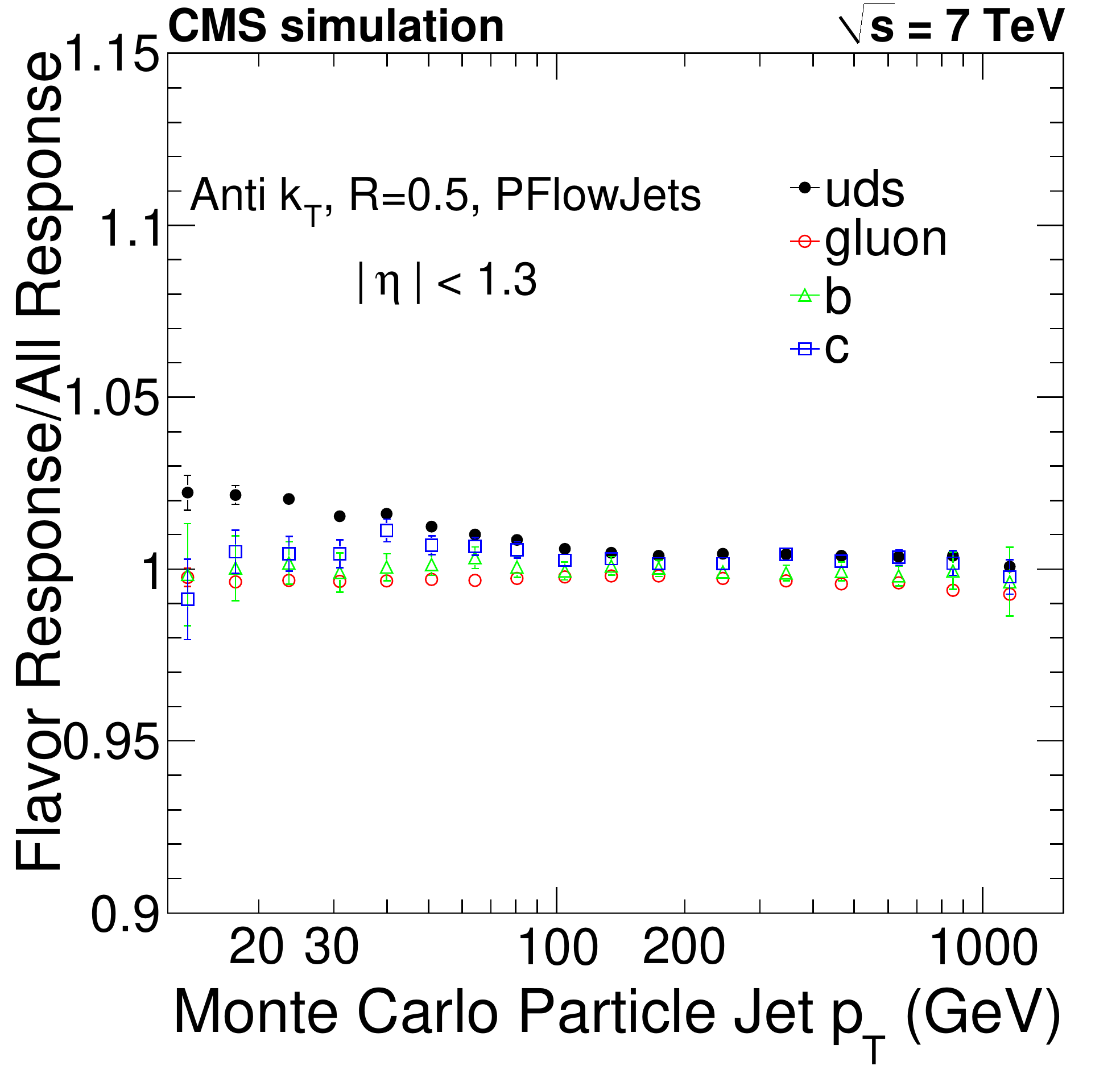}
    \caption{Simulated jet energy response, in {\sc PYTHIA6} Z2 tune, of different jet flavours normalized to the response of the QCD flavour mixture, as a function of the true particle jet \pt, in the region $|\eta|<1.3$ for the three jet types.}
    \label{fig:flavor}
  \end{center}
\end{figure}

\begin{figure}[ht!]
  \begin{center}
    \includegraphics[width=0.45\textwidth]{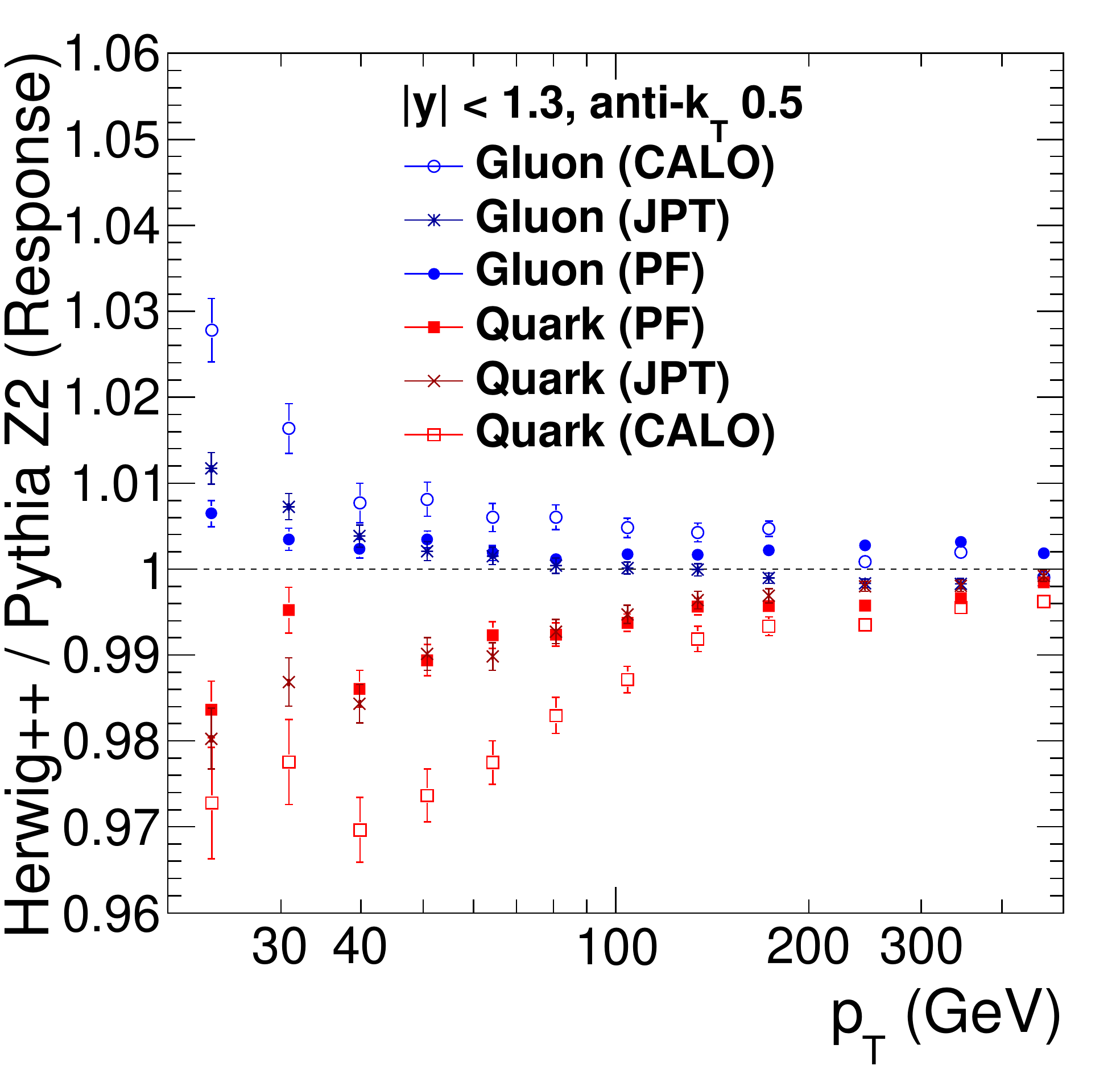}
    \caption{Response ratio predicted by {\sc Herwig++} and {\sc PYTHIA6} for jets originated by light quarks (uds) and gluons for the various jet types.}
    \label{fig:flavor_herwigpythia}
  \end{center}
\end{figure}

The default MC calibration is derived from the QCD sample and corresponds to a jet flavour composition enriched in low-\pt gluon jets. The jet energy response and resolution depend on the fragmentation properties of the initial parton: gluons and heavy-flavour quarks tend to produce more particles with a softer energy spectrum than light quarks. The investigation of the jet energy response of the various flavour types, for the different jet reconstruction techniques, is done with MC matching between the generated particle jet and the reconstructed jet. For each MC particle jet, the corresponding parton is found by spatial matching in the $\eta-\phi$ space. Figure~\ref{fig:flavor} shows the response of each flavour type (gluon, b-quark, c-quark, uds-quark), as predicted by {\sc PYTHIA6} (Z2 tune), in the region $|\eta|<1.3$, normalized to the average response in the QCD flavour mixture. The QCD flavour composition varies significantly with jet \pt, being dominated by gluon jets at low \pt and by quark jets at high \pt. Calorimeter jets show strong dependence on the flavour type with differences up to 10\%. This is attributed to the non-linear single-particle response in the calorimeters. For the track-based reconstructed jets, the flavour dependence is significantly reduced and not larger than 5\% and 3\% for JPT and PF jets respectively. The ability to measure precisely the charged particle momenta in the tracker reduces the contribution of calorimetry at low jet \pt. In all jet types, the jets originated from a light quark (u/d/s) have a systematically higher response than those from the other flavours, which is attributed to the harder spectrum of the particles that are produced in the fragmentation process. For comparison, Fig.~\ref{fig:flavor_herwigpythia} shows the flavour dependent response ratio of a different fragmentation model ({\sc Herwig++}) with respect to {\sc PYTHIA6}. 

\subsection{Relative Jet Energy Scale}

\subsubsection{Measurement} 

\begin{figure}[ht!]
  \begin{center}
    \includegraphics[width=0.45\textwidth]{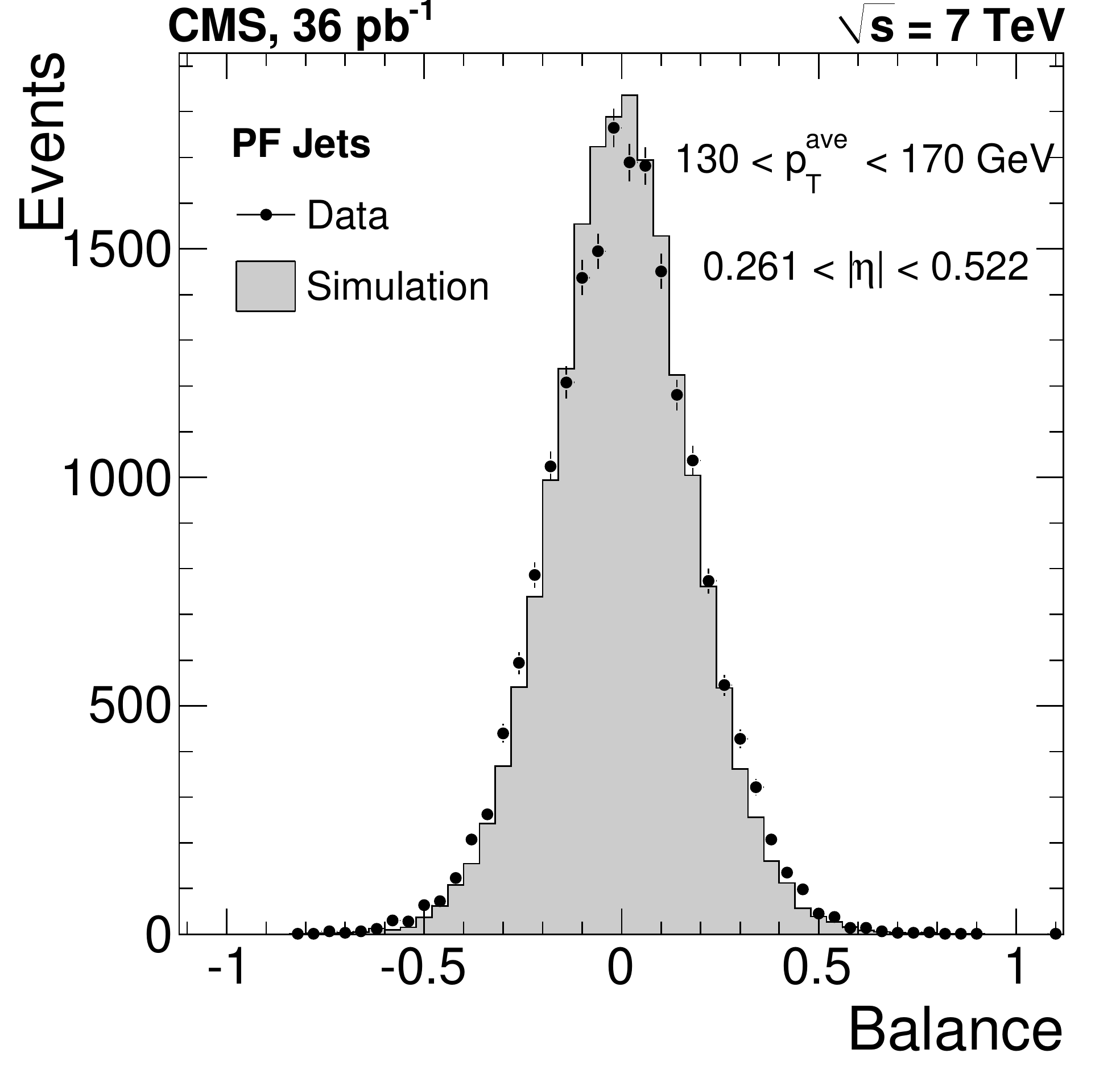}
    \includegraphics[width=0.45\textwidth]{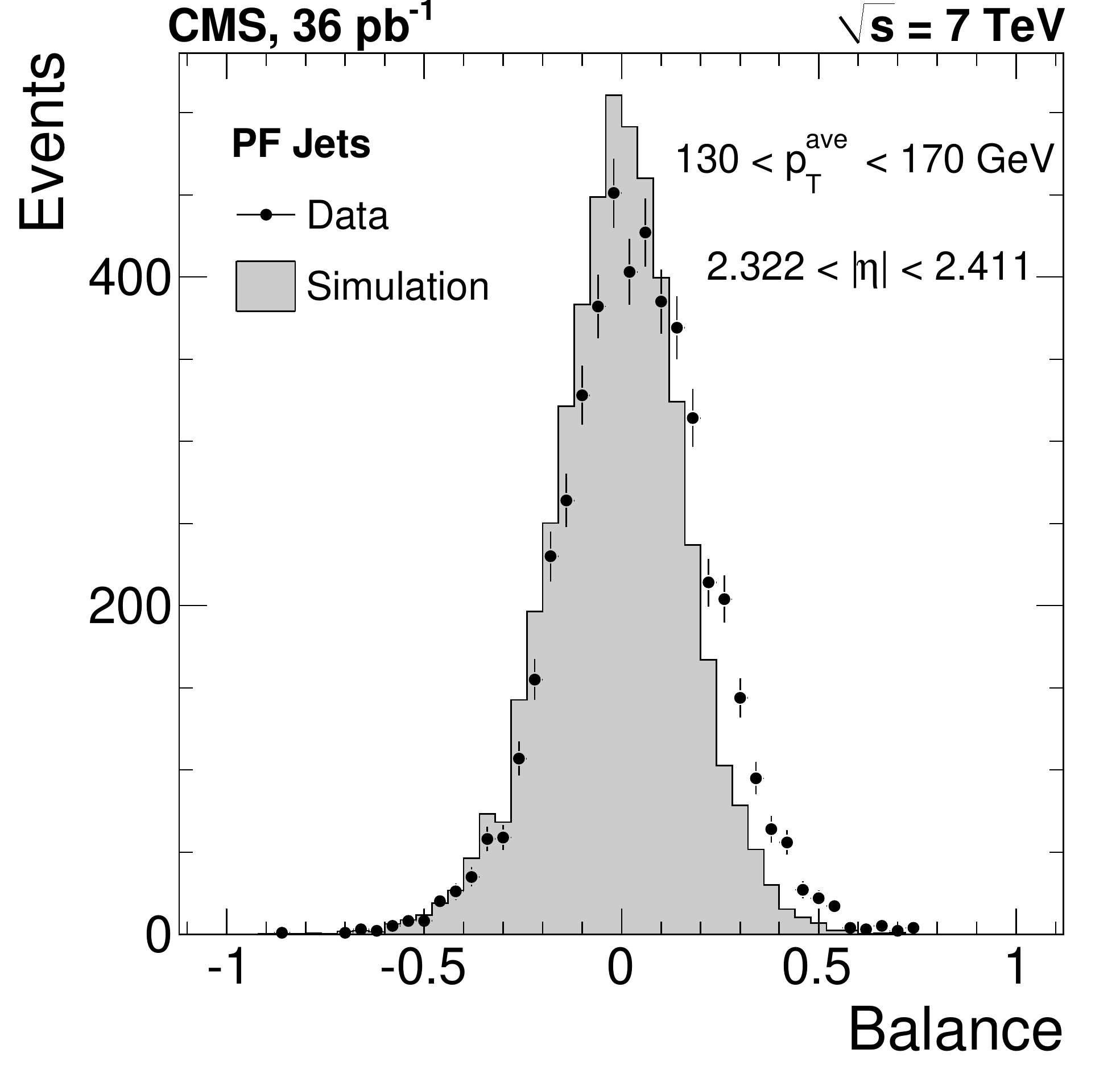}
    \caption{Example distributions of the dijet balance quantity for PF jets in two $\eta$ regions.}
    \label{fig:balance}
  \end{center}
\end{figure}

\begin{figure}[ht!]
  \begin{center}
    \includegraphics[width=0.45\textwidth]{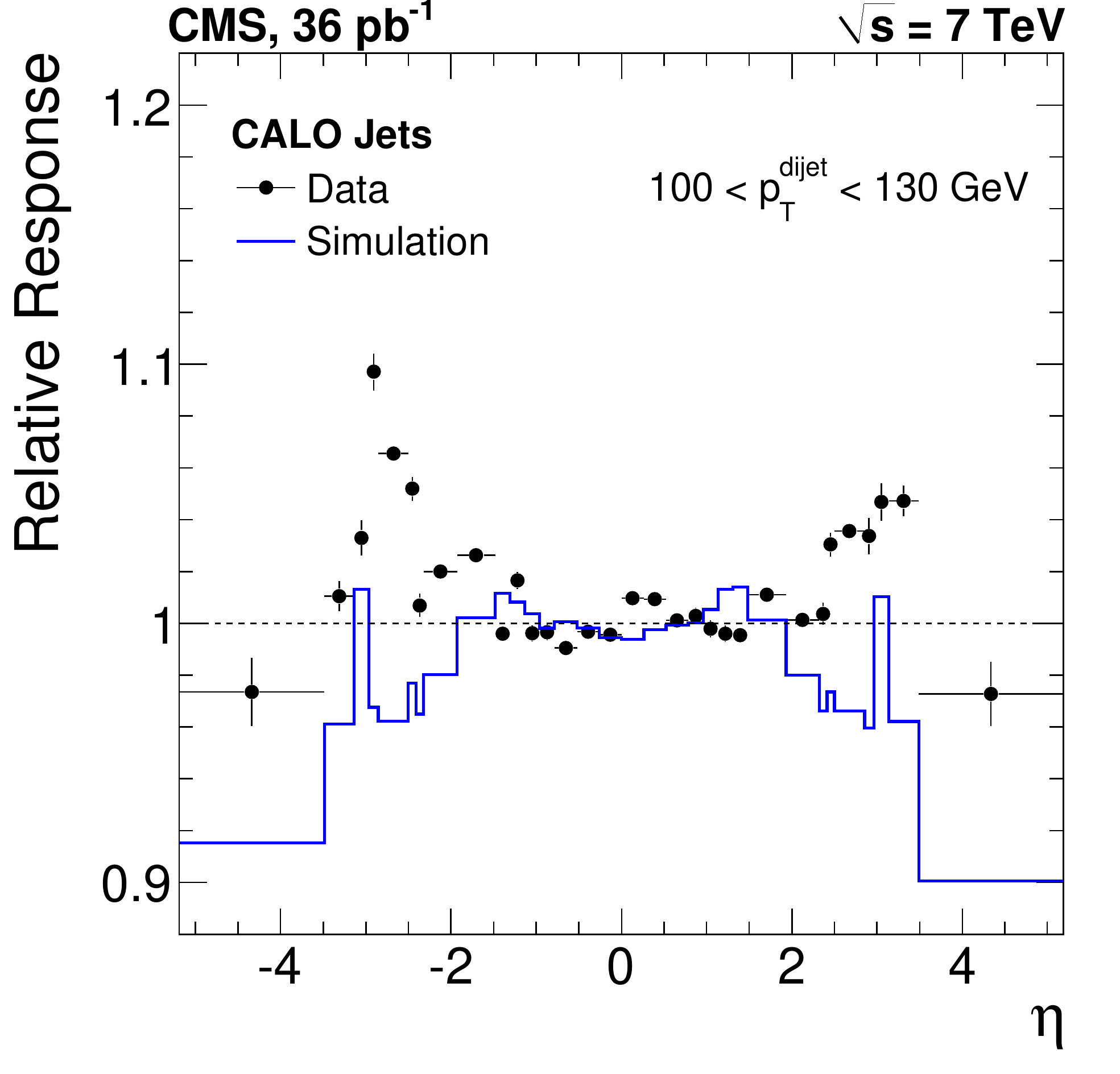}
    \includegraphics[width=0.45\textwidth]{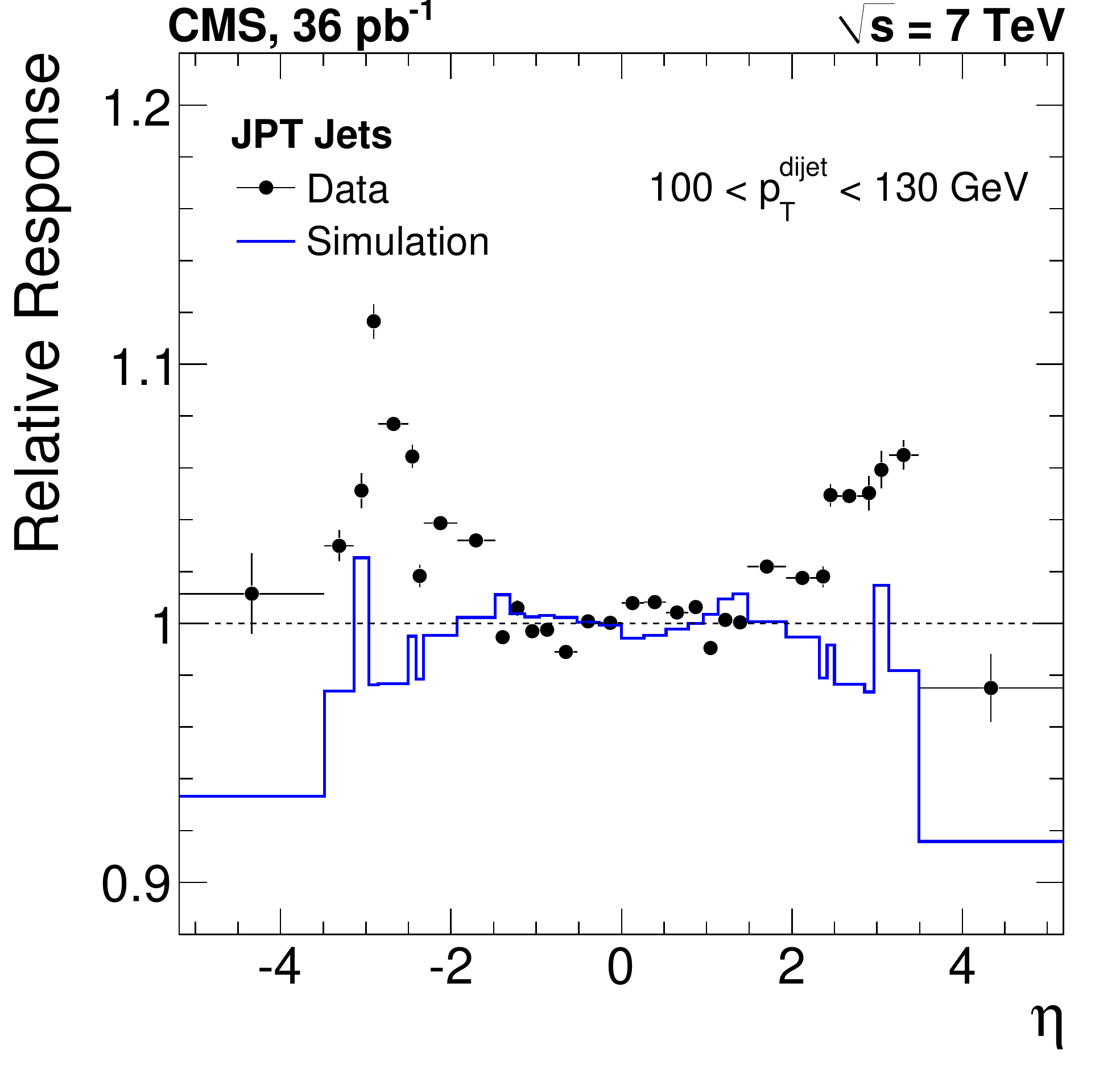}
    \includegraphics[width=0.45\textwidth]{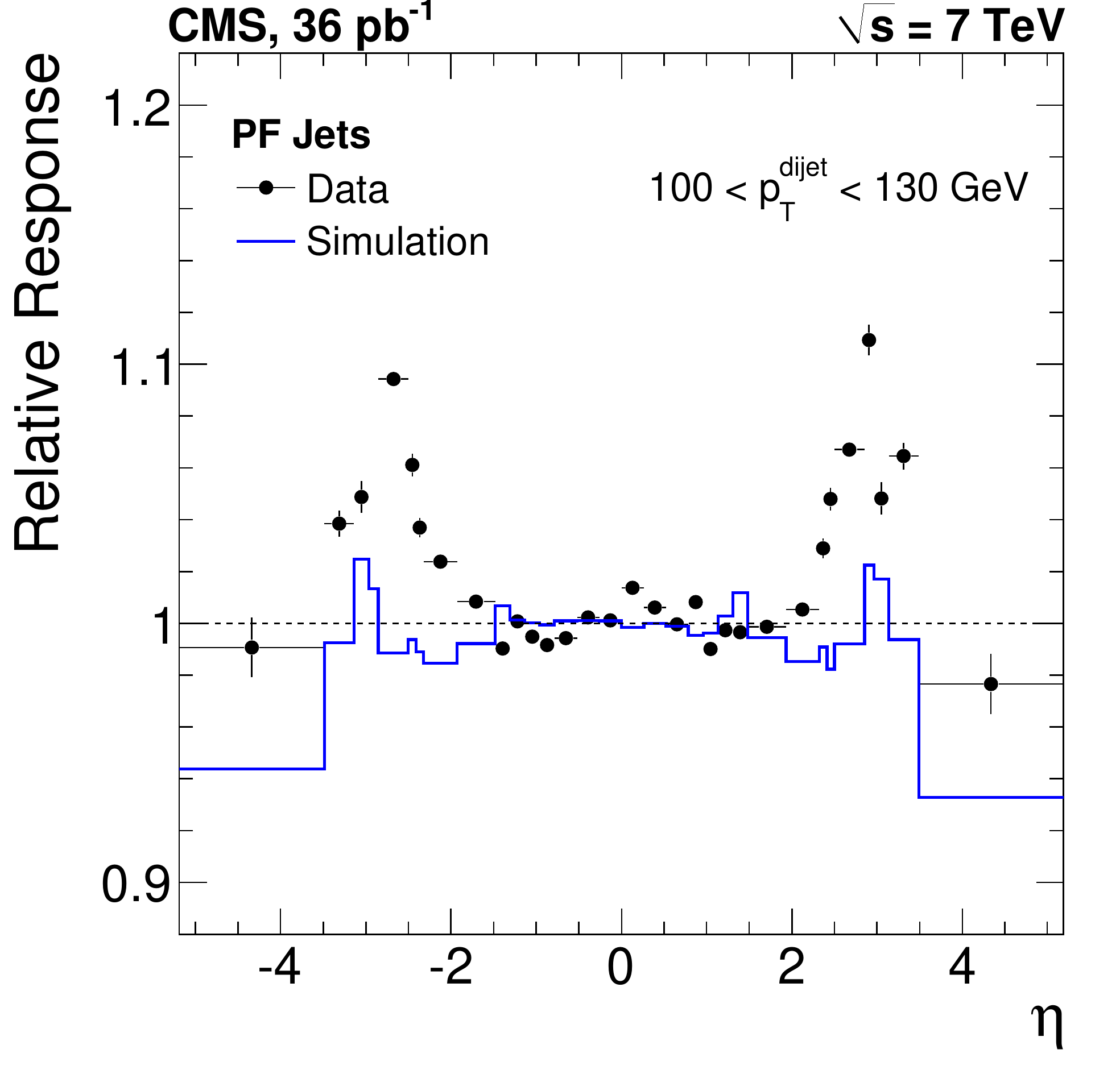}
    \caption{Relative jet energy response as a function of $\eta$, measured with the dijet balance method for CALO, JPT and PF jets respectively.}
    \label{fig:relrsp}
  \end{center}
\end{figure}

The dijet \pt-balance technique, described in Section~\ref{sec:methods}, is used to measure the response of a jet at any $\eta$ relative to the jet energy response in the region $|\eta|<1.3$. Figure~\ref{fig:balance} shows example distributions of the balance quantity $\mathcal{B}$ for PF jets in two pseudorapidity bins. Figure~\ref{fig:relrsp} shows the relative response as a function of $\eta$ in the range $100\GeV < \ptave < 130\GeV$. Ideally, the relative response of the corrected jets in the simulation should be equal to unity. However, because of the resolution bias effect (Section~\ref{sec:resbias}), the relative response in the simulation is found to deviate from unity by an amount equal to the resolution bias. The comparison of the data with the MC simulations implicitly assumes that the resolution bias in the data is the same as in the simulation. This assumption is the dominant systematic uncertainty related to the measurement of the relative response with the dijet balance method.

In order to reduce the radiation bias (Section~\ref{sec:radbias}), a selection is applied on the ratio $\alpha=\ptthird/\ptave$ and the nominal analysis value is $\alpha<0.2$. The residual relative correction calculation is done in three steps: first, the $\eta$-symmetric part, $C_\text{sym}$, is measured in bins of $|\eta|$, in order to maximize the available statistics, with the nominal requirement $\alpha<0.2$. Then, a correction factor $k_\text{rad}$ is applied to take care of the extrapolation to $\alpha=0$, and finally the asymmetry in $\eta$, $\mathcal{A}_R(|\eta|)$, is taken into account. The residual correction for the relative jet energy scale is formally expressed below:

\begin{equation}
  C_\text{rel}(\pm\eta) = \frac{k_\text{rad}(|\eta|)\cdot C_\text{sym}(|\eta|)}{1\mp \mathcal{A}_R(|\eta|)}.
\end{equation}

The $C_\text{sym}$ component is defined by comparing the relative response in data and MC simulations: 

\begin{equation}
  C_\text{sym}(|\eta|) = \left<\frac{R_{MC}^{\alpha<0.2}}{R_{data}^{\alpha<0.2}}\right>_{\pt},
\end{equation}

averaged over the entire \pt range. This is justified by the fact that no statistically significant \pt-dependence is observed in the comparison between data and simulation.

Since the additional radiation and the UE are not perfectly modeled in the simulation, a correction needs to be applied by extrapolating to zero third-jet activity, as discussed in Section~\ref{sec:radbias}. The radiation correction $k_{rad}$ is defined as: 

\begin{equation}
  k_\text{rad} = \lim_{\alpha\to 0}\left(\dfrac{\left<\frac{R_{MC}^\alpha}{R_{data}^\alpha}\right>_{\pt}}{\left<\frac{R_{MC}^{\alpha<0.2}}{R_{data}^{\alpha<0.2}}\right>_{\pt}}\right).
\end{equation}

\begin{figure}[ht!]
  \begin{center}
    \includegraphics[width=0.45\textwidth]{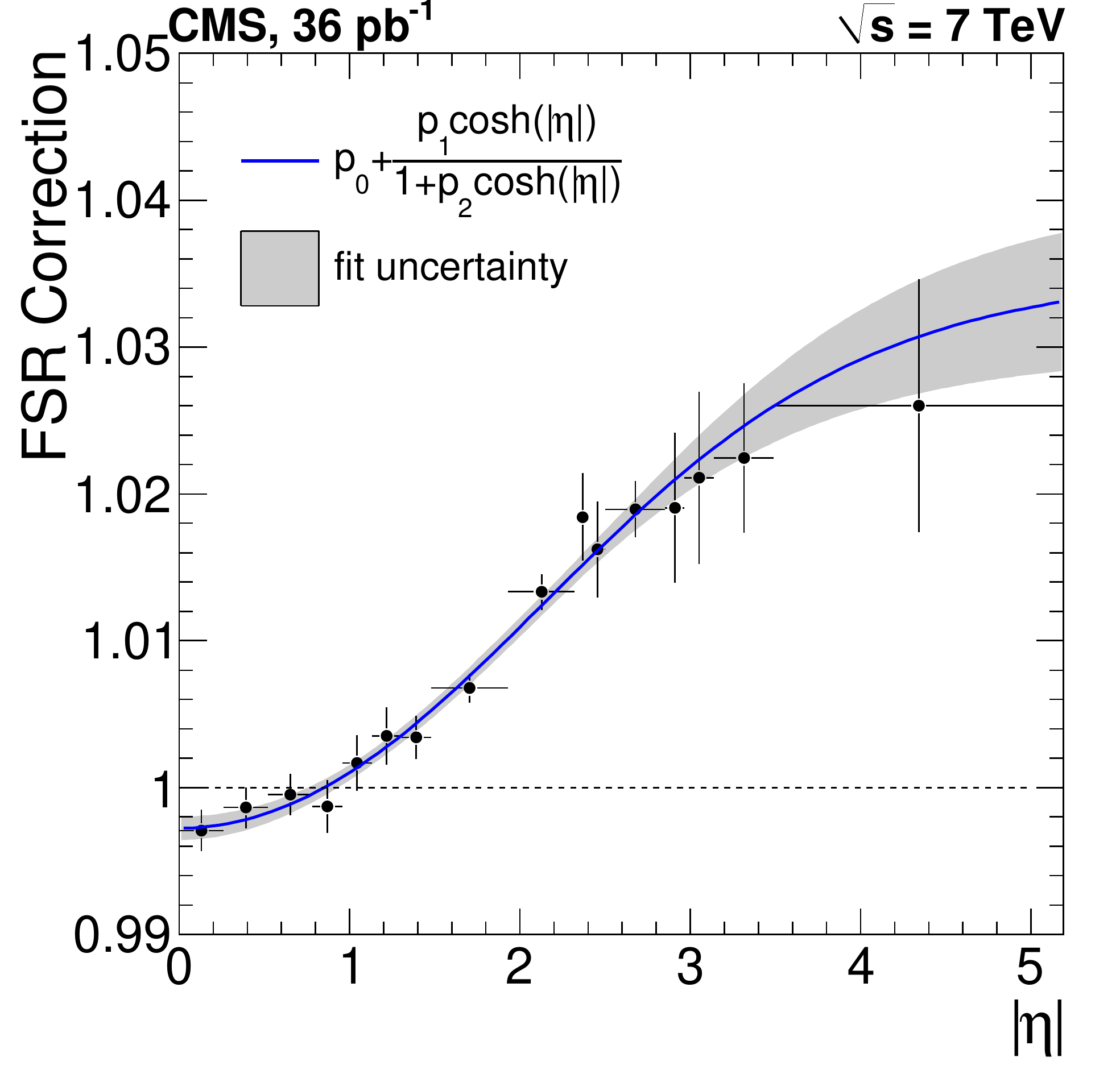}
    \includegraphics[width=0.45\textwidth]{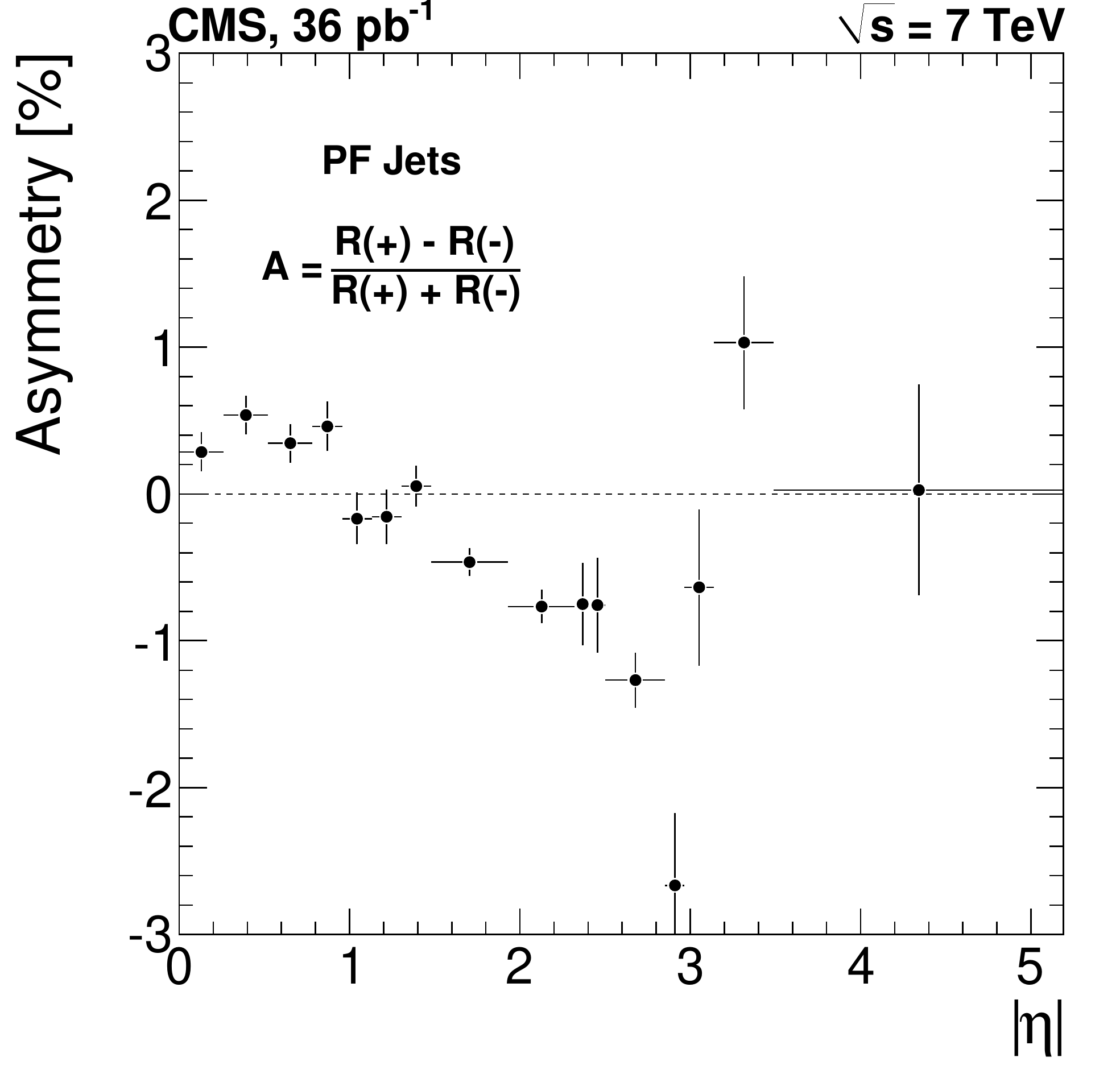}
    \caption{Left: correction $k_\text{rad}$ of the relative jet energy residual due to initial and final state radiation. Right: relative jet energy response asymmetry as a function of jet $|\eta|$, for $\alpha<0.2$.}
    \label{fig:FSR}
  \end{center}
\end{figure}

Figure~\ref{fig:FSR} (left) shows the radiation correction that needs to be applied to the measurement at the working point $\alpha<0.2$. The correction is negligible in the central region while it reaches the value of 3\% at larger rapidities.

The asymmetry of the response in $\eta$ is quantified through the variable $\mathcal{A}_R$:

\begin{equation}
  \mathcal{A}_R(|\eta|) = \frac{R(+|\eta|)-R(-|\eta|)}{R(+|\eta|)+R(-|\eta|)},
\end{equation}

where $R(+|\eta|)$ ($R(-|\eta|)$) is the relative response measured in the data at the detector part lying in the direction of the positive (negative) z-axis. Figure~\ref{fig:FSR} (right) shows the measured asymmetry. It is found to be similar for the different jet types.

\begin{figure}[ht!]
  \begin{center}
    \includegraphics[width=0.45\textwidth]{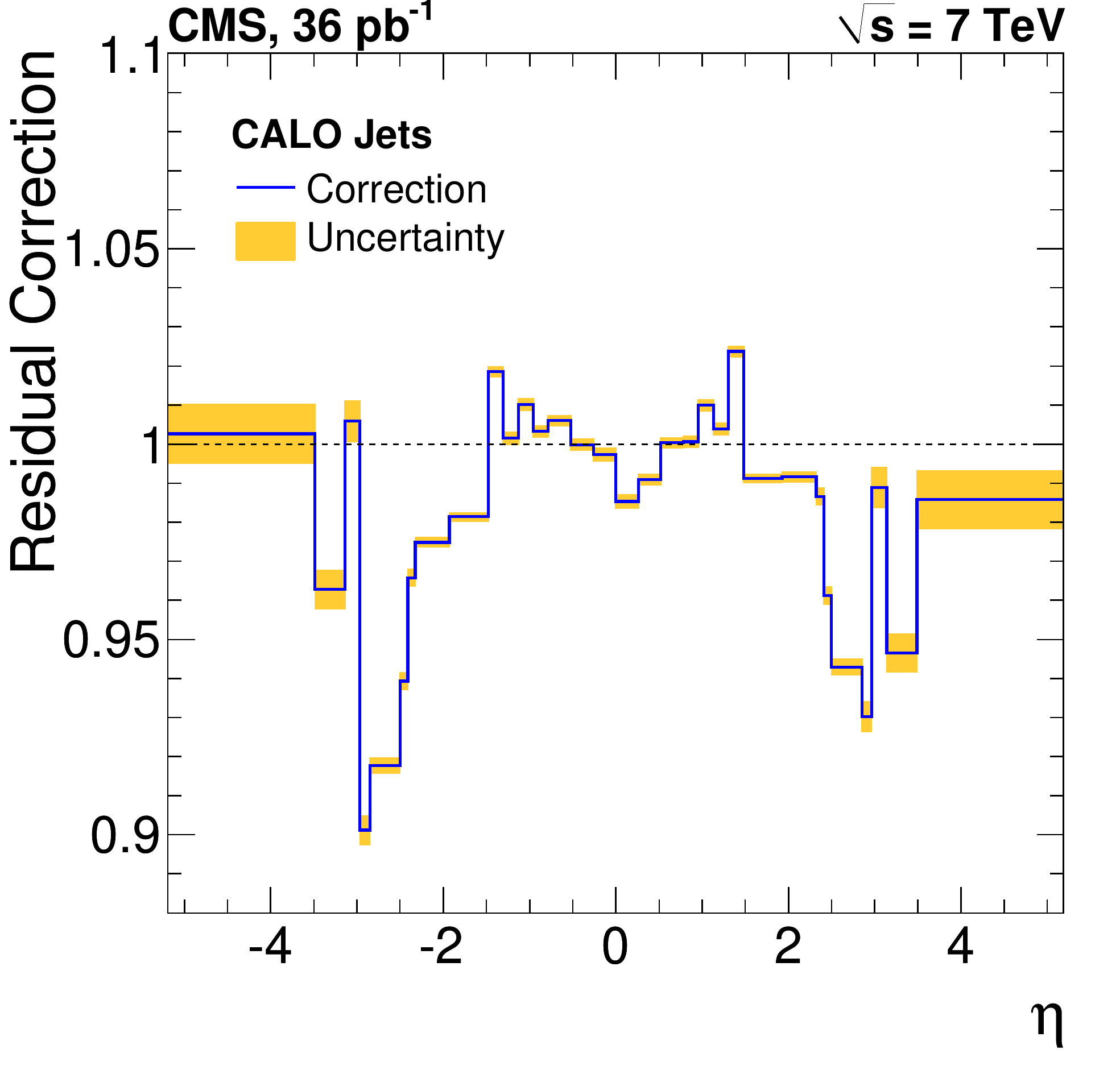}
    \includegraphics[width=0.45\textwidth]{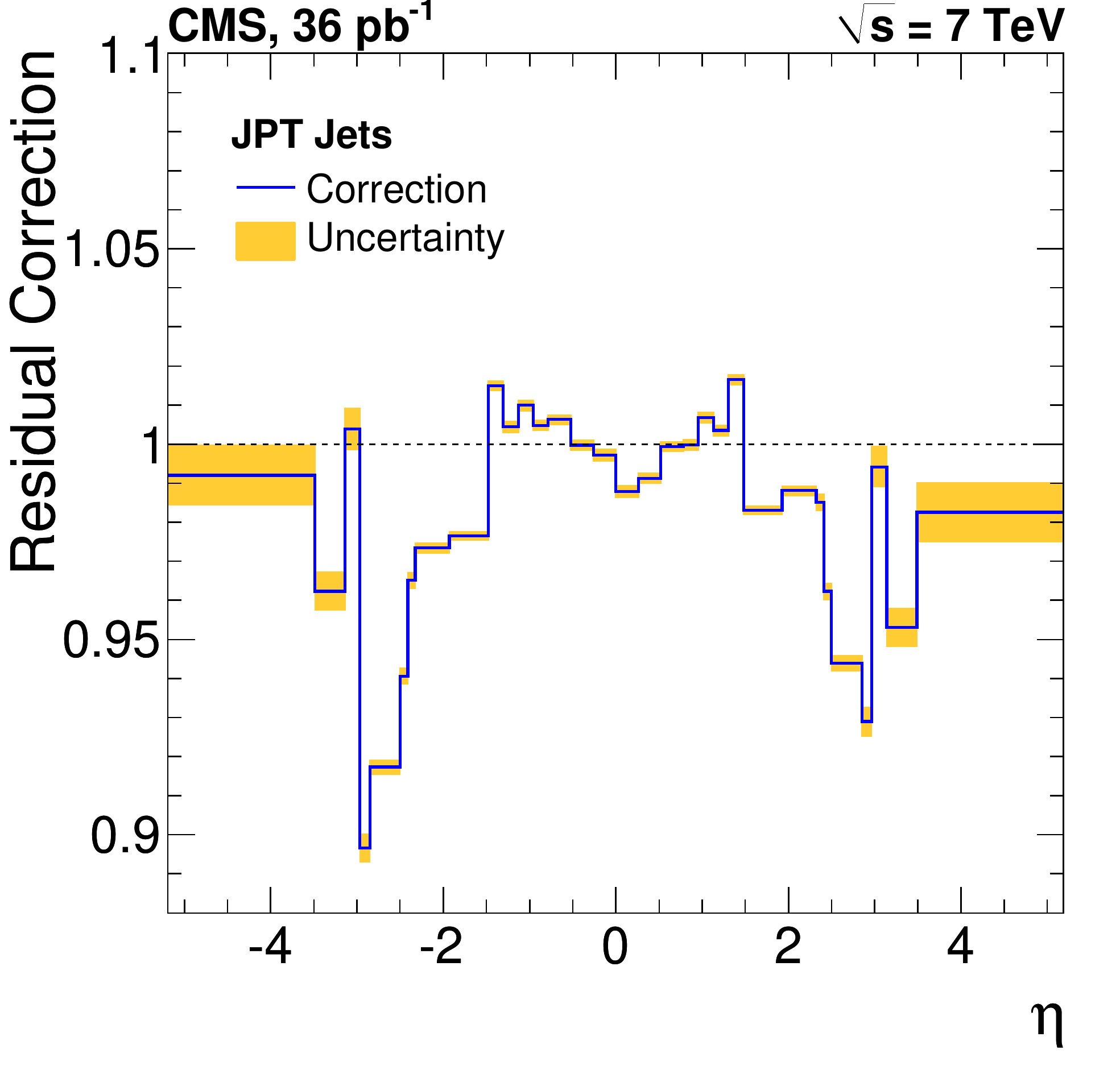}
    \includegraphics[width=0.45\textwidth]{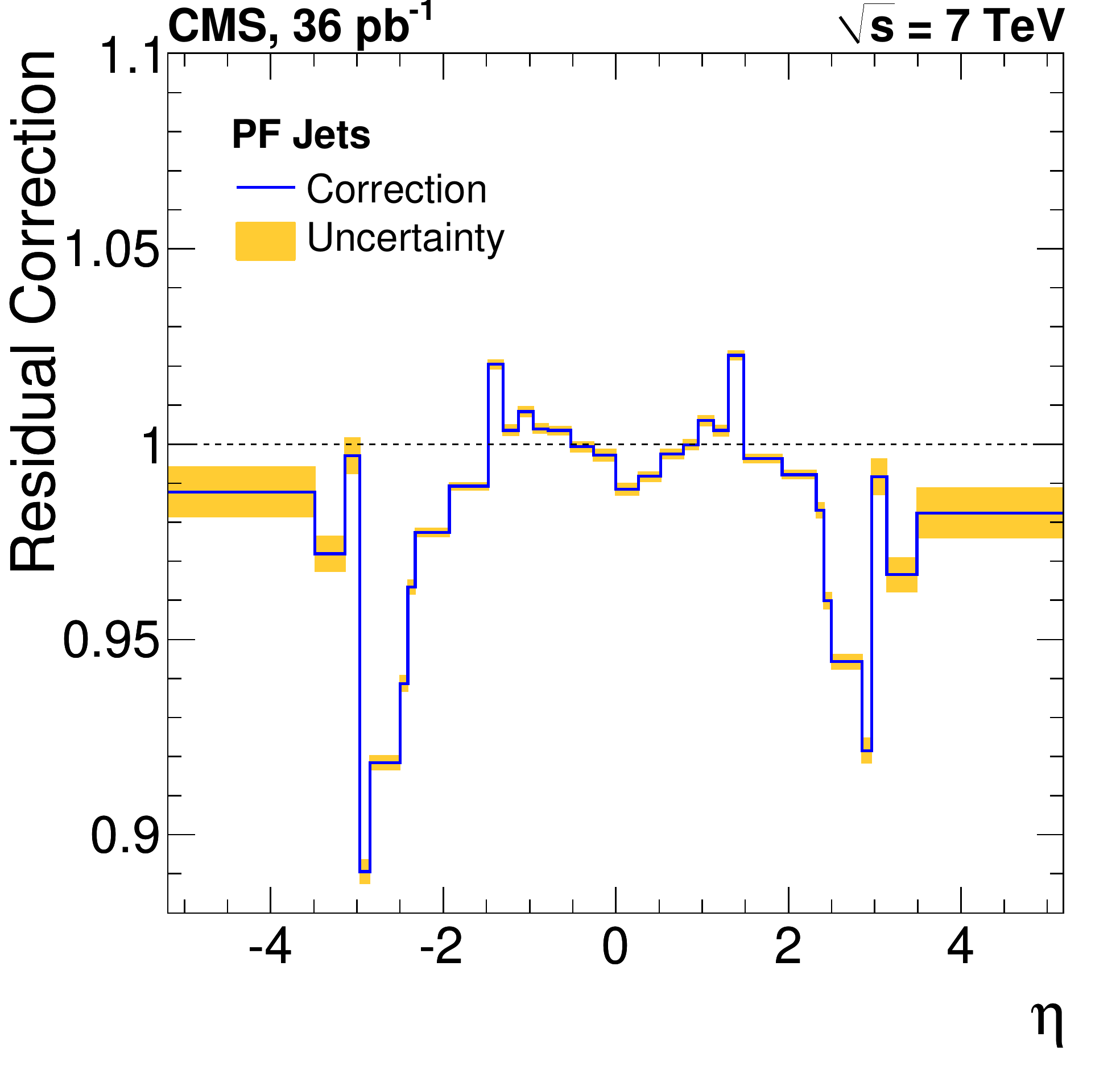}
    \caption{Relative jet energy residual correction as a function of jet $\eta$ for CALO, JPT and PF jets respectively. The band shows the uncertainty due to statistics, radiation corrections, and asymmetry in $\eta$.}
    \label{fig:final_residual}
  \end{center}
\end{figure}

Figure~\ref{fig:final_residual} shows the final residual correction, as a function of $\eta$, for all jet types. This correction is typically of the order of 2-3\%, with the exception of the region $2.5<|\eta|<3.0$ where it reaches the value of 10\%. The region where the larger discrepancy between data and MC simulations is observed (Fig.~\ref{fig:relrsp}), coincides with the border between the endcap and the forward calorimeters. It has also been observed~\cite{JME-10-008} that the single-particle response shows similar behavior in this region.

\begin{figure}[ht!]
  \begin{center}
    \includegraphics[width=0.45\textwidth]{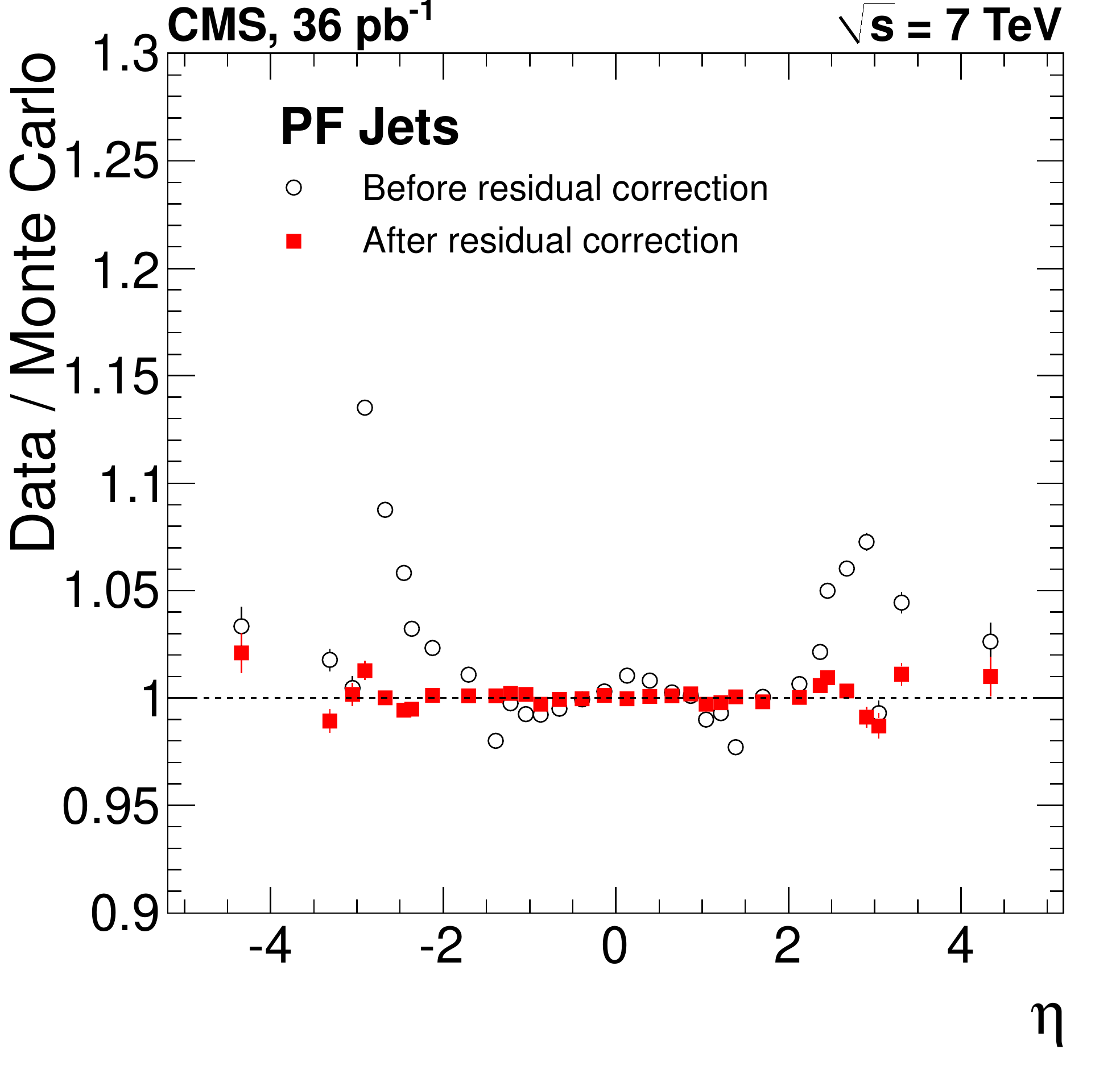}
    \caption{Relative response ratio between data and MC simulation before and after the residual correction.}
    \label{fig:L2Closure}
  \end{center}
\end{figure}

Finally, Fig.~\ref{fig:L2Closure} demonstrates that the derived residual correction establishes an almost perfect agreement between data and simulation.

\subsubsection{Uncertainty}

\begin{figure}[ht!]
  \begin{center}
    \includegraphics[width=0.45\textwidth]{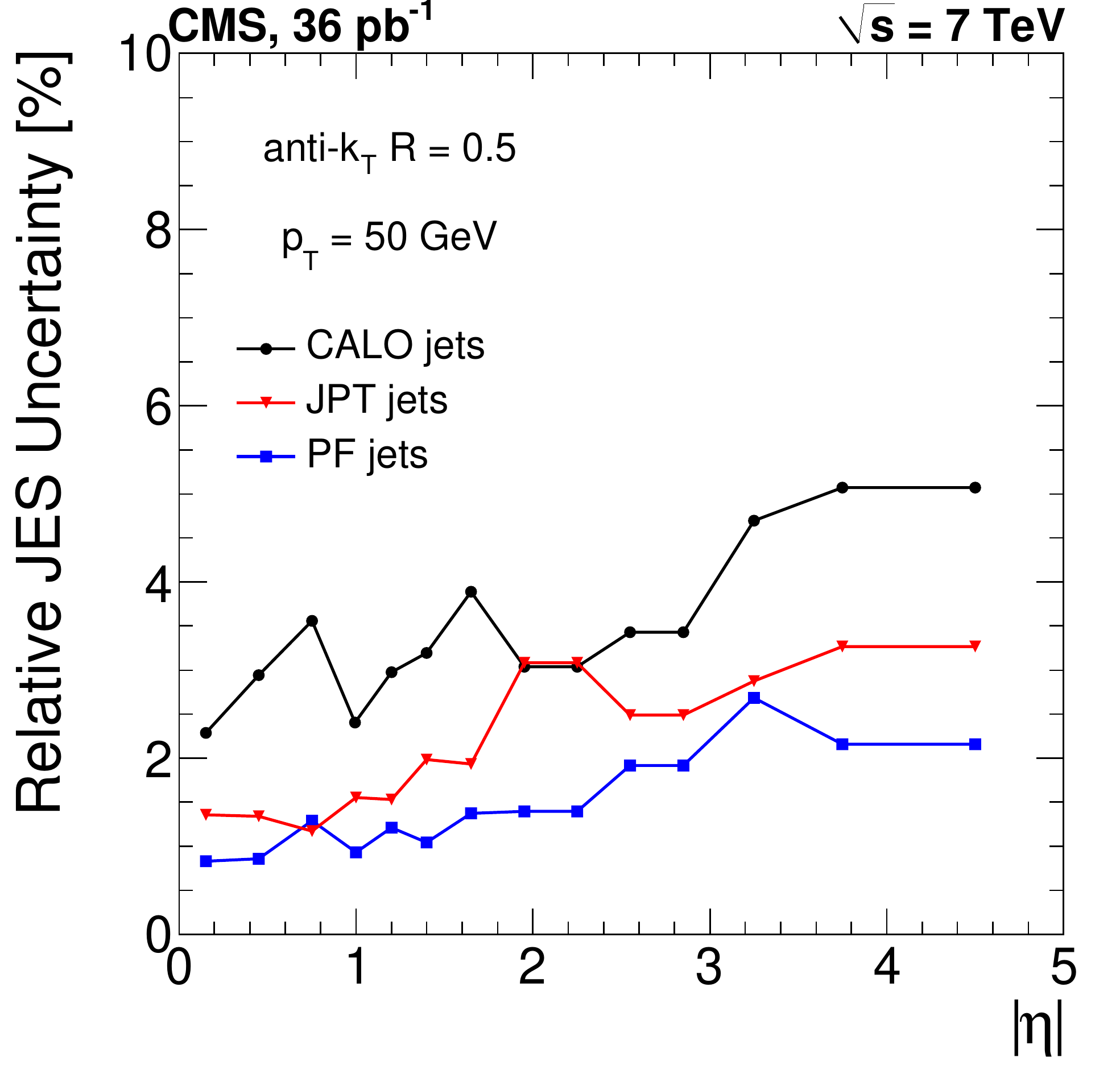}
    \includegraphics[width=0.45\textwidth]{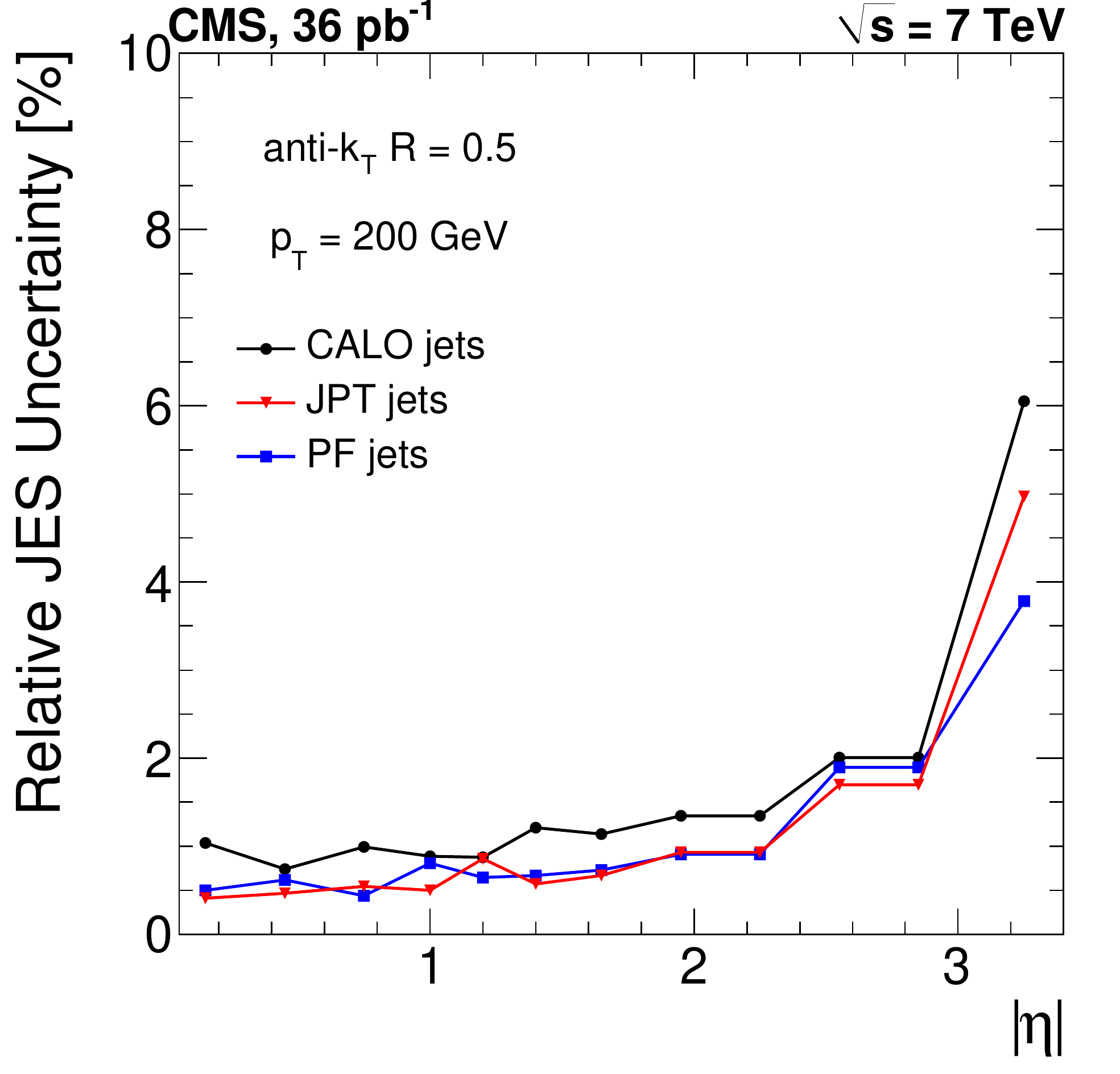}
    \caption{Relative jet energy residual correction uncertainty, as a function of $\eta$ for jet $\pt=50\GeV$ (left) and $\pt=200\GeV$ (right).}
    \label{fig:residual_unc}
  \end{center}
\end{figure}

The dominant uncertainty of the relative residual correction is due to the simulation of the jet energy resolution, which defines the magnitude of the resolution bias. The estimate of the systematic uncertainty is achieved by varying the jet \pt resolution according to the comparisons between data and MC simulations shown in Section~\ref{sec:res}. Other sources of uncertainty, such as lack of available events, radiation correction and asymmetry in $\eta$ are found to be smaller than 1\%. The total uncertainty of the relative jet energy scale is shown in Fig.~\ref{fig:residual_unc} as a function of the jet $|\eta|$ for two characteristic values of jet \pt (50\GeV, 200\GeV). The CALO jets have systematically larger uncertainty, as opposed to PF jets which have the smallest while the JPT jets uncertainty lies between the values for the other two jet types. This pattern is consistent with the behavior of the jet energy resolution. Also, it is observed that the relative scale uncertainty grows toward larger rapidities because of the larger resolution uncertainty. 

\clearpage

\subsection{Absolute Jet Energy Scale}
\subsubsection{Measurement}
\begin{figure}[ht!]
  \begin{center}
    \includegraphics[width=0.45\textwidth]{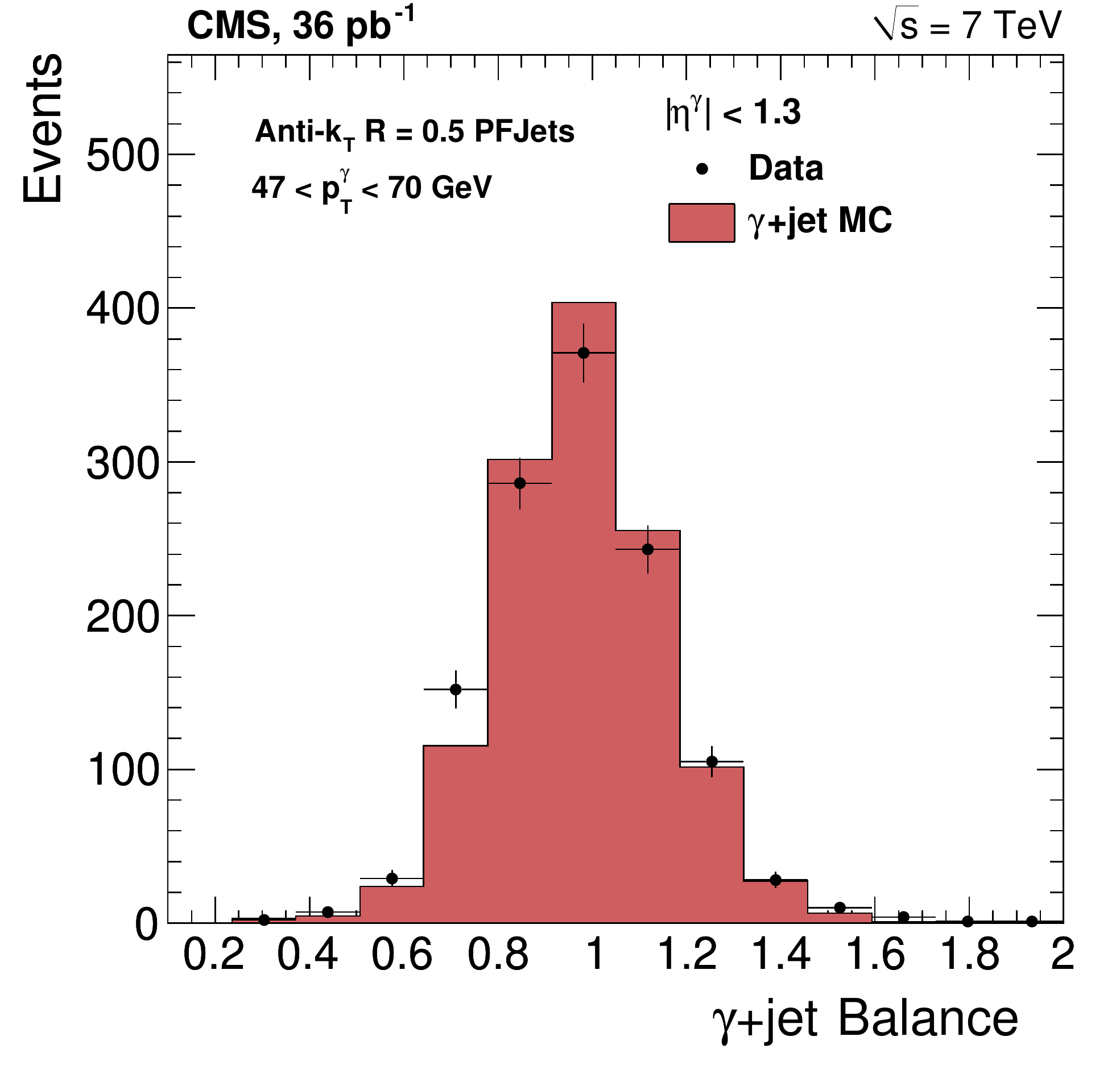}
    \includegraphics[width=0.45\textwidth]{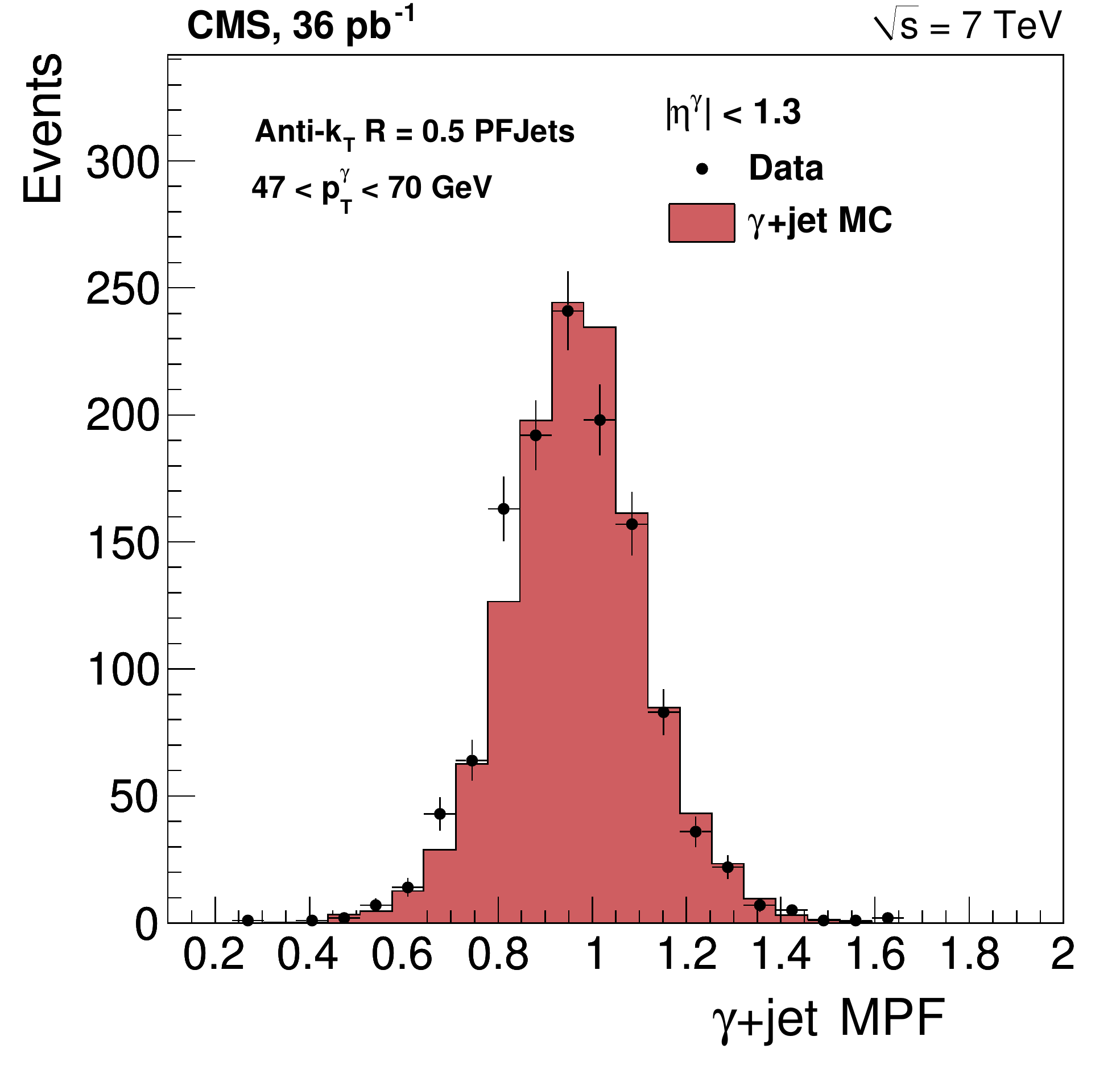}
    \caption{Example response distributions for PF jets from \pt-balancing (left) and MPF (right) in the $\gamma$+jets sample.}
    \label{fig:photonResponse}
  \end{center}
\end{figure}

The absolute jet energy response is measured in the reference region $|\eta|<1.3$ with the MPF method using $\gamma/Z$+jets events,  and the result is verified with the \pt-balancing method. The $\gamma$ or the Z are used as reference objects because their energy is accurately measured in ECAL (photon, $Z\rightarrow e^+e^-$) or in the tracker and muon detectors ($Z\rightarrow \mu^+\mu^-$). Figure~\ref{fig:photonResponse} shows example distributions of the MPF and \pt-balancing methods for PF jets in the $\gamma$+jet sample.

The actual measurement is performed only for PF jets because of the full consistency between the jet and the \vecmet reconstruction (both use the same PF candidates as inputs). The absolute energy scale of the remaining jet types (CALO, PF) is determined by comparison to the corresponding PF jet after jet-by-jet matching in the $\eta-\phi$ space.    

In the selected $\gamma$+jets sample, the presence of a barrel jet ($|\eta|<1.3$) recoiling against the photon candidate in azimuth by $\Delta \phi > 2.7$ is required. To reduce the effect of initial and final state gluon radiation that degrades the jet-photon \pt-balance, events containing additional jets with $\ptsecond > \alpha\cdot\pt^{\gamma}$ and outside the $\DeltaR=0.25$ cone around the photon direction are vetoed. The \pt-balance and MPF response measurements are performed in the same way with data and MC samples with different values of the threshold on $\alpha$ and the data/MC ratio is extrapolated to $\alpha=0$. This procedure allows the separation of the $\gamma$-jet intrinsic \pt-imbalance from the imbalance caused by hard radiation (Section~\ref{sec:radbias}). 

\begin{figure}[ht!]
  \begin{center}
    \includegraphics[width=0.45\textwidth]{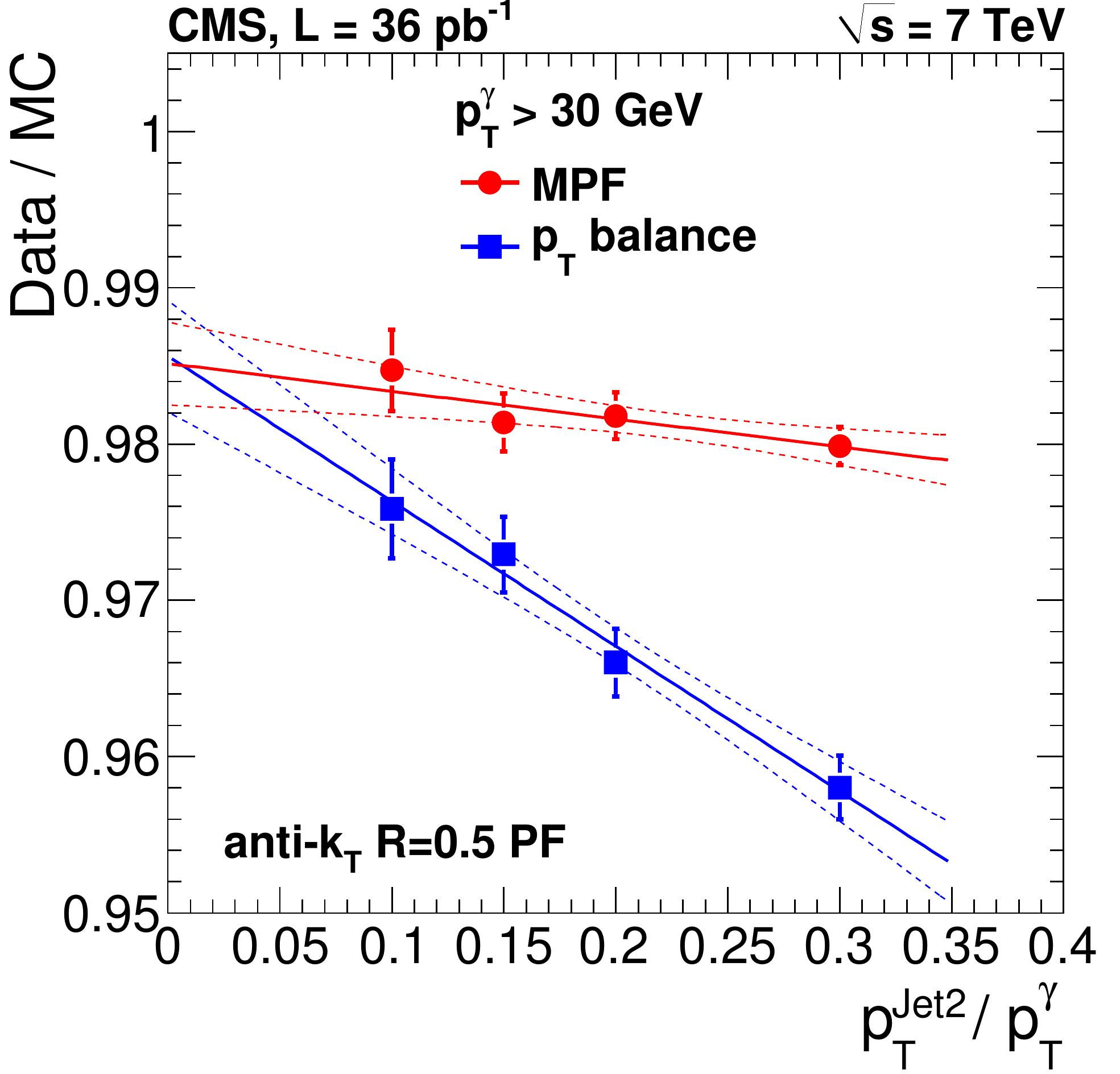}
    \includegraphics[width=0.45\textwidth]{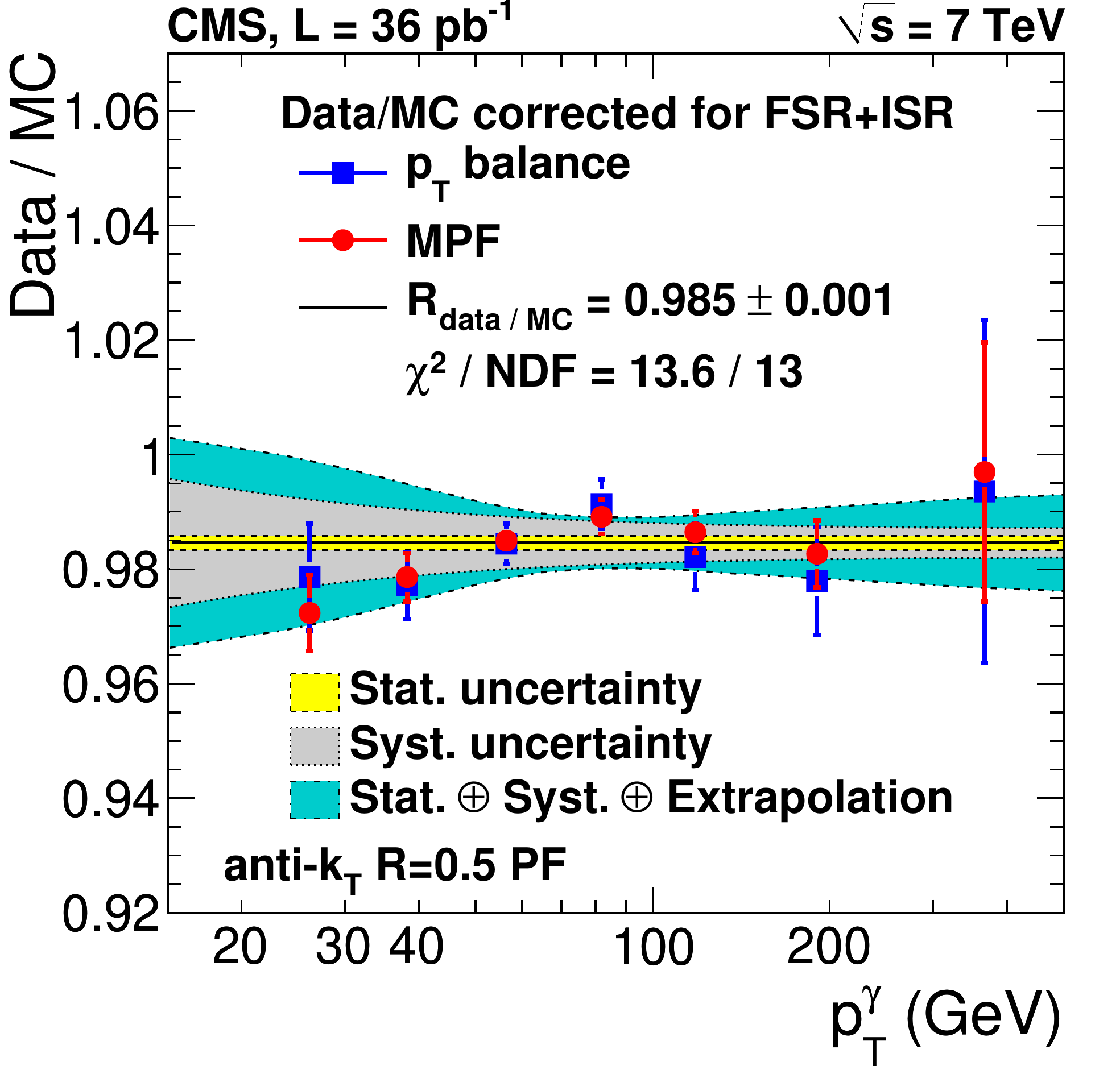}
    \caption{Left: dependence of the data/MC ratio of the jet energy response on the second jet \pt threshold. Right: data/MC ratio of the jet energy response, after extrapolation to zero second jet \pt, as a function of $\pt^\gamma$. Solid squares and solid circles correspond to the \pt-balancing and the MPF methods, respectively.}
    \label{fig:photon}
  \end{center}
\end{figure}

Figure~\ref{fig:photon} (left) shows the data/MC jet-energy-response ratio, relative to the $\gamma$ ECAL scale, extrapolated as a function of the threshold on the second jet \pt. In the \pt-balancing method, the secondary jet effect is more pronounced because it affects directly the transverse momentum balance between the photon and the leading jet. In the MPF method, the presence of the secondary jet(s) affects the measurement to a lesser extent, and mainly through the response difference between the leading jet and the secondary softer jet(s). For loose veto values, the ratio data/MC in both methods is lower than unity, while the agreement improves by tightening the veto. Figure~\ref{fig:photon} (right) shows the data/MC response ratio after the extrapolation to $\alpha=0$ for both MPF and \pt-balancing methods, as a function of $\pt^{\gamma}$. The two measurements are statistically uncorrelated to a good approximation and the two sets of points are fitted together with a constant value. The fit gives data/MC = $0.985 \pm 0.001$, relative to the $\gamma$ ECAL scale, which leads to an absolute response residual correction $C_\text{abs}=1/0.985=1.015$ (Eq.~(\ref{eq:jec_components})), constant in \pt.  

\begin{figure}[ht!]
  \begin{center}
    \includegraphics[width=0.45\textwidth]{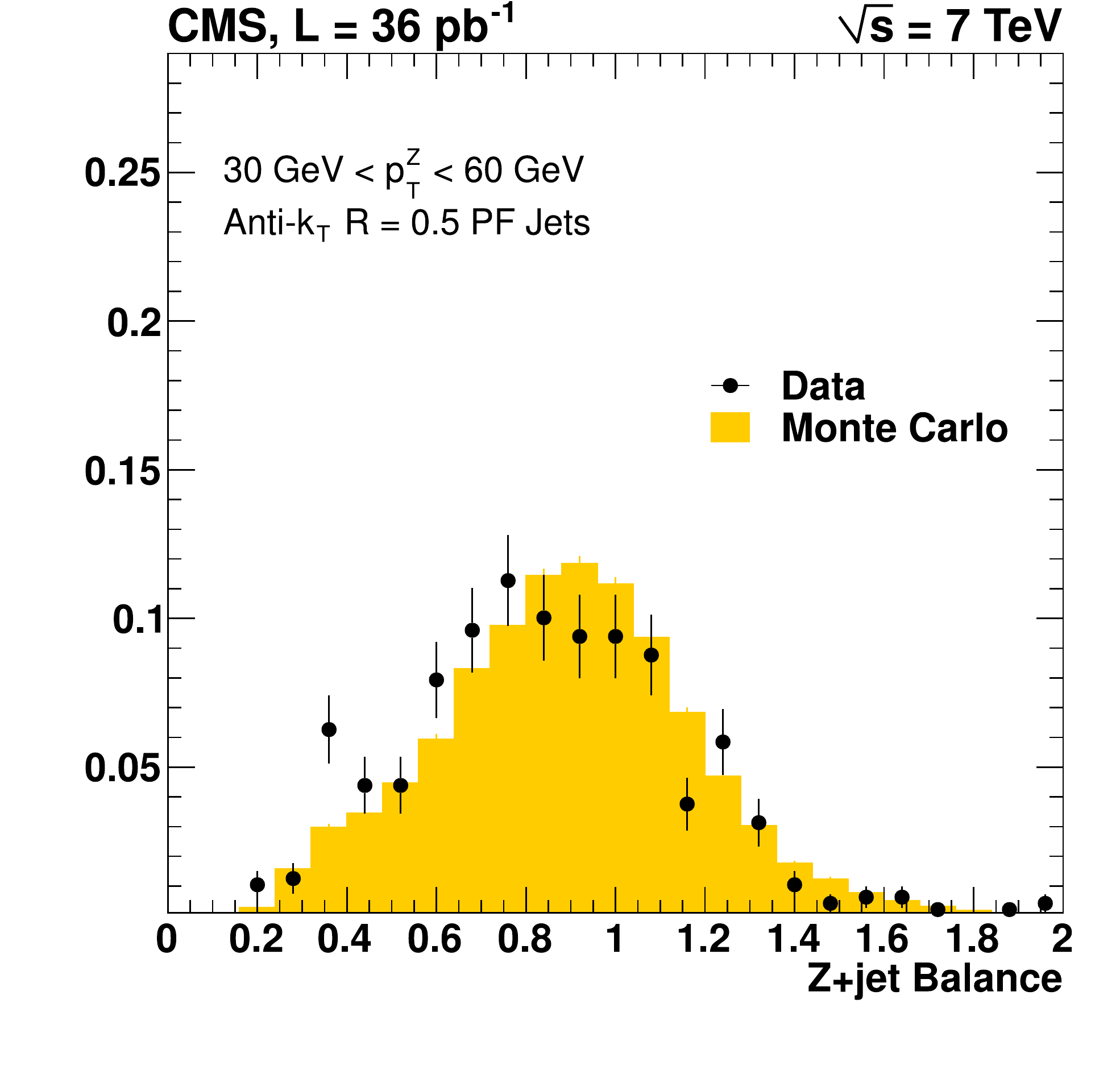}
    \includegraphics[width=0.45\textwidth]{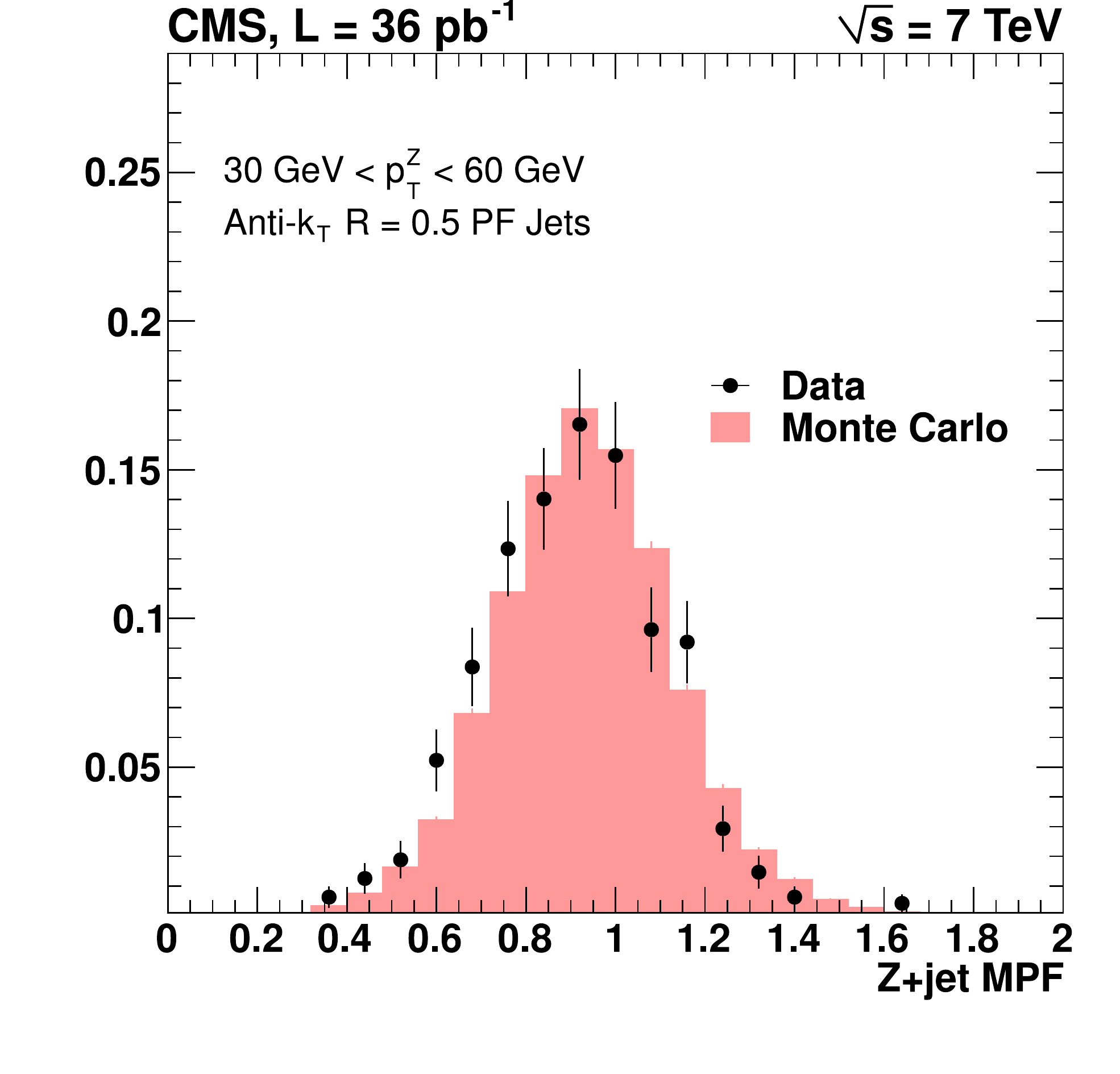}
    \caption{Left: jet energy response from $Z(\mu^+\mu^-)$+jets \pt-balancing in the bin $30<\pt^Z<60\GeV$. Right: jet energy response from $Z(\mu^+\mu^-)$+jets MPF in the bin $30<\pt^Z<60\GeV$.}
    \label{fig:ZJB}
  \end{center}
\end{figure}

In addition to the $\gamma$+jets sample, the absolute jet energy response is also measured from the Z+jets sample. Figure~\ref{fig:ZJB} shows two characteristic response distributions in the $30\GeV<\pt^Z<60\GeV$ bin, as an example, measured from the $Z(\mu^+\mu^-)$+jets sample with the \pt-balancing and the MPF methods. The Z+jets samples cover the $\pt^Z$ range from 20\GeV to 200\GeV. 

\begin{figure}[ht!]
  \begin{center}
    \includegraphics[width=0.45\textwidth]{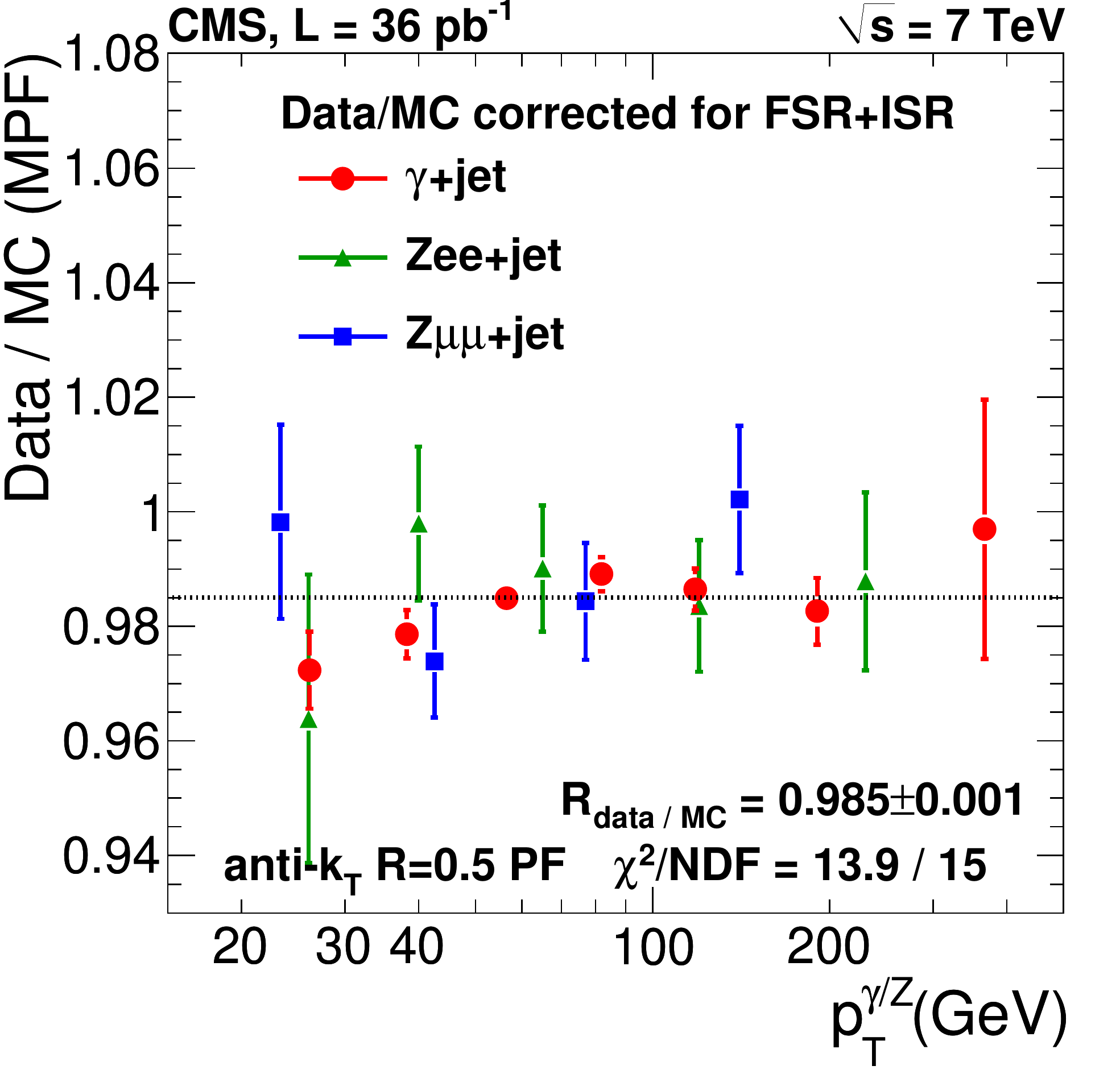}
    \caption{Ratio of data over MC for the MPF response, as a function of $\pt^{\gamma,Z}$ in the photon+jet sample (circles), $Z(e^+e^-)$+jet sample (triangles) and $Z(\mu^+\mu^-)$+jet sample (squares).}
    \label{fig:dataMCAll}
  \end{center}
\end{figure}

In order to combine the results from the photon+jet and Z+jet samples, the more precise MPF method is employed identically in all relevant samples. Figure~\ref{fig:dataMCAll} shows the data/MC ratio as a function of $\pt^{\gamma,Z}$ after correcting for the final and initial state radiation differences between data and simulation (extrapolation to $\alpha=0$). Although the size of the Z+jets data sample is smaller than the $\gamma$+jets sample, the results from all samples are in good agreement, within the corresponding statistical uncertainties.

\subsubsection{Uncertainty Sources}

The uncertainty of the absolute jet energy scale measurement has six components: uncertainty in the MPF method for PF jets, photon energy scale, MC extrapolation beyond the reach of the available dataset, offset due to noise and pile-up at low-\pt (as discussed in Section~\ref{sec:offset_unc}), MC residuals (the level of closure of the MC correction in the MC), and the jet-by-jet matching residuals for CALO and JPT jets.  

\begin{figure}[ht!]
  \begin{center}
    \includegraphics[width=0.45\textwidth]{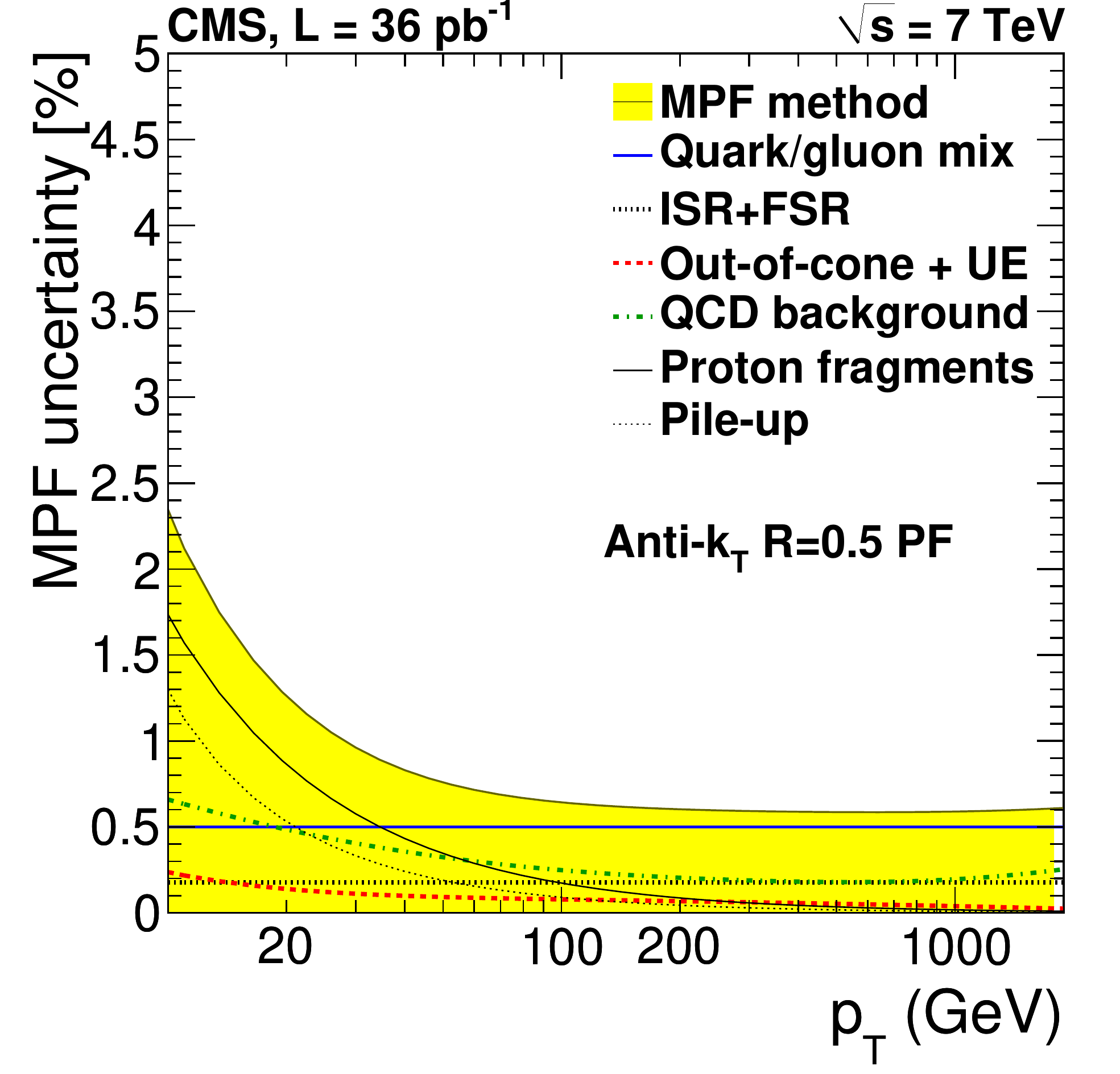}
    \caption{Jet energy scale uncertainty in the MPF method for PF jets.}
    \label{fig:jecuncert_mpf}
  \end{center}
\end{figure}

{\bf MPF Uncertainty for PF Jets.} The MPF method is affected by several small uncertainties that mainly contribute at low \pt: flavour mapping, parton-to-particle level sensitivity, QCD background, secondary jets, and proton fragments. The various contributions are shown in Fig.~\ref{fig:jecuncert_mpf}.

The flavour mapping uncertainty accounts for the response difference between jets in the quark-rich $\gamma$+jets sample used to measure the absolute jet energy scale, and those in the reference, gluon-rich QCD multijet sample. This is estimated from the average quark-gluon response difference between {\sc PYTHIA6} and {\sc Herwig++} (Fig.~\ref{fig:flavor_herwigpythia}) in the region $30-150\GeV$. The latter is chosen because it is the \pt region best constrained by the available data. For PF jets, the flavour mapping uncertainty amounts to $\sim 0.5\%$.

By definition, the MPF response refers to the parton level because the photon is perfectly balanced in the transverse plane, against the outgoing partons. However, the default jet energy response refers to the particle level, which includes the UE and the hadronization effects. The parton-to-particle level response interpretation therefore is sensitive to the UE and the out-of-cone showering (OOC). The corresponding uncertainty is estimated from the simulation by using jets reconstructed with larger size parameter ($R=0.7$, more sensitive to UE and OOC) and comparing the extrapolation to the zero secondary jet activity with respect to the nominal size parameter ($R=0.5$). The resulting uncertainty has a weak \pt-dependence and is smaller than $0.2\%$. 

The dominant background for $\gamma$+jets events is the QCD dijet production where one leading jet fragments into a hard isolated $\pi^0\to\gamma+\gamma$. Such events can alter the measured \pt-balance because the leading neutral $\pi^0$ carries only a fraction of the initial parton energy. The QCD background uncertainty is estimated by repeating the measurement, using a loose and a tight photon identification, and is found to be negligible compared to the current statistical precision.

The MPF response at low \pt is sensitive to the undetected energy that leaks outside the forward calorimeter acceptance at $|\eta|>5$. This results in an underestimation of the MPF response, compared to the true response. The uncertainty due to the undetected energy is taken from the simulation and is estimated to be $50\%$ of the difference between the MPF response and the true response. 

The secondary jet activity is found to be significantly different between data and MC, and it is corrected by extrapolating the data/MC ratio for the MPF and \pt-balance methods to zero secondary jet activity. The related uncertainty is estimated as half of the radiation bias correction applied to the MPF method.

{\bf Photon Energy Scale Uncertainty.} The MPF and \pt-balancing methods are directly sensitive to the uncertainty in the energy of the $\gamma$ used as a reference object. The $\gamma$ energy scale uncertainty is estimated to be $\sim 1\%$ based on studies presented elsewhere~\cite{EGM-10-003}.

{\bf Monte Carlo Extrapolation.} The in situ measurement of the absolute jet energy scale is feasible only in the \pt range where $\gamma$+jets data are available. For the current dataset this range extends to around $300\GeV$. However, the jet \pt range probed in the entire dataset is generally more than three times higher than in the $\gamma$+jets sample. In QCD dijet events, jets as high as $\pt = 1\TeV$ are observed. Because of the absence of data for direct response measurement at high \pt, the calibration relies on the simulation. Based on the data vs. MC comparison in the region of available $\gamma$+jets data, conclusions can be drawn for the extrapolation of the jet energy correction at the highest jet \pt.

\begin{figure}[ht!]
  \begin{center}
    \includegraphics[width=0.45\textwidth]{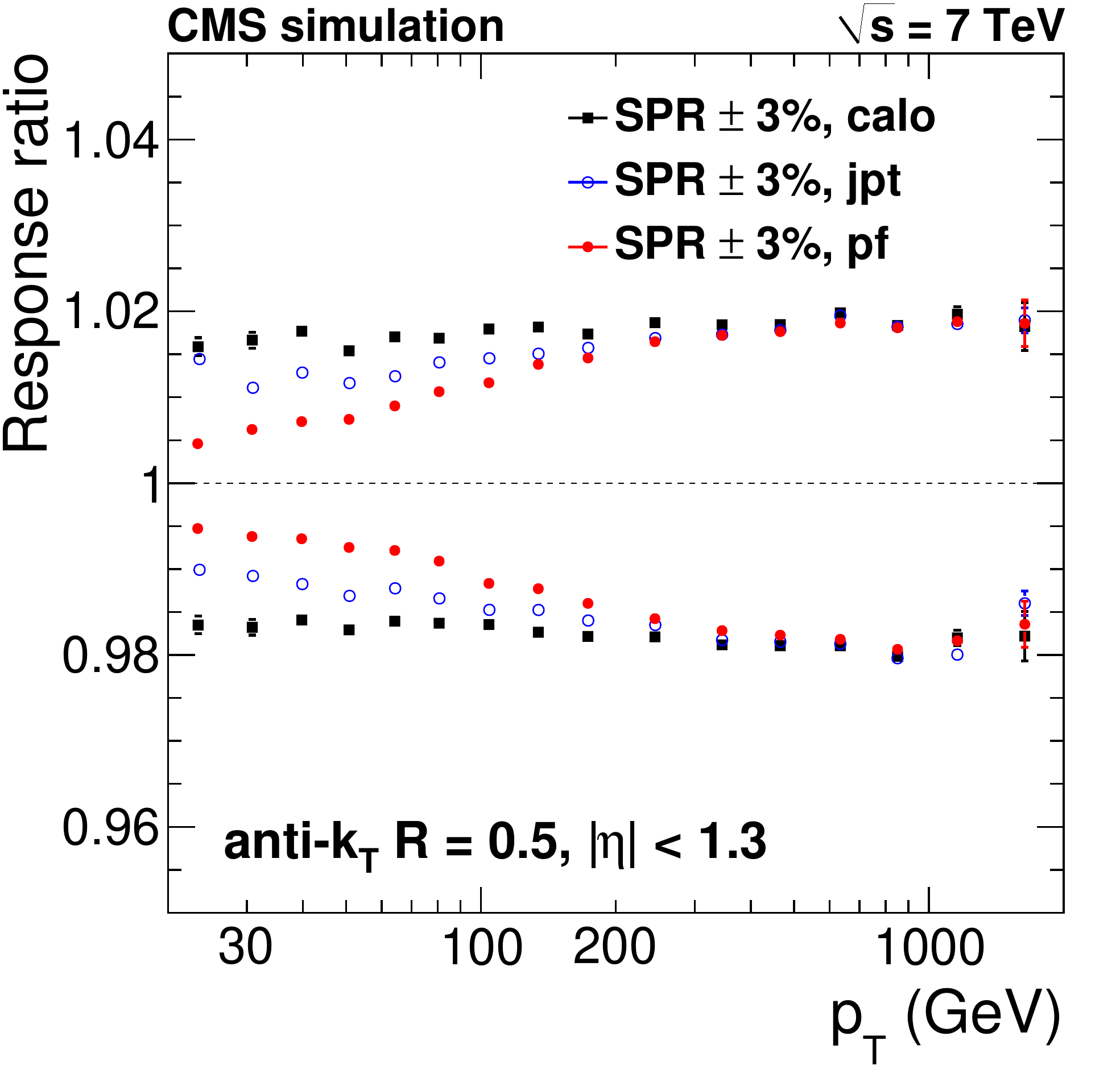} 
    \includegraphics[width=0.45\textwidth]{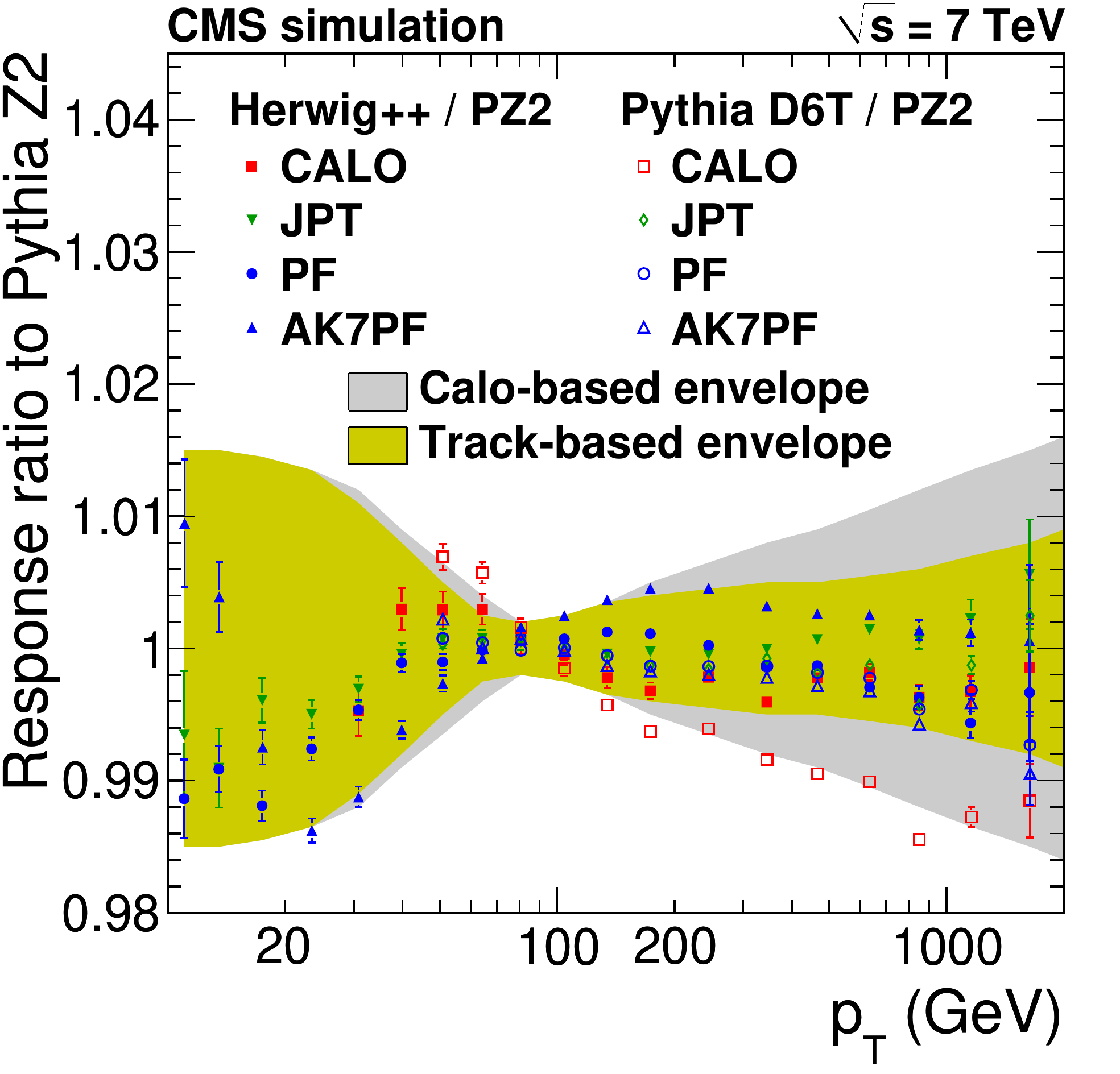}
    \caption{Left: sensitivity of the jet energy response in $|\eta|<1.3$ to the single-particle response (SPR) uncertainty. Right: dependence of the jet energy response on the fragmentation model. Here AK7PF stands for PF jets reconstructed with the anti-$k_T$ algorithm with size parameter $R=0.7$.}
    \label{fig:FragSpr}
  \end{center}
\end{figure}

The simulation uncertainty for the high-\pt jets arises from two main sources: the single-particle response (SPR) and the fragmentation modeling. The former is measured directly in data by using isolated tracks and comparing the energy deposited in the calorimeters with the momentum measured by the tracker. The currently available measurement~\cite{JME-10-008} indicates that the data/MC disagreement is less than 3\%. The SPR uncertainty is translated to a jet energy response uncertainty by modifying accordingly the simulation. Figure~\ref{fig:FragSpr} (left) shows the impact of the SPR uncertainty on the response of the different jet types, in the region $|\eta|<1.3$. For CALO jets, the induced uncertainty is roughly constant vs. \pt and approximately equal to 2\%. The track-based algorithms are less affected at low-\pt by the SPR uncertainty because the energy is primarily measured by the tracker. However, as the jet \pt increases and the track momentum measurement becomes less precise compared to the calorimetric measurement, the track-based jet types behave like CALO jets. The transition is smooth and is completed at jet $\pt\sim 300\GeV$. 

The other source of systematic uncertainty is related to the fragmentation properties, which include the parton shower and the hadronization simulation. Since jets are composite objects, realized as ``sprays" of highly collimated particles, and the calorimeter response is non-linear, the jet energy response depends on the number and the spectrum of the particles it consists of. The sensitivity to the fragmentation modelling is studied by generating QCD events from various MC generators which are then processed by the full simulation of the CMS detector. The MC generators employed are: {\sc PYTHIA6} (tunes D6T~\cite{D6T} and Z2) and {\sc Herwig++}\cite{HERWIG}. Figure~\ref{fig:FragSpr} (right) shows the response ratio of the various models with respect to {\sc PYTHIA6},with Z2 tune, which is the default. The differences between the models are negligible at $\pt\sim 80\GeV$, while they grow up to 1.5\% at low and high jet \pt. 

\begin{figure}[ht!]
  \begin{center}
    \includegraphics[width=0.45\textwidth]{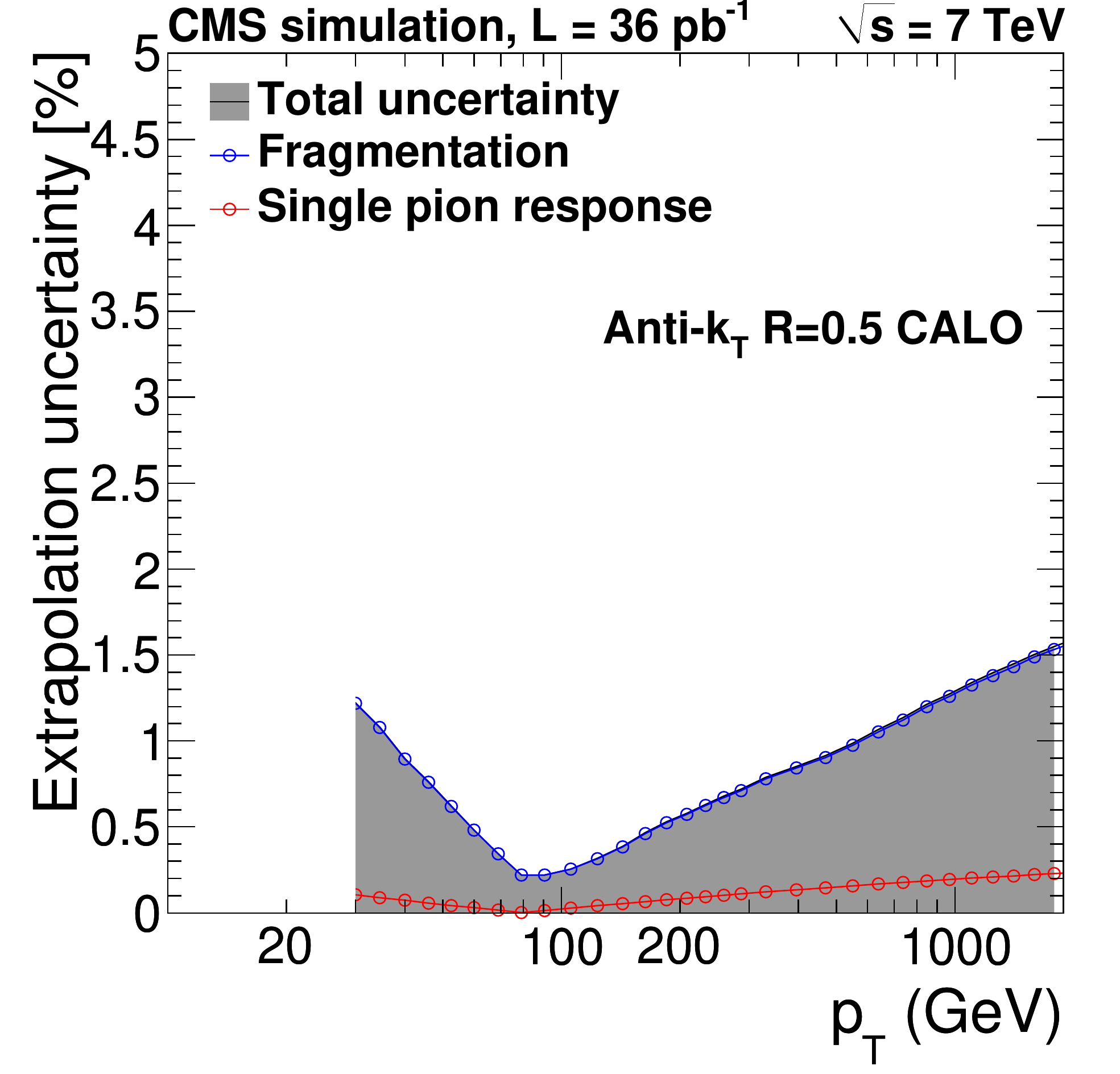}
    \includegraphics[width=0.45\textwidth]{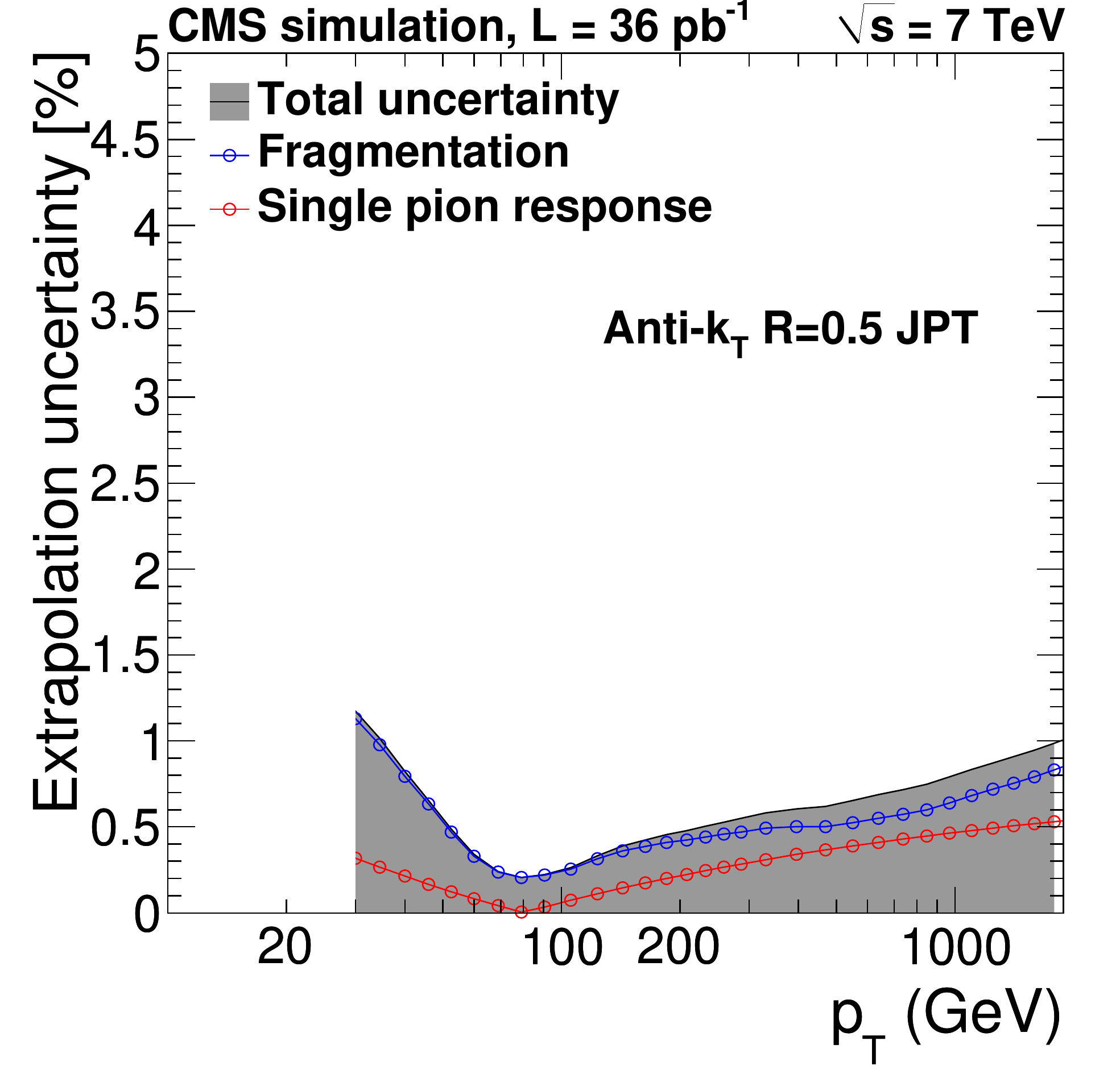}
    \includegraphics[width=0.45\textwidth]{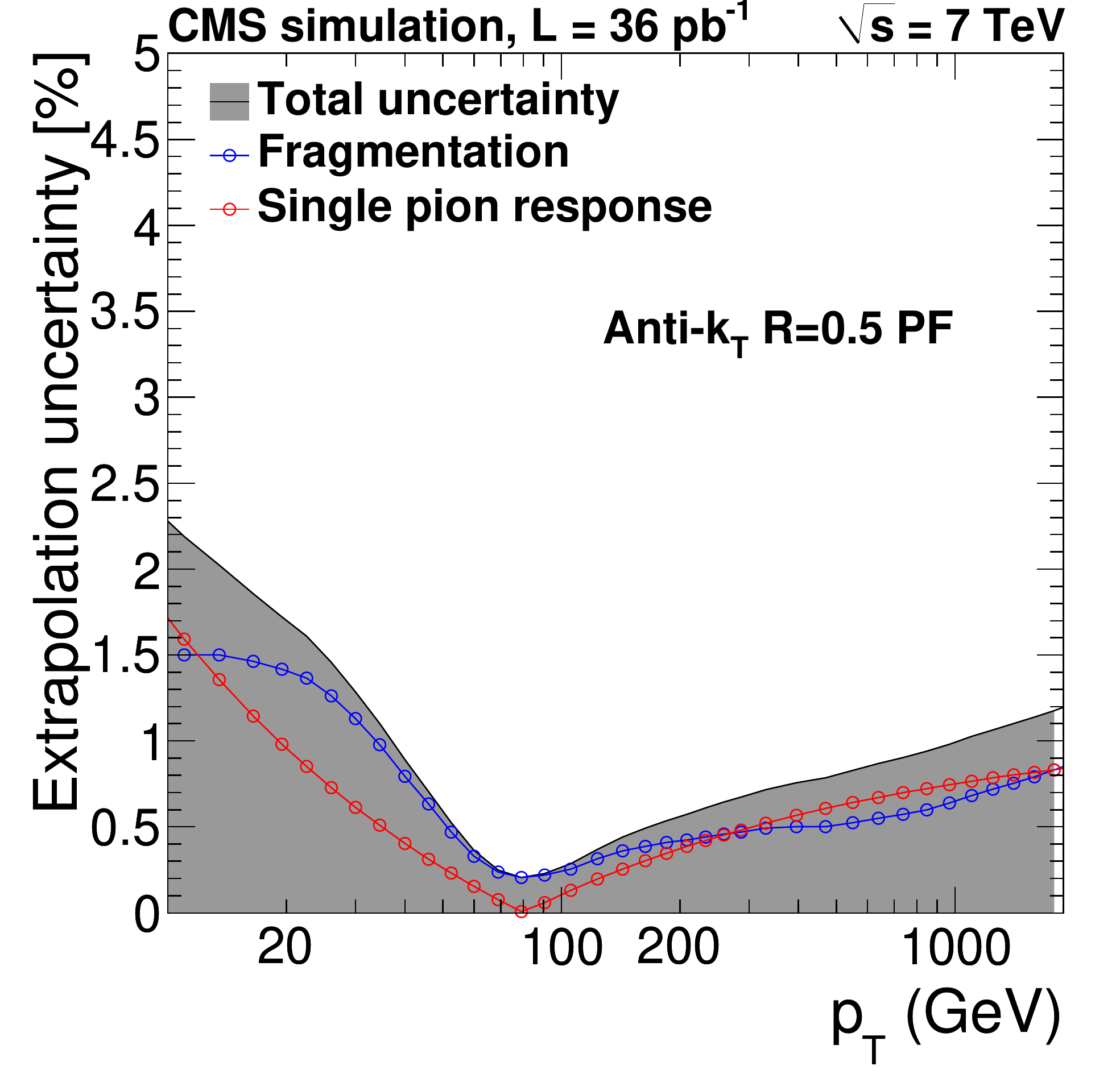}
    \caption{Uncertainty of the absolute jet energy response in the region $|\eta|<1.3$ related to the MC extrapolation for CALO, JPT and PF jets respectively. }
    \label{fig:highptUnc}
  \end{center}
\end{figure}

The combined MC uncertainty of the absolute jet energy response due to SPR and fragmentation is shown in Fig.~\ref{fig:highptUnc}.

The particle-flow algorithm reconstructs individual particles, prior to jet clustering. This allows the detailed study of the PF jet composition in terms of charged hadrons, photons and neutral hadrons. In particular, the jet energy response is closely related to the energy fraction carried by the three major composition species. The purpose of this study is to demonstrate that the MC simulation is able to describe accurately the PF jet composition observed in data and therefore can be trusted to predict the PF jet response in the kinematic regions where the in situ measurement is not possible.

\begin{figure}[ht!]
  \begin{center}
    \includegraphics[width=0.45\textwidth]{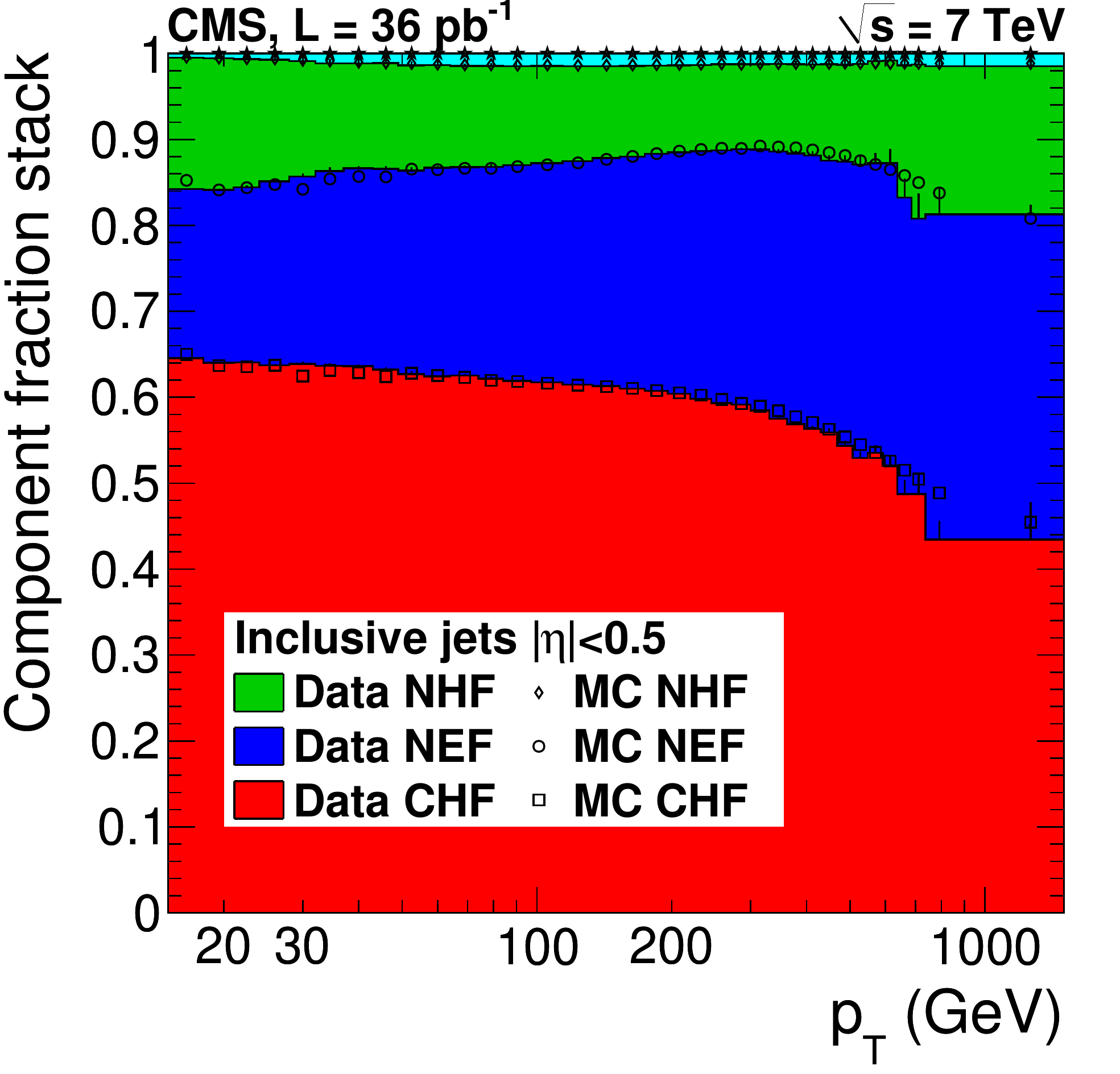}
    \includegraphics[width=0.45\textwidth]{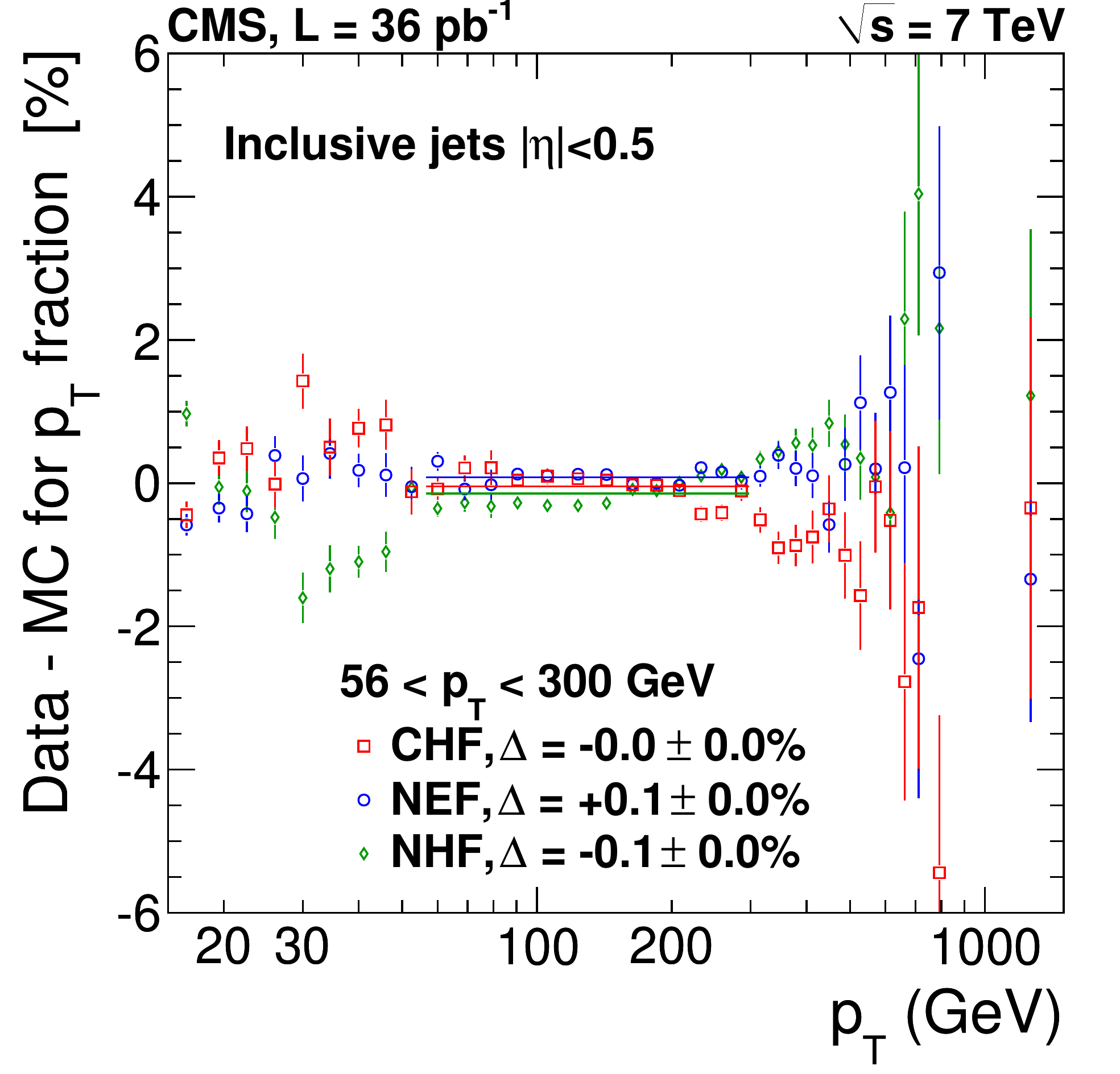}
    \caption{PF jet composition. Left: energy fraction carried by charged hadrons (CHF), photons (NEF), and neutral hadrons (NHF) as a function of jet \pt in the region $|\eta|<0.5$. The filled histograms and the markers represent the data and the simulation respectively. Right: \pt fraction difference between data and MC.}
    \label{fig:composition}
  \end{center}
\end{figure}

Figure~\ref{fig:composition} (left) shows the fraction of jet energy carried by the various particle types. Charged hadrons, photons, and neutral hadrons carry $\sim 65\%,\,20\%$, and $15\%$ of the jet energy respectively at low jet \pt, as expected from the general properties of the fragmentation process. As the jet \pt increases, charged hadrons become more energetic and more collimated, while the tracking efficiency and momentum resolution worsen. This increases the probability for a charged hadron to leave detectable energy only in the calorimeters and to be classified either as a neutral electromagnetic object (photon) or as a neutral hadron. Therefore, for higher jet \pt, the energy fraction carried by photons and neutral hadrons is increased. The excellent agreement between data and simulation quantified in Fig.~\ref{fig:composition} (right) proves that the simulation can be safely trusted to predict the absolute jet energy response. 

\begin{figure}[ht!]
  \begin{center}
    \includegraphics[width=0.45\textwidth]{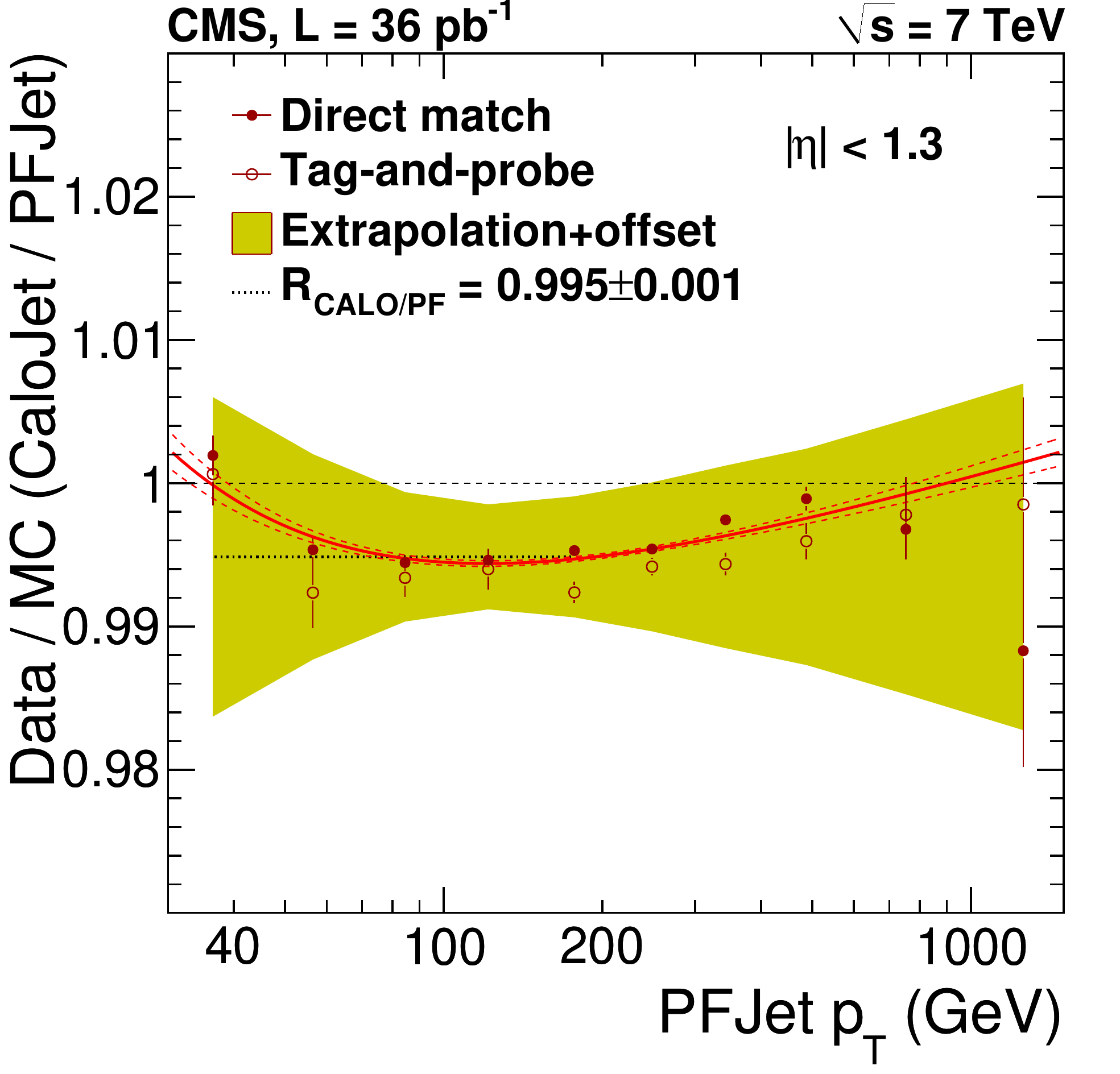}
    \includegraphics[width=0.45\textwidth]{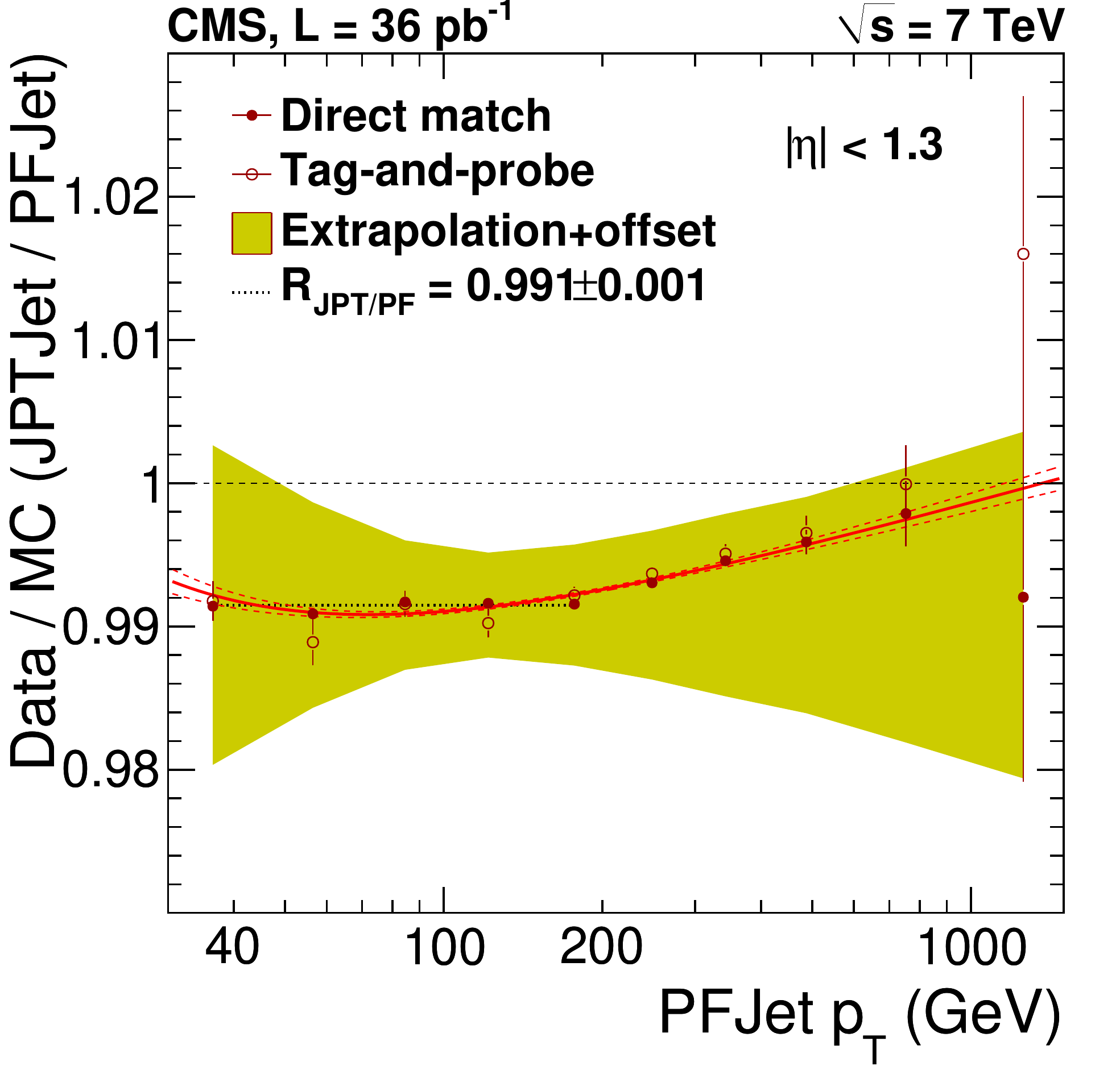}
    \caption{Left: CALO vs. PF jet \pt response ratio between data and MC simulation. Right: JPT vs. PF jet \pt response ratio between data and MC simulation. The solid circles correspond to direct matching in the $\eta-\phi$ space and the open circles correspond to a tag (PF jet) and probe (CALO/JPT jet) method.}
    \label{fig:caloVsPF}
  \end{center}
\end{figure}

{\bf Jet-by-Jet Matching.} Once the jet energy scale is established for PF jets, the estimated uncertainties are transfered to the other jet types. This is done by direct jet-by-jet comparison between different jet types in the QCD dijet sample. The PF and CALO (JPT) jets are spatially matched in the $\eta,\,\phi$ space by requiring $\DeltaR<0.25$. For the matched jet pairs the relative response of CALO (JPT) jets $\pt^{CALO}/\pt^{PF}$ ($\pt^{JPT}/\pt^{PF}$) is measured as a function of $\pt^{PF}$ (the study is described in detail in Ref.~\cite{JME-10-003}). A cross-check of the direct jet matching is done with a tag-and-probe method in dijet events, with the PF jet being the tag object and the CALO/JPT jets being the probe objects. The results are summarized in Fig.~\ref{fig:caloVsPF} where the response ratio data/MC of the CALO and JPT response relative to the PF jets is shown. The observed disagreement is at the level of $0.5\%$, indicating that the precision of the CALO and JPT calibration is comparable to that of the PF jets. The observed $0.5\%$ level of data/MC disagreement is taken into account as an additional systematic uncertainty for CALO and JPT jets.

\subsubsection{Uncertainty}

As described in the previous sections, the absolute jet energy response is measured in situ for PF jets with the MPF method in $\gamma$/Z+jets events. The systematic uncertainties related to the measurement itself are summarized in Fig.~\ref{fig:jecuncert_mpf}. The estimation of the systematic uncertainty in the kinematic region beyond the reach of the $\gamma$+jets sample is based on the simulation and its sensitivity to the single-particle response and the fragmentation models. In addition, the uncertainty on the $\gamma$ energy scale needs to be taken into account since the jet energy response is measured relative to the $\gamma$ scale. The direct jet-by-jet spatial matching, allows the transfer of the PF jet-energy- scale uncertainty to the other jet types (CALO, JPT). Finally, a flavour uncertainty is assigned from the response differences between the quark and gluon originated jets. These are taken from Fig.~\ref{fig:flavor_herwigpythia} and cover the absolute scale uncertainty in physics samples with a different flavour mixture than the reference QCD multijet sample. 

\begin{figure}[ht!]
  \begin{center}
    \includegraphics[width=0.45\textwidth]{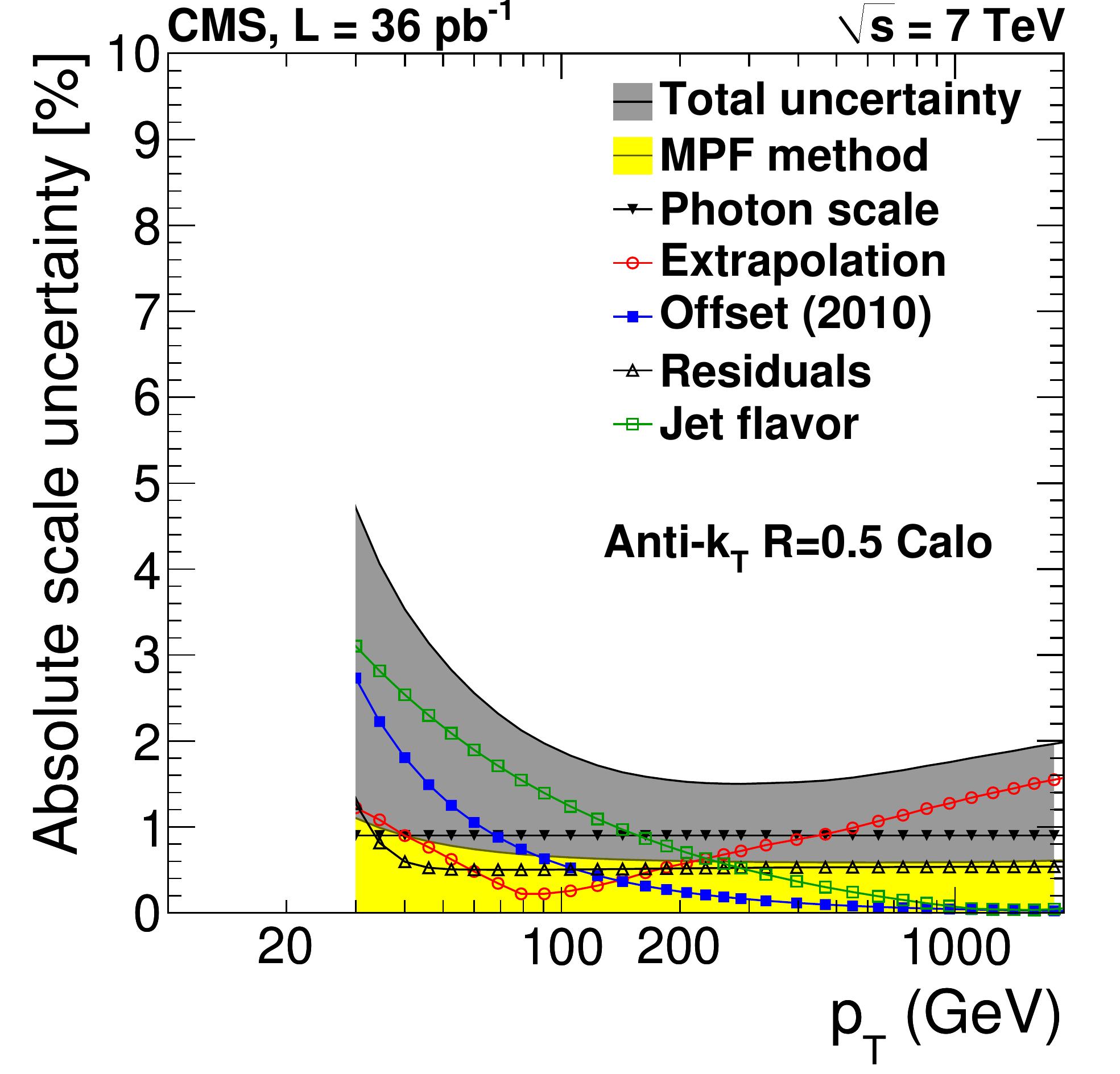}
    \includegraphics[width=0.45\textwidth]{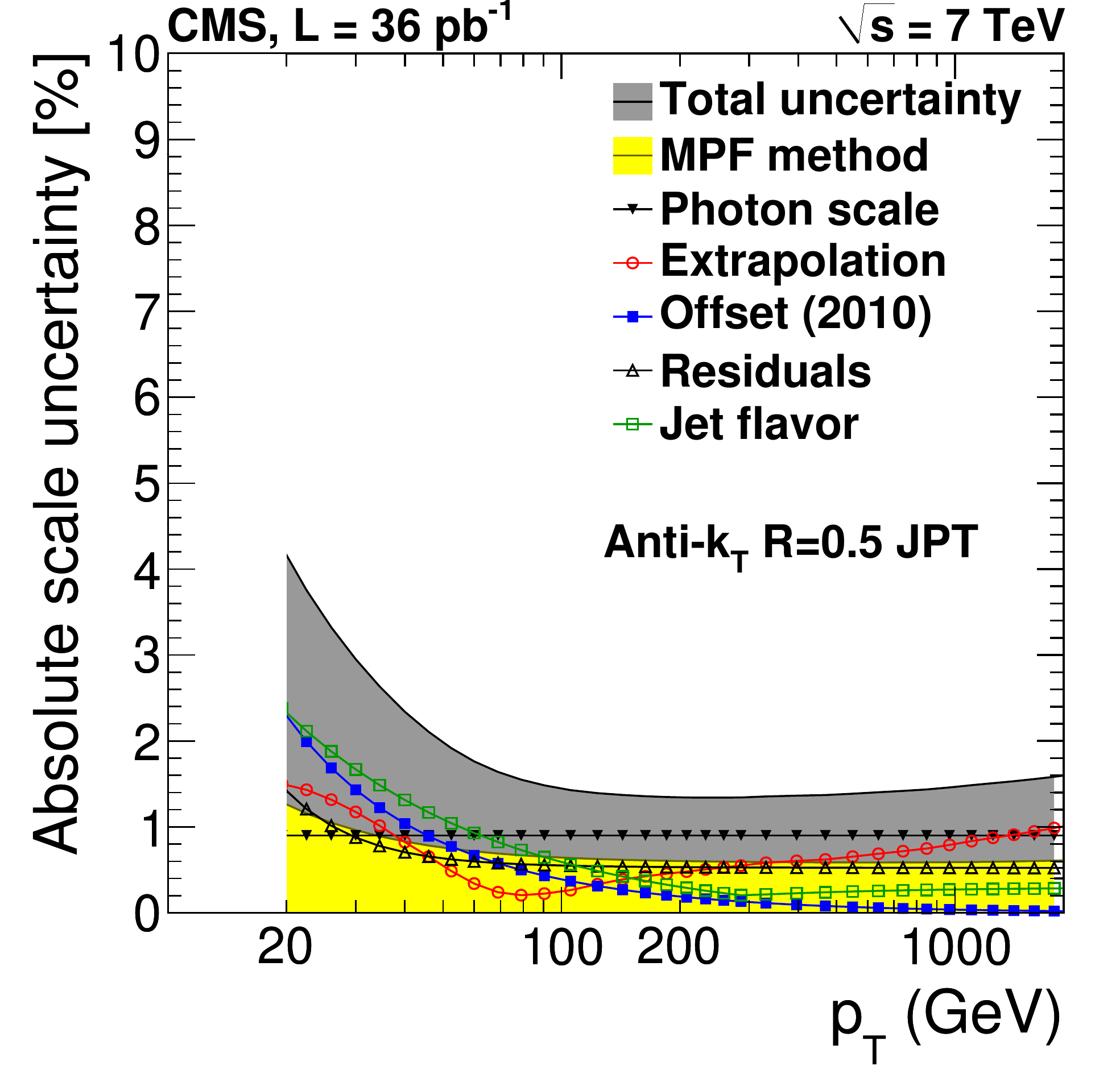}
    \includegraphics[width=0.45\textwidth]{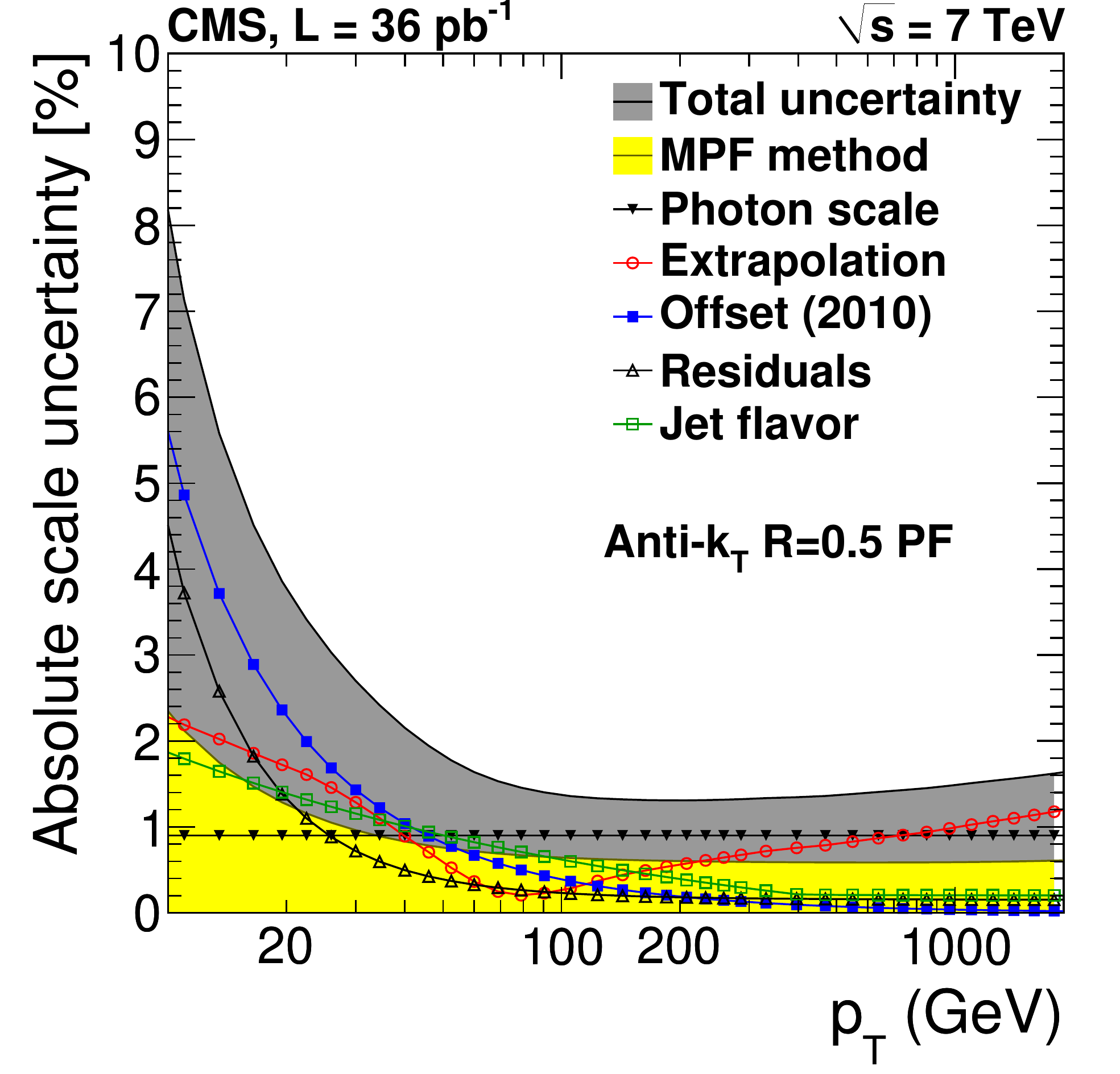}
    \caption{Absolute jet energy scale uncertainty as a function of jet \pt for CALO, JPT and PF jets respectively.}
    \label{fig:absunc}
  \end{center}
\end{figure}

Figure~\ref{fig:absunc} shows the absolute energy scale uncertainties for the three jet types, combined with the offset correction uncertainty corresponding to the average number of pile-up events in the datasets considered for this paper. The low jet \pt threshold indicates the minimum recommended \pt for each jet type: 30\GeV, 20\GeV, and 10\GeV for CALO, JPT, and PF jets respectively. At low jet \pt the offset uncertainty dominates with significant contribution from the MC truth and jet-by-jet matching residuals. At the intermediate jet \pt, where enough data for the in situ measurements are available, the $\gamma$ energy scale uncertainty dominates. At high jet \pt, the uncertainty due to the MC extrapolation is dominant. Overall, the absolute jet energy scale uncertainty for all jet types is smaller than 2\% for $\pt>40\GeV$. 

\clearpage

\subsection{Combined Jet Energy Correction}

\begin{figure}[ht!]
  \begin{center}
    \includegraphics[width=0.45\textwidth]{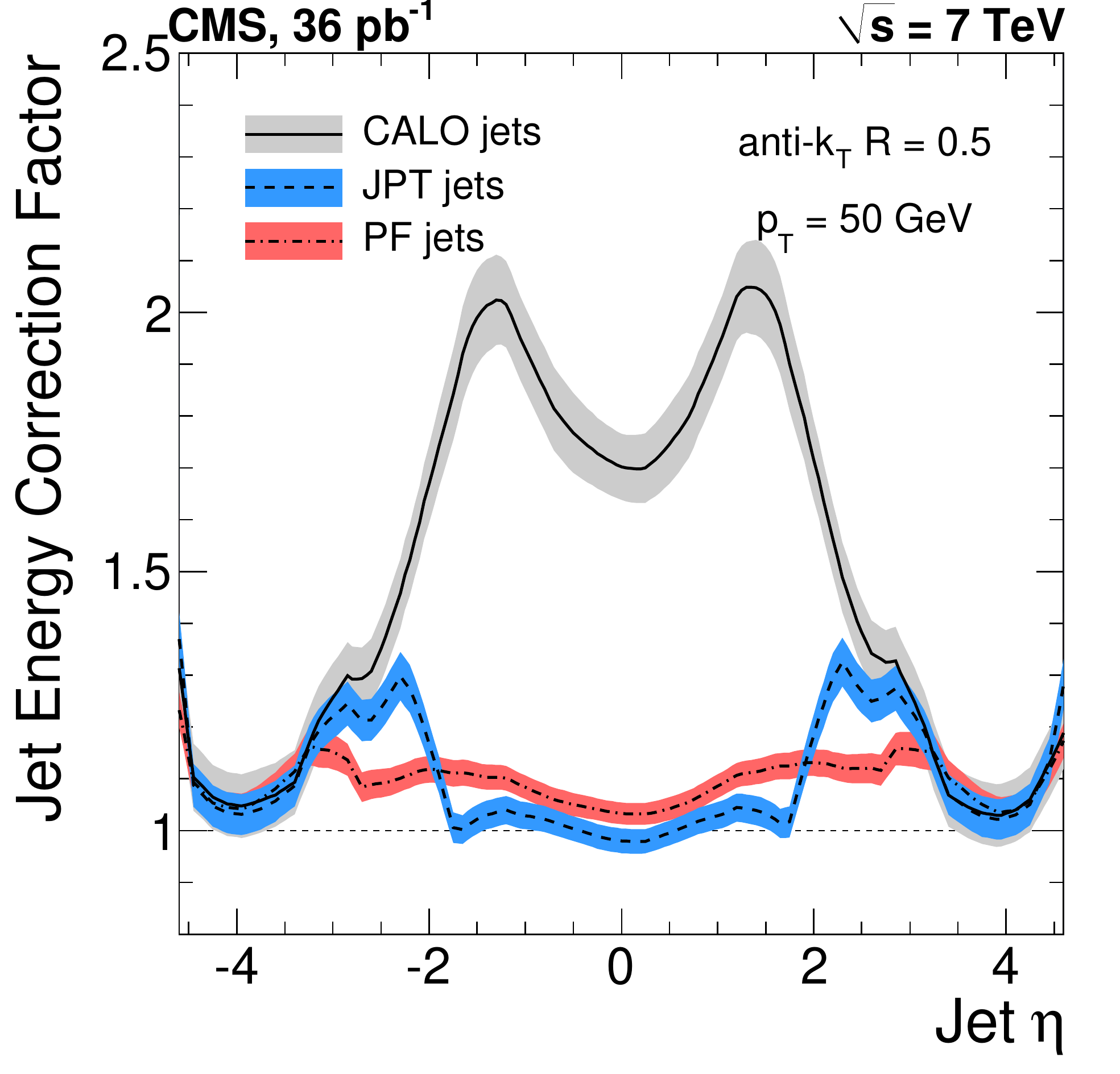}
    \includegraphics[width=0.45\textwidth]{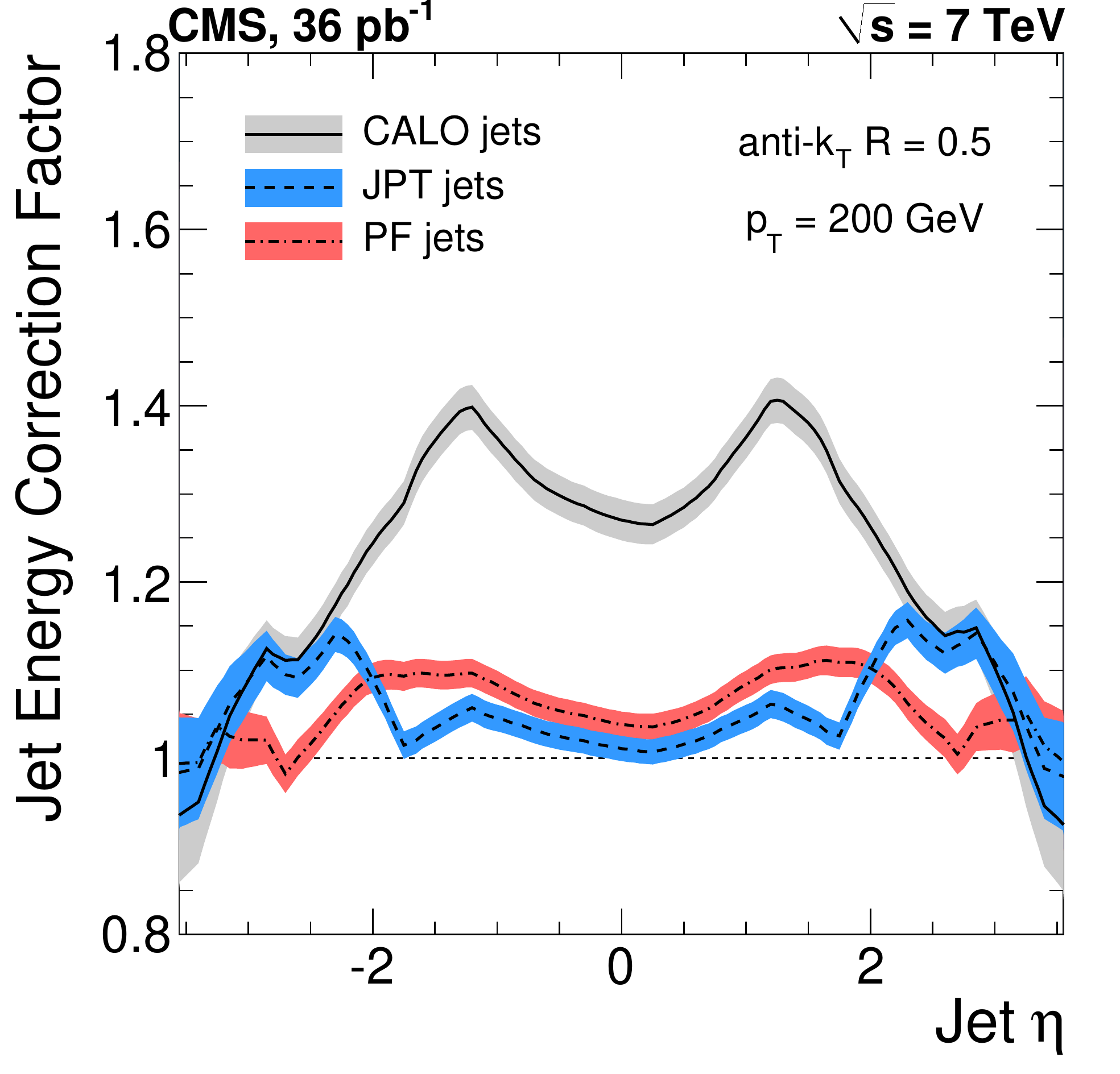}
    \caption{Total jet-energy-correction factor, as a function of jet $\eta$ for $\pt=50\GeV$ (left) and $\pt=200\GeV$ (right). The bands indicate the corresponding uncertainty.}
    \label{fig:finalJECvsEta}
  \end{center}
\end{figure}

\begin{figure}[ht!]
  \begin{center}
    \includegraphics[width=0.45\textwidth]{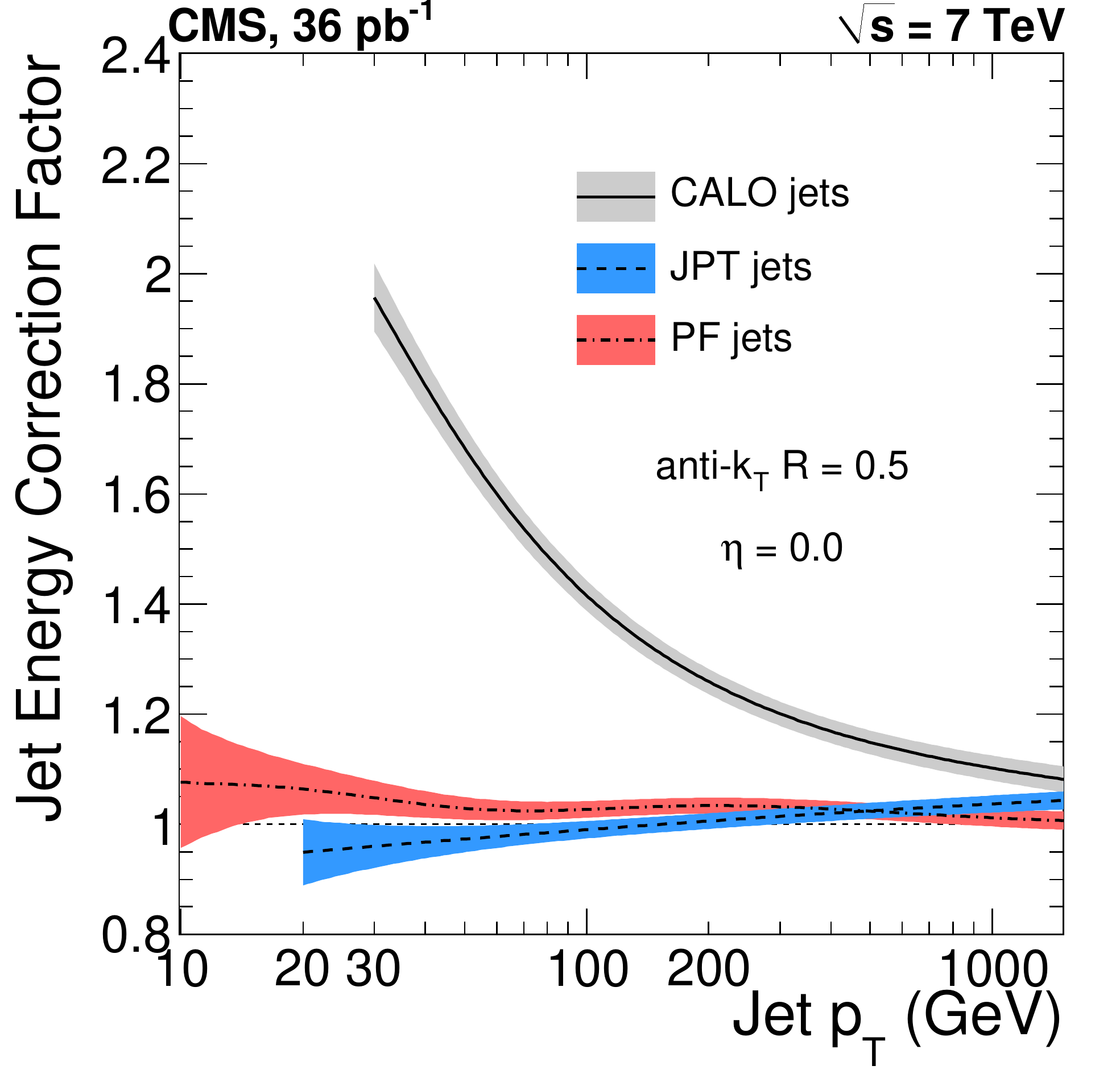}
    \includegraphics[width=0.45\textwidth]{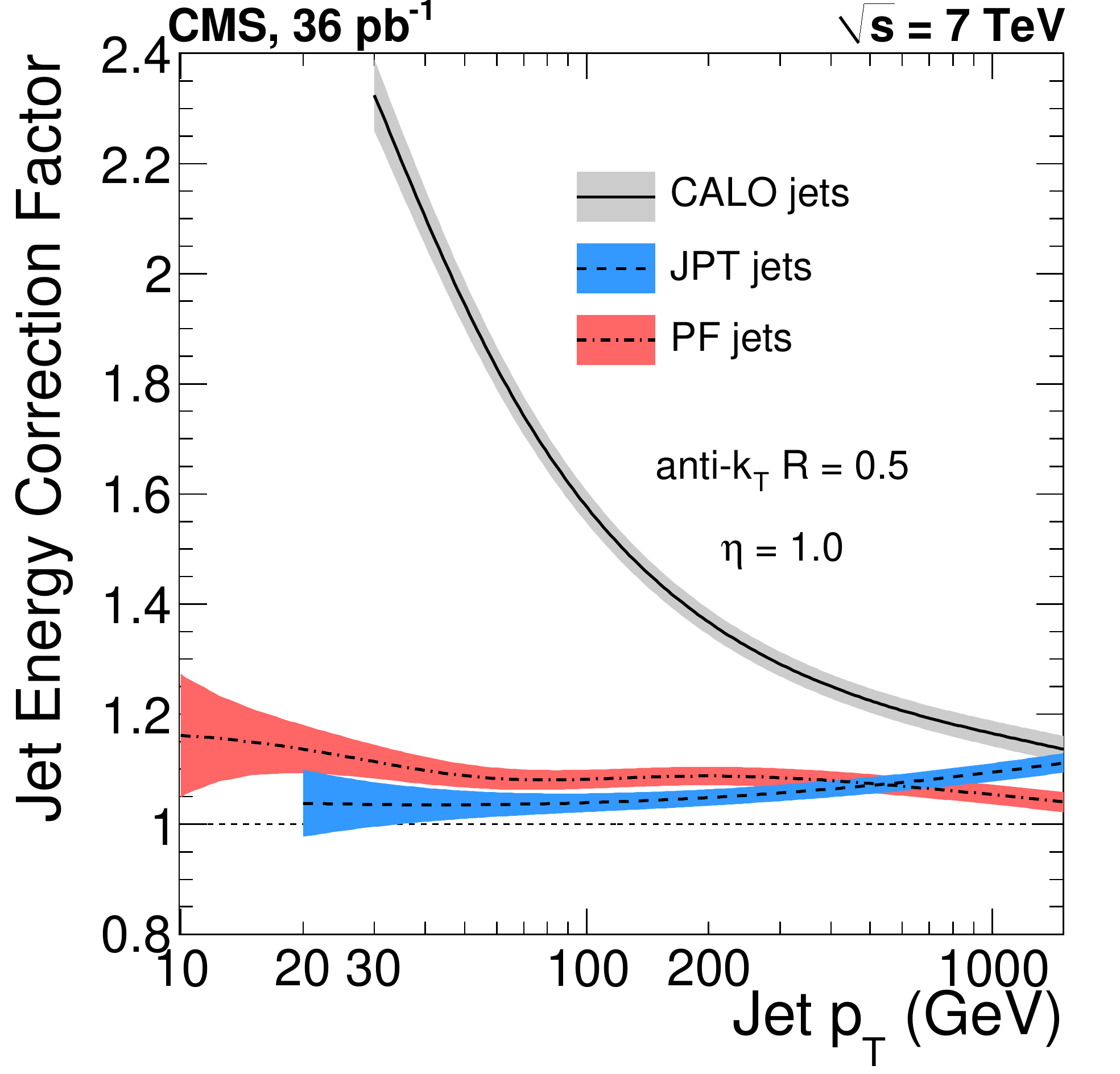}
    \includegraphics[width=0.45\textwidth]{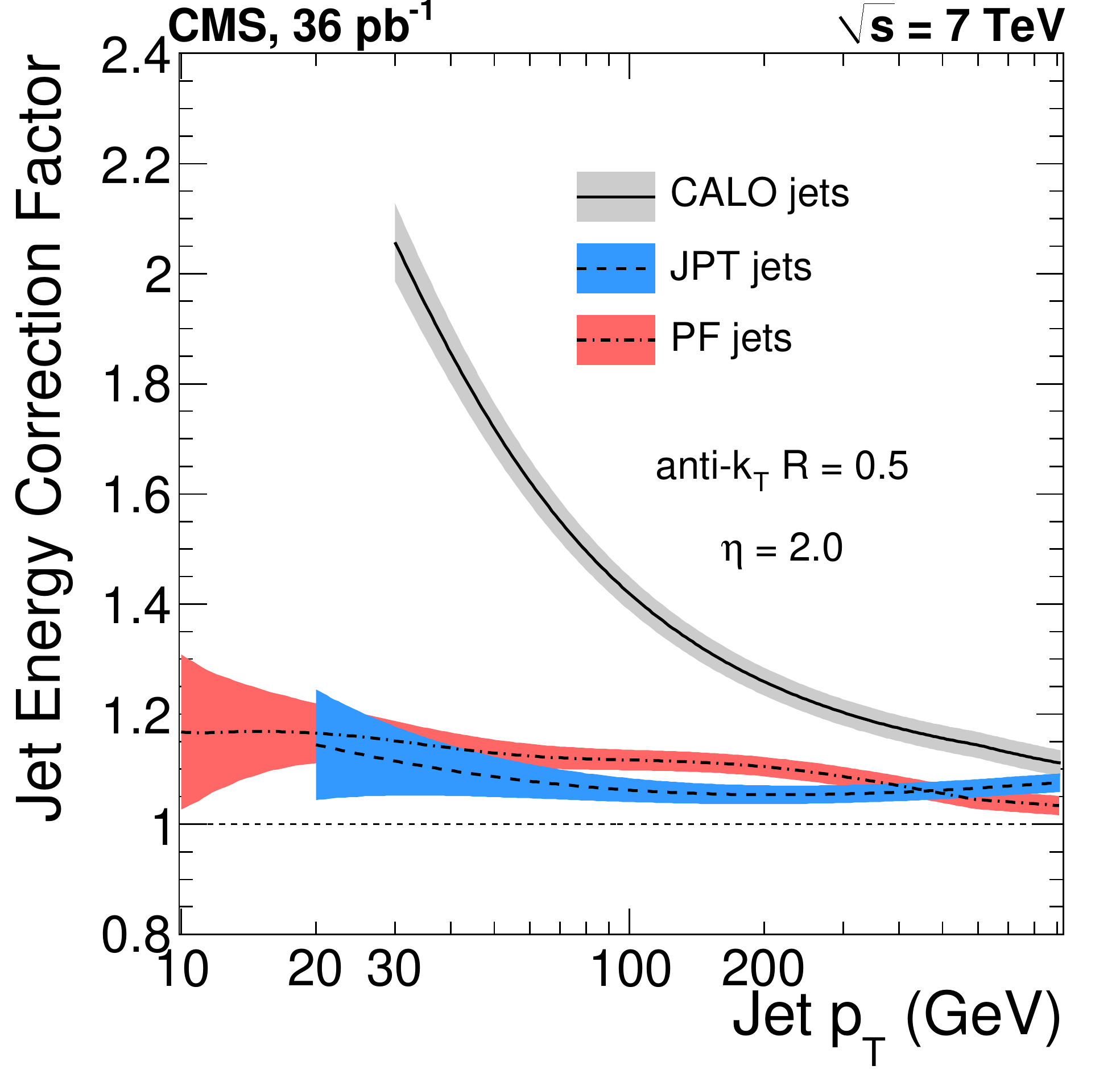}
    \includegraphics[width=0.45\textwidth]{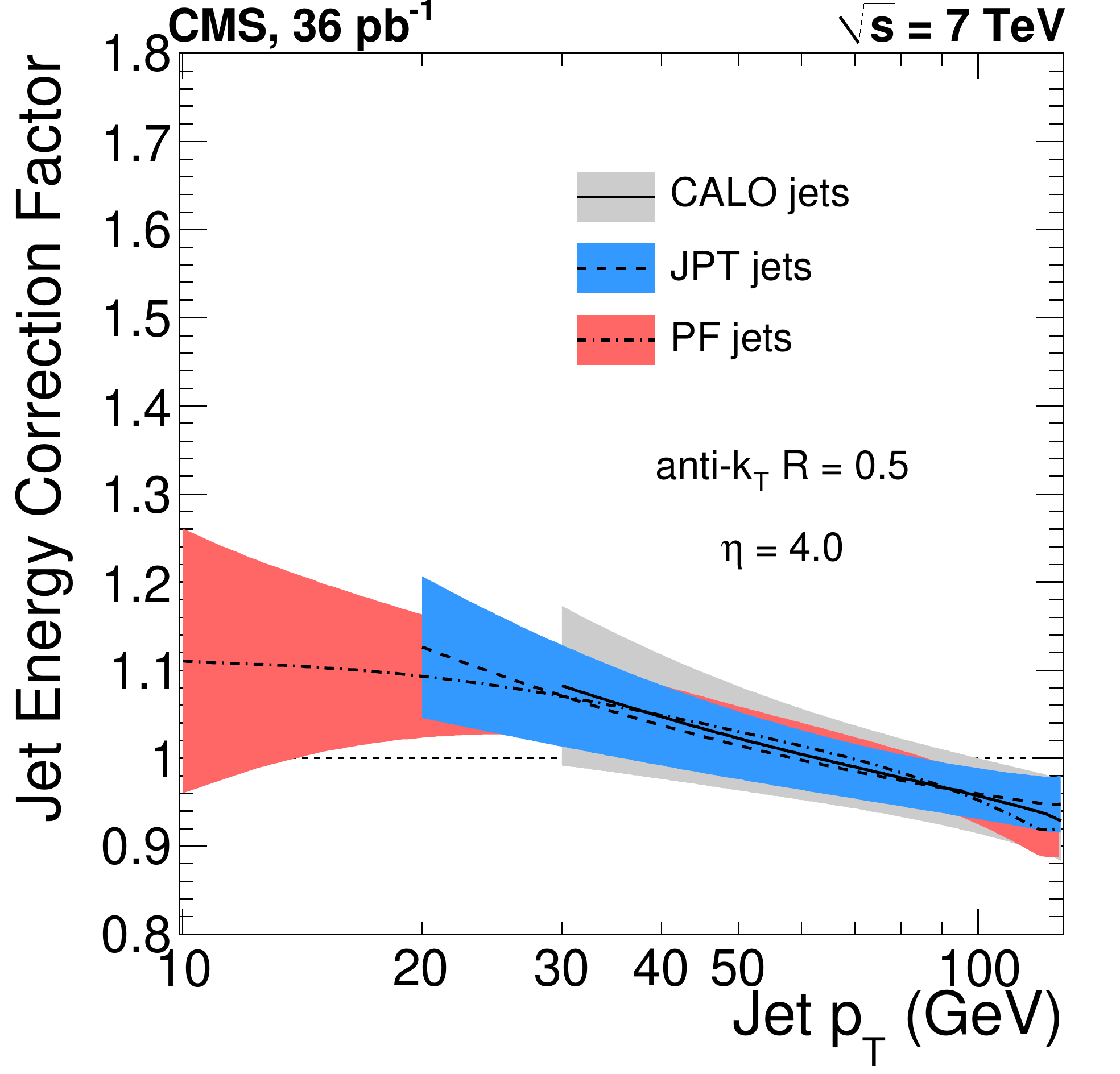}
    \caption{Total jet-energy-correction factor, as a function of jet \pt for various $\eta$ values. The bands indicate the corresponding uncertainty.}
    \label{fig:finalJECvsPt}
  \end{center}
\end{figure}

\begin{figure}[ht!]
  \begin{center}
    \includegraphics[width=0.45\textwidth]{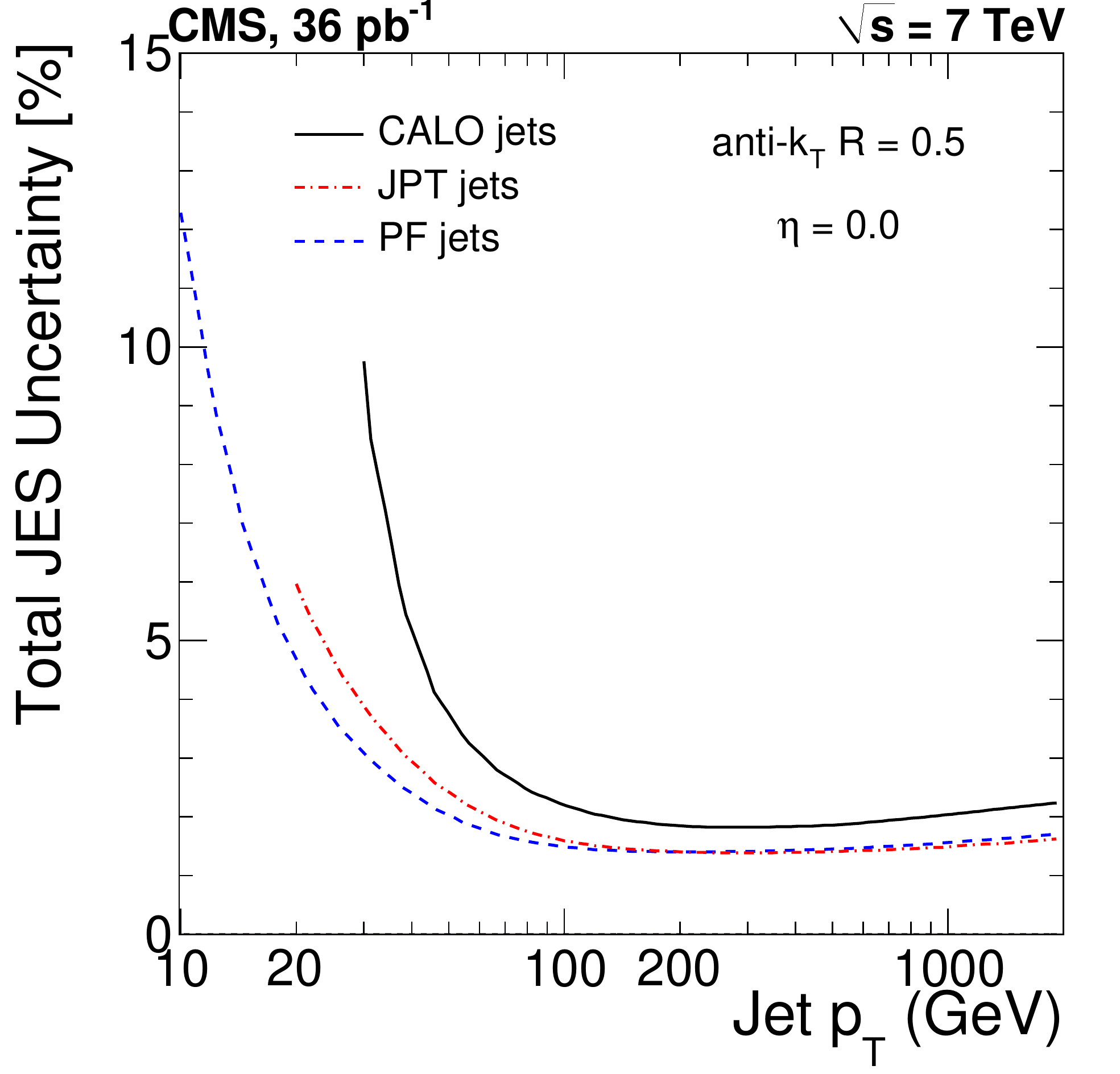}
    \includegraphics[width=0.45\textwidth]{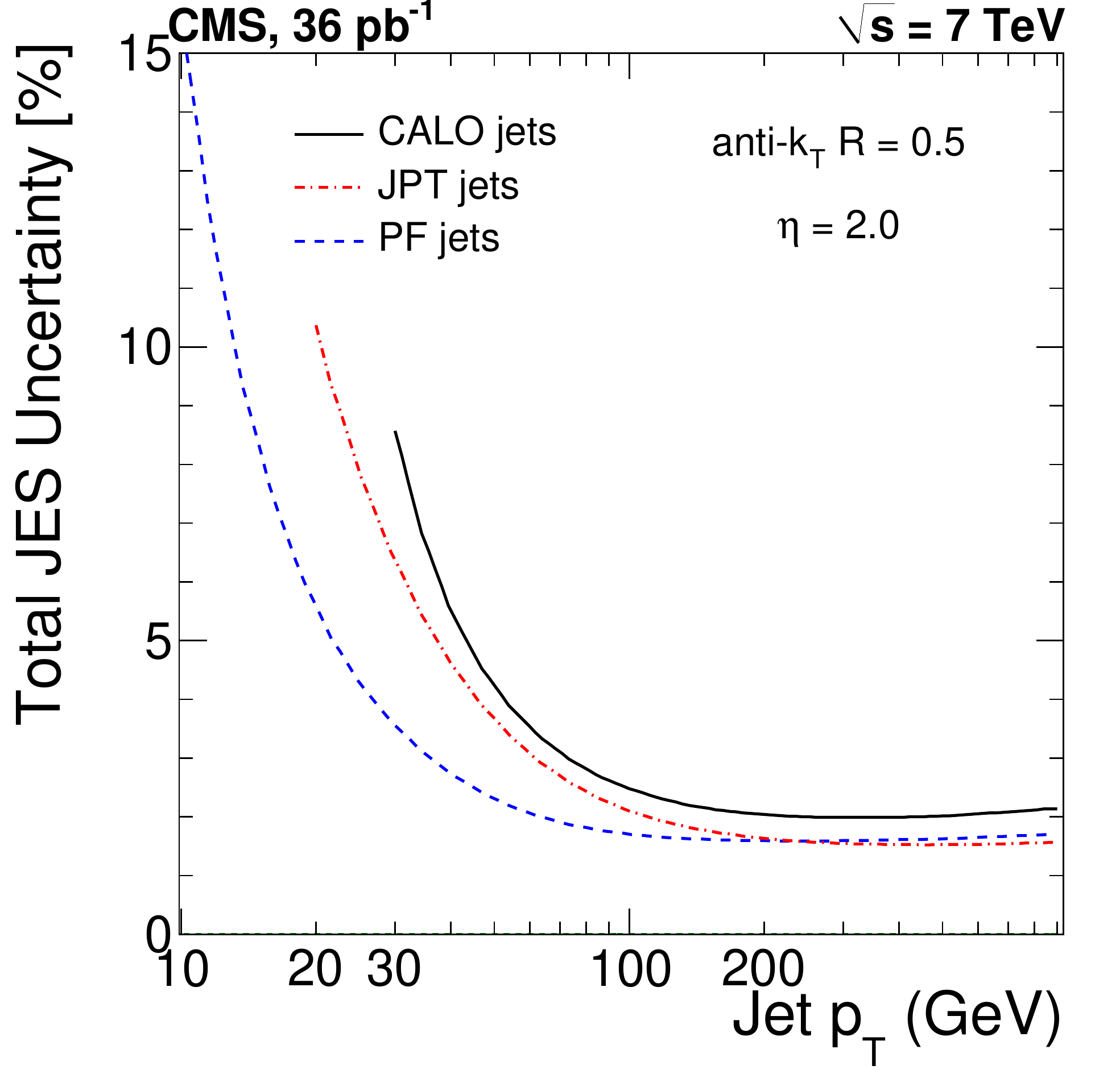}
    \includegraphics[width=0.45\textwidth]{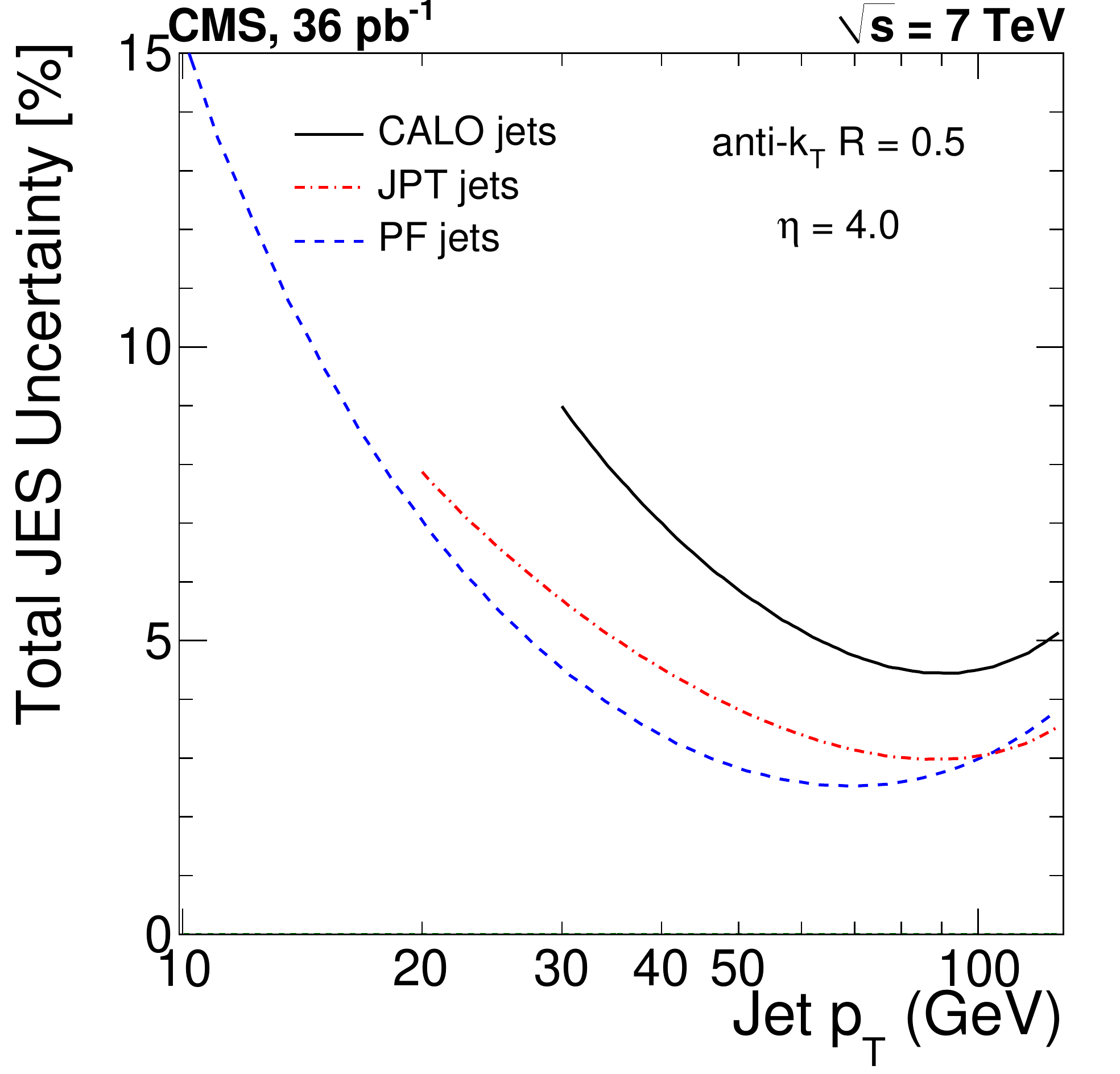}
    \caption{Total jet-energy-scale uncertainty, as a function of jet \pt for various $\eta$ values.}
    \label{fig:finalUncvsPt}
  \end{center}
\end{figure}

In this section, the combined MC and residual calibration is presented along with the total jet energy scale systematic uncertainty. Following Eq.~(\ref{eq:jec_components}), the residual corrections for the relative and absolute response are multiplied with the generator-level MC correction, while the corresponding uncertainties are added in quadrature. Figure~\ref{fig:finalJECvsEta} shows the combined calibration factor as a function of jet-$\eta$ for $\pt=50,\,200\GeV$. Because of the smallness of the residual corrections, the combined correction has the shape of the MC component, shown in Fig.~\ref{fig:mctruthVsEta}. The total correction as a function of jet \pt is shown in Fig.~\ref{fig:finalJECvsPt} for various $\eta$ values. Figure~\ref{fig:finalUncvsPt} shows the total jet energy scale uncertainty as a function of jet \pt. At low jet \pt the relative energy scale uncertainty makes a significant contribution to the total uncertainty while it becomes negligible at high \pt. In the forward region, the relative scale uncertainty remains significant in the entire \pt-range. In general PF jets have the smallest systematic uncertainty while CALO jets have the largest.

\clearpage

\section{Jet Transverse Momentum Resolutions}\label{sec:res}

In the following sections, results on jet \pt resolutions are presented, 
extracted from generator-level MC information, and measured 
from the collider data. Unless stated otherwise, CALO, PF and JPT jets 
are corrected for the jet energy scale, as described in the previous section.

The jet \pt resolution is measured from two different samples, 
in both data and MC samples, using methods described in 
Section~\ref{sec:methods}:

\begin{itemize}
\item The dijet asymmetry method, applied to the dijet sample,
\item The photon-plus-jet balance method, applied to the $\gamma+$jet 
sample.
\end{itemize}
 
The dijet asymmetry method exploits momentum conservation in the transverse plane of the dijet system and
is based (almost) exclusively on the measured kinematics of the dijet events. 
This measurement uses two ways  of describing  the jet resolution 
distributions in data and simulated events. The first method 
makes use of a truncated RMS to characterize the core of the distributions.
The second method  employs functional fitting  of the full jet resolution function, 
and is currently limited to a Gaussian approximation for the jet \pt 
probability density.

The $\gamma+$jet balance method exploits the balance in the transverse 
plane between the photon and 
the recoiling jet, and it uses the photon as a 
reference object whose \pt is accurately measured in ECAL. The width
of the $\pt/\pt^{\gamma}$ distribution provides information on 
the jet \pt resolution in a given $\pt^{\gamma}$ bin.  
The resolution is determined independently for both data and simulated events.
The results extracted from $\gamma$+jet \pt balancing provide
useful input for validating the CMS detector simulation, and serve as an independent and complementary cross-check
of the results obtained with the dijet asymmetry method.

In the studies presented in this paper, the resolution broadening from 
extra radiation activity is removed by extrapolating to the ideal case of a two-body process, both in data and in MC.
In addition, the data/MC resolution ratio is derived.

\subsection{Monte Carlo  Resolutions}
\label{sec:res_mctruth}
The jet \pt resolution derived from generator-level MC information   
information in the simulation, serves as a benchmark for the measurements 
of the jet resolution in collision data samples, 
using the methods introduced above and discussed in the following sections.

\begin{figure}[t]
\centering
\includegraphics[width=0.49\textwidth]{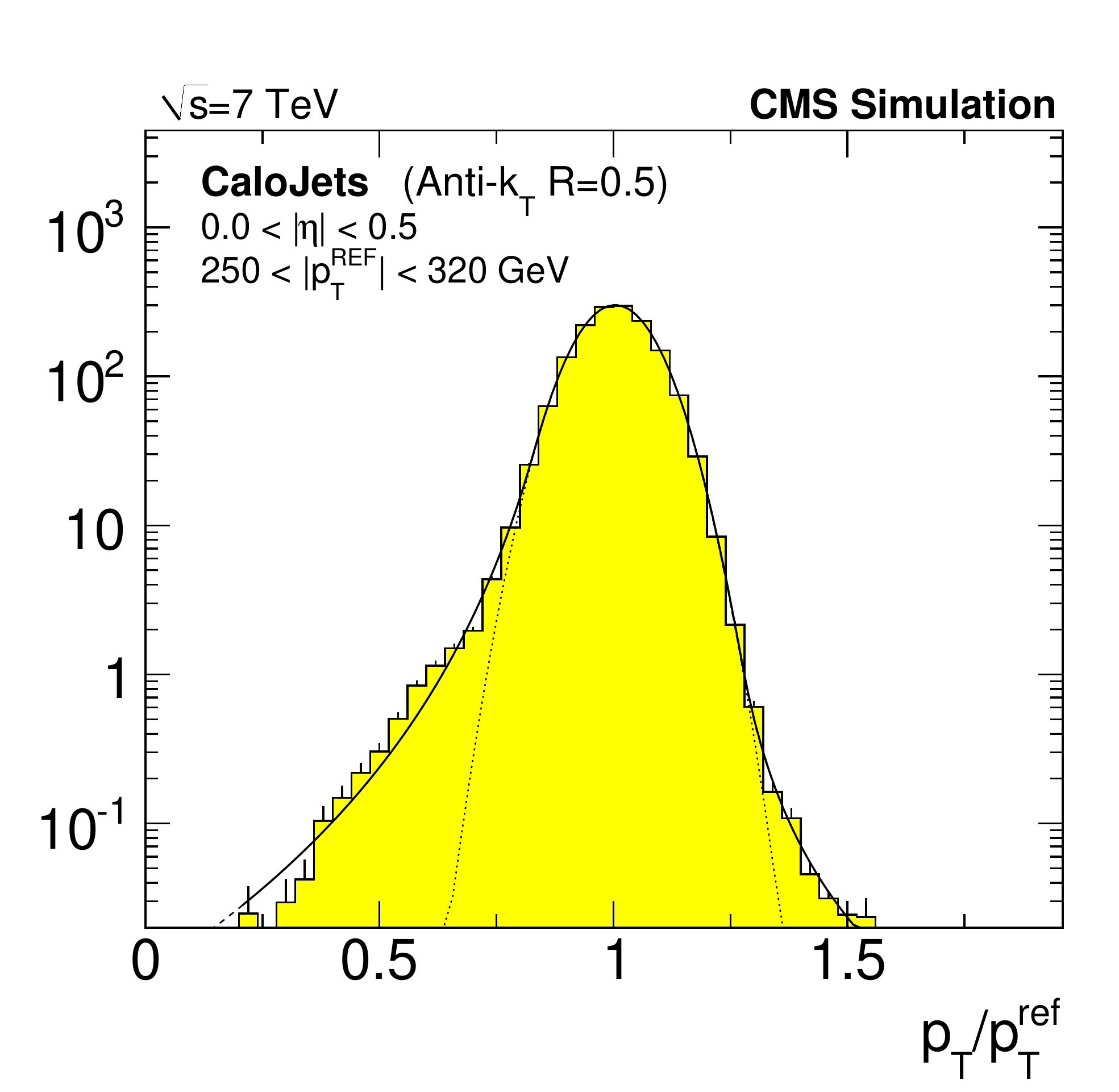}
\caption{Distribution of the simulated CALO jet response, $\pt^{reco}/\pt^{gen}$, in a particular $|\eta|$ and $\pt^{gen}$ range. Fit examples with a Gaussian and a double-sided Crystal-Ball function are shown.
}
\label{fig:fit-example}
\end{figure}

The measurement of the jet \pt resolution in the simulation is performed using {\sc pythia} QCD dijet
events. The MC particle jets are matched geometrically to the reconstructed jets (CALO, JPT, or PF) by requiring
their distance in $\eta-\phi$ space to be $\Delta R < \Delta R_{Max}$.

The jet \pt response is defined as the ratio $\pt^{reco}/\pt^{gen}$
where $\pt^{reco}$ and $\pt^{gen}$ refer to the transverse momenta  
of the reconstructed jet and its matched reference MC particle jet respectively.

The width of the jet \pt response distribution, in a given $|\eta|$ and
$\pt^{gen}$ bin, is interpreted as the generator-level MC jet \pt 
resolution.  Figure~\ref{fig:fit-example} shows an example of 
$\pt^{reco}/\pt^{gen}$ distribution for CALO jets
in $|\eta|<0.5$ and with $250<\pt^{gen}<320\GeV$.

\subsection{Dijet Measurements}
\label{sec:res_dijet}

The principles of the dijet asymmetry method for the measurement
of the jet \pt resolution were presented in Section~\ref{sec:methods}. Here, the results of the measurement are presented.

The idealized topology of two jets with exactly compensating
transverse momenta is spoiled in realistic collision events by the
presence of extra activity, e.g. from additional soft radiation or from
the UE. The resulting asymmetry distributions are broadened and
the jet \pt resolution is systematically underestimated.
Other effects can also cause jet imbalance.
For example, fragmentation effects cause some energy to be
showered outside the jet cone (``out of cone radiation'').
The width of the asymmetry distribution is thus a convolution of
these different contributions:

\begin{equation}
\sigma_\mathcal{A} = \sigma_{intrinsic}\oplus\sigma_{imbalance}
\label{eq:convolution}
\end{equation}

\begin{figure}[h]
\centering
\includegraphics[width=0.49\textwidth]{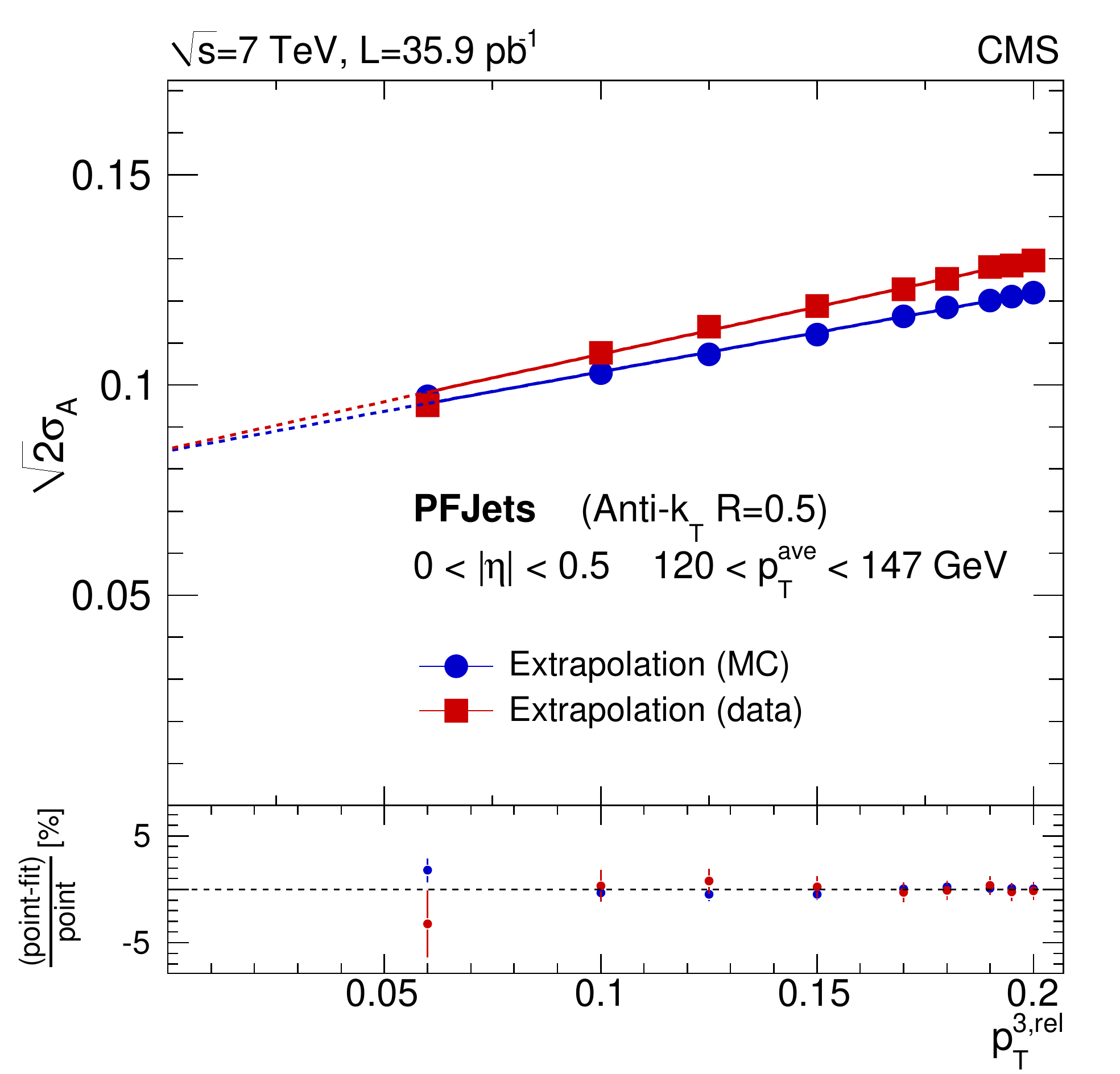}
\includegraphics[width=0.49\textwidth]{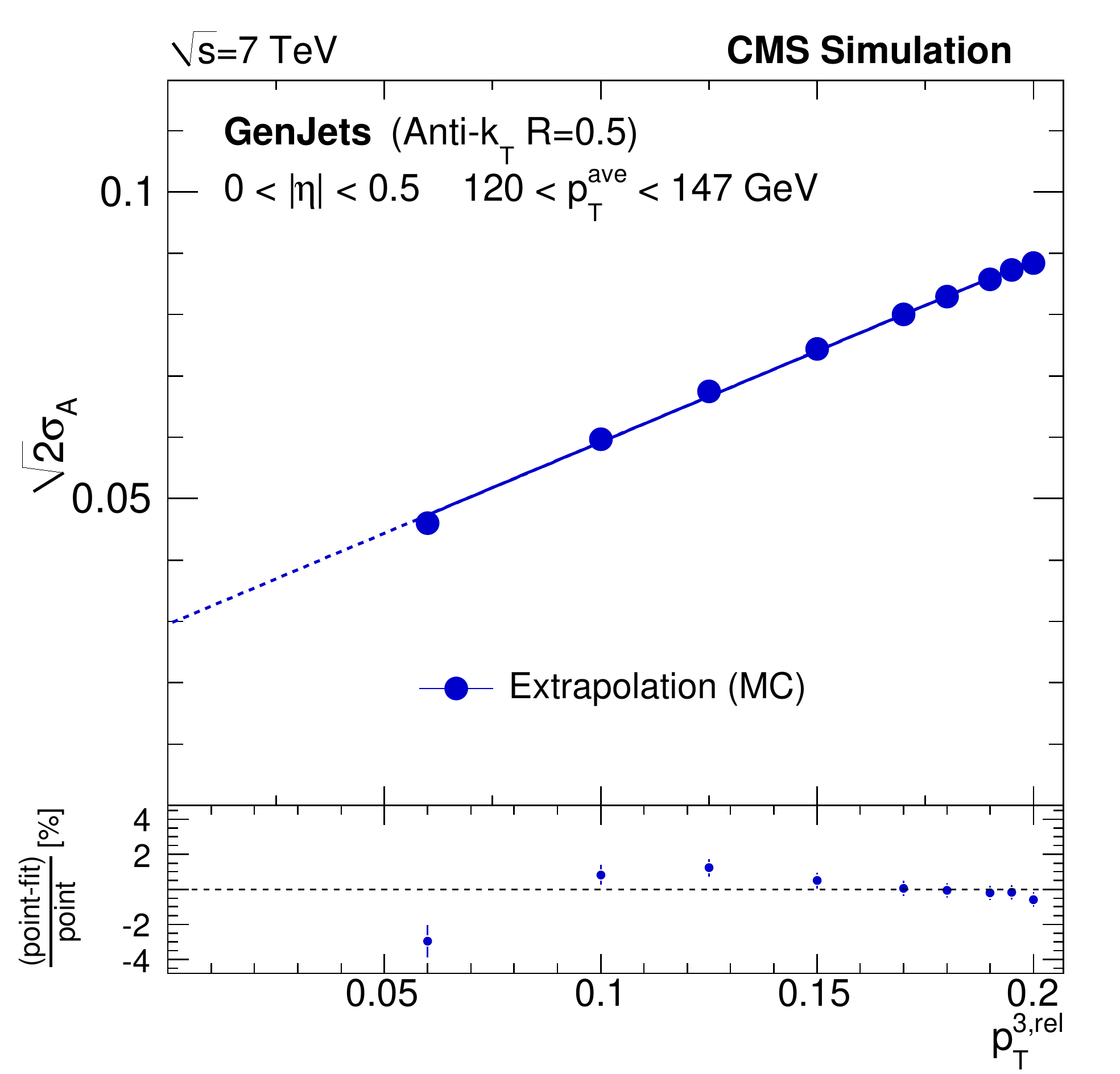}

  \caption{Examples of extrapolations of $\sqrt{2}\sigma_\mathcal{A}$
   as a function of
  \ptrelthree to zero for PF jets ($R=0.5$) in $|\eta|<0.5$ and $120<\ptave<147\GeV$ (left).
  Example of a corresponding extrapolation  for MC particle jets (right).}
\label{fig:extrap-example}
\end{figure}

\begin{figure}[h]
\centering
\includegraphics[width=0.45\textwidth]{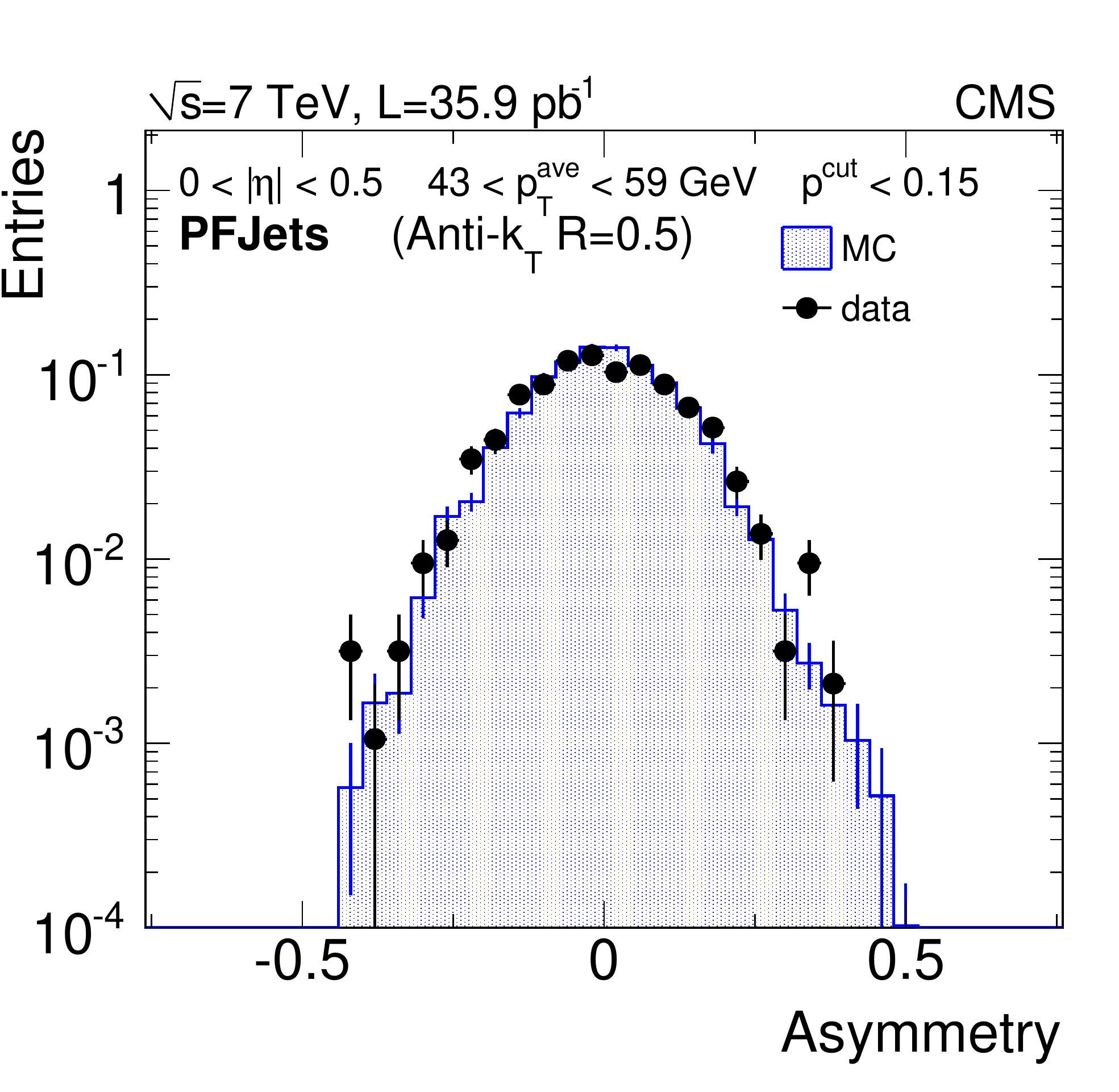}
\includegraphics[width=0.45\textwidth]{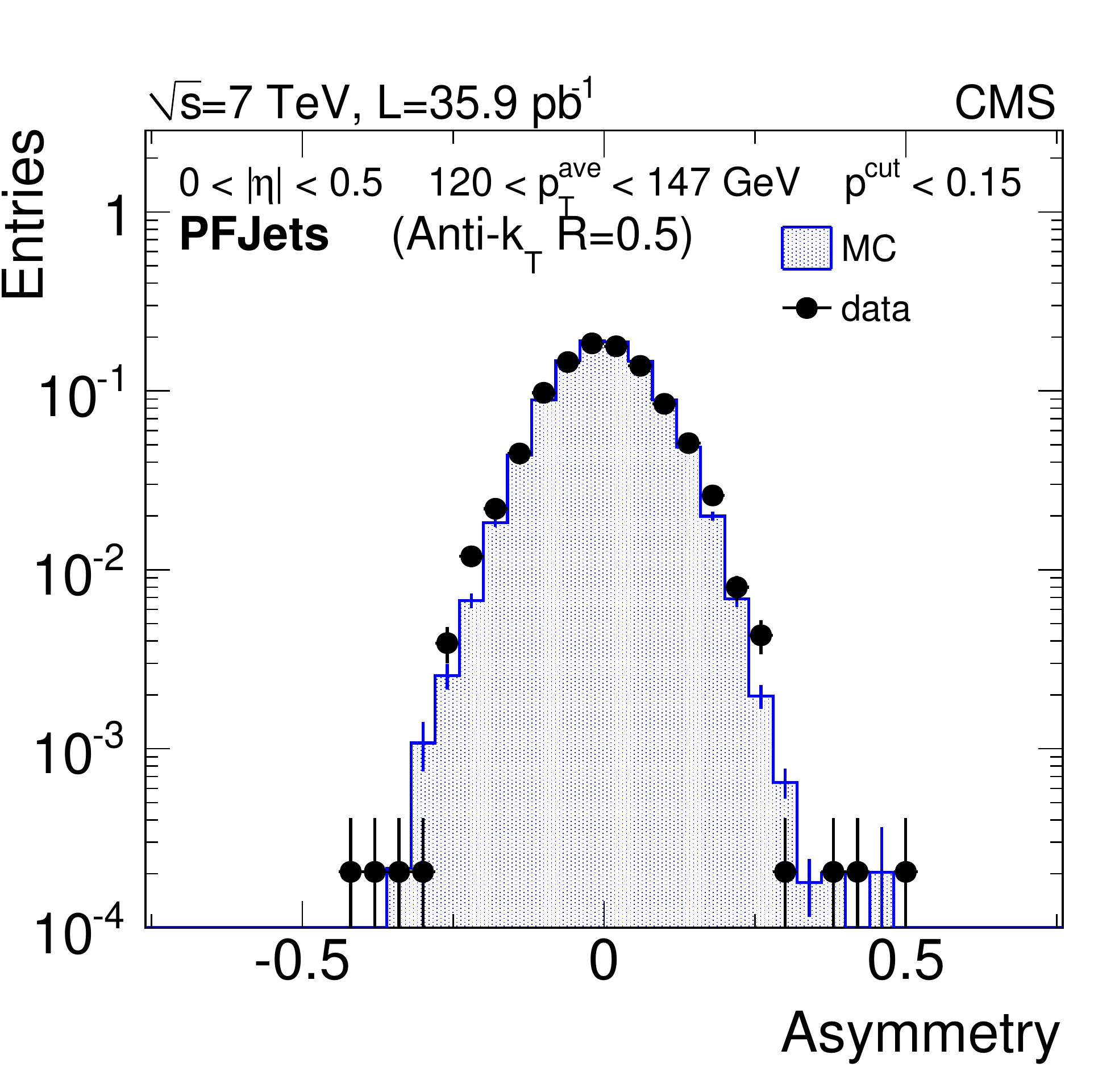}
\includegraphics[width=0.45\textwidth]{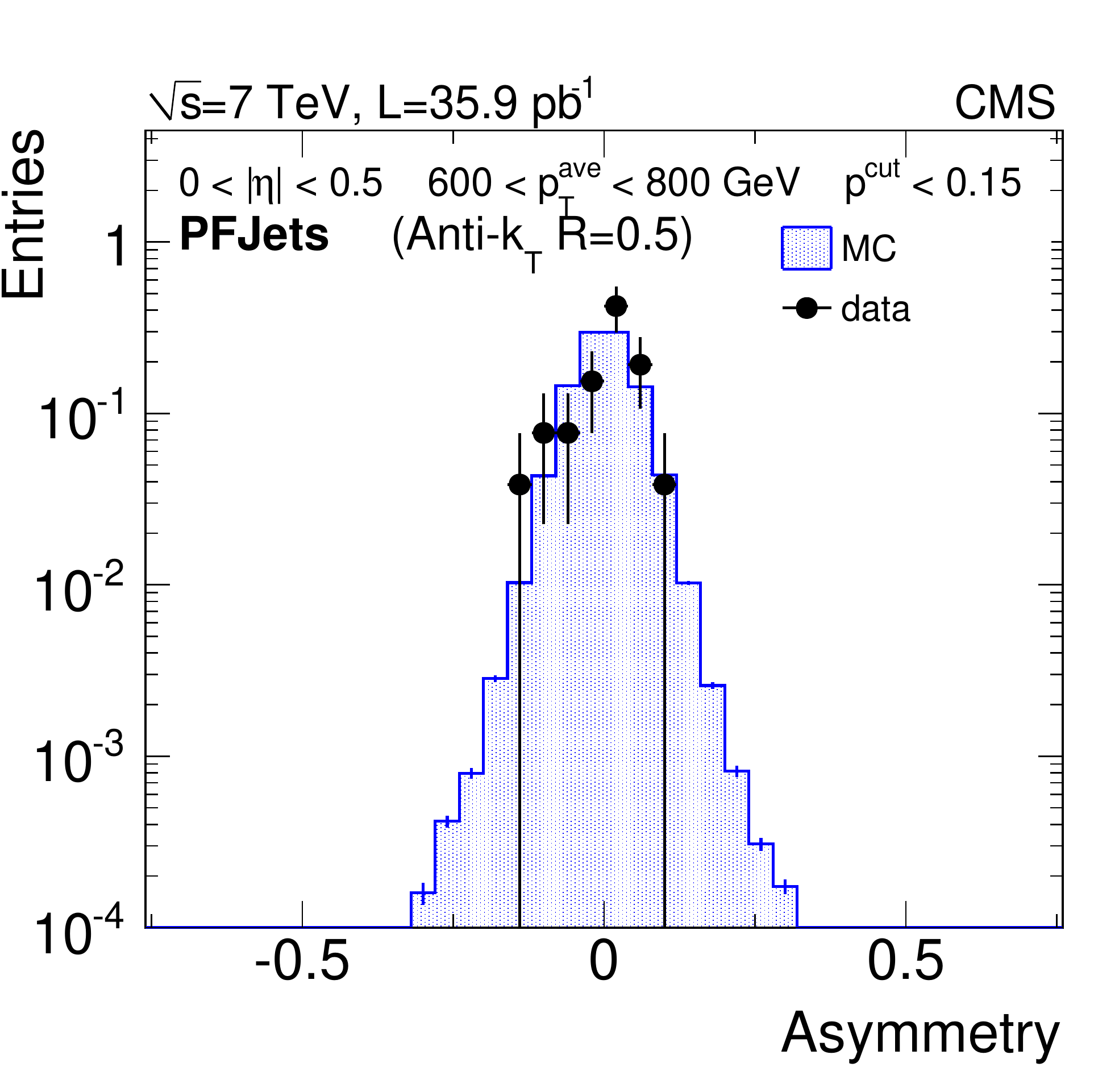}
\caption{Examples of PF jet asymmetry distributions for $|\eta|<0.5$ and a low-\ptave bin (top left), a medium-\ptave bin (top right) and a high-\ptave bin (bottom), determined from QCD simulation (blue histograms) and compared with the result from data (black dots).}
\label{fig:asym-mcdata-pf-eta00to05}
\end{figure}

To account for soft radiation in dijet events,
the measurement of the asymmetry in each $\eta$ and
$\ptave$ bin is carried out
multiple times, for decreasing amounts of extra activity, and the jet \pt
resolution is extracted by extrapolating the extra event activity to zero,
as discussed in Section \ref{sec:radbias}.
The ratio of the transverse momentum of the third jet in the event over
the dijet average \pt, $\pt^{Jet3, rel}=\ptrelthree$, is used as a measure of
the extra activity.
The extrapolation procedure is illustrated in
Fig.~\ref{fig:extrap-example} (left) for the
$120 < \ptave < 147\GeV$ bin of PF jets
and for the corresponding bin of MC particle jets (right).
The width of each asymmetry distribution $\sigma_\mathcal{A}$, as well as the resolutions obtained
using generator-level MC information, are derived based on the RMS of the
corresponding  distributions.
Some characteristic example distributions for the raw asymmetry are shown
for PF jets in Fig.~\ref{fig:asym-mcdata-pf-eta00to05}.

\begin{figure}[h]
  \centering
\includegraphics[width=0.45\textwidth]{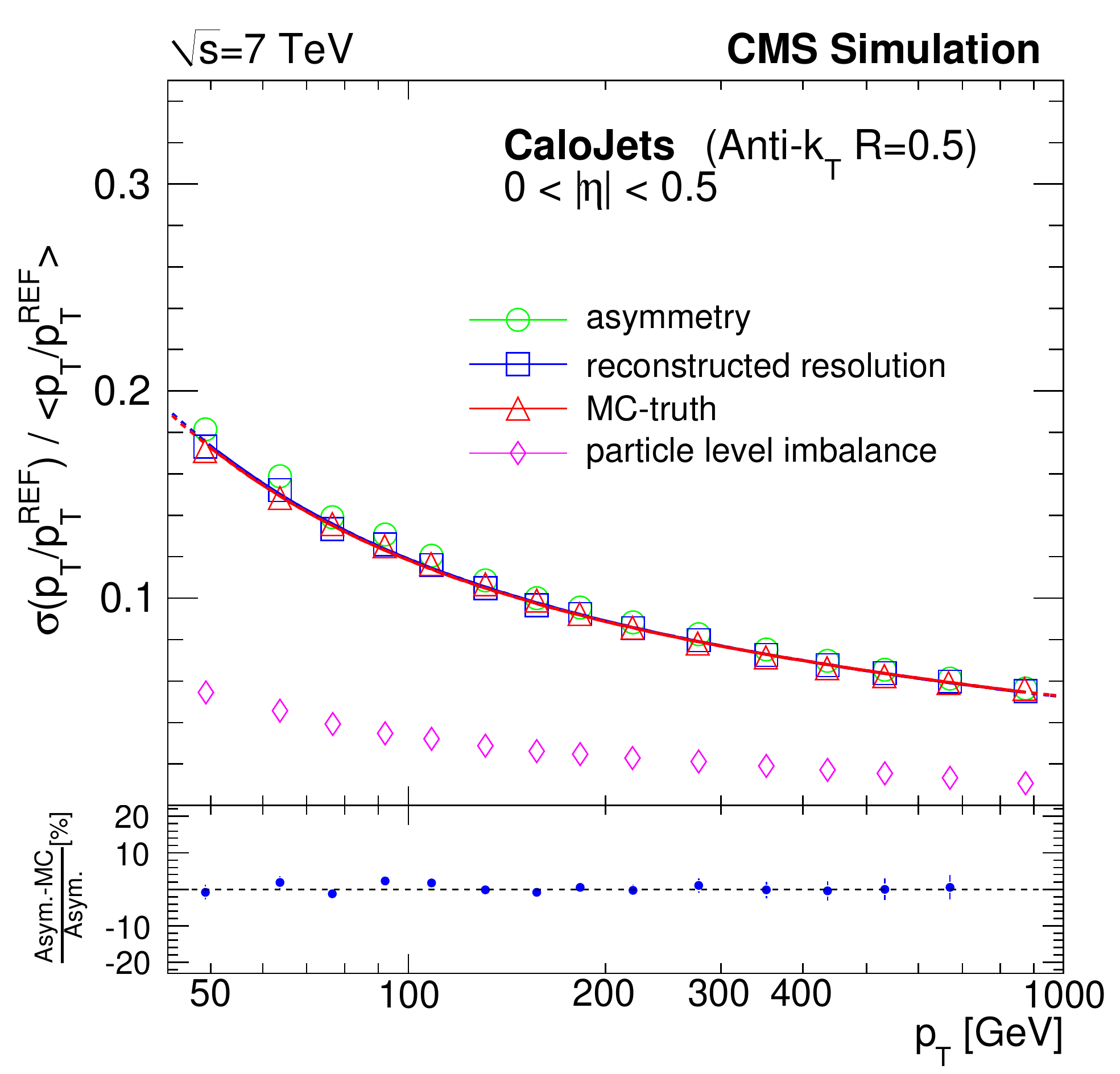}
\includegraphics[width=0.45\textwidth]{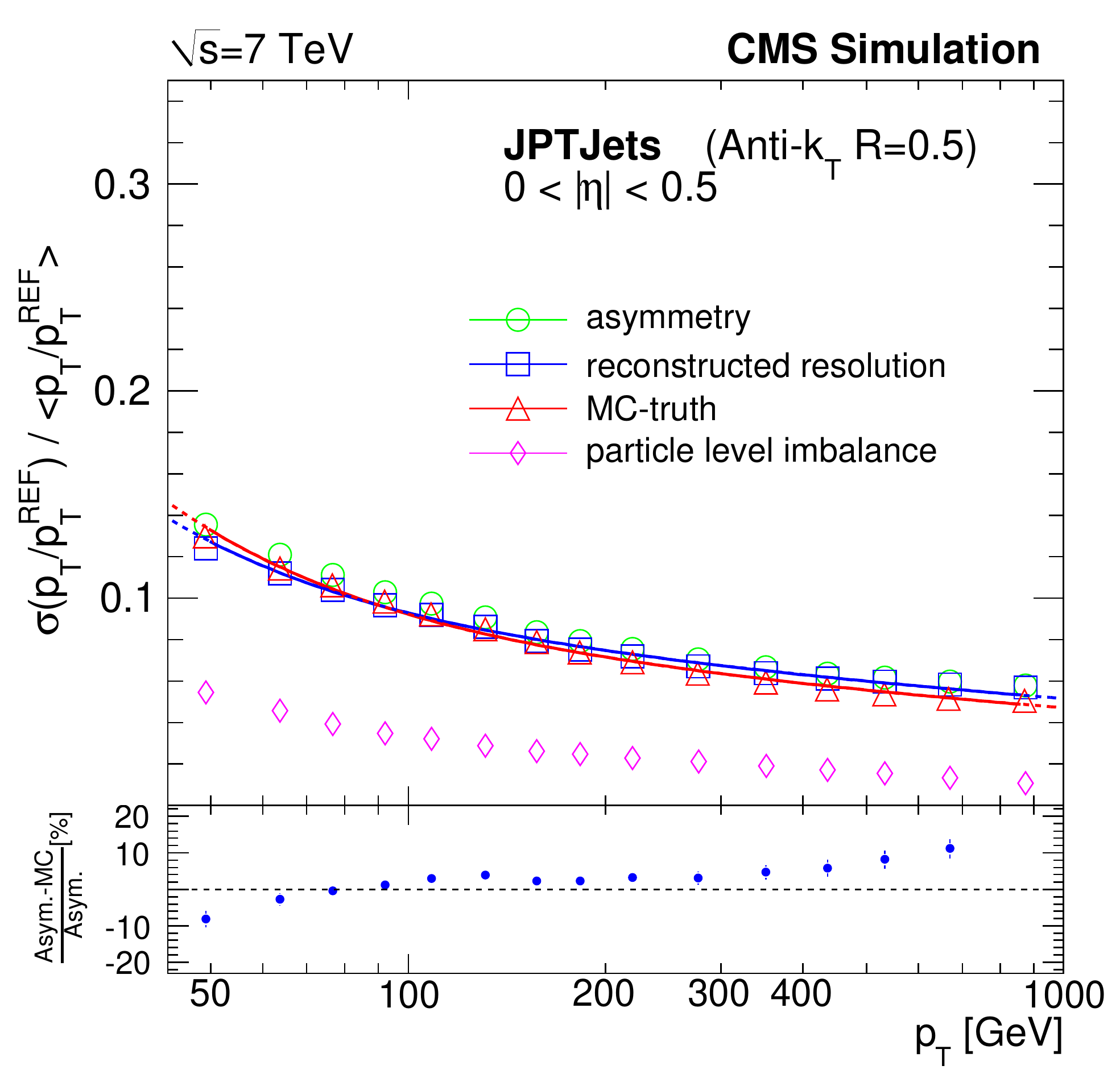}
\includegraphics[width=0.45\textwidth]{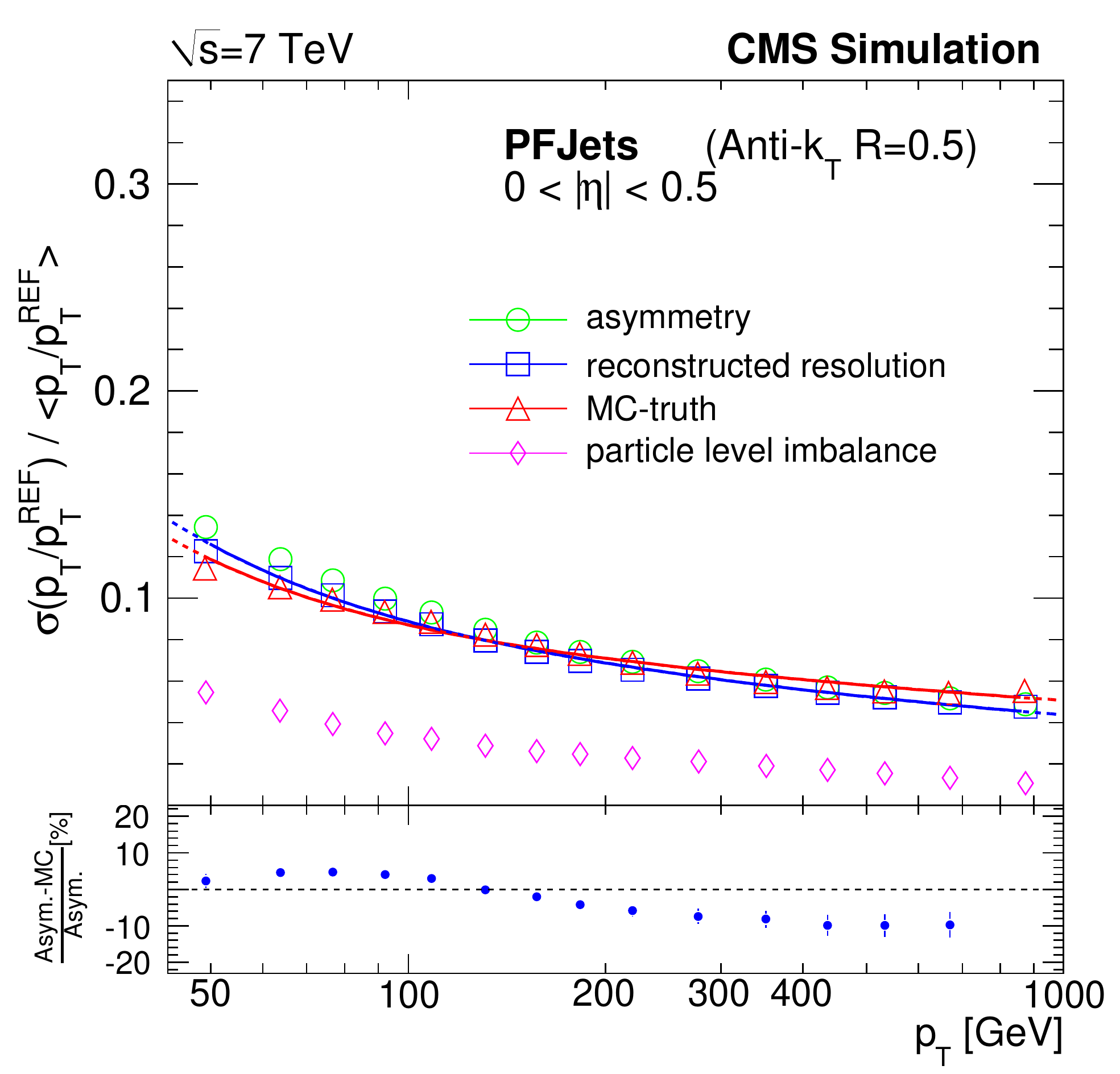}
  \caption{Application of the asymmetry method to simulated CALO (top left), JPT (top right), and PF jets (bottom) in $|\eta|<0.5$.
The reconstruction-level (green circle) and particle-level (magenta diamond) results are shown together with the final measurement (blue square), compared to the generator-level MC (denoted as MC-truth) derived resolution (red triangle).}
  \label{fig:reso-mc-00to05}
\end{figure}

To account for the particle-level imbalance contribution to the measured jet \pt resolution, the asymmetry method is applied to the generated MC particle jets. Then the extrapolated particle-level resolution is subtracted in quadrature from the measurement.
Figure~\ref{fig:reso-mc-00to05} illustrates the different steps of the asymmetry procedure for CALO, JPT, and PF jets respectively.
The total \pt resolution derived from the extrapolation of the reconstructed asymmetry is shown in green circle, the estimation of the particle-level imbalance resolution from the application to MC particle jets is shown in magenta diamond, and the quadrature subtraction to the final asymmetry result is shown in blue square. All three can be described by a fit to a variation of the standard
formula for calorimeter-based resolutions,

\begin{equation}
  \frac{\sigma(\pt)}{\pt} =
  \sqrt{\mathrm{sgn}(N)\cdot\left(\frac{N}{\pt}\right)^2+S^2\cdot
    \pt^{(M-1)}+C^2},
\end{equation}

where, $N$ refers to the "noise", $S$ to the "stochastic",
and $C$ to the "constant" term. The additional parameter $M$ is
introduced, and the negative sign of the noise term is allowed,
to improve the fits to the jet \pt resolution vs. \pt,
for jets that include tracking information (JPT, PF),
while retaining a similar functional form as
the one used for CALO jets. The resolution estimated from  generator-level
MC information is shown in red triangles, and good agreement with the result
of the  asymmetry method is observed. The ratio
$\mathrm{\frac{MC(generator-level)}{MC(asymmetry)}}$ is obtained as a
function of \pt, for each jet type and in each $\eta$-bin and is later
applied to the data measurement as a bias correction.

Several sources of systematic uncertainties are identified:

The linear extrapolation at half-the-distance between the standard working point (at $\ptrelthree=0.15$) and zero is evaluated, and the difference from the full extrapolation to zero is assigned as an uncertainty. The size of the particle-level imbalance is varied by $25\%$ and the impact of the measurement is studied when subtracting $75\%$ and $125\%$ of the original particle jet \pt resolution in quadrature.

\begin{figure}[h]
  \centering

\includegraphics[width=0.45\textwidth]{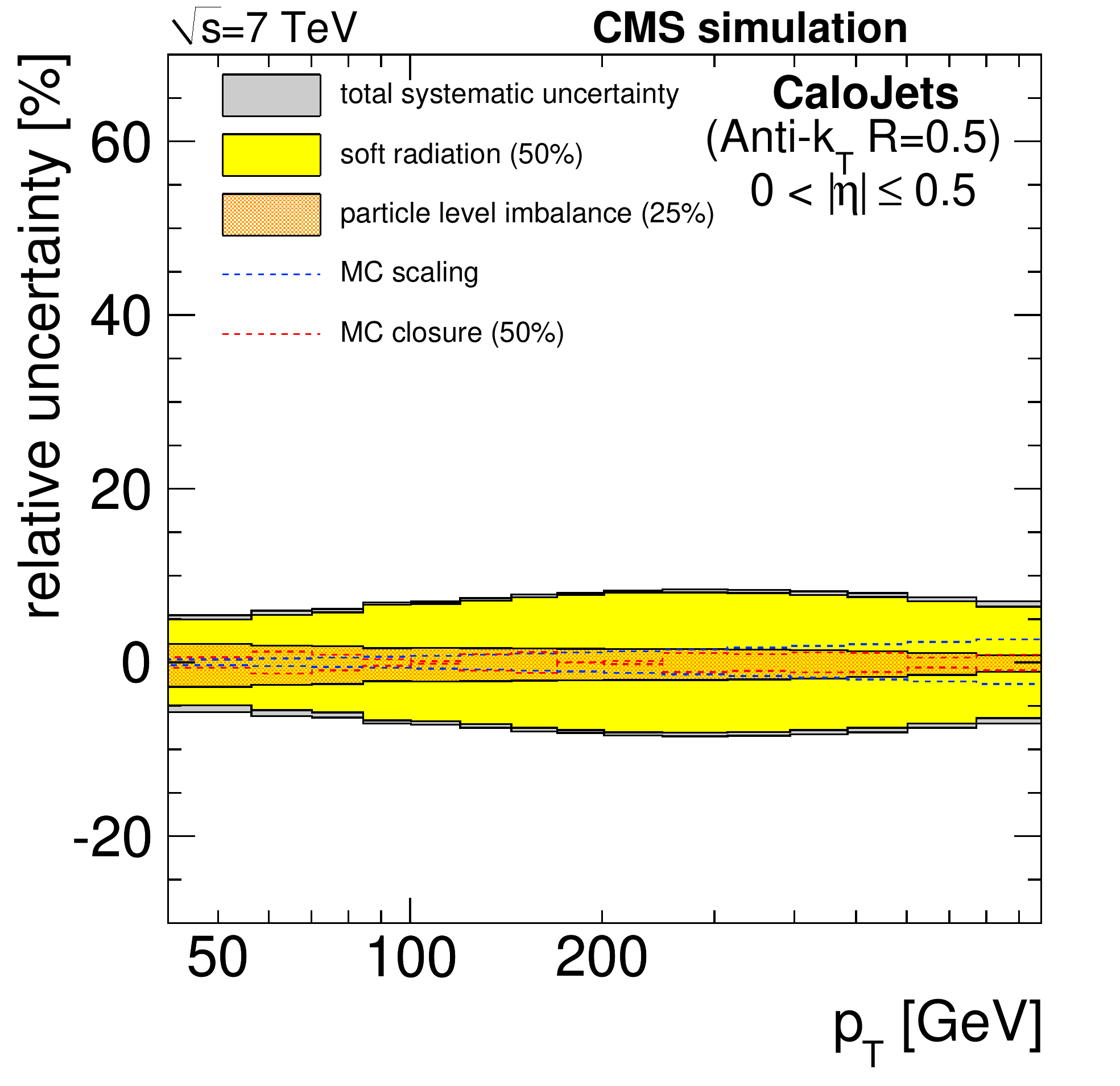}
\includegraphics[width=0.45\textwidth]{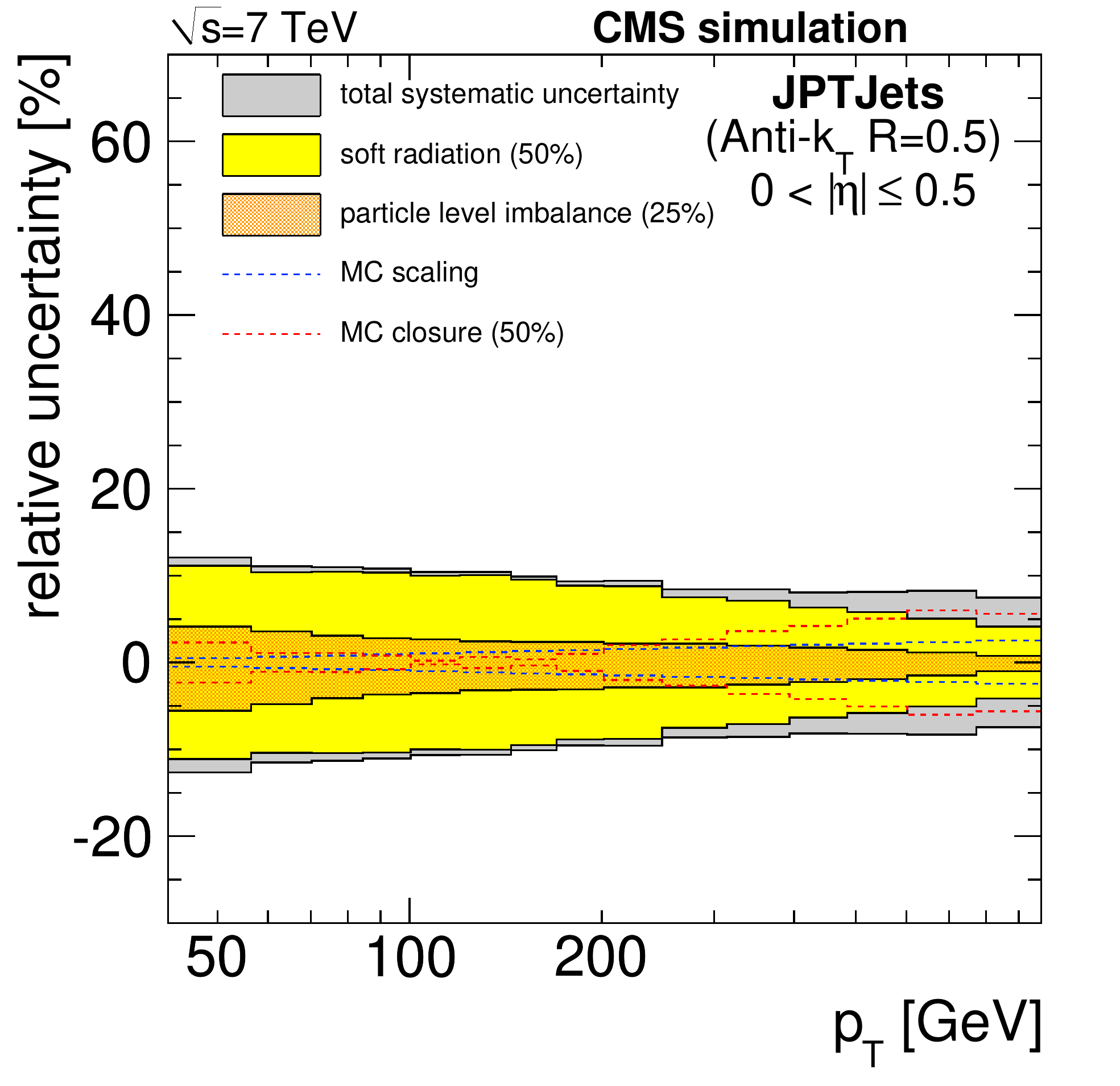}
\includegraphics[width=0.45\textwidth]{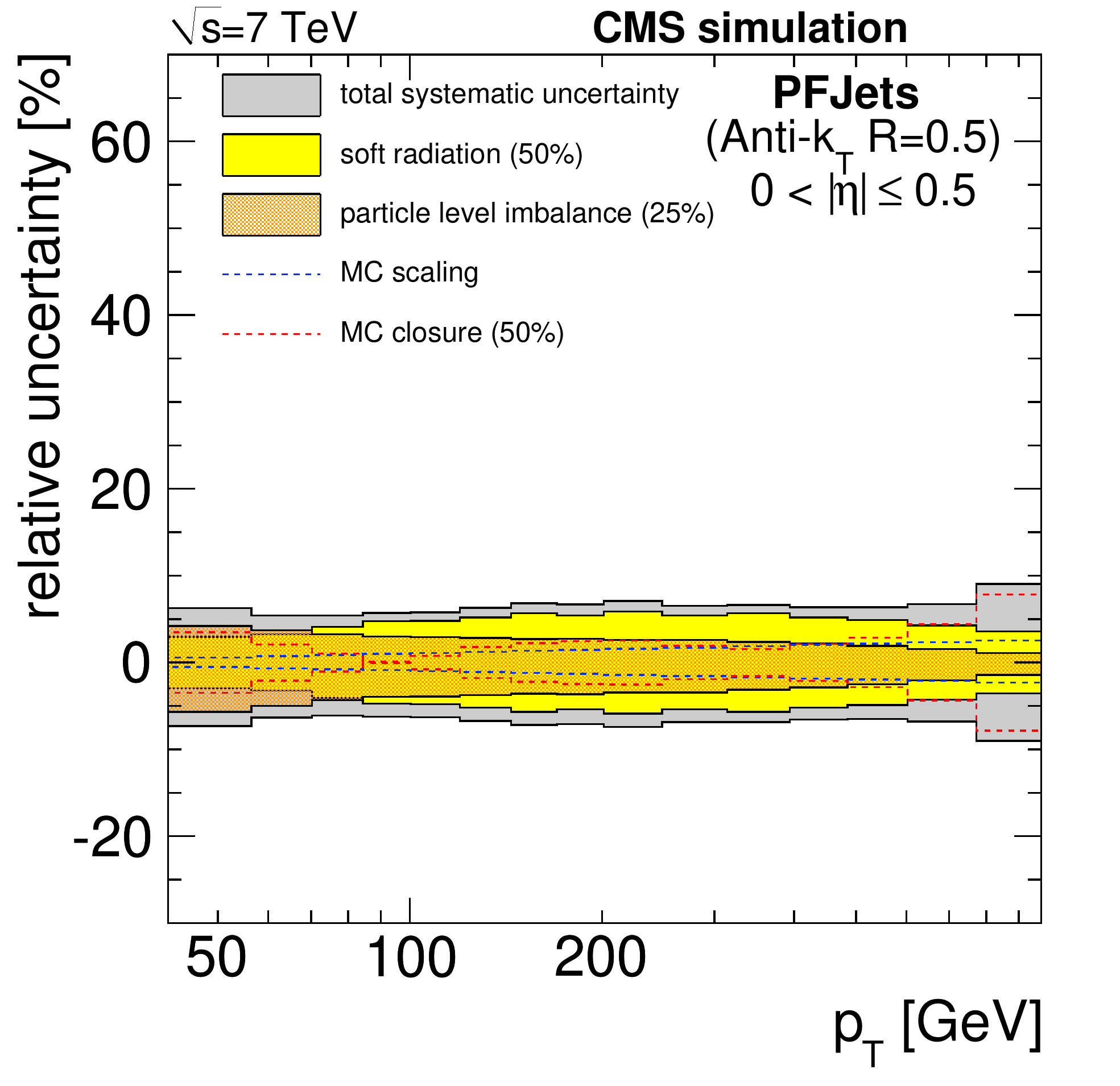}
  \caption{Relative systematic uncertainty of the asymmetry method to simulated CALO (top left), JPT (top right), and PF jets (bottom) for $|\eta|<0.5$.}
  \label{fig:urel-mc-00to05}
\end{figure}

Performing the analysis on simulated events, we observe deviations (biases) from the obtained and expected values, referred to as ``MC
closure  residuals''. A conservative $50\%$ of the MC closure residuals
$\frac{\mathrm{MC}(\mathrm{generator-level})-\mathrm{MC}(\mathrm{asymmetry})}{\mathrm{MC}(\mathrm{asymmetry})}$
is taken as an additional relative systematic uncertainty, corresponding to the bias correction. By comparing the asymmetry measured in data with the expectation from MC simulations, an  additional constant term is fitted, describing the observed discrepancy  between data and simulation, as  described below. The statistical  uncertainty from the  fit of the constant term is assigned as a systematic uncertainty. Figure~\ref{fig:urel-mc-00to05} shows the size of the different systematic uncertainties as a function of \ptave and for a central $\eta$ bin, for the three jets types. The particle-level imbalance uncertainty is shown in opaque orange, the solid yellow contribution corresponds to the uncertainty from the soft radiation variation, and the dashed-red line depicts the impact from the remaining differences in the MC closure. The relative uncertainty due to particle-level imbalance is larger for JPT and PF jets than for CALO jets because the absolute values of the raw resolutions are significantly smaller for JPT and PF, and thus more sensitive to the imbalance subtraction, than in the CALO jet case. The dashed blue line shows the contribution of the uncertainty on the additional constant term. The total systematic uncertainty for each resolution measurement is obtained by summing all individual components in quadrature, and is represented by the grey filled area in Fig.~\ref{fig:urel-mc-00to05}. The sensitivity of the method to the presence of additional collisions due to pile-up has been assessed by applying the measurement to the subsample of the data where exactly one primary vertex candidate is reconstructed, and no significant deviations from the inclusive measurement are observed.

The presented measurements of the jet \pt resolution, obtained by applying
the asymmetry method to data, yield systematically poorer resolution
compared  to the simulation. This discrepancy is quantified by taking the
fits to the MC asymmetry results, fixing all parameters, and adding in quadrature an additional constant term, as the only free parameter in a subsequent fit to the data asymmetry. The fitted additional constant term provides a good characterization of the discrepancy, which was verified by several closure tests based on MC. A likely source of the discrepancy is an imperfect intercalibration of the CMS calorimeters, which affects analyses based on the corresponding datasets.

\begin{figure}[h]
  \centering
\includegraphics[width=0.45\textwidth]{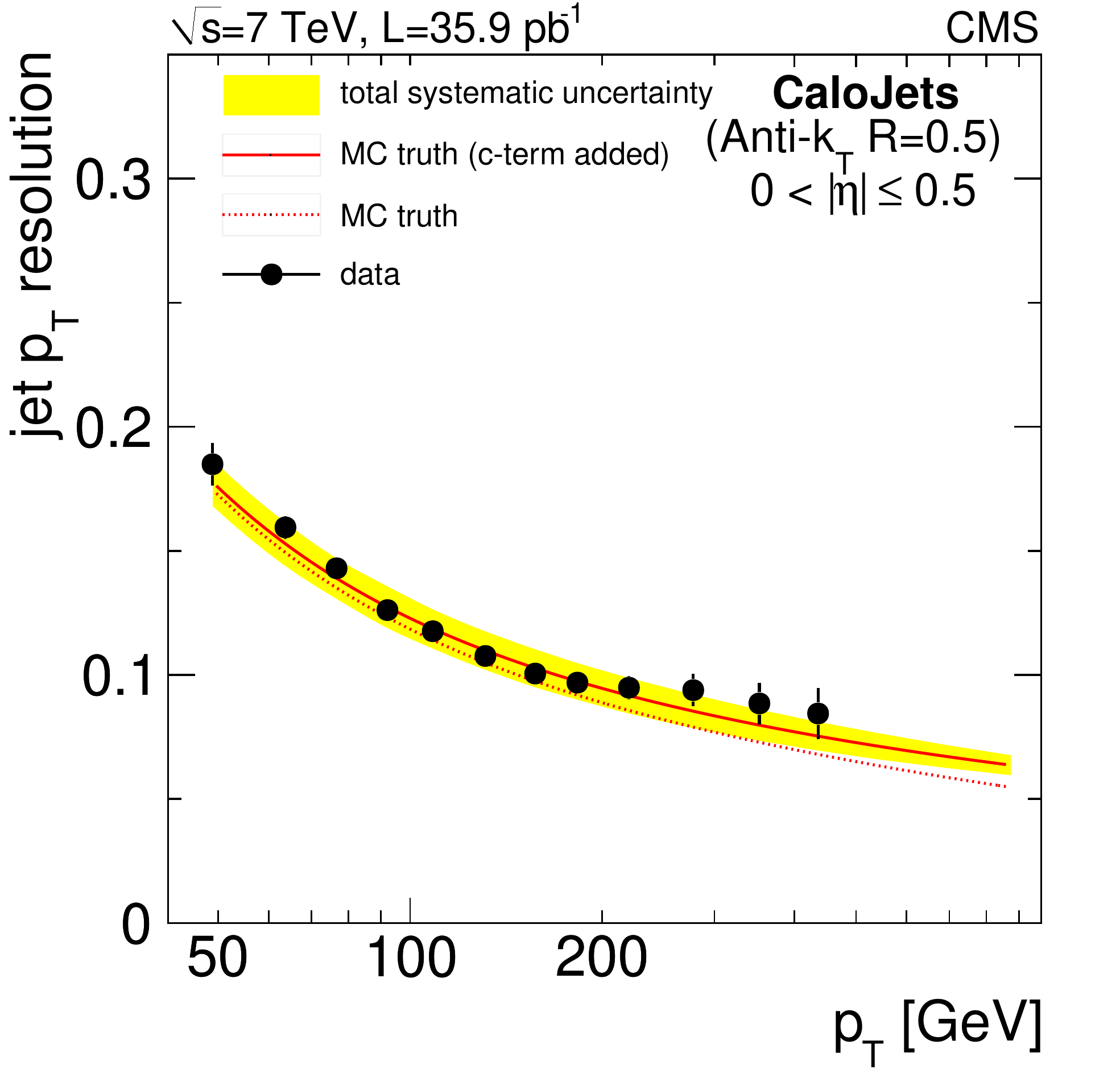}
\includegraphics[width=0.45\textwidth]{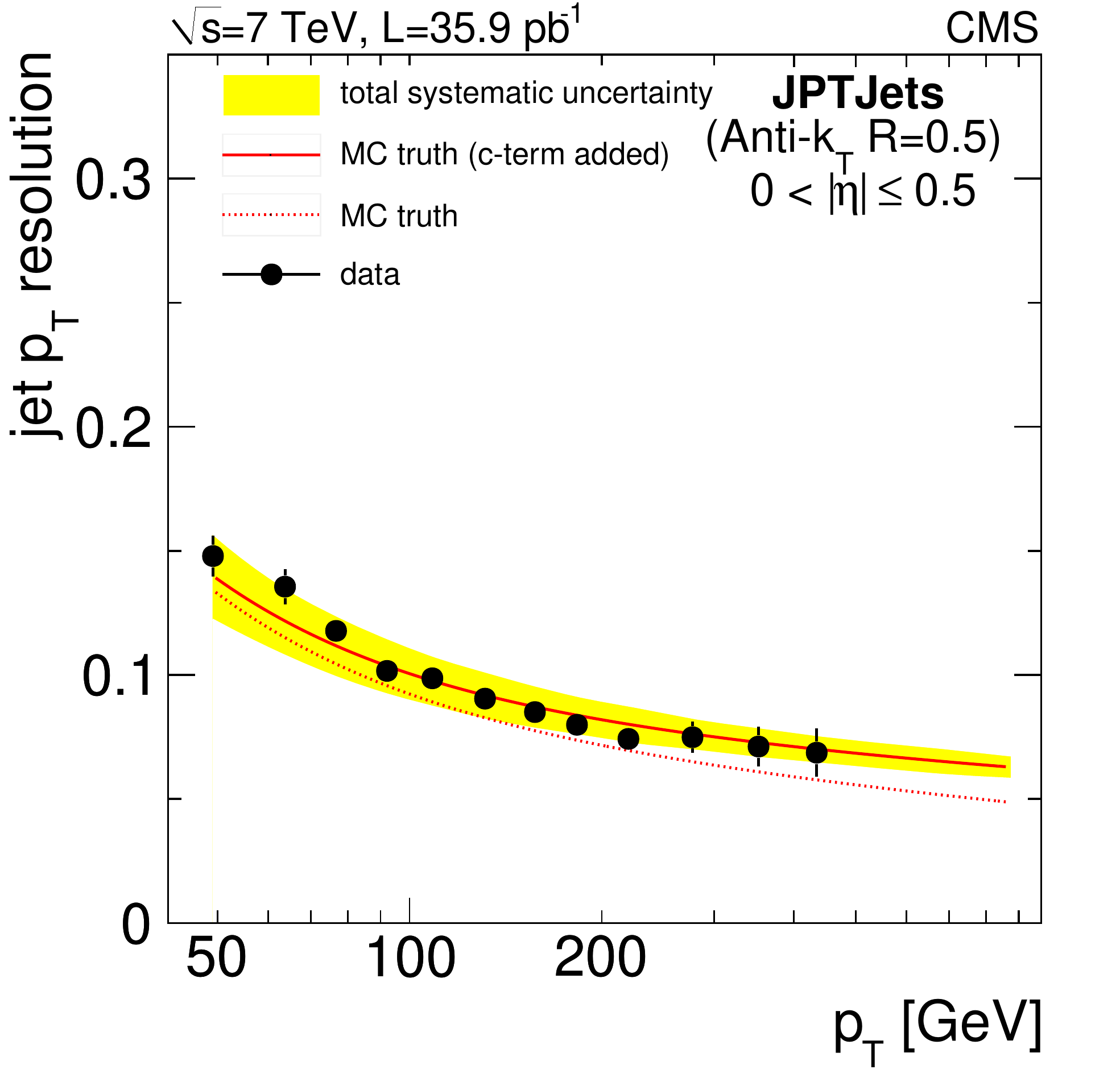}
\includegraphics[width=0.45\textwidth]{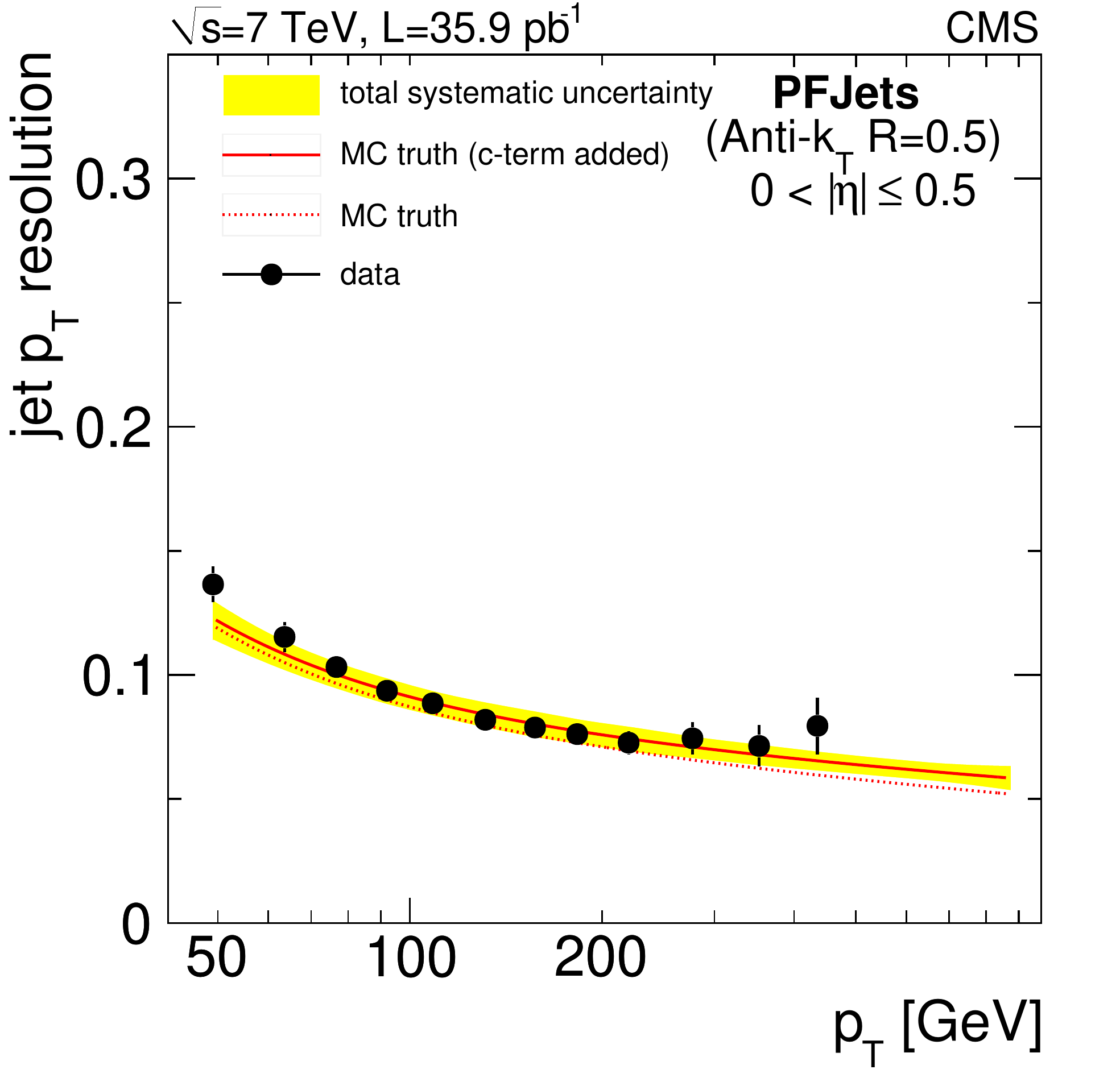}
  \caption{Bias-corrected data measurements, compared to the generator-level MC (denoted as MC-truth) \pt resolution before (red-dashed line) and after correction for the measured discrepancy between data and simulation (red-solid line) for CALO (top left), JPT (top right), and PF jets (bottom) in $|\eta|<0.5$.}
  \label{fig:uabs-mcdata-00to05}
\end{figure}

\begin{figure}[h]
  \centering

\includegraphics[width=0.45\textwidth]{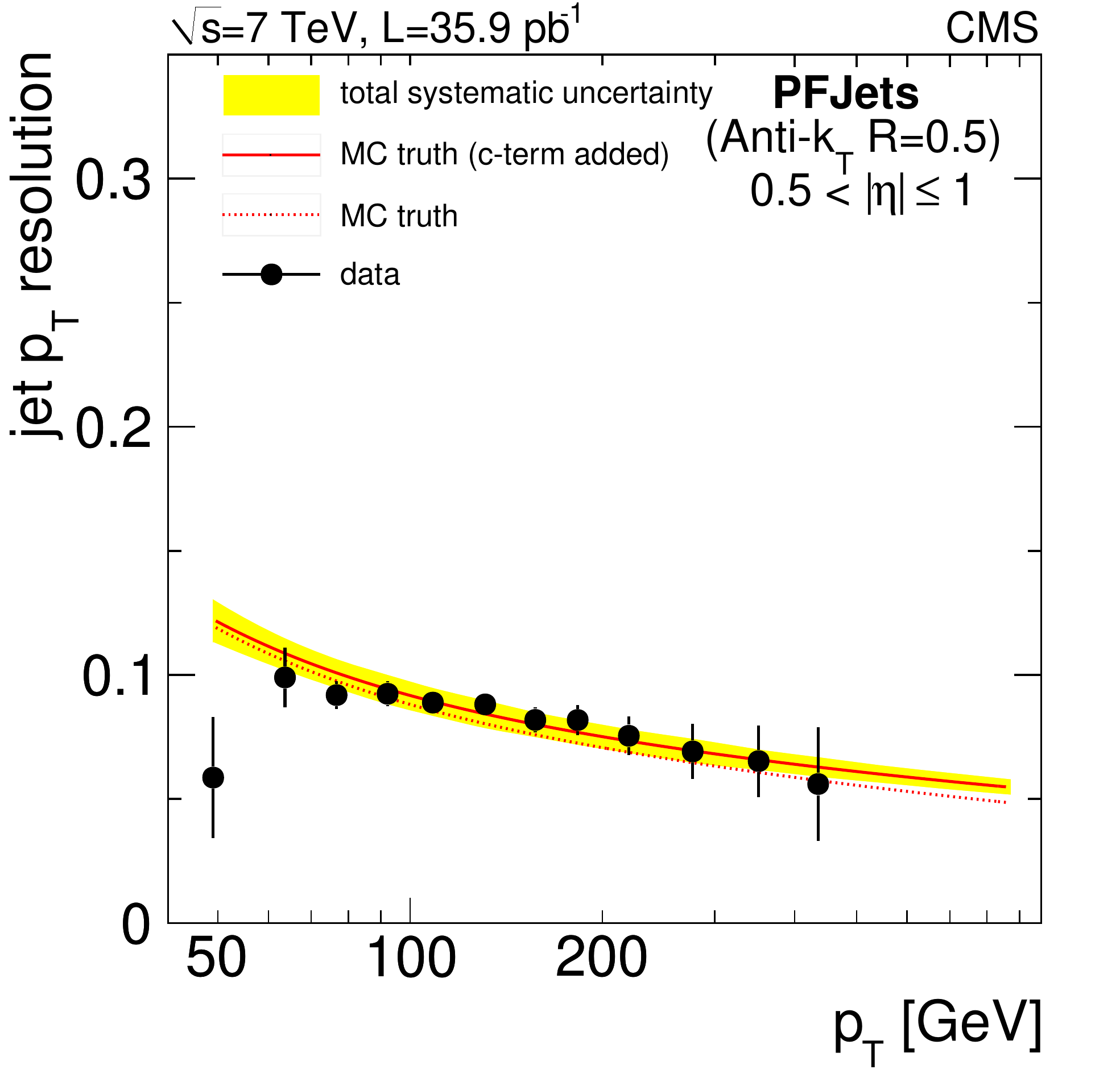}
\includegraphics[width=0.45\textwidth]{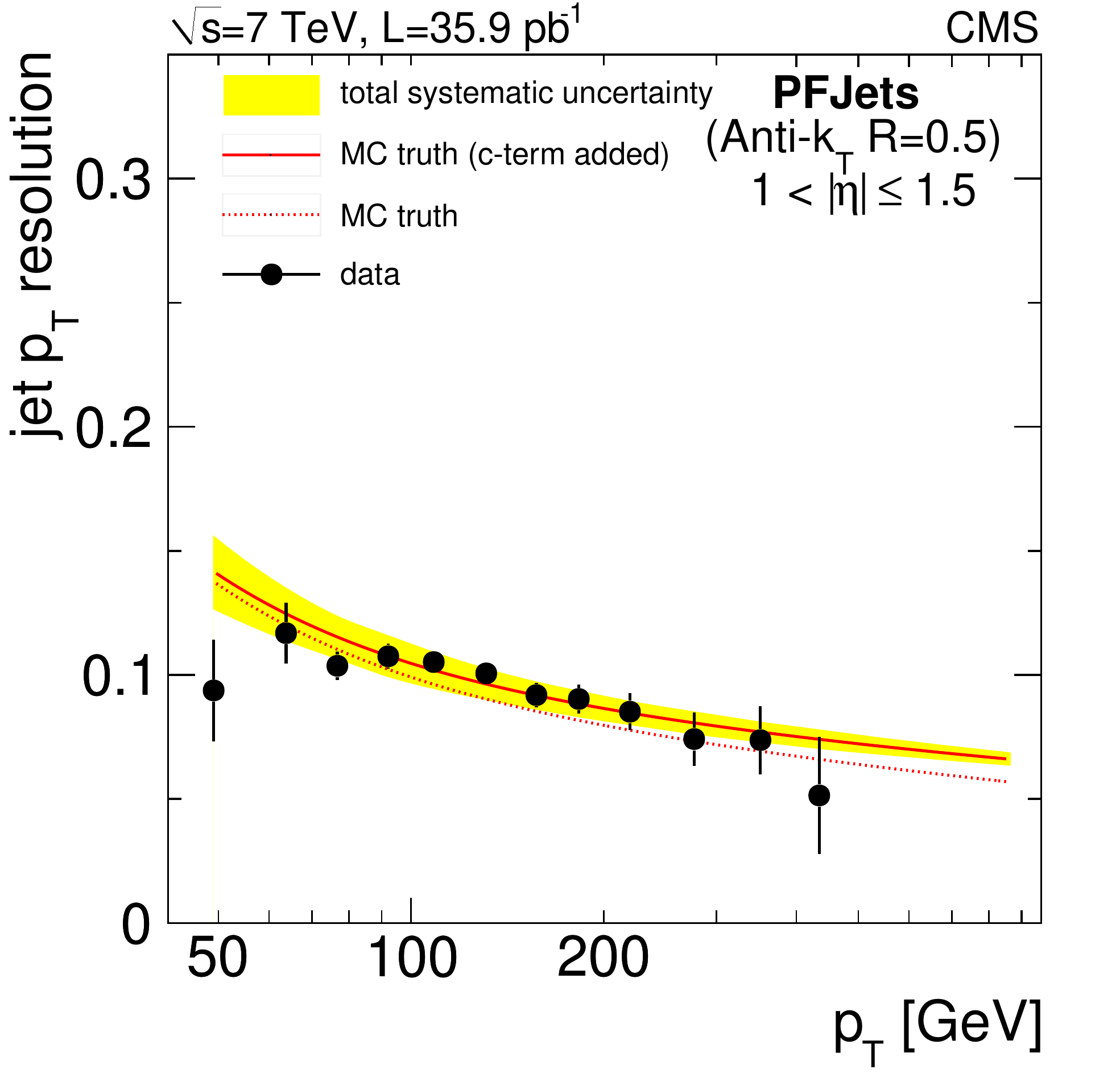}
\includegraphics[width=0.45\textwidth]{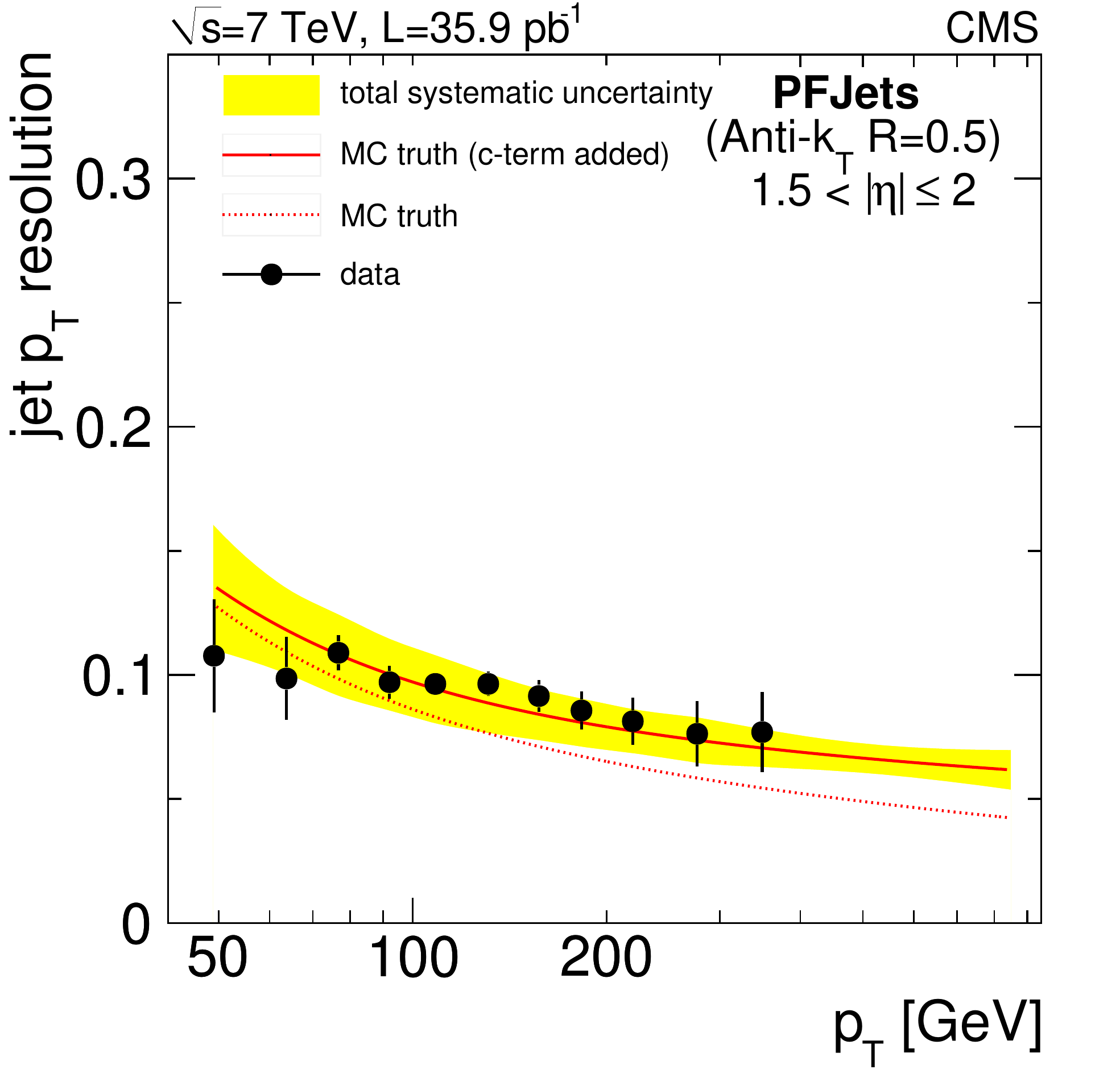}
\includegraphics[width=0.45\textwidth]{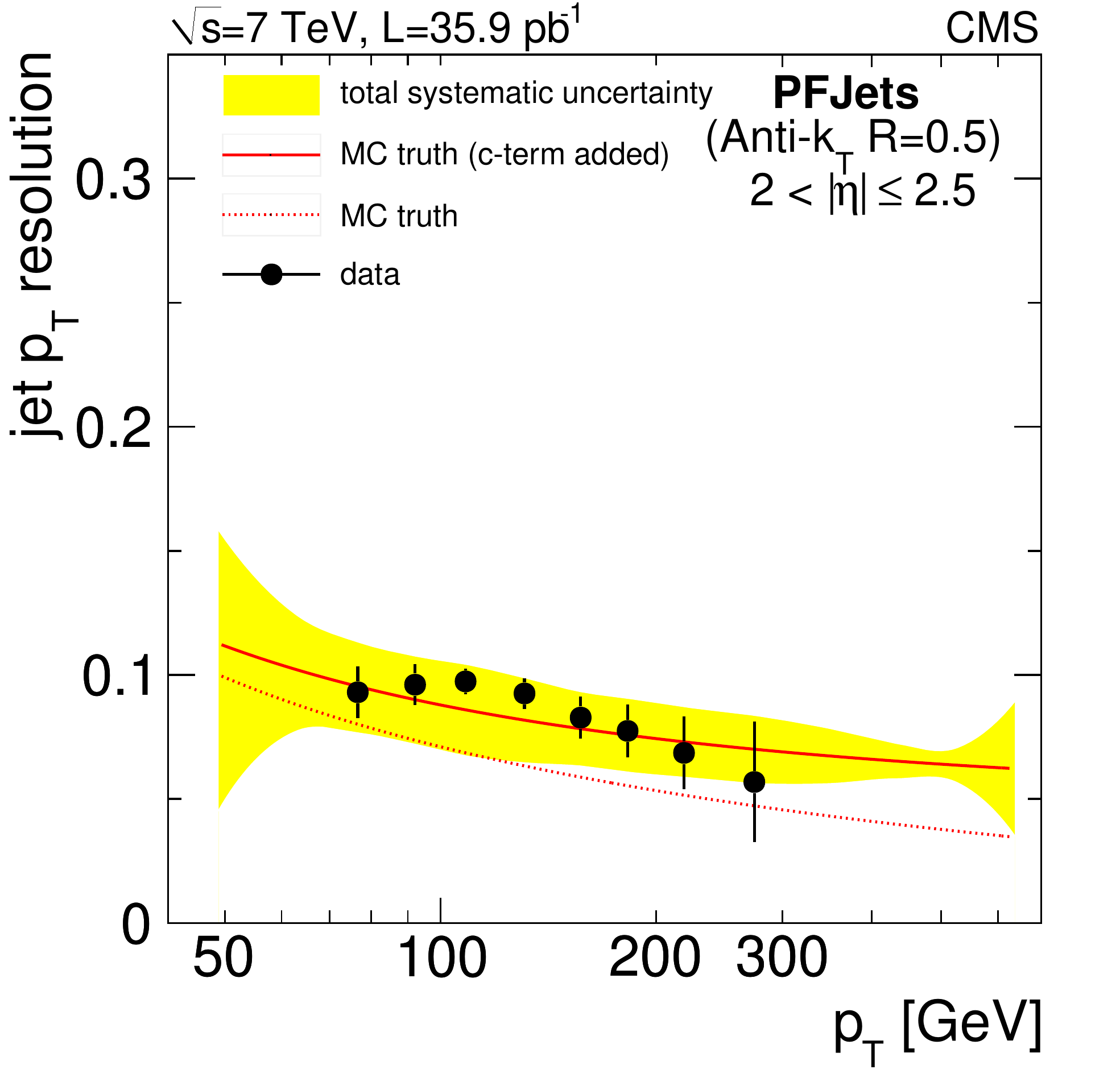}
\includegraphics[width=0.45\textwidth]{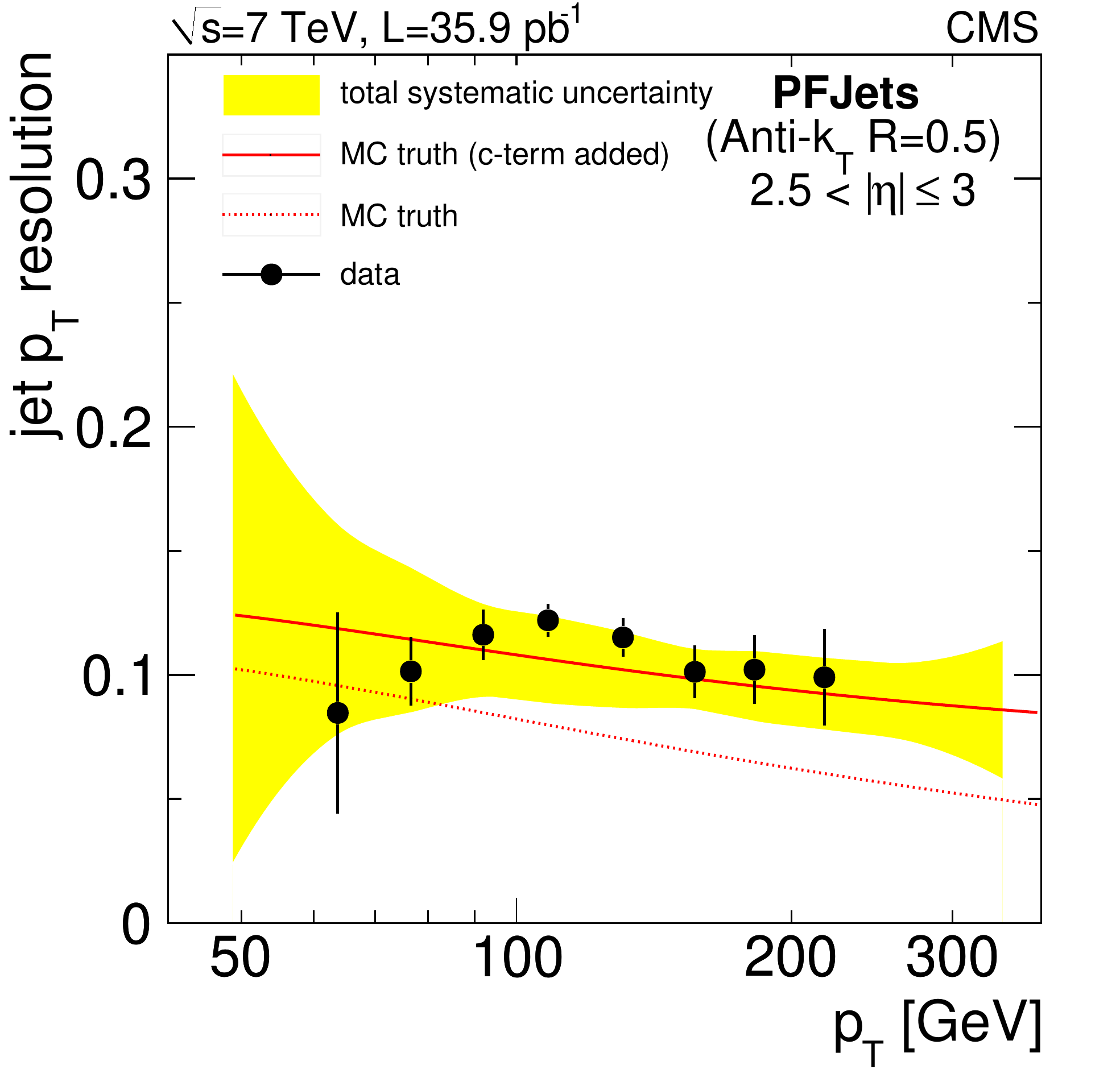}
\includegraphics[width=0.45\textwidth]{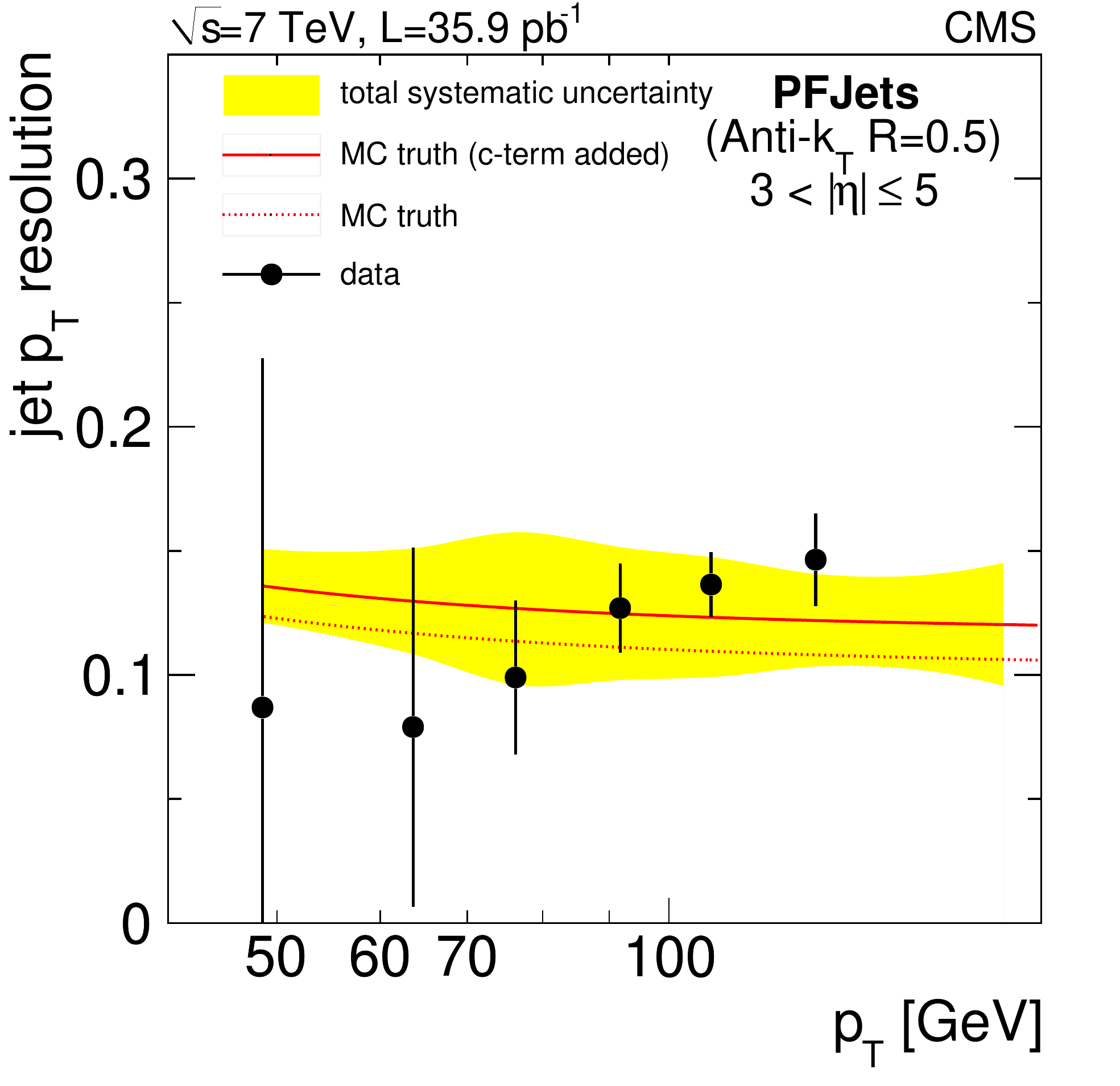}
  \caption{Bias-corrected data measurements, compared to the generator-level MC (denoted as MC-truth) resolution before (red-dashed line) and after correction for the measured discrepancy between data and simulation (red-solid line), compared to data, for PF jets in different $\eta$ ranges.}
  \label{fig:uabs-mcdata-05to50}
\end{figure}

The final results are presented in Figs.~\ref{fig:uabs-mcdata-00to05} (for all three types of jets, in the central region) and \ref{fig:uabs-mcdata-05to50} (for PF jets in all remaining $\eta$ bins). In each case, the solid red line depicts the resolution from generator-level MC, corrected for the measured discrepancy between data and simulation (constant term), and represents the best estimate of the jet \pt resolution in data. Consequently, it is central to the total systematic uncertainty band, drawn in yellow. The uncorrected generator-level MC resolution is shown as a red-dashed line for reference. The black dots are the bias-corrected data measurements, which are found to be in good agreement with the discrepancy-corrected  generator-level MC, within the statistical and systematic uncertainties. Note in particular that the agreement with the uncorrected generator-level MC resolution is considerably worse.

The dijet data are also investigated within the framework of the unbinned likelihood fit to the jet \pt resolution parameterization. This approach is developed in order to provide a cross-check of the results. It also serves  as a tool for the determination of the full jet \pt resolution  function, once larger collider data samples become available. This method directly takes into account biases in the event selection caused by the jet \pt resolution and the steeply falling jet \pt spectrum.

At the present stage, the jet \pt probability densities are approximated
by a truncated Gaussian, providing direct correspondence with the binned
fits discussed above. The resulting determination of the widths of the jet
\pt resolution (as function of \pt and $\eta$) is also affected by the
soft-radiation and hadronization (out-of-cone) effects. The fitted
resolution values are thus extrapolated to zero-radiation activity. The MC
particle-level imbalance is subtracted in quadrature to correct for
effects of hadronization. The method is applied to both data and MC, and
the results are consistent with the previously discussed binned fits to
the asymmetry distributions. Namely, poorer resolutions are observed in
data compared to the simulation.

\newpage
\clearpage 
\subsection{$\gamma$ + Jet Measurements}
\label{sec:res_gjet}

As for dijets, the measurement of the jet \pt resolution using the $\gamma+$jet \pt-balancing, involves an extrapolation of the event 
topology  to the ideal case of zero secondary hadronic activity, as described in detail in Section~\ref{sec:radbias}. To measure the jet \pt resolution from data, the observable $\sigma(\pt^{Jet}/\pt^{\gamma})$ is expressed as: 

\begin{equation}
   \sigma_{total}(\pt^{Jet}/\pt^{\gamma}) =  \sigma_{intrinsic}(\pt^{Jet}/\pt^{gen}) \oplus \sigma_{imbalance}(\pt^{gen}/\pt^{\gamma}),
\end{equation}

where the first term $\sigma_{intrinsic}(\pt^{Jet}/\pt^{gen})$ is the intrinsic (generator-level MC) resolution of interest. The second term $\sigma_{imbalance}(\pt^{gen}/\pt^{\gamma})$ is the ``imbalance'' term, arising from the presence of secondary jets in an event and from other effects, such as hadronization. The effect of extra jet activity is studied as a function of $\pt^{Jet2}/\pt^{\gamma}$. 

The jet \pt resolution is measured using two methods. The ``direct'' method measures the \pt resolution separately for data and MC, while the ``ratio'' method is specialized for the data/MC ratio. In the direct method, the intrinsic resolution is  taken to be independent of $\pt^{Jet2}/\pt^{\gamma}$, while the width of the imbalance is assumed to be a first-order polynomial in the 
$\pt^{Jet2}/\pt^{\gamma}$ fraction. The intercept ``$q$'' gives the limit of zero secondary jet activity and the parameter ``$m$'' describes the soft-radiation effects:

\begin{equation}
   \sigma_{intrinsic}(\pt^{Jet2}/\pt^{\gamma}) = c,
\end{equation}

\begin{equation}\label{eq:imb}
   \sigma_{imbalance} (\pt^{Jet2}/\pt^{\gamma}) = q + m \cdot \pt^{Jet2}/\pt^{\gamma},
\end{equation}

\begin{equation} \label{eq:totalRes}
   \sigma_{total} (\pt^{Jet2}/\pt^{\gamma}) = \sqrt{(c^2 + q^2 + 2qm \pt^{Jet2}/\pt^{\gamma}+m^2(\pt^{Jet2}/\pt^{\gamma})^2)}\,.
\end{equation}

In order to correct for the particle-level imbalance, the measured \pt resolution, using reconstructed jets and photons, is fitted with the functional form in Eq.~(\ref{eq:totalRes}), while keeping the parameter $q$ fixed to the value obtained from the fit to the imbalance in MC, using the functional form in Eq.~(~\ref{eq:imb}). A non-zero value of the intercept $q$ is expected, which represents the irreducible imbalance due to, e.g. hadronization and photon resolution effects. The parameter $c$ is the measured intrinsic resolution from the reconstructed quantities. The same functional form is used in fits to both MC and data samples.

\begin{figure}[htbp]
\begin{center}
\includegraphics[angle=90,width=0.48\textwidth]{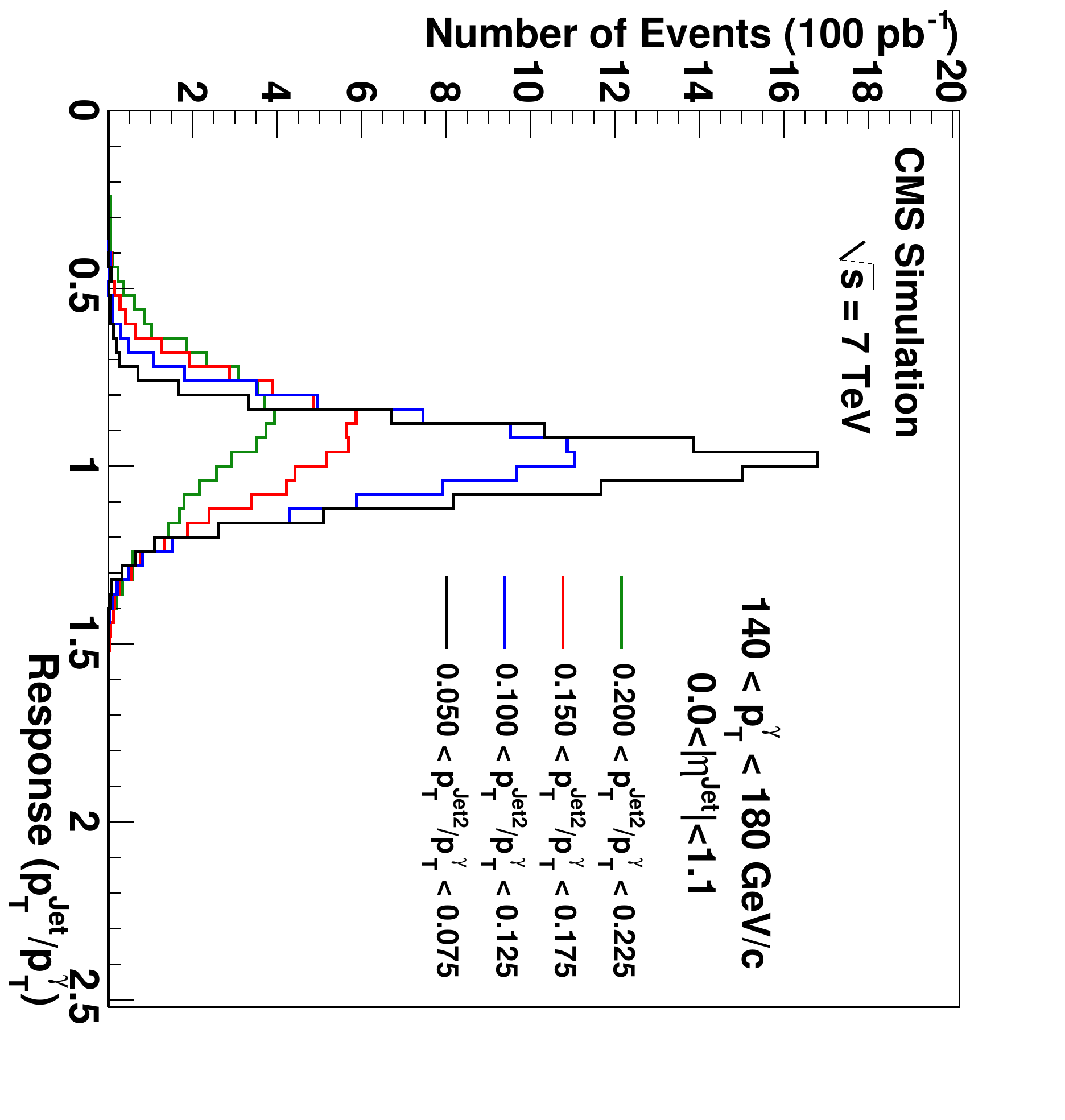}
\includegraphics[width=0.48\textwidth]{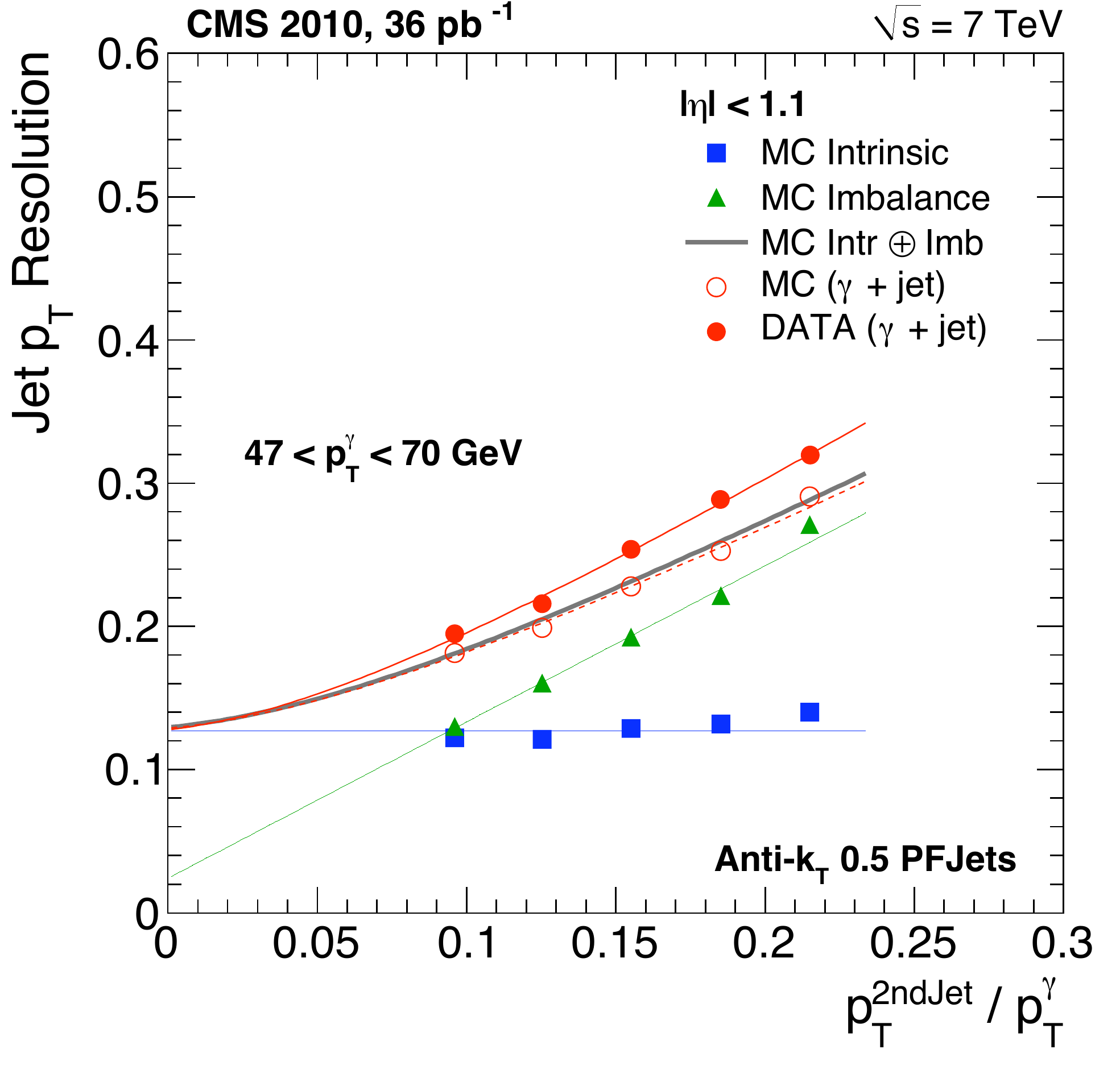}
\end{center}
\caption[]{Response distributions in different bins of $\pt^{Jet2}/\pt^{\gamma}$, and for $140<\pt^{\gamma}<180\GeV$ (left), components of the jet \pt resolution, as a function of $\pt^{Jet2}/\pt^{\gamma}$ for PF jets in data and MC samples (right).}
\label{fig:GenPtRecPhotRes}
\end{figure}

Figure~\ref{fig:GenPtRecPhotRes} (left) shows $\pt^{jet}/\pt^{\gamma}$ distributions, for different bins of $\pt^{Jet2}/\pt^{\gamma}$, in MC simulation. The effect of tightening the secondary jet activity is clearly visible as a narrowing of the spread of the  distributions. The intrinsic resolution on the other hand, is independent of any other activity in the event. The evolution of the 
RMS of the \pt-balance distribution in bins of $\pt^{Jet2}/\pt^{\gamma}$ is shown in Fig.~\ref{fig:GenPtRecPhotRes} (right) as the red circular points, for both data and MC samples. The blue square points show the intrinsic resolution of PF  jets in MC simulation, measured as the RMS of the $\pt^{PFJet}/\pt^{gen}$. The grey thicker line shows the total expected resolution defined as the quadrature sum of intrinsic (blue square points) and imbalance component (green triangle points), as a function of $\pt^{Jet2}/\pt^{\gamma}$. The dotted-red line is the result of the fit, applied to MC simulation, and the thinner solid-red line is the functional fit to the data measurement.

The $\gamma+$jet sample has significant contamination from QCD dijet background with a jet misidentified as a photon. To pass the photon isolation and cluster shape requirements, such a jet must be composed of a leading $\pi^0$ (with $\pi^0 \to\gamma\gamma$) accompanied by a very low hadronic activity. These misidentified ``photons'' have energy-scale similar to the genuine prompt photons, a good energy resolution, and can therefore serve as valid reference objects for this analysis. In the selected photon sample, the presence of a jet is required, with $|\eta|<1.1$, recoiling against the photon candidate in azimuth within $\Delta \phi > 2\pi/3$. 

\begin{figure}[t]
\centering
\includegraphics[width=0.49\textwidth]{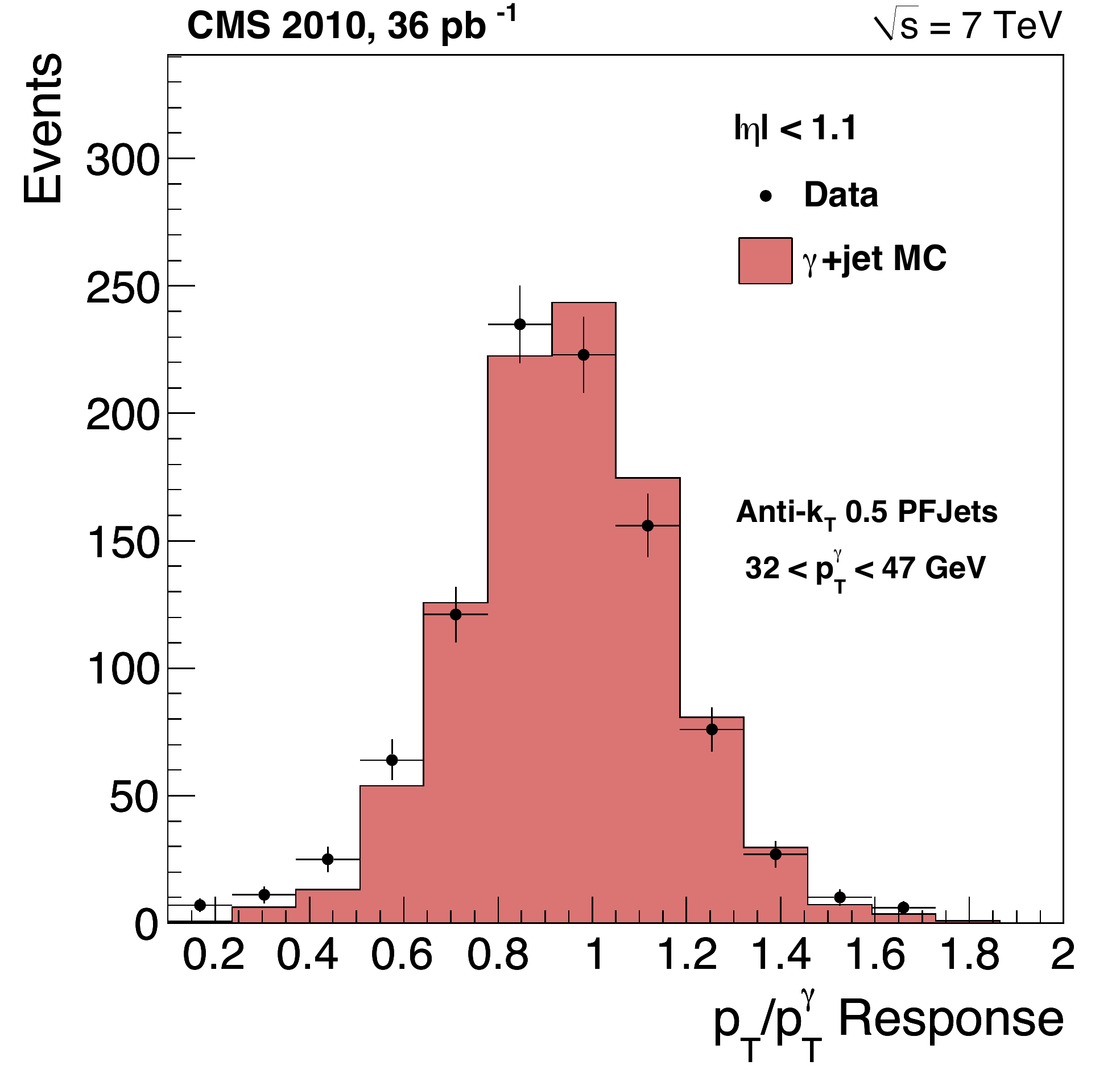}
\includegraphics[width=0.49\textwidth]{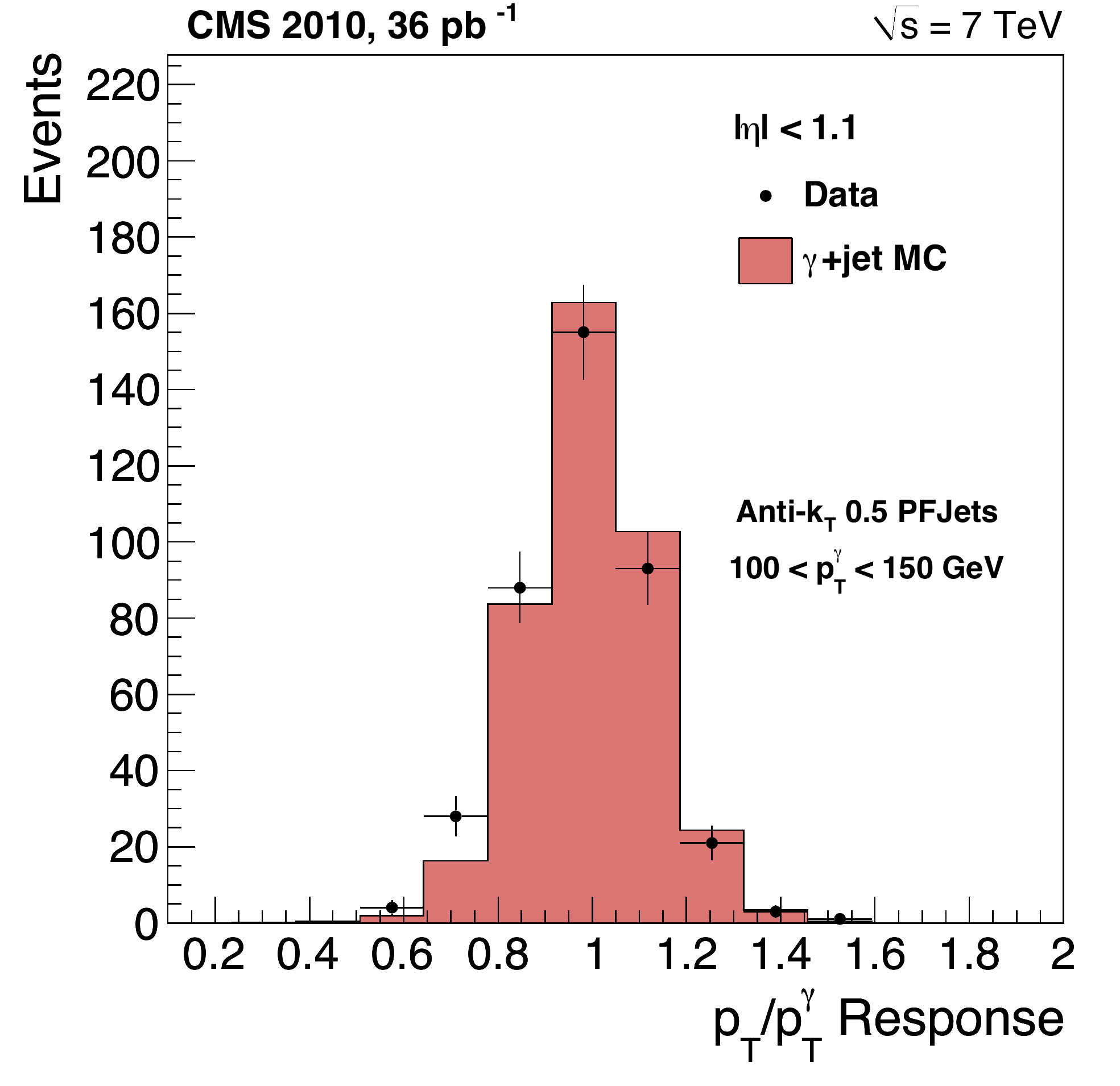}
\caption{Distributions of $\pt/\pt^{\gamma}$ in data and MC sample for PF jets in two representative $\pt^{\gamma}$ bins.}
\label{fig:response}
\end{figure}

Distributions of the $\pt/\pt^{\gamma}$ variable for PF jets are shown in Fig.~\ref{fig:response} for data and MC. The distributions are not centered around the response of 1.0, because of the impact of soft radiation, as illustrated in Fig.~\ref{fig:GenPtRecPhotRes}.

\begin{figure}[t]
\centering
\includegraphics[width=0.45\textwidth]{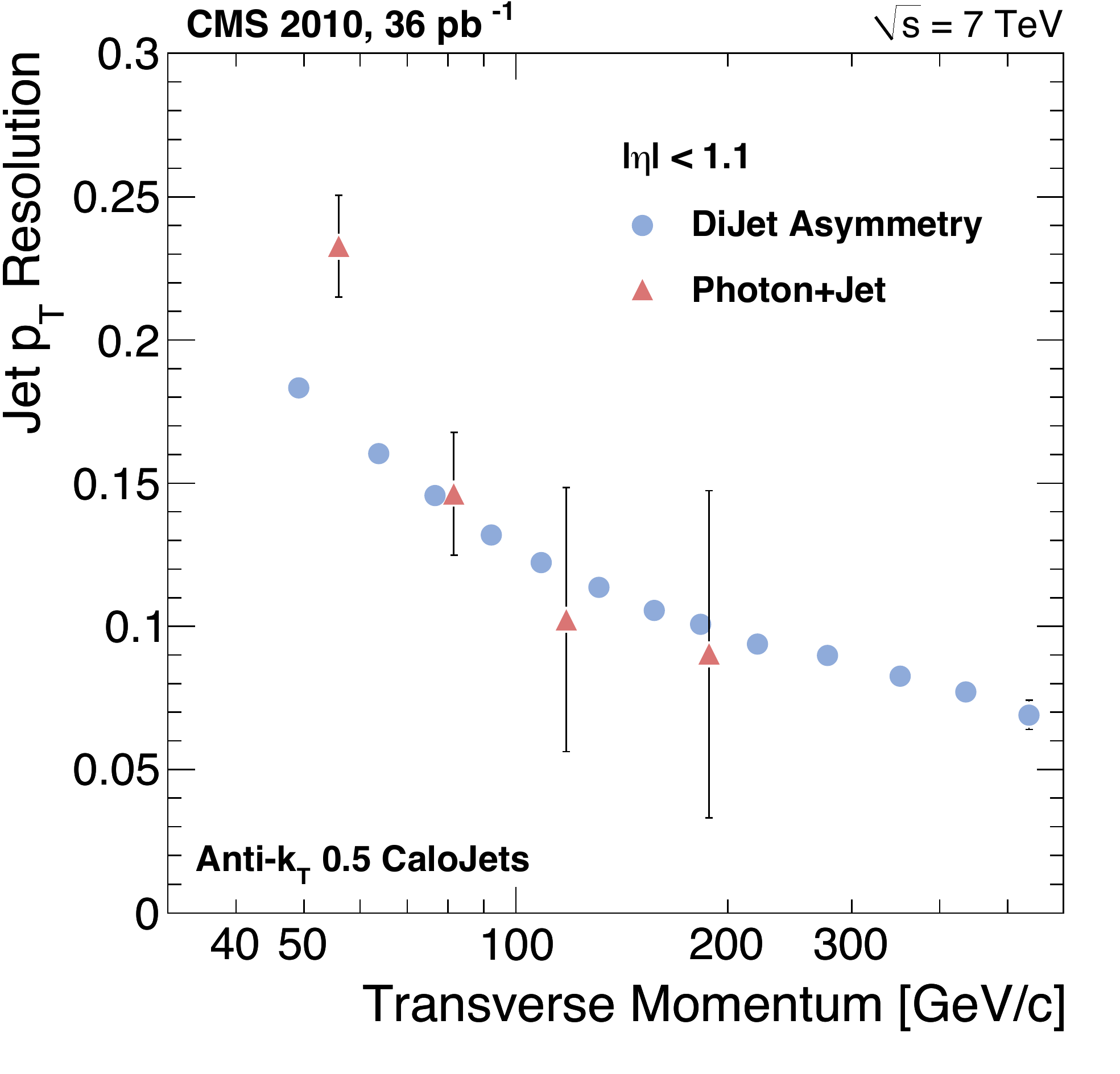}
\includegraphics[width=0.45\textwidth]{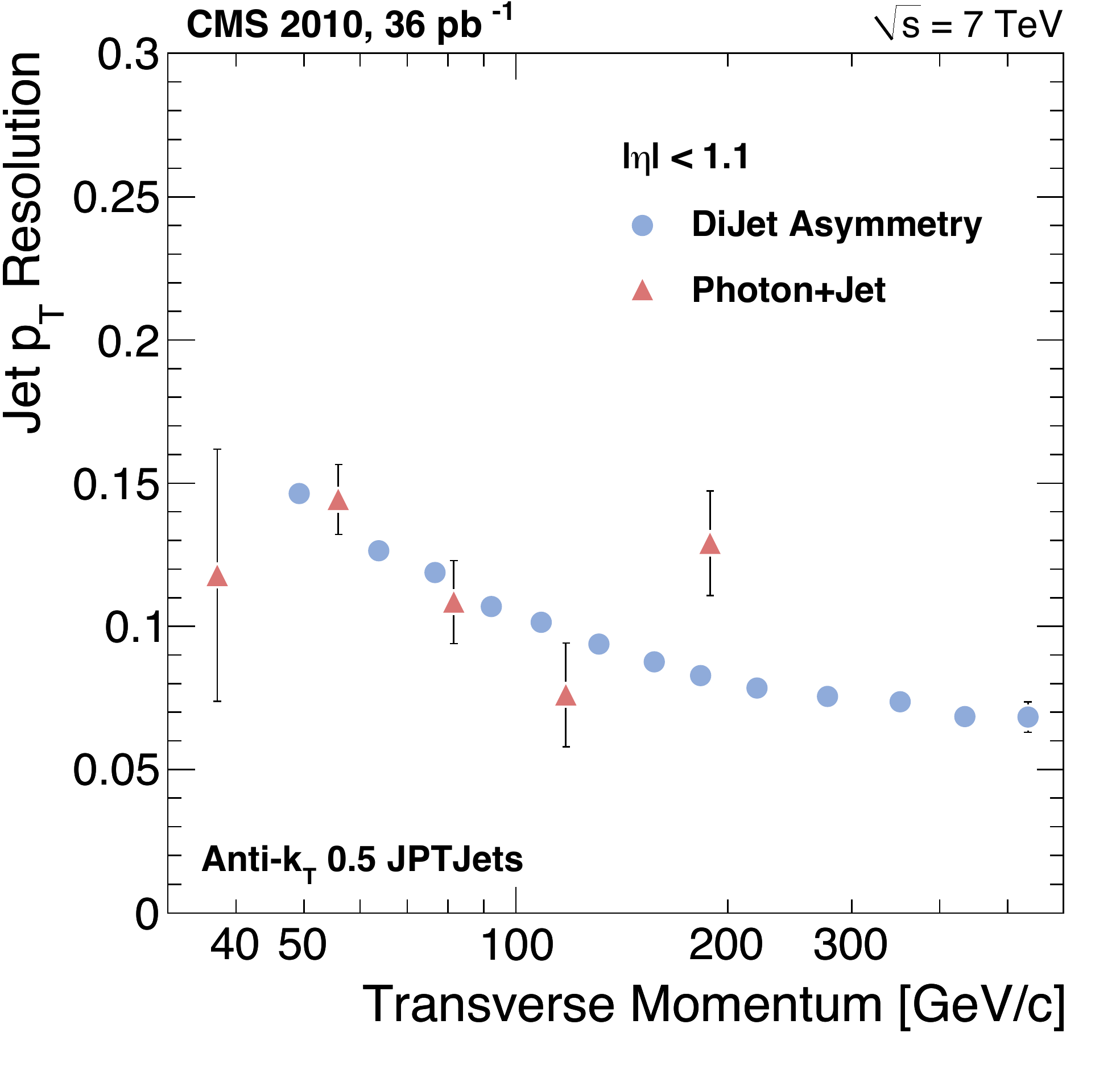}
\includegraphics[width=0.45\textwidth]{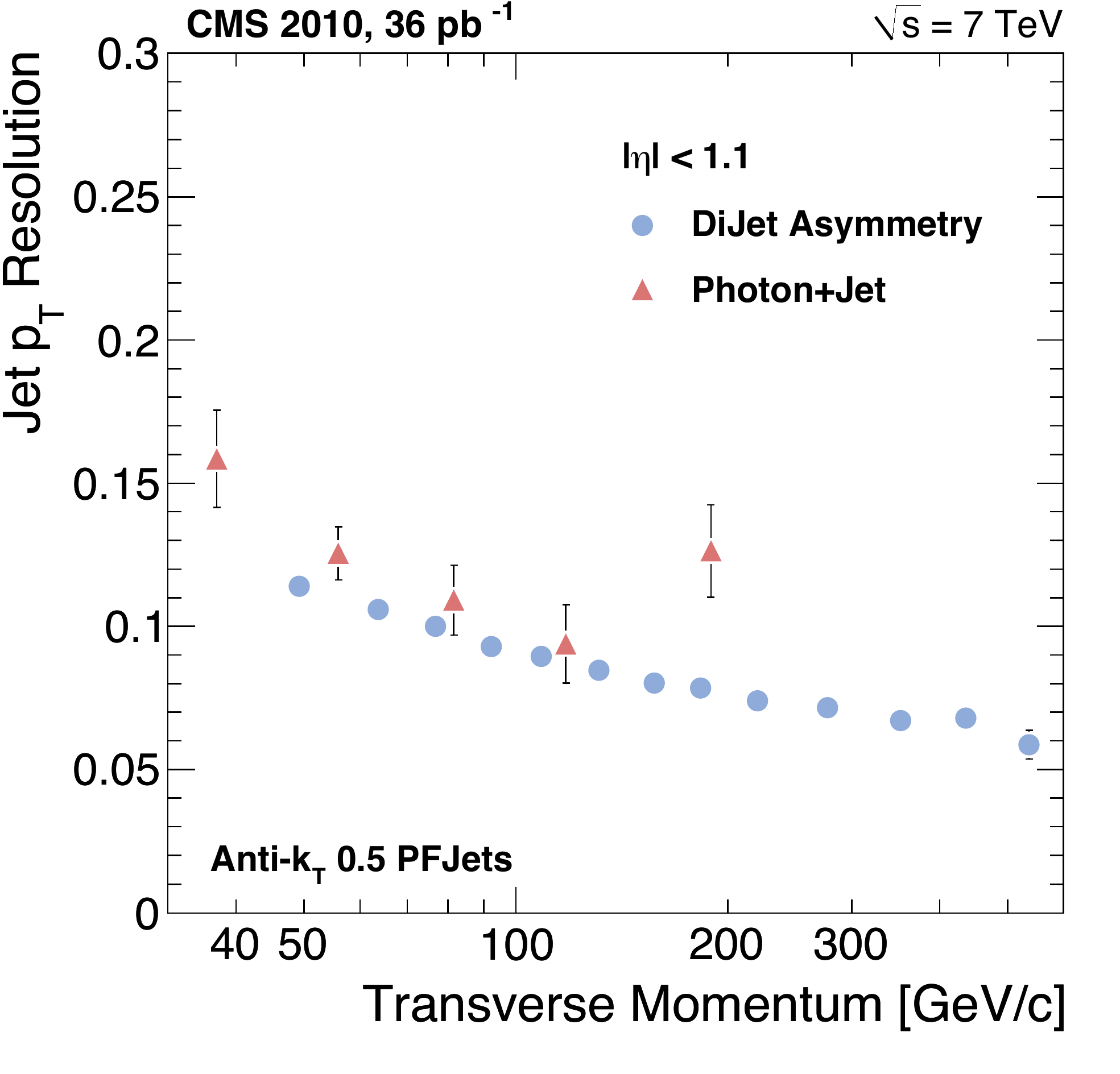}
\caption{Jet \pt resolutions from $\gamma+$jet (red triangular points) and dijet asymmetry (blue circular points) measurements, for CALO (top left), JPT (top right), and PF jets (bottom).}
\label{fig:gammajet_dijet}
\end{figure}

The jet \pt resolution obtained from the $\gamma+$jet analysis, 
provides a cross-check of the dijet asymmetry results, and a 
reasonable agreement is observed between the two measurements, as 
shown in Fig.~\ref{fig:gammajet_dijet}. At the current level of statistics, 
the $\gamma+$jet method  yields poorer resolutions for $\pt>150\GeV$.

The complementary ``ratio'' method is based on taking the ratio of the data and MC intrinsic resolutions versus $\pt^{Jet2}/\pt^{\gamma}$ before the extrapolation. The intrinsic resolutions are first derived in data and MC, by subtracting in quadrature, from the total measured resolutions, the imbalance predicted in the simulation. The strength of the method is that the extrapolation fit is performed only once, and that the fitted observable, as estimator of the ratio of data and MC intrinsic resolutions, is expected to be a constant function of $\pt^{Jet2}/\pt^{\gamma}$. The intrinsic resolution derived in data is confirmed to be flat vs. $\pt^{Jet2}/\pt^{\gamma}$, as expected, providing a test of the procedure. Any deviation from a flat dependence would have indicated a limitation of the simulation to model the imbalance in the data. The results for the ratio are consistent with the constant $\pt^{Jet2}/\pt^{\gamma}$ behaviour and therefore a simple constant fit has been performed in bins of \pt and $\eta$. Systematic uncertainties due to variation of the extrapolation fit range and the uncertainty of jet energy corrections have been evaluated, adding up to $\pm (3-4)\%$ for the ratio. An additional uncertainty of $\pm (2-4)\%$ is assigned to the ratio, due to the assumption that the MC simulation correctly models the imbalance in data. This  uncertainty is evaluated by varying the relevant features at generator level, namely, the modeling of hadronization, treatment of multiple-parton final states and modeling of the $k_T$-kick. The direct and ratio methods are consistent with each other within uncertainties. The statistical errors on the results from the ratio method are smaller than 
from the direct method, because of the fact that in the ratio method the imbalance is fixed to the MC-based result, while in  the direct method the parameter $m$, describing the imbalance part of the resolution, is free in the fits to data and MC samples.

\begin{figure}[htbp]
\begin{center}
\includegraphics[width=0.49\textwidth]{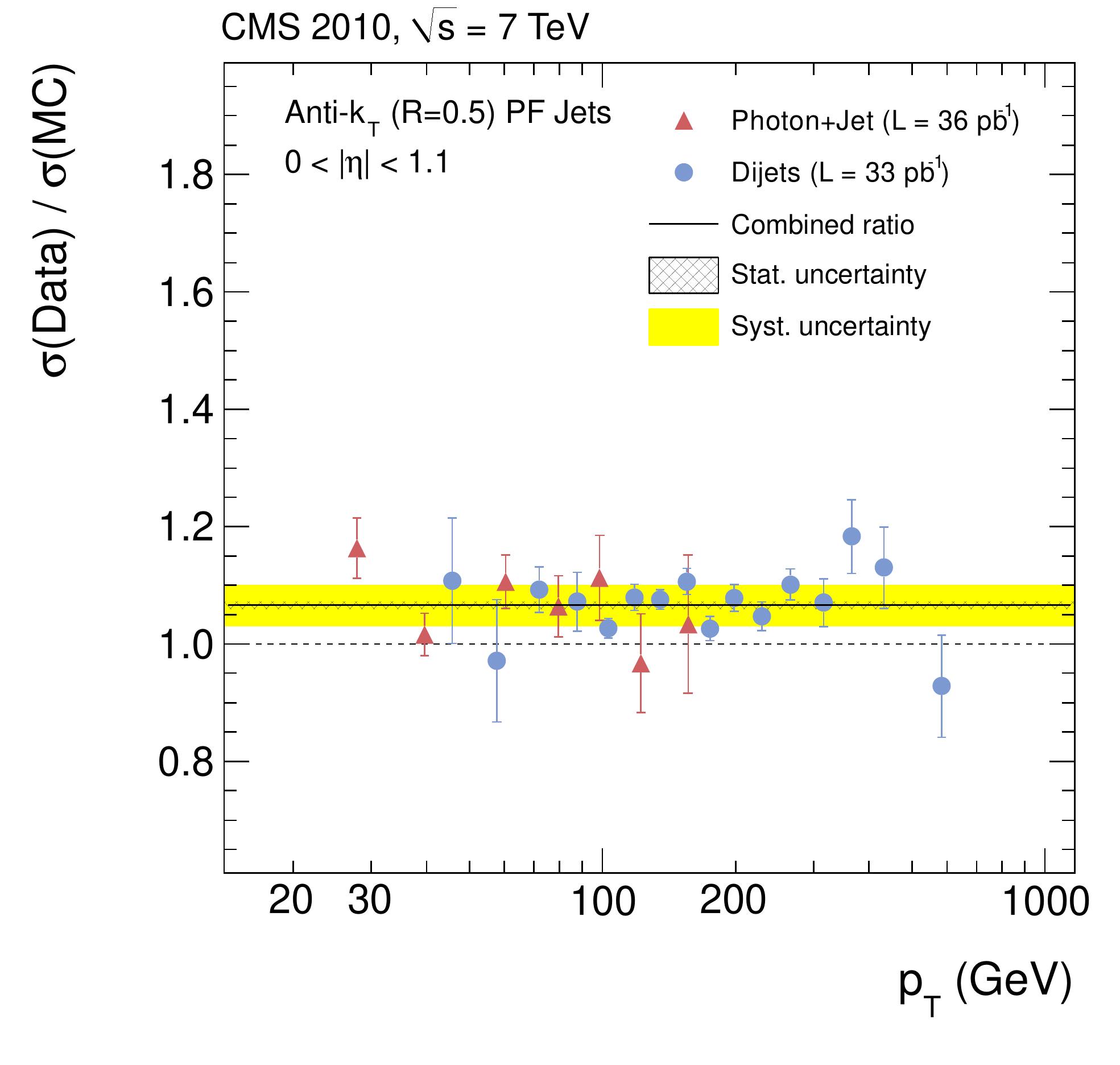}
\includegraphics[width=0.49\textwidth]{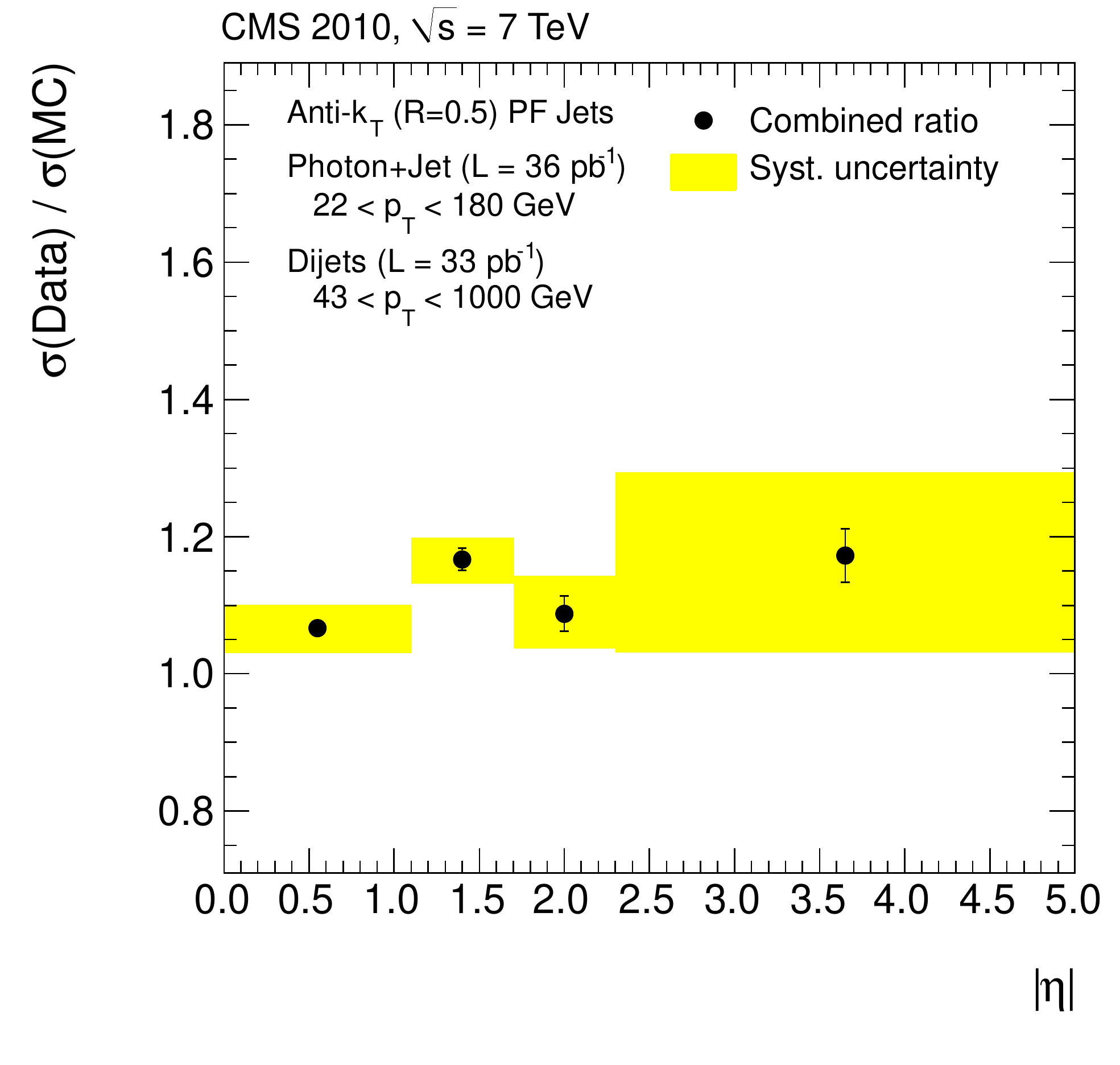}
\end{center}
\caption[]{The ratio of jet \pt resolutions in data and MC samples vs. \pt, in $|\eta|<1.1$, from dijet and $\gamma+$jet samples; a combination fit to both  data sets is also shown (left). Results from the combination  fits vs. \pt in various $\eta$ ranges (right).}
\label{fig:ResolutionRatioSum}
\end{figure}

The results from the ratio method are compared in Fig.~\ref{fig:ResolutionRatioSum} (left) to the results from the dijet samples (using the unbinned likelihood fits), as a function of \pt, in the barrel region. Since the two samples have different jet flavour 
compositions, the generator-level MC resolutions from the corresponding MC samples  have been compared and verified that the difference is within $3\%$. The dependence of the data/MC ratio on the flavour difference is expected to be $\sim 1\%$ under a conservative assumption of $30\%$ uncertainty on the modeling of the flavour composition in the simulation.  This component is included in the comparison of the data/MC ratio between the dijet and $\gamma +$jet samples.

Therefore, the two results are compared directly, without applying additional jet composition corrections. Both sets of points are fully consistent and have been combined in fits, using a constant function of \pt for each $\eta$ range. The $\eta$ dependence of the measured intrinsic resolution ratios is also shown in Fig.~\ref{fig:ResolutionRatioSum} (right).

The results from the ratio method, as well as the results  from the dijet samples (using the unbinned likelihood fits), integrated over \pt and for all four different $|\eta|$ regions are illustrated in Table \ref{tab:ResFit:ScalingFactors}. As before, good agreement is observed between the two methods, within the statistical and systematic uncertainties. The quoted uncertainties for the dijet analysis combine the systematic uncertainties assigned to the extrapolation procedure,  the particle-level imbalance correction, the jet-energy-scale correction, and the particle-level dijet \pt spectrum. 

\begin{table}[ht]
  \caption{Ratios of the resolution measured in data and simulation for different jet $\eta$ ranges using the unbinned likelihood fits in dijet samples; the last column gives the ratio for PF jets from $\gamma$+jet samples for comparison. Stated are the statistical uncertainties from the fit and the upper and lower systematic uncertainties.}
  \begin{center}
\begin{tabular}[ht]{ccccc}
\hline
$|\eta|$ bin & Ratio CALO Jets & Ratio JPT Jets & Ratio PF Jets & Ratio PF Jets in $\gamma$+jet\\
\hline
\raisebox {0mm} [5mm] [3mm] {
0.0 -- 1.1} & $1.088 \pm 0.007 ^{+ 0.076}_{- 0.075}$ & $1.087 \pm 0.006^{+0.080}_{-0.078}$ &$1.066 \pm 0.007 ^{+
0.074}_{ - 0.072}$& $1.07\pm 0.020^{+0.024}_{- 0.033}$ \\
\raisebox {0mm} [3mm] [3mm] {
1.1 -- 1.7} & $1.139 \pm 0.019 ^{+ 0.084}_{ - 0.084}$ & $1.213 \pm 0.015^{+0.081}_{-0.080}$ &$1.191 \pm 0.019 ^{+
0.064}_{ - 0.062}$& $1.10 \pm 0.031^{+0.031}_{-0.039}$ \\
\raisebox {0mm} [3mm] [3mm] {
1.7 -- 2.3} & $1.082 \pm 0.030 ^{+ 0.140}_{ - 0.139}$ & $1.018 \pm 0.021^{+0.071}_{-0.071}$ &$1.096 \pm 0.030 ^{+
0.089}_{ - 0.085}$& $1.07 \pm 0.048^{+0.056}_{-0.047}$ \\
\raisebox {0mm} [3mm] [3mm] {
2.3 -- 5.0} & $1.065 \pm 0.042 ^{+ 0.237}_{ - 0.235}$ & $1.068 \pm 0.036^{+0.139}_{-0.139}$ &$1.166 \pm 0.050 ^{+
0.198}_{ - 0.199}$& $1.18 \pm 0.062^{+0.043}_{-0.072}$ \\
\hline
\end{tabular}
  \end{center}
  \label{tab:ResFit:ScalingFactors}
\end{table}

\subsection{Measurement of Jet Resolution Tails}
\label{sec:res_tails}

One of the most promising signatures of physics beyond the standard model involves events with multiple jets and a large missing transverse energy \met. A background to this signal is expected from QCD multijet production where \met can originate, e.g. from fluctuations in the detector response to jets. One way to estimate the QCD background in the high \met signal region is to smear particle-level multijet events with parameterizations of the full jet \pt resolution functions that model both the Gaussian core and the tails of the distributions. It is therefore important to quantify the non-Gaussian 
component of the jet \pt resolution in order to predict accurately the QCD background.

Two complementary studies of resolution tails are presented, using dijet and 
$\gamma+$jet  events. For these studies, the focus is on the PF jet 
reconstruction, since it provides the best jet \pt resolution and is adopted in the primary physics analyses most sensitive to the impact of the jet \pt resolution tails.

\subsubsection{Dijet Asymmetry Measurement}

The full resolution functions can be derived using the generator-level MC 
information in 
the simulation. To validate the MC simulation description of the 
\pt-resolution  tails with the currently available data samples, the 
fractional number of events in the tail regions of the dijet \pt asymmetry 
distributions is compared between data and simulation.

\begin{figure}[ht]
 \centering
  \begin{tabular}{cc}
    \includegraphics[width=0.45\textwidth]{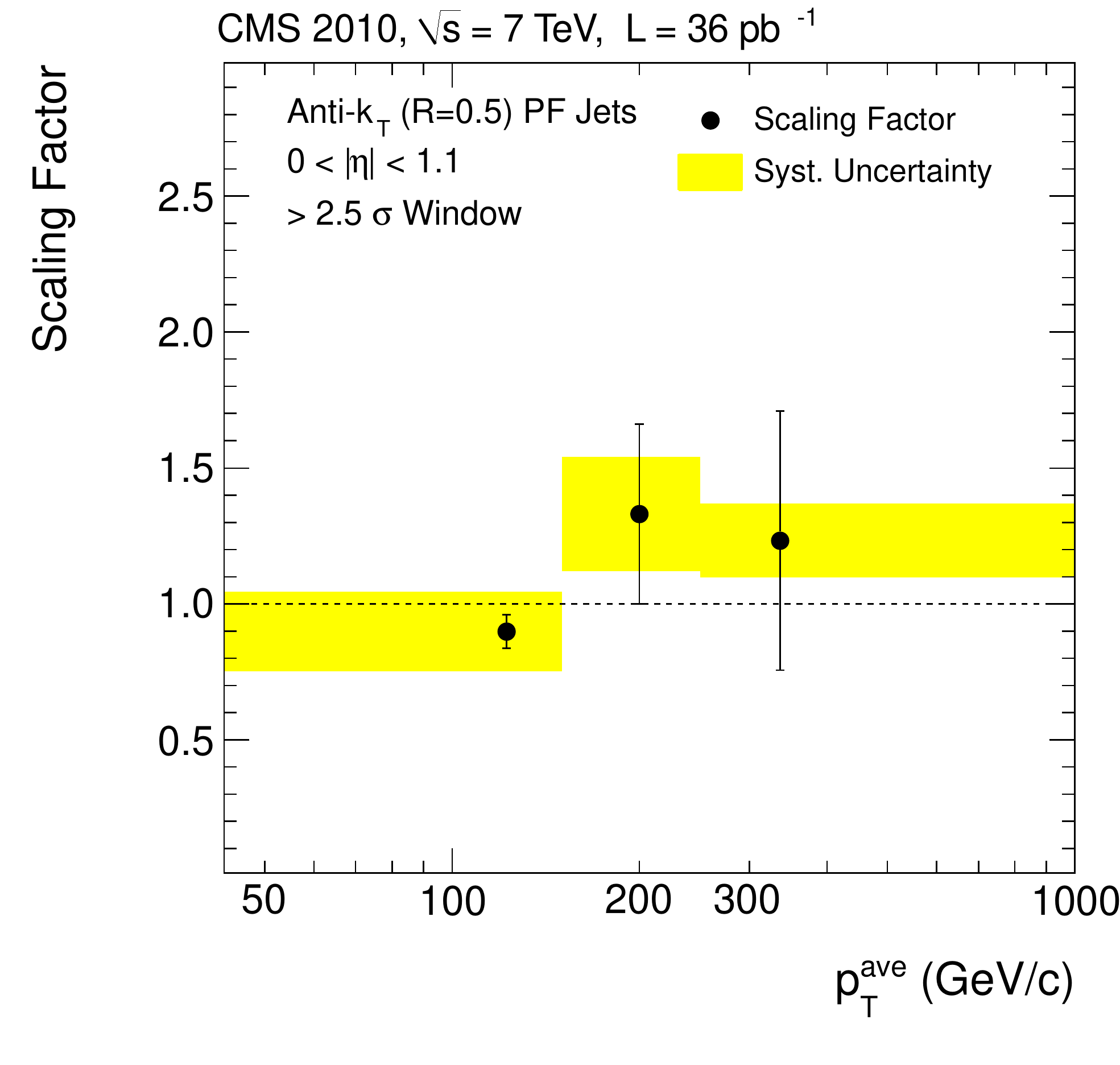} &
    \includegraphics[width=0.45\textwidth]{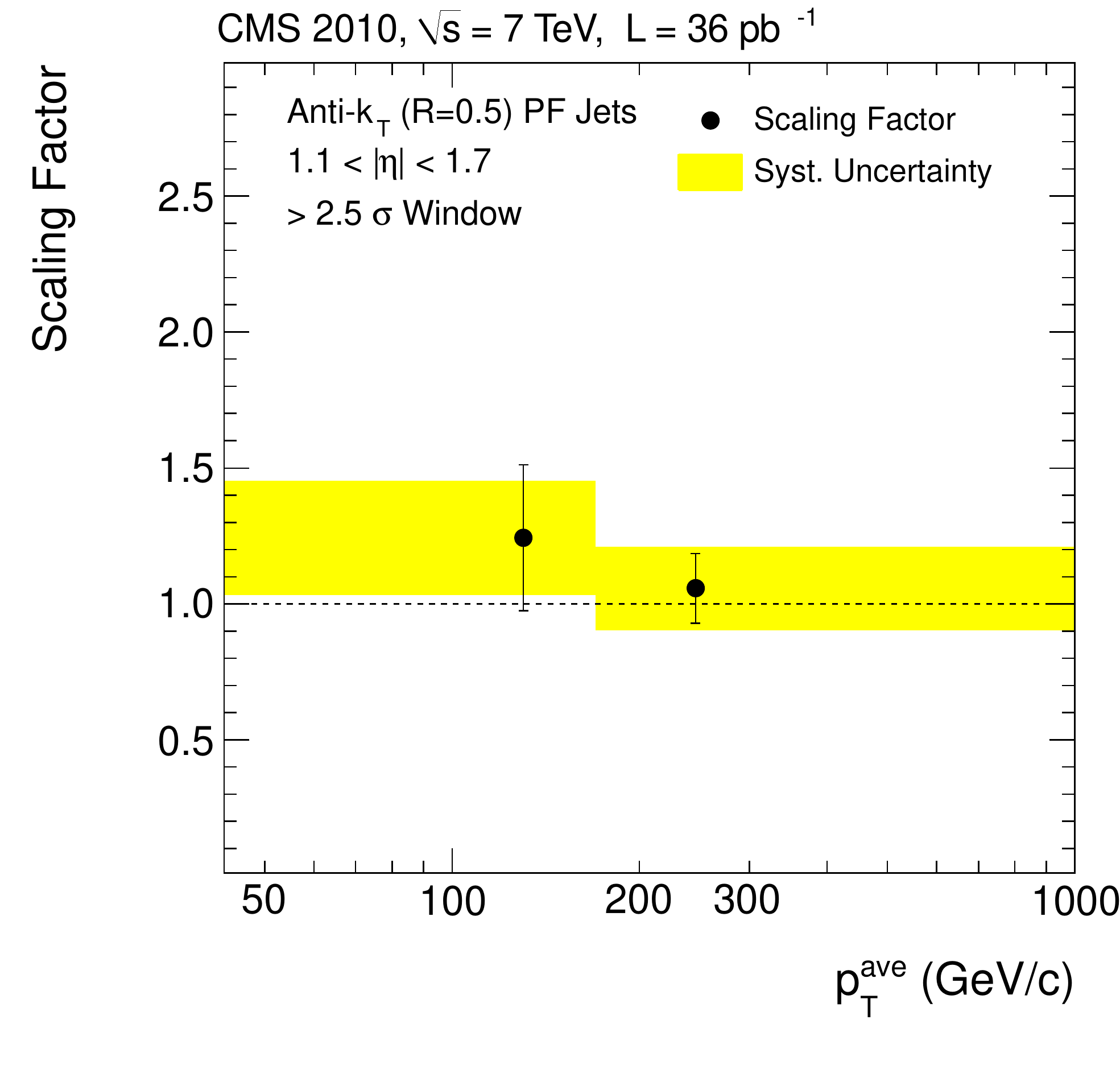} \\
 \end{tabular}
  \caption{The data/MC scaling factors for the tails of the resolutions observed in the dijet 
samples for different $\eta$ and \pt bins, using the $>2.5\sigma$ window.}
  \label{fig:Asym:ScalingFactors}
\end{figure}

As shown before (Table ~\ref{tab:ResFit:ScalingFactors}), the 
central-core widths of the response and asymmetry 
distributions differ between data and MC samples. The adopted strategy is 
to  adjust the MC response distributions to have the same core resolutions 
in MC simulations as in data. Then, the fraction of events in a given 
asymmetry  window in the tail of the distribution is calculated with both 
data and MC samples. Different tail regions have been studied; here the 
results 
for the window \mbox{$2.5\sigma - \infty$} are presented. These fractions 
are observed to depend on the threshold on the third-jet \pt, and are 
therefore extrapolated to zero.
The measured ratio between data and MC fractions from asymmetry is used to 
correct the fraction from generator-level MC  in the form of a scaling 
factor. Since the  dijet asymmetry distribution is symmetric by 
construction, the measured tail-scaling factors average over the low and 
the 
high response tails. To validate the method and to quantify biases caused 
by the event selection, the extrapolated fractions in the MC simulation have 
been compared to the expectation from the asymmetry  
from the generator-level response. Small deviations in the MC closure 
are taken  as a source of systematic uncertainty. Other systematic 
uncertainties have been estimated from the extrapolation procedure and 
from the scaling of the central-core widths between data and MC samples.
The final results for the scaling factors are presented in 
Fig.~\ref{fig:Asym:ScalingFactors}. These results demonstrate that, 
given the current data statistics, the observed data over MC ratios of 
the resolutions tails are within a factor of 1.5.

\subsubsection{$\gamma$ + Jet Measurement}

\begin{figure}[htbp]
\begin{center}
\includegraphics[width=0.45\textwidth]{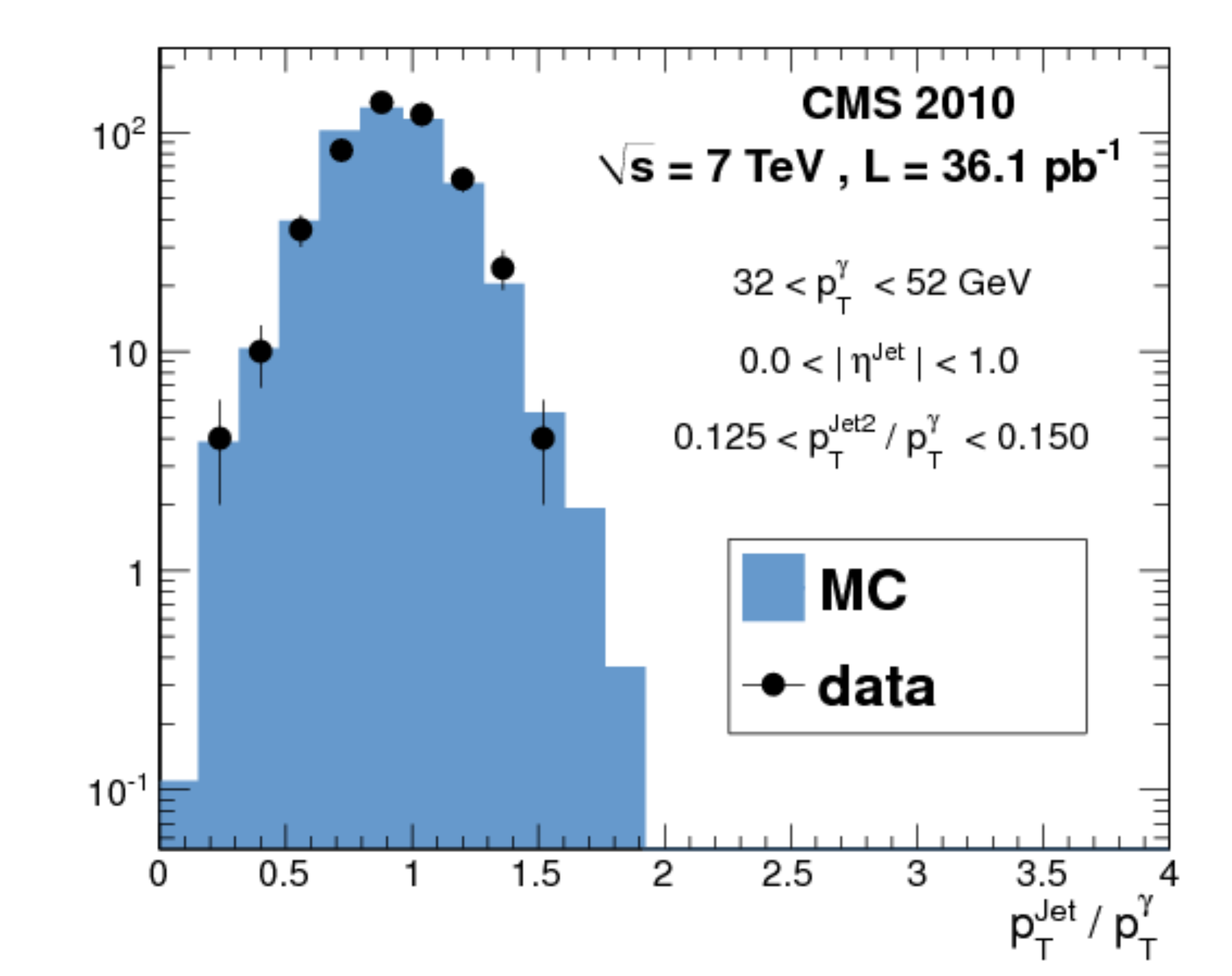}
\includegraphics[width=0.45\textwidth]{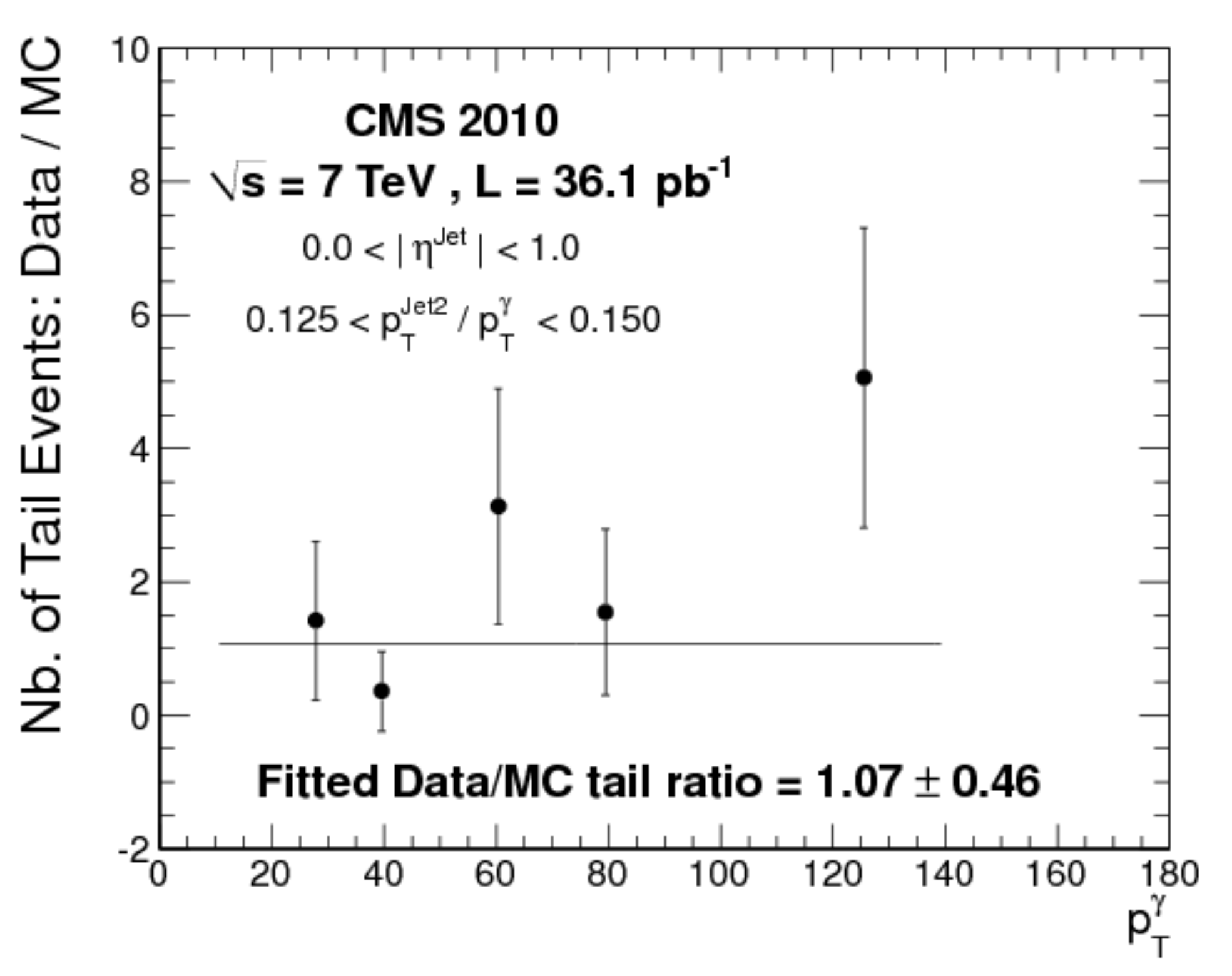}
\end{center}
\caption[]{An example jet \pt resolution function for the $32<\pt^{\gamma}<52\GeV$ bin (left); ratio of the number of tail 
events  in data and MC samples vs. $\pt^\gamma$ (right).} 
\label{fig:tailFits}
\end{figure}

This method uses the \pt balance in $\gamma$+jet events to estimate the 
non-Gaussian tails of the resolution. It can be used separately for the 
low- and high-response tails because the response distribution 
is not symmetric  as in the case of dijet measurement. The number of 
events  outside 2.5$\sigma$ range is counted in bins of $\pt^{\gamma}$, to 
compare the resolution tails 
from the MC simulation to the level observed in data. An example 
distribution is shown 
in Fig.~\ref{fig:tailFits} (left) for the  $32<\pt^{\gamma}<52\GeV$ bin and central $\eta$, and the ratio of the number of tail events in data and MC
samples  vs. $\pt^{\gamma}$ is shown on the right, as function of 
$\pt^{\gamma}$. 
The available statistics in data do not allow to measure the resolution tails
with precision, but a constant fit  to the ratio is consistent within 
uncertainties with the study using the dijet sample, and thereby provides 
a cross-check.

\clearpage
\section{Summary}\label{sec:summary}

A study of the jet energy response and the \pt-resolution in the CMS detector has been presented. The various measurements were performed using the 2010 dataset of proton-proton collisions at $\sqrt{s}=7\TeV$ corresponding to an integrated luminosity of $36\pbinv$. Three different jet reconstruction methods have been examined: calorimeter jets, jet-plus-track jets, and particle-flow jets, clustered with the anti-$k_T$ algorithm with a distance parameter $R=0.5$.

The jet energy response of all jet types is well understood and good agreement between data and simulation has been observed. The calibration is based on MC simulations, while residual corrections are needed to account for the small differences between data and simulation. The calibration chain also includes an offset correction, which removes the
additional energy inside jets due to pile-up events. Various in situ measurements, which utilize the transverse momentum balance, have been employed to constrain the systematic uncertainty of the jet energy scale. For all jet types, the total energy scale uncertainty is smaller than 3\% for $\pt>50\GeV$ in the region $|\eta|<3.0$. In the forward region $3.0<|\eta|<5.0$, the energy scale uncertainty for calorimeter jets increases to 5\% (Fig.~\ref{fig:finalUncvsPt}).

The jet \pt-resolution has been studied, using the dijet and $\gamma+$jets
samples in both data and simulation. For PF jets in the region
$|\eta|<0.5$ with a $p_T$ of 100 GeV the measured resolution in the data
is better than  $10\%$ (Figs.~\ref{fig:uabs-mcdata-00to05}-\ref{fig:uabs-mcdata-05to50}). The core as well as the tails  of  the jet
\pt-resolution function  have been estimated, and close agreement  is
observed between the  $\gamma+$jets and dijet samples. The core of the
measured  jet  \pt-resolution in data is broader than the one obtained
from  the  simulation, by $10\%$ in the central region and up
to $20\%$ in the forward region. The resolution tails are in
agreement with the simulation within statistical uncertainty.

\section*{Acknowledgements}
\hyphenation{Bundes-ministerium Forschungs-gemeinschaft Forschungs-zentren} We wish to congratulate our colleagues in the CERN accelerator departments for the excellent performance of the LHC machine. We thank the technical and administrative staff at CERN and other CMS institutes. This work was supported by the Austrian Federal Ministry of Science and Research; the Belgium Fonds de la Recherche Scientifique, and Fonds voor Wetenschappelijk Onderzoek; the Brazilian Funding Agencies (CNPq, CAPES, FAPERJ, and FAPESP); the Bulgarian Ministry of Education and Science; CERN; the Chinese Academy of Sciences, Ministry of Science and Technology, and National Natural Science Foundation of China; the Colombian Funding Agency (COLCIENCIAS); the Croatian Ministry of Science, Education and Sport; the Research Promotion Foundation, Cyprus; the Estonian Academy of Sciences and NICPB; the Academy of Finland, Finnish Ministry of Education and Culture, and Helsinki Institute of Physics; the Institut National de Physique Nucl\'eaire et de Physique des Particules~/~CNRS, and Commissariat \`a l'\'Energie Atomique et aux \'Energies Alternatives~/~CEA, France; the Bundesministerium f\"ur Bildung und Forschung, Deutsche Forschungsgemeinschaft, and Helmholtz-Gemeinschaft Deutscher Forschungszentren, Germany; the General Secretariat for Research and Technology, Greece; the National Scientific Research Foundation, and National Office for Research and Technology, Hungary; the Department of Atomic Energy and the Department of Science and Technology, India; the Institute for Studies in Theoretical Physics and Mathematics, Iran; the Science Foundation, Ireland; the Istituto Nazionale di Fisica Nucleare, Italy; the Korean Ministry of Education, Science and Technology and the World Class University program of NRF, Korea; the Lithuanian Academy of Sciences; the Mexican Funding Agencies (CINVESTAV, CONACYT, SEP, and UASLP-FAI); the Ministry of Science and Innovation, New Zealand; the Pakistan Atomic Energy Commission; the State Commission for Scientific Research, Poland; the Funda\c{c}\~ao para a Ci\^encia e a Tecnologia, Portugal; JINR (Armenia, Belarus, Georgia, Ukraine, Uzbekistan); the Ministry of Science and Technologies of the Russian Federation, the Russian Ministry of Atomic Energy and the Russian Foundation for Basic Research; the Ministry of Science and Technological Development of Serbia; the Ministerio de Ciencia e Innovaci\'on, and Programa Consolider-Ingenio 2010, Spain; the Swiss Funding Agencies (ETH Board, ETH Zurich, PSI, SNF, UniZH, Canton Zurich, and SER); the National Science Council, Taipei; the Scientific and Technical Research Council of Turkey, and Turkish Atomic Energy Authority; the Science and Technology Facilities Council, UK; the US Department of Energy, and the US National Science Foundation.
 Individuals have received support from the Marie-Curie programme and the European Research Council (European Union); the Leventis Foundation; the A. P. Sloan Foundation; the Alexander von Humboldt Foundation; the Associazione per lo Sviluppo Scientifico e Tecnologico del Piemonte (Italy); the Belgian Federal Science Policy Office; the Fonds pour la Formation \`a la Recherche dans l'Industrie et dans l'Agriculture (FRIA-Belgium); the Agentschap voor Innovatie door Wetenschap en Technologie (IWT-Belgium); and the Council of Science and Industrial Research, India.

\clearpage

\bibliography{auto_generated}   

\providecommand{\href}[2]{#2}\begingroup\raggedright\begin{thebibliography}{10}%
\makeatletter
\providecommand{\hrefCMSnoop }[0]{\@secondoftwo}%
\makeatother

\bibitem{CMS}
\hrefCMSnoop {} {{ CMS} Collaboration, ``The CMS experiment at the CERN LHC'',}
  \textit{ JINST} \textbf{ 03} (2008) S08003.
  \href{http://dx.doi.org/10.1088/1748-0221/3/08/S08004}{\texttt{
  doi:10.1088/1748-0221/3/08/S08004}}.

\bibitem{PTDRI}
\href {http://cdsweb.cern.ch/record/706847} {{ CMS} Collaboration, ``CMS TRIDAS
  Project Technical Design Report, Volume 1, The Trigger Systems'',} CMS TDR
  {CERN/LHCC 2000-38}, (2000).

\bibitem{HLT}
\hrefCMSnoop {} {{ CMS} Collaboration, ``The CMS High Level Trigger'',}
  \textit{ Eur. Phys. J. C} \textbf{ 46} (2006) 605.
  \href{http://dx.doi.org/10.1140/epjc/s2006-02495-8}{\texttt{
  doi:10.1140/epjc/s2006-02495-8}}.

\bibitem{AKT}
\hrefCMSnoop {} {M.~Cacciari, G.~P. Salam, and G.~Soyez, ``The anti-kt jet
  clustering algorithm'',} \textit{ JHEP} \textbf{ 04} (2008) 063.
  \href{http://dx.doi.org/10.1088/1126-6708/2008/04/063}{\texttt{
  doi:10.1088/1126-6708/2008/04/063}}.

\bibitem{JME-09-002}
\href {http://cdsweb.cern.ch/record/1190234} {{ CMS} Collaboration, ``Jet Plus
  Tracks Algorithm for Calorimeter Jet Energy Corrections in {CMS}'',} \textit{
  CMS Physics Analysis Summary} \textbf{ CMS-PAS-JME-09-002} (2009).

\bibitem{PFT-09-001}
\href {http://cdsweb.cern.ch/record/1194487} {{ CMS} Collaboration,
  ``Particle--Flow Event Reconstruction in {CMS} and Performance for Jets,
  Taus, and {\MET}'',} \textit{ CMS Physics Analysis Summary} \textbf{
  CMS-PAS-PFT-09-001} (2009).

\bibitem{PFT-10-002}
\href {http://cdsweb.cern.ch/record/1279341} {{ CMS} Collaboration,
  ``Commissioning of the Particle-Flow Reconstruction in Minimum-Bias and Jet
  Events from {\Pp\Pp} Collisions at 7 {TeV}'',} \textit{ CMS Physics Analysis
  Summary} \textbf{ CMS-PAS-PFT-10-002} (2010).

\bibitem{JME-09-008}
\href {http://cdsweb.cern.ch/record/1259924} {{ CMS} Collaboration,
  ``Calorimeter Jet Quality Criteria for the First {CMS} Collision Data'',}
  \textit{ CMS Physics Analysis Summary} \textbf{ CMS-PAS-JME-09-008} (2010).

\bibitem{JME-10-003}
\href {http://cdsweb.cern.ch/record/1279362} {{ CMS} Collaboration, ``Jet
  Performance in pp Collisions at $\sqrt{s}$=7 {TeV}'',} \textit{ CMS Physics
  Analysis Summary} \textbf{ CMS-PAS-JME-10-003} (2010).

\bibitem{EGM-10-005}
\href {http://cdsweb.cern.ch/record/1279143} {{ CMS} Collaboration, ``Photon
  reconstruction and identification at $\sqrt{s}$ = 7 {TeV}'',} \textit{ CMS
  Physics Analysis Summary} \textbf{ CMS-PAS-EGM-10-005} (2010).

\bibitem{EWK-10-002}
\hrefCMSnoop {} {{ CMS} Collaboration, ``Measurements of inclusive W and Z
  cross sections in pp collisions at sqrt(s)=7 TeV'',} \textit{ JHEP} \textbf{
  01} (2011) 080. \href{http://dx.doi.org/10.1007/JHEP01(2011)080}{\texttt{
  doi:10.1007/JHEP01(2011)080}}.

\bibitem{spps}
\hrefCMSnoop {} {{ UA2} Collaboration, ``Measurement of Production and
  Properties of Jets at the CERN anti-p p Collider'',} \textit{ Z. Phys. C}
  \textbf{ 20} (1983) 117. \href{http://dx.doi.org/10.1007/BF01573214}{\texttt{
  doi:10.1007/BF01573214}}.

\bibitem{jes_d0}
\hrefCMSnoop {} {{ D0} Collaboration, ``Determination of the absolute jet
  energy scale in the D0 calorimeters'',} \textit{ Nucl. Inst. Meth. A}
  \textbf{ 424} (1999) 352.
  \href{http://dx.doi.org/10.1016/S0168-9002(98)01368-0}{\texttt{
  doi:10.1016/S0168-9002(98)01368-0}}.

\bibitem{jes_cdf}
\hrefCMSnoop {} {{ CDF} Collaboration, ``Determination of the jet energy scale
  at the Collider Detector at Fermilab'',} \textit{ Nucl. Inst. Meth. A}
  \textbf{ 566} (2006) 375.
  \href{http://dx.doi.org/arXiv:hep-ex/0510047}{\texttt{
  doi:arXiv:hep-ex/0510047}}.

\bibitem{d0-asymmetry}
\hrefCMSnoop {} {{ D0} Collaboration, ``High \pt jets in $p\bar{p}$ collisions
  at $\sqrt{s} = 630\GeV$ and $1800\GeV$'',} \textit{ Phys. Rev. D} \textbf{
  64} (2001) 032003.
  \href{http://dx.doi.org/10.1103/PhysRevD.64.032003}{\texttt{
  doi:10.1103/PhysRevD.64.032003}}.

\bibitem{JME-09-007}
\href {http://cdsweb.cern.ch/record/1194493} {{ CMS} Collaboration, ``Jet
  Reconstruction Performance at {CMS}'',} \textit{ CMS Physics Analysis
  Summary} \textbf{ CMS-PAS-JME-09-007} (2009).

\bibitem{PU_JET_AREAS}
\hrefCMSnoop {} {M.~Cacciari and G.~P. Salam, ``Pileup subtraction using jet
  areas'',} \textit{ Phys. Lett. B} \textbf{ 659} (2007) 119.
  \href{http://dx.doi.org/10.1016/j.physletb.2007.09.077}{\texttt{
  doi:10.1016/j.physletb.2007.09.077}}.

\bibitem{JET_AREAS}
\hrefCMSnoop {} {M.~Cacciari, G.~P. Salam, and G.~Soyez, ``The Catchment Area
  of Jets'',} \textit{ JHEP} \textbf{ 04} (2007) 005.
  \href{http://dx.doi.org/10.1088/1126-6708/2008/04/005}{\texttt{
  doi:10.1088/1126-6708/2008/04/005}}.

\bibitem{KT1}
\hrefCMSnoop {} {S.~Catani, Y.~L. Dokshitzer, and B.~R. Webber, ``The
  K-perpendicular clustering algorithm for jets in deep inelastic scattering
  and hadron collisions'',} \textit{ Phys. Lett. B} \textbf{ 285} (1992) 291.
  \href{http://dx.doi.org/10.1016/0370-2693(92)91467-N}{\texttt{
  doi:10.1016/0370-2693(92)91467-N}}.

\bibitem{KT2}
\hrefCMSnoop {} {S.~Catani {et~al.}, ``Longitudinally invariant K(t) clustering
  algorithms for hadron hadron collisions'',} \textit{ Nucl. Phys. B} \textbf{
  406} (1993) 187.
  \href{http://dx.doi.org/10.1016/0550-3213(93)90166-M}{\texttt{
  doi:10.1016/0550-3213(93)90166-M}}.

\bibitem{KT3}
\hrefCMSnoop {} {D.~Ellis and D.~E. Soper, ``Successive combination jet
  algorithm for hadron collisions'',} \textit{ Phys. Rev. D} \textbf{ 48}
  (1993) 3160. \href{http://dx.doi.org/10.1103/PhysRevD.48.3160}{\texttt{
  doi:10.1103/PhysRevD.48.3160}}.

\bibitem{PYTHIA}
\hrefCMSnoop {} {T.~Sj{\"o}strand, S.~Mrenna, and P.~Skands, ``PYTHIA 6.4
  physics and manual'',} \textit{ JHEP} \textbf{ 05} (2007) 026.
  \href{http://dx.doi.org/10.1088/1126-6708/2006/05/026}{\texttt{
  doi:10.1088/1126-6708/2006/05/026}}.

\bibitem{D6T}
\hrefCMSnoop {} {R.~Field, ``Early LHC Underlying Event Data-Findings and
  Surprises'',} (2010). \href{http://www.arXiv.org/abs/1010.3558}{\texttt{
  arXiv:1010.3558}}.

\bibitem{GEANT4}
\hrefCMSnoop {} {S.~Agostinelli {et~al.}, ``Geant 4 -- A Simulation Toolkit'',}
  \textit{ Nucl. Inst. Meth.} \textbf{ A506} (2003) 250.
  \href{http://dx.doi.org/10.1016/S0168-9002(03)01368-8}{\texttt{
  doi:10.1016/S0168-9002(03)01368-8}}.

\bibitem{JME-10-008}
\href {http://cdsweb.cern.ch/record/1279141} {{ CMS} Collaboration,
  ``Single-Particle Response in the {CMS} Calorimeters'',} \textit{ CMS Physics
  Analysis Summary} \textbf{ CMS-PAS-JME-10-008} (2010).

\bibitem{EGM-10-003}
\href {http://cdsweb.cern.ch/record/1279350} {{ CMS} Collaboration,
  ``Electromagnetic calorimeter calibration with 7 TeV data'',} \textit{ CMS
  Physics Analysis Summary} \textbf{ CMS-PAS-EGM-10-003} (2010).

\bibitem{HERWIG}
\hrefCMSnoop {} {M.~Bahr {et~al.}, ``Herwig++ Physics and Manual'',} \textit{
  Eur. Phys. J. C} \textbf{ 58} (2008) 639.
  \href{http://dx.doi.org/10.1140/epjc/s10052-008-0798-9}{\texttt{
  doi:10.1140/epjc/s10052-008-0798-9}}.

\end{thebibliography}\endgroup

\cleardoublepage \appendix\section{The CMS Collaboration \label{app:collab}}\begin{sloppypar}\hyphenpenalty=5000\widowpenalty=500\clubpenalty=5000\textbf{Yerevan Physics Institute,  Yerevan,  Armenia}\\*[0pt]
S.~Chatrchyan, V.~Khachatryan, A.M.~Sirunyan, A.~Tumasyan
\vskip\cmsinstskip
\textbf{Institut f\"{u}r Hochenergiephysik der OeAW,  Wien,  Austria}\\*[0pt]
W.~Adam, T.~Bergauer, M.~Dragicevic, J.~Er\"{o}, C.~Fabjan, M.~Friedl, R.~Fr\"{u}hwirth, V.M.~Ghete, J.~Hammer\cmsAuthorMark{1}, S.~H\"{a}nsel, M.~Hoch, N.~H\"{o}rmann, J.~Hrubec, M.~Jeitler, W.~Kiesenhofer, M.~Krammer, D.~Liko, I.~Mikulec, M.~Pernicka, B.~Rahbaran, H.~Rohringer, R.~Sch\"{o}fbeck, J.~Strauss, A.~Taurok, F.~Teischinger, P.~Wagner, W.~Waltenberger, G.~Walzel, E.~Widl, C.-E.~Wulz
% ...
\vskip\cmsinstskip
% ...
1:~~Also at CERN, European Organization for Nuclear Research, Geneva, Switzerland\\
% ...

\end{sloppypar}
\end{document}